\documentclass[12pt,a4paper]{article}
\usepackage{jheppub}
\usepackage{chemarrow}
\usepackage{amsmath,amssymb,amsfonts,pstricks,mathtools}
\usepackage{float}
\usepackage{subfig}
\usepackage{scalefnt}
\usepackage{wrapfig}
\usepackage{fancybox}
\usepackage{ifsym}
\newcommand{\bea}{\begin{eqnarray}\displaystyle}
\newcommand{\eea}{\end{eqnarray}}

\newlength{\arrow}
{\setlength{\fboxsep}{15pt}
\setlength{\mylength}{\linewidth}%
\addtolength{\mylength}{-2\fboxsep}%
\addtolength{\mylength}{-2\fboxrule}%
\Sbox
\minipage{\mylength}%
\setlength{\abovedisplayskip}{0pt}%
\setlength{\belowdisplayskip}{0pt}%
\equation}%
{\endequation\endminipage\endSbox
\[\fbox{\TheSbox}\]}

\def \Z {{\mathbb Z}}
\def \C {{\mathbb C}}
\def \R {{\mathbb R}}

\def\sp{\mathrm{Sp}}
\def\Dsl{\,\raise.15ex\hbox{/}\mkern-13.5mu D}

\begin{document}
\title{$tt^*$ Geometry in 3 and 4 Dimensions}

\author[\S]{Sergio Cecotti,}
\author[\dag]{Davide Gaiotto,}
\author[\ast]{Cumrun Vafa}

\affiliation[\S]{International School of Advanced Studies (SISSA),
via Bonomea 265, 34136 Trieste, Italy}

\affiliation[\dag]{Perimeter Institute for Theoretical Physics, Waterloo, Ontario, Canada N2L 2Y}

\affiliation[\ast]{Jefferson Physical Laboratory, Harvard University, Cambridge, MA 02138, USA}
\abstract{We consider the vacuum geometry of supersymmetric theories with 4 supercharges,
on a flat toroidal geometry. The 2 dimensional vacuum geometry is known to be captured by the $tt^*$ geometry.
In the case of 3 dimensions, the parameter space is $(T^{2}\times \R)^N$ and the vacuum
geometry turns out to be a solution to a generalization of monopole equations in $3N$ dimensions where the relevant
topological ring is that of line operators.  We compute the generalization of the 2d cigar amplitudes, which
lead to $S^2\times S^1$ or $S^3$ partition functions which are distinct from the
supersymmetric partition functions on these spaces, but reduce to them in a certain limit.
  We show the sense in which these amplitudes generalize the structure of 3d Chern-Simons theories and 2d  RCFT's.
  In the case of 4 dimensions the parameter space is of the form $(T^3\times \R)^M\times T^{3N}$,
and the vacuum geometry is a solution to a mixture of generalized monopole equations and generalized instanton equations
(known as hyper-holomorphic connections).  In this case the topological rings are associated to surface operators. 
We discuss the physical meaning of the generalized Nahm transforms which act on all of these geometries. }
\maketitle

\section{Introduction}
Supersymmetric quantum field theories are rich with exactly computable quantities.  These have various degrees of complexity
and carry different information about the underlying supersymmetric theory.
Recently many interesting amplitudes have been computed on $S^d$ or $S^{d-1}\times S^1$ for various
dimensions and for theories with various amounts of supersymmetry.  These geometries
are particularly relevant for the conformal limit of supersymmetric theories, where conformal transformations
can flatten out the spheres.  Away from the conformal fixed point, one can still formulate and compute these
supersymmetric partition functions,
but this involves adding unphysical terms to the action to preserve the supersymmetry, which in particular are not compatible with unitarity.

It is natural to ask whether away from conformal points one can compute supersymmetric amplitudes
without having had to add unphysical terms to the action, and in particular study non--trivial amplitudes in flat space.
A prime example of this would be studying the geometry of the supersymmetric theory on flat toroidal geometries.
In particular we can consider $T^{d-1}$ flat torus as the space, with periodic boundary conditions for supercharges.
Supersymmetric theories have a number $N$ of vacua and in this context one can ask what is the geometry
of the Berry's $U(N)$ connection of the vacuum states as a function of parameters of the underlying theory.
This question has been answered in the case of 2 dimensions for theories with $(2,2)$ supersymmetry
which admit deformations with mass gap \cite{ttstar} leading to what is called the $tt^*$ geometry.
The equations characterizing $U(N)$ connection on the $k$-complex dimensional parameter space
are known as the $tt^*$ equations.  In the case $k=1$ these reduce to $U(N)$ Hitchin equations, which
in turn can be viewed as the reduction of self-dual Yang-Mills equations from 4 to 2 dimensions. 

It is natural to try to generalize these results to supersymmetric theories in higher dimensions which admit mass gap.
The interesting theories, by necessity, would have up to 4 supercharges: they would include in 3 dimensions the ${\cal N}=2$
theories and in 4 dimensions the ${\cal N}=1$ supersymmetric models. \footnote{Our considerations also apply to $d$-dimensional half-BPS defects in $(d+2)$-dimensional 
field theories with eight supercharges}  
Some evidence that such a generalization should be possible, at least in the case of ${\cal N}=2$, $d=3$, has
been found in  \cite{pasquettietal,GG}.
The strategy to determine the higher dimensional $tt^*$ geometries
is rather simple:  We can view their toroidal compactification as a 2d theory with infinitely many fields.  Therefore
the $tt^*$ equations also apply to these theories as well.  The $S^1$ and $T^2$ compactifications of three and four-dimensional gauge theories 
gives 2d theories analogous to (infinite dimensional) gauged linear sigma models with twisted masses. 
These 2d theories have infinitely many vacua similar to the $|n\rangle$ vacua of QCD.  It is natural to consider
the analog of $|\theta \rangle$ vacua which corresponds
to turning on twists by flavor symmetries as we go around the $tt^*$ compactification circle. 
These extra parameters lead to $tt^*$ equations formulated on a higher dimensional space.  

In the case of $d=3$ the equations capturing the $tt^*$ geometry
live in the $3r$ dimensional space $(T^{2}\times\R)^r$, where $r$ is the rank of flavor symmetry.  This space
arises by choosing $2r$ flavor symmetry twists around the cycles of $T^2$ and $r$ twisted masses associated
to flavor symmetries.  In the case of $r=1$ the $tt^*$ equations coincide with the Bogomolny monopole equations, which
can be viewed as the reduction of self-dual Yang-Mills from 4 to 3 dimensions.  The more general $r\geq 1$ case
can be viewed as a generalization of monopole equations to higher dimensions.  The chiral operators of the 2d theory
lift to line operators of the 3d theory.

Similarly one can consider ${\cal N}=1$ theories in $d=4$.  In this case, if the flavor symmetry has rank
$r$, the $tt^*$ parameter space will be $T^{3r}$, corresponding to the $3r$ twist parameters for the flavor symmetry, where
$T^r$ can be viewed as the Cartan torus of the flavor group.   In this case the $tt^*$ equations are again a generalization of
monopole equations but now triply periodic.
 If, in addition, we have $m$  $U(1)$ gauge symmetries \footnote{The $tt^*$ geometry is independent of the 4d gauge couplings, and unaffected by the Landau pole. Later in the paper, we will also show how the Landau pole can be avoided by appropriate UV completions which do not modify the $tt^*$ geometry itself} the parameter space has an extra factor $(T^{3}\times \R^1)^m$ corresponding to turning on  $3m$, $\int B_i\wedge F_i$ type terms
and $m$ FI-parameters.  In the case of $m=1$ the $tt^*$ equations are the self-dual Yang-Mills equations.  For higher
$m$ they describe hyper-holomorphic connections (or certain non-commutative deformations of them).  These are connections which are holomoprhic in any
choice of complex structure of the hyper-K\"ahler space $T^{3m}\times \R^m$.  In fact the generalized
monopole equations or the original 2d $tt^*$ equations can be viewed as reductions of the hyper-holomorphic structure
from $4m$ dimensions to $3m$ or $2m$ dimensions, respectively. Then the hyperholomorphic geometry is a unified framework for all $tt^*$ geometries. The chiral operators of 2d theory lift to surface operators of the 4d theory.

There are also operations that one can do on quantum field theories.  In particular, we can gauge a flavor symmetry
or ungauge a gauge symmetry.   More generally, we consider extensions of these actions on the space
of field theories to  $\sp(2g,{\C})$ actions on 
 $2d$ theories with $(2,2)$ supersymmetry or  $\sp(2g,{\Z})$ actions on $3d$ theories \cite{W3} with ${\cal N}=2$ supersymmetry.
 At the level of the $tt^*$ geometry, as we shall show, these turn out to correspond
to generalized Nahm transformations on the space of hyper-holomorphic connections or their reductions.

The derivation of $tt^*$ equations for the vacuum geometry in 2 dimensions involved studying
topologically twisted theories on cigar or stretched $S^2$ geometries.  It is natural to ask what is the relation
of this to supersymmetric partition functions on $S^2$. It has been shown recently \cite{2dpart1,2dpart2} that in the case
of conformal theories they are the same, but in the case of the mass deformed ones, they differ, and
the $tt^*$ amplitude is far more complicated.   We explain in this paper how one can recover the supersymmetric
partition functions from the $tt^*$ amplitues by taking a particular limit.

For the case of the 3 dimensional theories, one can still define and compute the amplitudes
on stretched $S^2\times S^1$ or $S^3$ (depending on how we fill the $T^2$ on either side).
These involve some novel ideas which are not present in the case of 2d $tt^*$ geometry.  In particular
the realization of modular transformations on $T^2$ as gauge transformations on the 
$tt^*$ geometry plays a key role and gives rise to the $S^3$ partition function.  Moreover the line
operators inserted on the two ends of $S^3$ give rise to a matrix which is a generalization of the $S$-matrix
for the $\tau\rightarrow -1/\tau$ modular transformation of rational conformal field theories in 2d,
while the line operator ring plays the role of the Verlinde algebra \cite{verlinealg}.
In fact in special IR limits the 3d theory typically reduces to a product of topological Chern-Simons theories
in 3d and in this case the $tt^*$ $S$-matrix reduces to the usual $S$--matrix of 2d RCFT's as was shown in \cite{wittenCS}.
Thus the $tt^*$ geometry gives an interesting extension of the Verlinde structure which includes UV degrees
of freedom of the theory. We show that, just as in the 2d case, these partition functions
agree with the supersymmetric partition functions on $S^3_b$ and $S^2\times S^1$ at the conformal point,
but differ from them when we add mass terms.  The $tt^*$ partition functions are more complicated functions
but in a certain limit, these partition functions reduce to the supersymmetric partition functions.
 Similarly, one can extend these ideas to the 4d theory and compute $tt^*$ partition functions
on the elongated spaces $S^2\times T^2$ and $S^3\times S^1$.
\medskip

The plan of this paper is as follows:  In section 2 we review the $tt^*$ geometry  in 2 dimensions.  In section 3 we show how this
can be extended to the cases where flavor symmetries give rise to infinitely many vacua, and how
the monopole equations, self-dual Yang-Mills equations, and more generally the hyper-holomorphic connections
can arise.
In section 4 we introduce the notion of an $\sp(2g,\mathbb{A})$ action on these QFTs, which changes
the theory, and show how this transformation acts on the $tt^*$ geometry as generalized Nahm transformations.
In section 5 we apply these ideas to 3 dimensional ${\cal N}=2$ theories.  In section 6 we give
examples of the $tt^*$ geometry in 3 dimensions.  In section 7 we study the case of $tt^*$ geometry for ${\cal N}=1$ theories in 4d.
In section 8 we give some examples in the 4d case.  In section 9 we take a preliminary step for the interpretation
of the CFIV index \cite{CFIV} as
applied to higher dimensional theories and in particular to $d=3$.
Some of the technical discussions are postponed to the appendices.

\section{Review of $tt^*$ geometry in 2 dimensions}\label{sec:rev2d}

In this section we review $tt^*$ geometry in 2 dimensions \cite{ttstar}.  We consider $(2,2)$
supersymmetric theories in 2 dimensions which admit supersymmetric deformations which introduce a mass gap and
preserve an $SO(2)_R$ charge.
The deformations of these theories are divided to superspace type deformations,
involving F-terms, and D-term variations. \footnote{There may be twisted F-term deformations as well, but for our purposes they behave as D-term deformations} The D-term variations are known not to affect the vacuum geometry, and so
we will not be interested in them.  The F-term deformation space is a complex space with complex coordinates $t^i$, whose tangent is parameterized by chiral fields $\Phi_i$ and correspond to deformations of the theory by F-terms
$$\int d^2\theta d^2z \ \delta W+c.c.=\int d^2\theta d^2z\ \delta t_i\Phi^i+c.c.$$
The chiral operators form a commutative ring:
$$\Phi_i \Phi_j={C_{ij}}^k\, \Phi_k,$$
and similarly for the anti-chiral operators:
$${\overline \Phi_i} {\overline  \Phi_j}={\overline C_{ij}}^k\;{\overline  \Phi_k}.$$  The ${C_{ij}}^k$ are only
a function of $t^i$ and ${\overline C_{ij}}^k$ are only a function of $\overline{t^i}$. 

\begin{figure}
\centering
\includegraphics[width=.8\textwidth,height=0.4\textwidth]{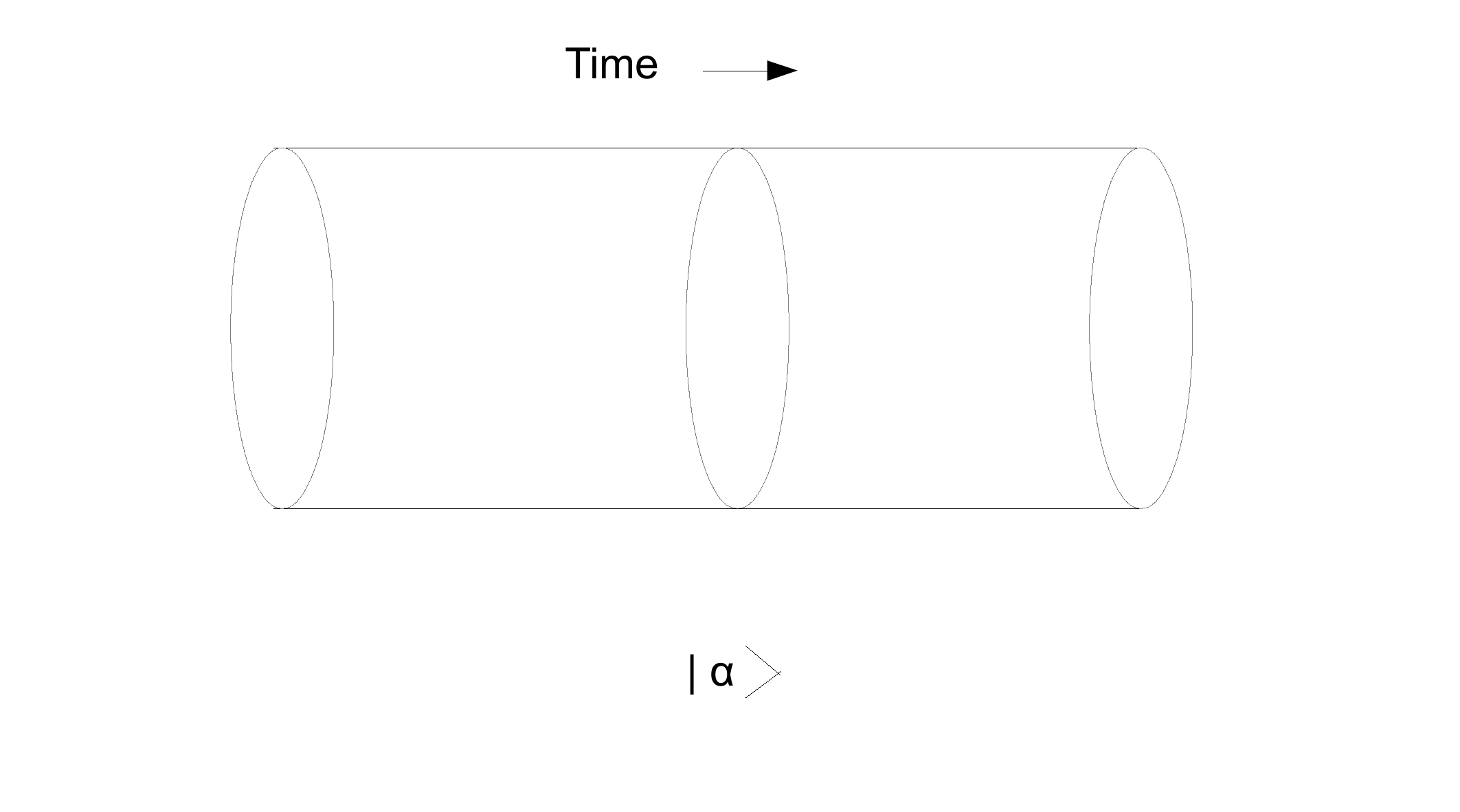}
\caption{1+1 dimensional geometry with circle of length $\beta$ as the space and vacuum $|\alpha\rangle$.}
\label{fig:1}
\end{figure}

Consider the theory on a circle with supersymmetric periodic boundary conditions for fermions. 
Let $|\alpha\rangle$ denote the ground states of the theory (see Fig. \ref{fig:1}).
As we change the parameters of the theory the ground states vary inside the full
Hilbert space of the quantum theory.  Let us denote this by $|\alpha(t,{\overline t})\rangle$.  Then we can
define Berry's connection, as usual, by\footnote{\ Here and in the following equations, equalities of states
signify equalities up to projection onto the ground state subsector.}
$${\partial\over \partial{t^i}}|\alpha(t,{\overline t})\rangle =(A_i)_\alpha^\beta\ |\beta(t,{\overline t})\rangle$$
$${\partial\over \partial{{\overline t^i}}}|\alpha(t,{\overline t})\rangle =({\overline A_{i}})_\alpha^\beta \ |\beta(t,{\overline t})\rangle.$$
It is convenient to define covariant derivatives $D_i,{\overline D_i}$ by
$$D_i=\partial_i  -A_i\qquad {\overline D_i}={\overline \partial_i} -{\overline A_i}$$
In other words over the moduli space of the theory, parameterized by $t_i$ we naturally get a connection of rank $N$ bundle where $N$ denotes the number of vacua
of the $(2,2)$ theory.\footnote{\ We have enough supersymmetry to guarantee that the number of vacua does
not change as we change the parameters of the theory.}  Note that the length
of the circle where we consider the Hilbert space can be chosen to be fixed, say 1, and the radius dependence
can be obtained by the RG flow dependence of the parameters of the theory.  For $(2,2)$ theories this corresponds to
\begin{equation}\label{adsorbingbeta}
W\rightarrow \beta W
\end{equation}
where $\beta$ is the length of the circle.

\begin{figure}
\centering
\includegraphics[width=.8\textwidth]{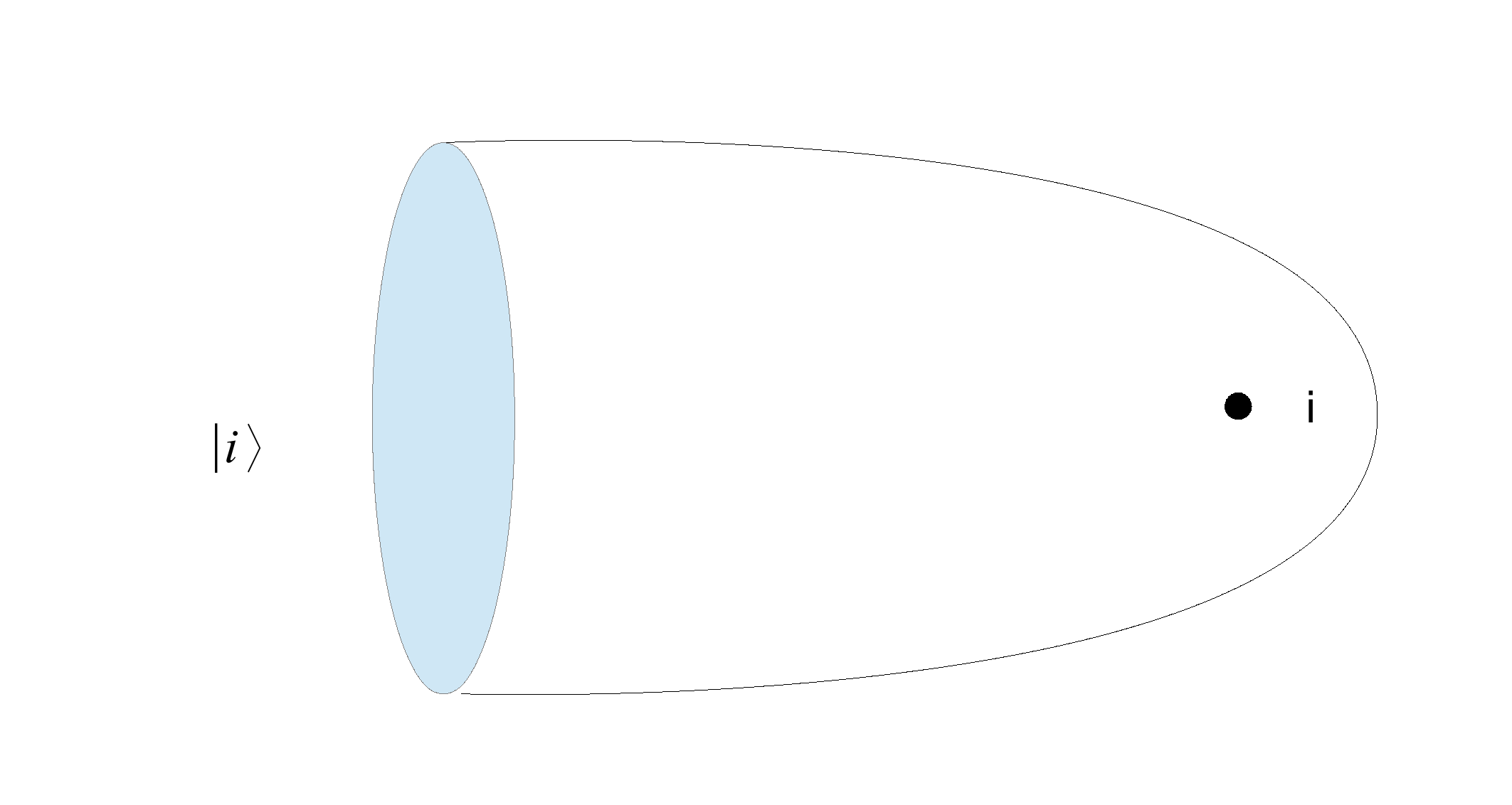}
\caption{A holomorphic basis for states can be produced by topologically twisted
path-integral on an infinitely long cigar, with chiral fields inserted at the tip of the cigar.}
\label{fig:2}
\end{figure}

The ground states of the theory form a representation of the chiral ring:
$$\Phi_i |\alpha\rangle={(C_i)_\alpha}^{\beta} |\beta \rangle$$
and similarly for ${\overline \Phi_i}$.  It turns out that there are two natural
bases for the vacua, which are obtained as follows:  Since the $(2,2)$ theory enjoys an $SO(2)$ R-symmetry,
one can consider a topological twisted version of this theory \cite{wittenTFT}.  In particular we consider a cigar
geometry with the topological twisting on the cigar. 
We consider a metric on the cigar which involves a flat metric
sufficiently far away from the tip of the cigar.  The topologically twisted theory is identical to the physical untwisted theory on flat
space. Path-integral determines
a state in the physical Hilbert space.   We next consider the limit where the length of the cigar $L\rightarrow \infty$. 
In this limit the path-integral projects the state to a ground state of theory.  Since chiral operators are among the BRST
observables of the topologically twisted theory, we can insert them anywhere in the cigar and change the state we get
at the boundary.  Consider the path-integral where $\Phi_i$ is inserted at the tip of the cigar.  The resulting
ground state will be labeled by $|i\rangle$ (see Fig. \ref{fig:2}).

In particular there is a distinguished state among
the ground states when we insert no operator (or equivalently when we insert the `chiral field' 1 at the tip
of the cigar) which we denote by $|0\rangle$.  
In this basis of vacua, the action of the $\Phi_i$ coincides with the ring coefficients:
$$\Phi_i |j\rangle ={C_{ij}}^k |k\rangle$$
In other words ${(C_i)_j}^k={C_{ij}}^k$.   Note that
$$|i\rangle =\Phi_i |0\rangle.$$
 Moreover this basis for the vacuum bundle exhibits the holomorphic structure
of the bundle.   Namely, in this basis $(A_{\overline i})_j^k=0$.  Similarly, when we topologically twist
in a complex conjugate way, we get a distinguished basis of vacua corresponding to anti-chiral fields
$|{\overline i}\rangle$.  These form an anti-holomorphic section of the vector bundle for which
$(A_{i})_{\overline j}^{\overline k}=0$.

\begin{figure}
\centering
\includegraphics[width=.8\textwidth]{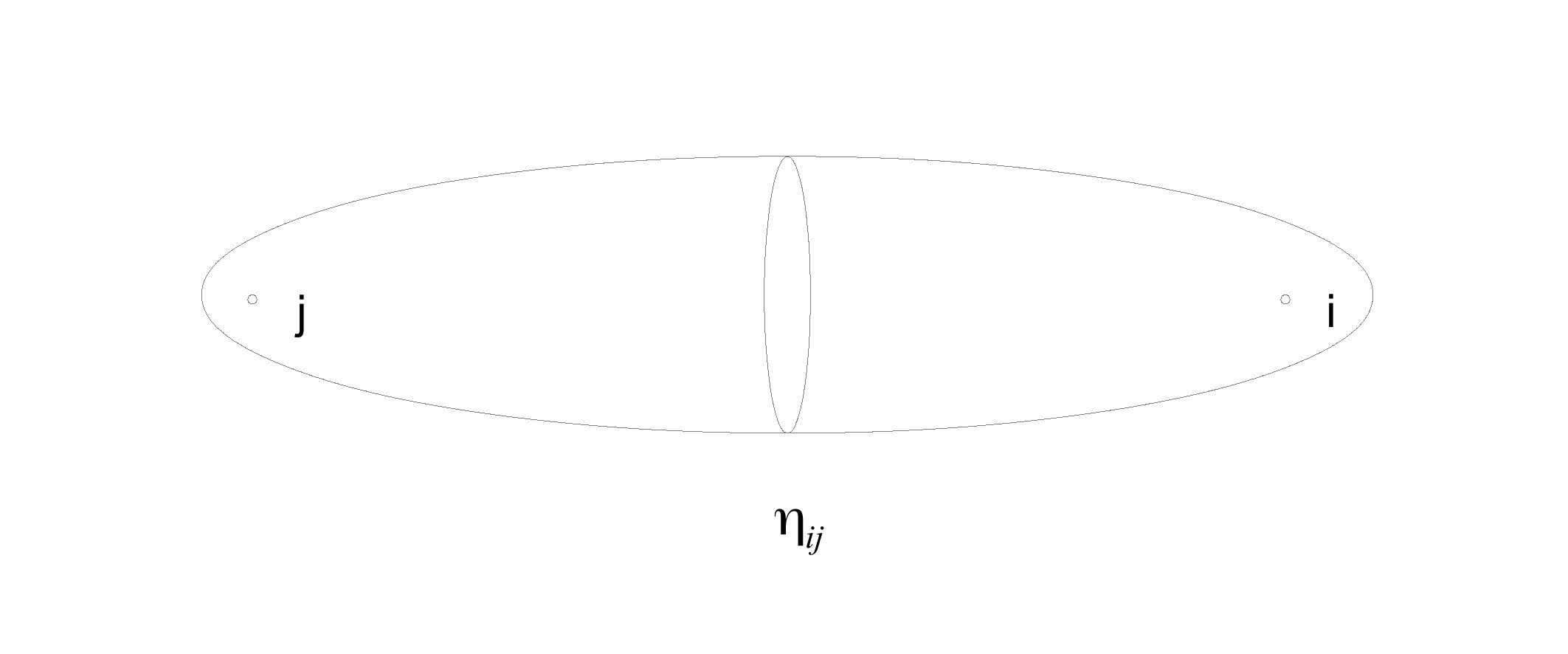}
\caption{The topologically twisted two point function $\eta_{ij}$ can be computed by topologically twisted
path-integeral on $S^2$, where we insert the chiral operators on the two ends of the sphere.  The path-integral
respects supersymmetry for arbitrary choice of metric on $S^2$.}
\label{fig:eta}
\end{figure}

Given these two distinguished bases for the ground states, it is natural to ask how they are related
to one another.  One defines
$$\eta_{ij}=\langle i|j \rangle\qquad g_{i{\overline j}}=\langle {\overline j}|i\rangle$$
and similarly for the complex conjugate quantities.  $\eta$ is a symmetric pairing
and it can be formulated purely in terms of the topologically twisted theory on the sphere  (see Fig. \ref{fig:eta}).
It only depends on holomorphic parameters.  It is convenient (and possible) to choose
a basis for chiral fields such that $\eta$ is a constant matrix.

\begin{figure}
\centering
\includegraphics[width=.8\textwidth]{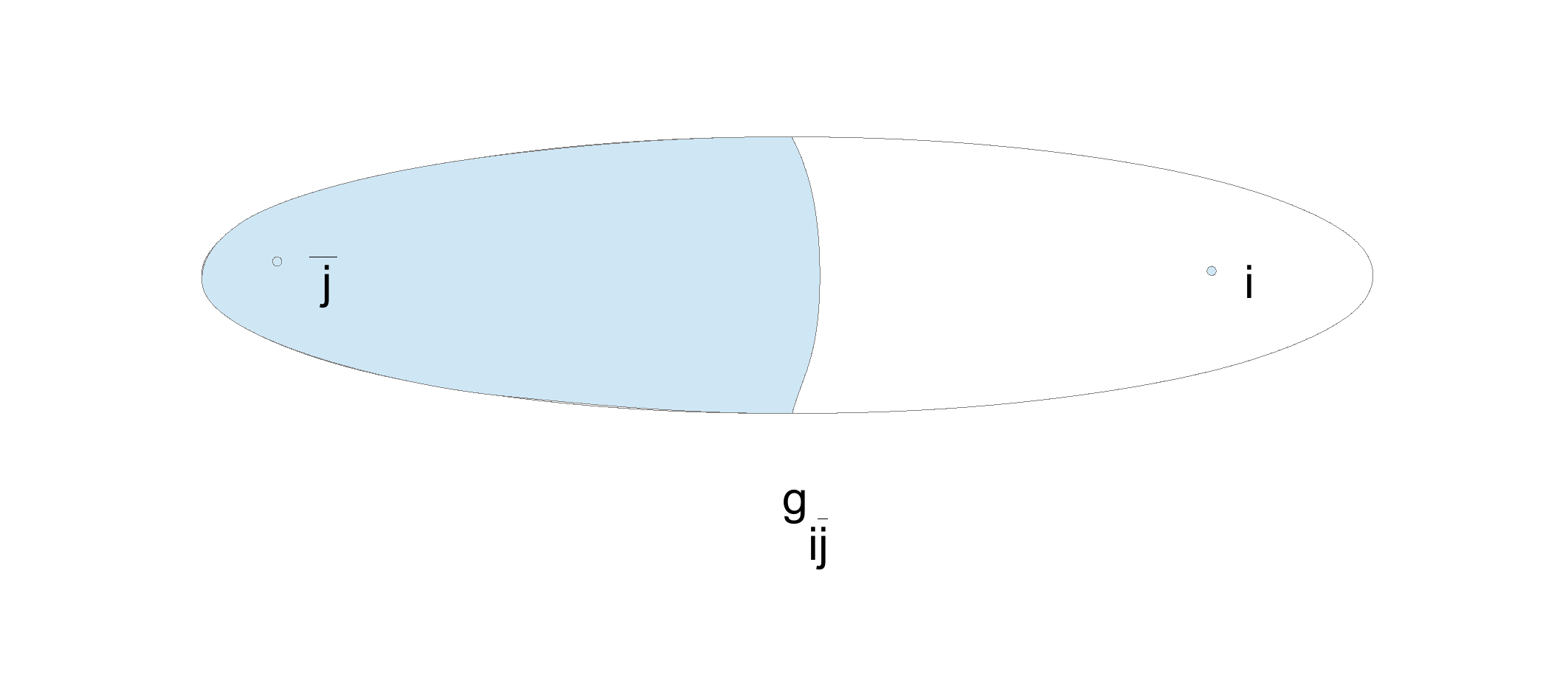}
\caption{The hermitian metric, which is induced from the hermitian inner product on the
Hilbert space in the ground states of the theory can be obtained by path-integral on an infinitely elongated
$S^2$ where on one half we have a topologically twisted theory with chiral fields inserted and on the other the anti-topological twisted
theory with anti-chiral fields inserted.}
\label{fig:g}
\end{figure}

On the other hand, $g$ is a hermitian
metric depending on both $t$ and ${\overline t}$ and 
is far more complicated to compute.  It can be formulated as a path integral
on a sphere composed of two cigars connected to one another, where we do topological
twisting on one side and anti-topological twisting on the other side.  Furthermore we take
the limit in which the length of the cigar goes to infinity.  For any finite length of
the cigar the path integral does not preserve any supersymmetry, and it is crucial to take the $L\rightarrow \infty$
to recover a supersymmetric amplitude (see Fig. \ref{fig:g}).

Note that the holomorphic and anti-holomorphic bases span the
same space so they are related by a matrix $M$:
$$|{\overline i}\rangle =M_{\overline i}^j|{j}\rangle.$$
$M$ can be computed in terms of $g, \eta$ as
$$M=\eta^{-1}g.$$
Furthermore, since $M$ represents the CTP operator acting on the ground states we
must have $MM^*=1$; this implies that
\begin{equation}\label{reality}
(\eta^{-1}g)(\eta^{-1}g)^*=1.
\end{equation}
Since $g$ is the usual inner product in the Hilbert space, it is easy to see that it is
covariantly constant with respect to the connections we have introduced:
$$D_ig_{k\overline l}={\partial_i} g_{k{\overline l}}-A_{ik}^jg_{j\overline l}-A_{i\overline l}^{\overline j} g_{k{\overline j}}=0={\overline D_i}g.$$
The $tt^*$ geometry gives a set of equations which characterize the curvature of the vacuum bundle.
They are given by
$$[{\overline D_i},C_j]=0=[D_i,{\overline C_j}]$$
$$[D_i,D_j]=0=[{\overline D_i},{\overline D_j}]$$
$$[D_i,C_j]=[D_j,C_i]\qquad  [{\overline D_i},{\overline C_j}]=[{\overline D_j},{\overline C_i}]$$
Furthermore the non-vanishing curvature of the Berry's connection is captured by the equations
$$[D_i,{\overline D_j}]=-[C_i,{\overline C_j}]$$
These equations can be summarized as the flatness condition for the following family
of connections parameterized by a phase $\zeta =e^{i\alpha}$.  Consider
\begin{equation}\label{laxconnection}
\nabla_i=D_i+\zeta C_i \qquad {\overline \nabla_i}={\overline D_i}+\zeta^{-1}{\overline C_i}
\end{equation}
The $tt^*$ equations can be summarized by the condition of flatness of $\nabla^{\alpha}$ and $\overline{\nabla^\alpha}$:
$$[\nabla _i,\nabla_j]=[{\overline \nabla_i},{ \overline \nabla_j}]=[\nabla_i,{\overline \nabla_j}]=0$$
for arbitrary phase $\alpha$, and in fact for all complex numbers $\zeta\in \C^*$.
Note that on top of these equation we have to impose the reality structure given by $MM^*=1$,
as an additional constraint. 

We shall refer to the flat connection \eqref{laxconnection} as the $tt^*$ \textit{Lax connection} (with spectral parameter $\zeta$). It is also known as the $tt^*$ Gauss--Manin connection.

For the case of one variable, the $tt^*$ equations become equivalent to the Hitchin equations \cite{Hitchin},
which itself is the reduction of instanton equations from 4 dimensions to 2 dimensions.  In that
context, if we represent the flat 4d space by two complex coordinates $(t,u)$ and reduce along $u$
the system on $t$ space will become the $tt^*$ system:
$$A_{\overline u} \leftrightarrow C$$
$$A_{ u}\leftrightarrow -{\overline C}$$
in which case the two non-trivial parts of the $tt^*$ read as
$$[{\overline D},C]=F_{{\overline t}{\overline u}}=0=F_{tu}=[D,{\overline C}]$$
$$F_{t{\overline t}}=[D,{\overline D}]=-[C,{\overline C}]=-F_{u{\overline u}}.$$
Thus $tt^*$ geometry with more parameters can be viewed as a dimensional reductions of a generalization of instanton equations.
As we will discuss later, and will be relevant for the generalizations of $tt^*$ geometry
to higher dimensions, the more general case can be viewed as a reduction
of tri-holomorphic connections on hyperK\"ahler manifolds.

The massive $(2,2)$ theories we consider will typically have a set of massive vacua 
in infinitely long space \footnote{\ Not to be confused with the states $|\alpha \rangle$ on the circle of finite length $\beta$.}, 
say corresponding to the critical points of the superpotential in a LG theory. 
In the topological gauge, the matrices $C_i$ are independent of the length $\beta$ of the compactification 
circle, and thus can be computed in terms of the properties of the theory on flat space. 

More precisely, the eigenvalues $p_{i,a}$ of the $C_i$ matrices should correspond to the vevs of the corresponding chiral 
operators $\Phi_i$ in the various massive vacua of the theory in flat space. In turn, these vevs can be expressed in terms of the 
low energy effective superpotentials $W^{(a)}[t_i]$ in each vacuum $a$ of the theory in flat space:
\begin{equation}
p_{i,a} = \partial_{t_i} W^{(a)}
\end{equation}
In a LG theory, the $W^{(a)}$ coincide with the values of the superpotential at the critical points. 

Although solving the $tt^*$ equations is generally hard, the solutions can be readily labelled by holomorphic data,
as will be discussed in section 4,
by trading in the usual way the $[D,D]=[C,C]$ equation for a complexification of the gauge group. 
Then the solution is labelled by the higher dimensional Higgs bundle defined by the pair $\overline D_i$ and 
$C_i$. For generic values of the parameters $t_i \in {\cal T}$ we can simultaneously diagonalize the $C_i$, 
and encode them into the Lagrangian submanifold ${\cal L}$ in $T^*{\cal T}$ defined by the pair $(t_i, p_i)$.
The corresponding eigenline defines a line bundle $L$ on ${\cal L}$. The pair $({\cal L},L)$ gives the spectral data which 
labels a generic solution of the $tt^*$ equations. 
For one-dimensional parameter spaces, this is the standard spectral data for a Hitchin system on ${\cal T}$. 
We refer to section 4 for further detail and generalizations. 

\subsection{Brane amplitudes}

The flat sections of the Lax connection over the parameter space have a physical
interpretation that will be important for us \cite{ttstar,onclassification,HIV}. 
The mass-deformed $(2,2)$ theory may admit supersymmetric boundary conditions (``D-branes''). 
Consider in particular some half-BPS boundary conditions which also preserve $SO(2)_R$.
There is a certain amount of freedom in picking which two supercharges will be preserved by the boundary condition.
The freedom is parameterized by a choice of a phase  given by a complex number $\zeta$ 
with norm 1.  Roughly, if we denote the $(2,2)$ supercharges as $Q^\pm_{L,R}$, where $L,R$ denote left- or right-moving, 
and $\pm$ the R-charge eigenvalue, a brane will preserve
\begin{equation}
Q^+_L + \zeta\, Q^+_R \qquad Q^-_L + \zeta^{-1}\, Q^-_R.
\end{equation}
A given half-BPS boundary condition $D$ in massive $(2,2)$ theories is typically a member of a 1-parameter family of 
branes $D^\zeta$ which preserve different linear combinations of supercharges. We will usually suppress the $\zeta$ superscript. 

We can use a brane $D$ to define states $|D \rangle$ or $\langle D|$ in the Hilbert space for the theory on a circle, 
even though this state will not be normalizable, as is familiar
in the context of D-brane states. We can project the states onto the supersymmetric ground states, i.e. 
we consider inner products such as 
$$\Pi[D,\zeta] = \langle D| \alpha\rangle .$$
Such a ``brane amplitude'' is a flat section of the $tt^*$ Lax connection with spectral parameter $\zeta$ \cite{HIV}. 

\begin{figure}
\centering
\includegraphics[width=.8\textwidth]{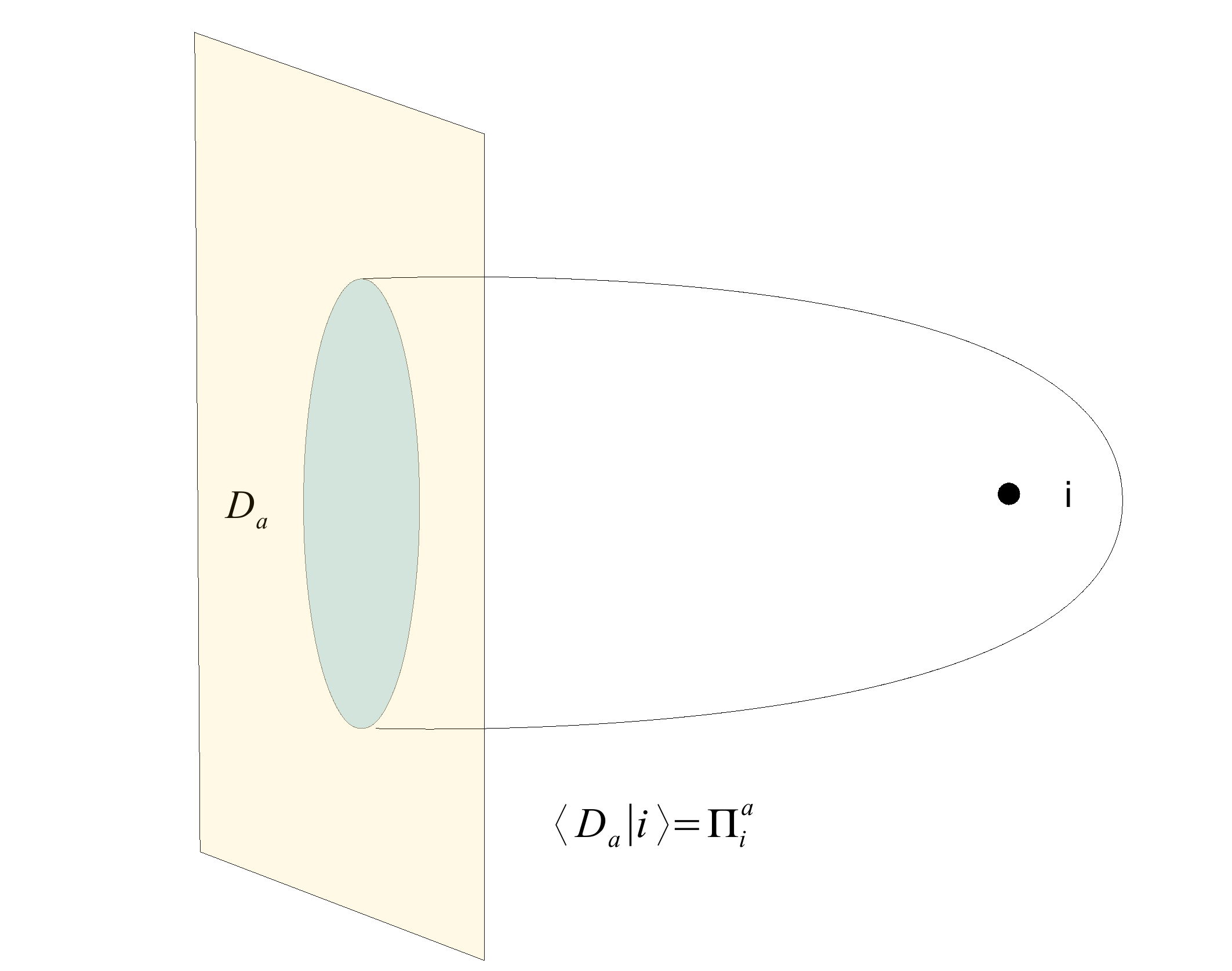}
\caption{The overlap of the vacuum states with the D-branes give rise to $\Pi^a_i$ which are flat
sections of the improved $tt^*$ connection.}
\label{fig:pi}
\end{figure}

It is useful to consider the brane amplitudes in the holomorphic gauge
$$\Pi_i=\langle D|i\rangle$$
which can be defined by a topologically twisted partition function on the semi-infinite cigar (see Fig. \ref{fig:pi}).

We can also define
$${\hat \Pi}_i[D] =\langle i |D\rangle$$
Looking at the BPS conditions, one notices that the ``same'' boundary condition can be used to define a
left boundary condition of parameter $\zeta$ or a right boundary condition of parameter $-\zeta$. 
Thus ${\hat \Pi}_i[D]$ is a left flat section for the $tt^*$ Lax connection of spectral parameter $-\zeta$. 
%
%
Using CTP, we can see
$$\langle {\overline i}| D\rangle =\Pi^\dagger_{\overline i}[D] \qquad \qquad \langle D|{\overline i}\rangle ={\hat \Pi}^\dagger_{\overline i}[D].$$
This is consistent with the observation that if $\zeta$ is a phase, the hermitean conjugate of standard flat section for the 
Lax connection of parameter $\zeta$ is a left flat section for the Lax connection of parameter $-\zeta$. \footnote{\ For general $\zeta$, that would be $- \overline \zeta^{-1}$.}

\begin{figure}
\centering
\includegraphics[width=.8\textwidth]{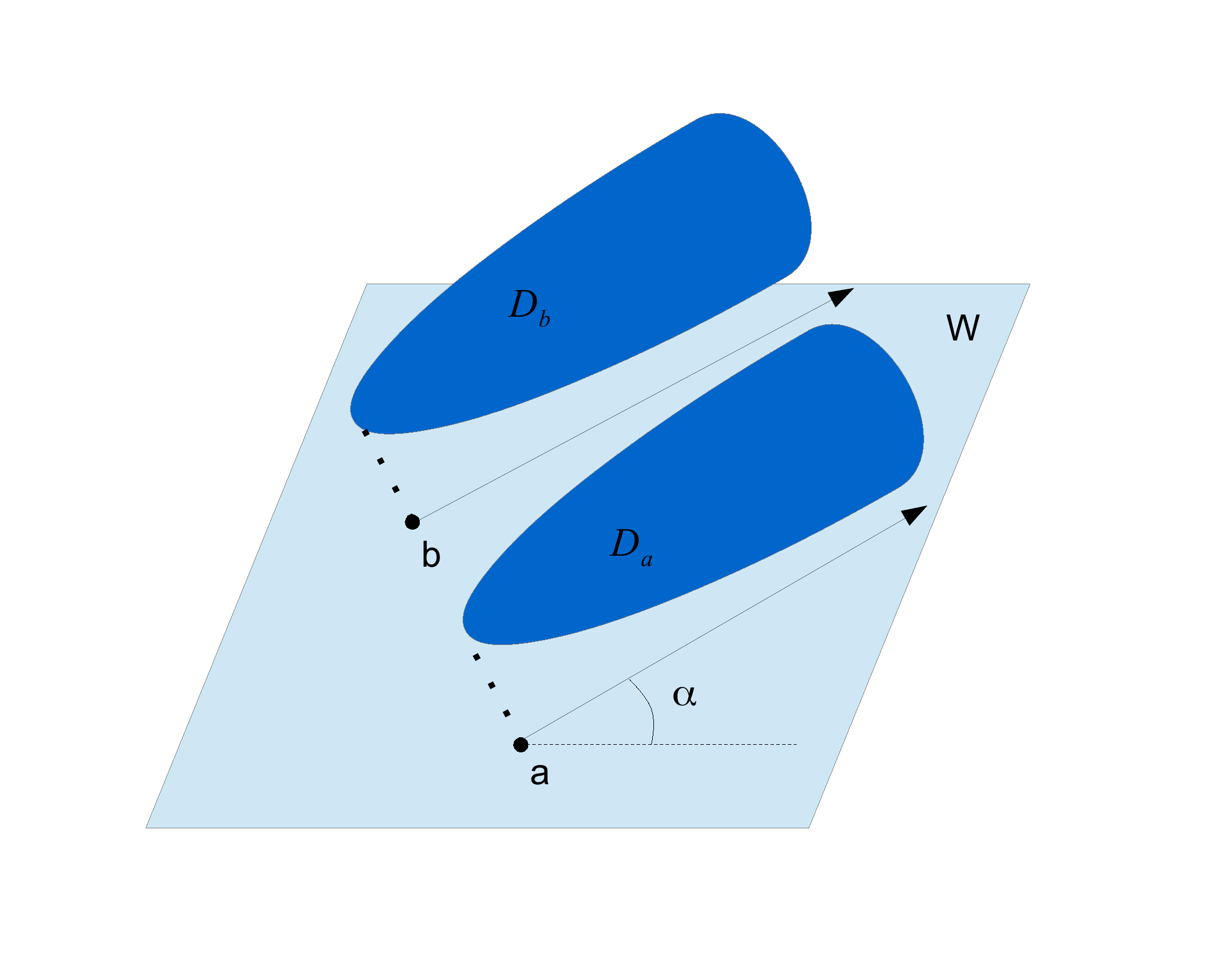}
\caption{D-branes in the LG description of $(2,2)$ theories are Lagrangian submanifolds which project
to straight lines in the $W$-plane emanating from critical points of $W$.  These objects are also known as \emph{`Lefschetz thimble branes'}. The slope $\alpha$ determines the combination of supercharges which the
D-brane preserves.  In the massive phases there is one D-brane per vacuum (and angle $\alpha$). }
\label{fig:3}
\end{figure}

For simplicity we will limit our discussion here
to the case of $(2,2)$ Landau-Ginzburg models, characterized by some superpotential
$W$.  The vacua are in one to one correspondence with critical points of $W$.
In the LG case there is a particularly nice class of branes \cite{HIV}, represented
by special mid-dimensional Lagrangian subspaces in field space sometimes called ``Lefschetz thimbles'', 
which are defined as contours of steepest descent for an integral of $e^{-\zeta W}$. 
They have the property that the value of the superpotential $W$ on that subspace is on a straight line, emanating from the critical value,
with slope given by the phase $\zeta$ (see Fig. \ref{fig:3}). \footnote{The D-branes introduced in \cite{HIV} project to straight-lines on W-plane.  This can be relaxed
to D-branes that at the infinity in field space approach straight lines and are more relaxed in the interior regions \cite{Herbst:2008jq}.  In this paper we will not need this extension and take the D-branes simply to project to the straight lines in $W$-plane.} 

Let $a$ denote a critical point of $W$.  The corresponding D-brane emanating from it will be denoted by $D_a$. 
Note that the $D_a$ are piece-wise continuous as a function of $\zeta$, 
with jumps at special values of $\zeta$ which are closely related to the BPS spectrum of the theory.
We will denote the corresponding brane amplitudes as $\Pi^a_i$.

The thimbles $D_a$ defined at $\zeta$ and $U_a$ defined at $-\zeta$ form a dual basis of Lagrangian cycles, and 
the inner product between the corresponding states is given by
$$\langle  D_a|U_b\rangle =\delta_{ab}$$
Note that since there are as many $|D_a\rangle$ as the ground state vacua, we can use them
to compute the ground state inner products, using the decomposition 
$1=\sum_a |U_a\rangle \langle D_a|$, which is valid acting on the ground states:
\begin{equation}\label{deco}
\begin{gathered}
\eta_{ij}=\langle j |i\rangle = \sum_a \langle j | U_a\rangle \langle D_a | i\rangle ={\hat \Pi}_j^a \Pi_i^a \\
g_{i{\overline j}}=\langle {\overline j}|i\rangle =\sum_a \langle{\overline  j} | U_a\rangle \langle D_a| i\rangle ={\Pi}^{a\dagger}_j\Pi_{ i}^a
\end{gathered}
\end{equation}

The thimble brane amplitudes $\Pi^a = \Pi[D_a]$ give a fundamental basis of flat sections for the $tt^*$ Lax connection \cite{onclassification,GMN2d4d}. 
Any other brane amplitude can be rewritten as a linear combination 
\begin{equation}
\Pi[D] = \sum_a n_a[D] \Pi^a
\end{equation}
with integer coefficients $n_a$ which coincide with the framed BPS degeneracies defined in \cite{GMN2d4d}
and can be computed as $\langle D|U^a \rangle$.

The brane amplitudes $\Pi[D,\zeta]$ can be analytically continued to any value of $\zeta$ in (the universal cover of) $\mathbb{C}^*$ so that 
$\Pi[D,\zeta]$ is holomorphic in $\zeta$. The general analysis of \cite{onclassification,GMN2d4d} shows that $\Pi[D,\zeta]$ will have essential singularities at $\zeta=0$ and $\zeta = \infty$, with interesting Stokes phenomena associated to the BPS spectrum of the theory. 

It is clear from the form of the Lax connection and of the eigenvalues $p_{i,a}$ of $C_i$ 
that the asymptotic behaviour as $\zeta \to \infty$ of a flat section should be 
\begin{equation}
\Pi[D,\zeta] \sim \sum_a e^{- \zeta W^{(a)}} v_a
\end{equation}
where $v_a$ are simultaneous eigenvectors of the $C_i$.

The thimble brane amplitudes have the very special property that $\Pi^a \sim e^{- \zeta W^{(a)}}$ for the analytic continuation of 
$\Pi^a[\zeta]$ to a whole angular sector of width $\pi$ around the value of $\zeta$ at which the thimbles are defined. 
This property, together with the relation between the jumps of the basis of thimble branes as we vary the reference 
value of $\zeta$ and the BPS spectrum of the theory \cite{HIV} allow one to reconstruct the $\Pi^a[\zeta]$ from their 
discontinuities by the integral equations described in \cite{onclassification,GMN2d4d}. \smallskip

Although a full review of these facts would bring us too far from the purpose of this paper, these is a simplified setup which 
captures most of the structure and will be rather useful to us. 
The $tt^*$ geometry has a useful ``conformal limit'', $\beta \to 0$. 
Although in this limit one would naively expect the dependence on relevant deformation parameters $t_i, \bar t_i$ 
to drop out, the behaviour of the brane amplitudes as a function of $\beta$, $\zeta$ is somewhat more subtle. 

More precisely, if we look at $\Pi_i[\zeta]$ and focus our attention on the region of large $\zeta$, \textit{i.e.}\! we keep $\zeta \beta$ finite as $\beta \to 0$,
only the $\overline t_i$ dependence really drops out and we are left with interesting functions of the holomorphic parameters 
$t_i$. The converse is true for the amplitudes $\Pi_{\overline i}[\zeta]= \langle D|{\overline i}\rangle$ in the anti-holomorphic gauge, for finite $\zeta^{-1} \beta$. 

In the LG case they are given by period integrals \cite{HIV}
$$\Pi_i=\int_{D} \Phi_i\; {\rm exp}\Big(-\zeta\,\beta\; W(X^\alpha, t)\Big)\;dX^1\wedge\cdots\wedge dX^n$$
$$\Pi_{\overline i}=\int_{D}{\overline \Phi_i}\; {\rm exp}\Big(- \zeta^{-1}\,\beta\;  {\overline W({\overline X}^{\alpha},{\overline t}})\Big)\;d\overline{X}^1\wedge\cdots \wedge d\overline{X}^n,$$
where, for later convenience, we reintroduced the explicit dependence on the $S^1$ length $\beta$, see eqn.\eqref{adsorbingbeta}.
The flatness under the $tt^*$ Lax connection reduces to the obvious facts that 
\begin{equation}
\partial_{t_j} \Pi_i + \beta \zeta C_{ij}^k \Pi_k =0  \qquad \qquad \partial_{\overline t_j} \Pi_i =0
\end{equation}

Due to the relation between thimble Lagrangian manifolds and steepest descent contours, it is obviously true that 
the thimble brane amplitudes $\Pi_i^a$ have the expected asymptotic behaviour at $\zeta \to \infty$. 
Furthermore, it is also clear that for an A-brane defined by some Lagrangian submanifold $D$ 
the integers $n_a[D]$ are simply the coefficients for the expansion of $D$ into the thimble cycles. 
The Stokes phenomena for the $tt^*$ geometry reduce to the standard Stokes phenomena 
for this class of integrals. 

There is another observation which will be useful later. 
Let us introduce one additional parameter $P_\alpha$ for each chiral field $X^\alpha$, deform the superpotential  $W\rightarrow W-X^\alpha P_\alpha$, and consider $\Pi^a$ for this deformed
$W$.  We can view $P_\alpha$ as part of the parameter space of $W$.  Let's focus on $\Pi^a_0$, \text{i.e.}\! the integral 
without insertion of chiral fields. 
$$\Pi^a_0\bigg|_{\overline \beta \rightarrow 0}=\int_{D_a}{\rm exp}\Big(- \zeta\,\beta\; W(X^\alpha, t)-\zeta\beta X^\alpha P_\alpha\Big)\;dX^1\wedge\cdots\wedge dX^n.$$
Now
consider the insertion of $\partial W/\partial X^\gamma-P_\gamma$ in the above integral, and use
integration by parts to conclude its vanishing:
$$\int_{D_a} dX^{\alpha}\; \left({\partial W\over \partial X^{\gamma}}-P_{\gamma}\right)\; \exp\!\Big(- \zeta\beta\, W(X^\alpha, t)-\zeta\beta X^\alpha P_\alpha\Big)=0$$
Which can be rewritten as
$$\left[\partial_\gamma W\! \left(-{1 \over \zeta}{\partial \over \partial P_\alpha}\right) -P_\gamma\right]\int_{D_a} dX^{\alpha}\;  \exp\!\Big(-\zeta\, \beta\, W(X^\alpha, t)-\zeta\,\beta X^\alpha P_\alpha\Big)=0.$$
In other words
 \begin{equation}\label{diffeq}
\left[\partial_\gamma W\! \left(-{1 \over \zeta\beta}\,{\partial \over \partial P_\alpha}\right) -P_\gamma\right]\Pi^a_0\bigg|_{\overline \beta\rightarrow 0}=0.
\end{equation}
Note that, replacing $\zeta\beta \rightarrow i/\hbar$, the above formula is suggestive of a quantum
mechanical system where $(X^\alpha, P_\alpha)$ form the phase space.  At this point however,
it appears that they are not on the same footing as $X^\alpha$ is a field but $P_\alpha$ is a parameter.
In section 4 we will see that we can in fact consider a dual LG system where $P_\alpha$ can be promoted to play
the role of fields.  More generally we will
see that we can have an $\sp(2g,{\C})$ transformation where one chooses a different
basis in which parameters are promoted to fields.  Here $g$ denotes the number of chiral fields.
  Indeed this structure also appeared for 2d (2,2) theories which arise by Lagrangian D-brane probes of Calabi-Yau in the context of topological strings \cite{AKV}
(which can be interpreted as codimension 2 defects in the resulting 4 or 5 dimensional theories), and the choice of parameters versus fields depends on the boundary data
of the Lagrangian brane.  We will see that this correspondence is not an accident.

\section{Extended $tt^*$ geometries and hyperholomorphic bundles}\label{section:periodictt*}

The basic assumption of the standard $tt^*$ analysis in the previous section is that the F-term deformation parameters $t^i$ are dual to well defined chiral operators of the theory,
such as single-valued holomorphic functions on the target manifold $\mathcal{M}$ of a LG model. 
This is not the only kind of deformation parameter which may appear in the F-terms of the theory. 
There are more general possibilities which lead to more general $tt^*$ geometries and new phenomena. 
In this section we discuss mostly situations which give rise to various dimensional reductions of the equations for a
hyper-holomorphic connection. We will briefly comment at the end on a more extreme 
situation in which the $tt^*$ geometry is based on \emph{non--commutative} spaces.

The simplest extension of $tt^*$ is generically associated to the existence of flavor symmetries in the theory. 
In a mirror setup, where one looks at the twisted F-terms for, say, a gauged linear sigma model, 
such deformation parameters are usually denoted as twisted masses.
If a flavor symmetry is present, which acts on the chiral fields of the GLSM, one can introduce the twisted masses as the vevs of the scalar field in a background 
gauge multiplet coupled to that flavor symmetry. In the presence of twisted masses, the low-energy twisted effective superpotential for the theory is a multi-valued function over the space of vacua, 
defined up to integral shifts by the twisted masses. 

\begin{figure}
\centering
\includegraphics[width=.8\textwidth]{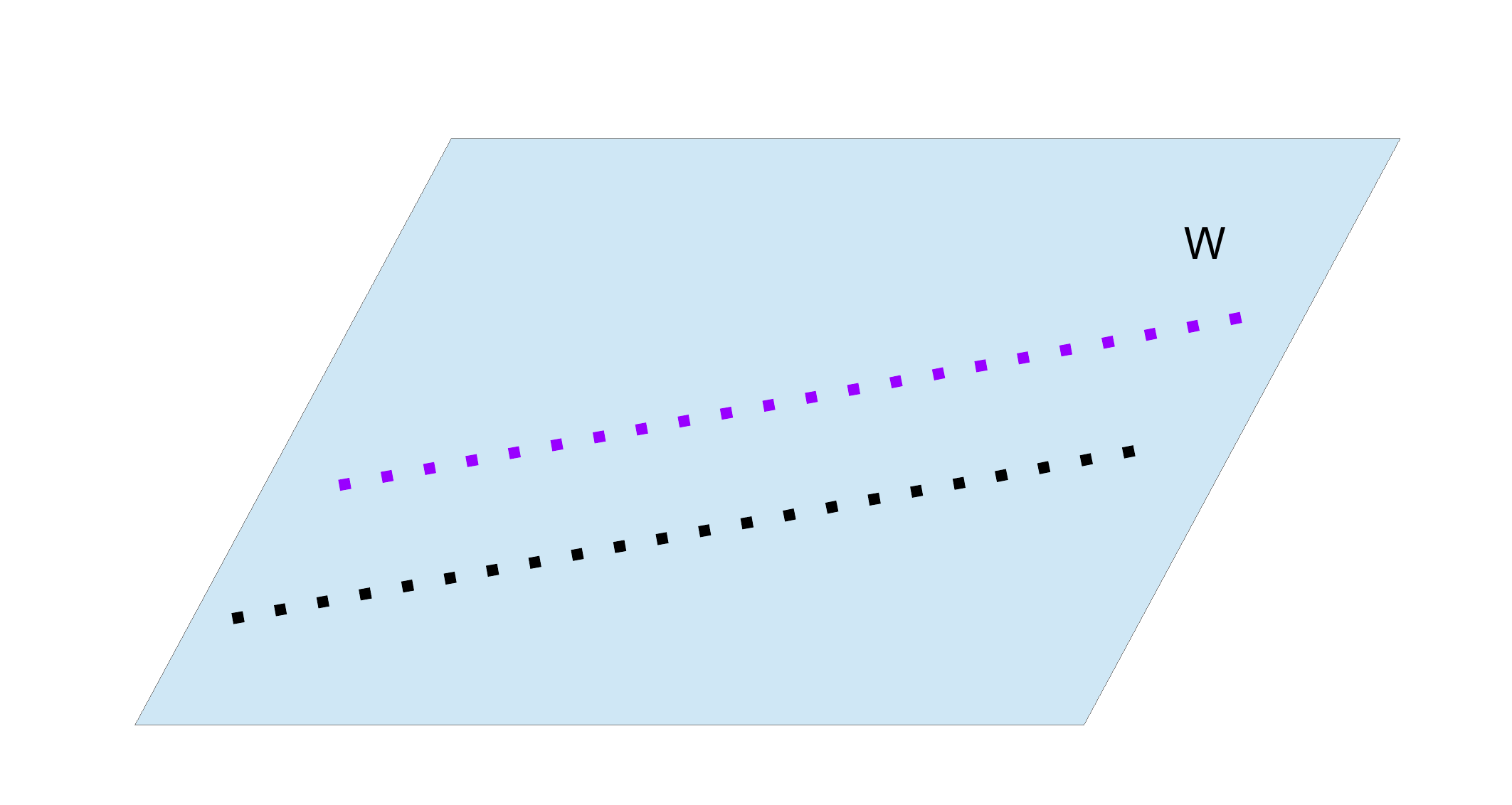}
\caption{In a 2d theory with one flavor symmetry each vacuum has infinitely many copies linearly shifted  in the $W$-plane by an amount $2\pi i\mu$.}
\label{fig:Wplane}
\end{figure}

In the context of LG theories, one can consider a non-simply connected target space $\mathcal{M}$, 
and superpotential deformations such that the holomorphic $1$--form $dW$ is closed but not exact on $\mathcal{M}$. 
The periods of $dW$ over 1-cycles of $\mathcal{M}$ give the ``twisted mass'' deformation parameters. 
In order to see the associated flavor symmetry, we can pull  the closed 1-form $dW$ 
to space-time, and thus define a conserved current. The simplest possibility, which occurs in the mirror of GLSM \cite{HoriV} and appears to be typical for the effective LG descriptions of 
UV-complete 2d field theories, is modelled on a poly-cylinder: a collection of LG fields with periodicity $Y_a\sim Y_a+2\pi i\,n_a$. A superpotential which includes the general linear term $\sum_a \mu_a Y_a$ 
will have discontinuities 
\begin{equation}\label{per:disc}
W(Y_a+2\pi i n_a)=W(Y_a)+ 2\pi i\sum_a n_a \mu_a,
\end{equation} 
with generic values for the complex twisted masses $4\pi i\mu_a$.\footnote{\ To understand the normalization, remember the BPS bound $M \geq 2|\Delta W|$.} We will denote these models as ``periodic''  (see Fig.\;\ref{fig:Wplane} for the case of one periodic field).

There is a second possibility which one encounters, for example, for 2d systems which occur as a surface defect in a 4d ${\cal N}=2$ gauge theory \cite{AKV,GGS}: the twisted mass parameters may not be all independent. We can model this 
occurrence by a LG theory with coordinates valued in an Abelian variety\footnote{\ At this level, it suffices that the $Y_a$'s take value in a complex torus of positive algebraic dimension. However, we may always reduce, without loss of generality, to the Abelian variety with the same field of meromorphic functions.},
\begin{equation}
Y_a\sim Y_a+2\pi i\, n_a+2\pi i\, \Omega_{ab}\,m_b.
\end{equation}
 A superpotential which includes the general linear term\footnote{\ This corresponds to the case of $dW$ a holomorphic differential on the Abelian variety. More generally, we may take $dW$ to be a closed \emph{meromorphic} differential. If $dW$ is a meromorphic differential of the second kind \cite{griffiths}, eqn.\eqref{abeliandisc} remains true with $\Omega_{ab}$ replaced by the relevant period matrix $\Lambda_{ab}$.} $\sum_a \mu_a Y_a$ 
will have discontinuities 
\begin{equation}\label{abeliandisc}
W(Y_a+2\pi i\, n_a+2\pi i\, \Omega_{ab}\,m_b)=W(Y_a)+2\pi i \sum_a \mu_a\! \left(n_a + \sum_{b} \Omega_{ab}\,m_b \right).
\end{equation} 
Thus each twisted mass parameter is associated to two flavor symmetries, whose conserved charges arise from the pull-back of $dY_a$ and $d \bar Y_a$. 
We will denote these models as ``doubly-periodic''.
For surface defects in 4d systems, the relation takes the form 
\begin{equation}
\Delta W= \frac{1}{2} \sum_i \Big(n_i\, a_i[u] + m_i\, a^D_i[u]\Big)
\end{equation} 
where $u$ are the Coulomb branch parameters in the bulk 4d theory, $(a_i, a_i^D)$ the Seiberg-Witten periods of the 4d theory, and the two conserved charges are the electric and magnetic charges 
for the bulk 4d gauge fields. \medskip

\paragraph{Periodic $tt^*$ geometries.} It was already observed in \cite{ttstar,onclassification,GMN2d4d} that the $tt^*$ geometries associated to a standard ``twisted mass'' deformation will be three-dimensional, rather than two-dimensional. Besides the mass parameters $\mu_a$ and $\bar \mu_a$, 
one has an extra angular parameter $\theta_a = 2 \pi x_a$. The angle $\theta_a$ has a direct physical interpretation: it is the flavor Wilson line parameter which appears when the 2d theory is quantized on a circle. 

There is an alternative point of view which is very useful in deriving the $tt^*$ equations for a periodic system. 
We can make the superpotential single-valued by lifting it to an universal cover $\widetilde{\mathcal{M}}$ of $\mathcal{M}$. 
Thus each vacuum $i$ of the original theory is lifted to infinitely many copies $(i,n_a)$, each associated to a sheet of the universal cover. 
We can define vacuum Bloch--waves labelled by the angles $\theta_a = 2 \pi x_a$ (\textit{i.e.}\! by the characters of the covering group $\widetilde{\mathcal{M}}\rightarrow\mathcal{M}$)
\begin{equation}\label{blockbasis}
\big|i;x\big\rangle=\sum_{n_a}e^{2\pi i n_a x_a}\,\big|i;n_a\big\rangle.
\end{equation} 

To describe the $tt^*$ geometry in the $\mu_a$ directions we need to compute the action of $\partial_{\mu_a} W$, 
which is not single-valued. This is simple, as the multi-valuedness of $\partial_{\mu_a} W$ is precisely controlled by the $n_a$. 
Let $n$ be the number of vacua in a reference sheet. Let $B_a$ be the $n\times n$ matrix
\begin{equation}\label{whatLLLs}
B_a= \text{diag}\!\left(\partial_{\mu_a} W(Y)\Big|_{Y=i\text{--th vacuum}\atop \text{in reference sheet}}\right).
\end{equation}
From eqn.\eqref{per:disc} we see that, acting on the $|i;n_a\rangle$ basis,
\begin{equation}
C_{\mu_a}=2\pi i\cdot \boldsymbol{1}\otimes N_a+ B_a\otimes \boldsymbol{1},
\end{equation}
where $N_a$ acts by multiplication by $n_a$. On the Bloch basis \eqref{blockbasis} this becomes the differential operator \cite{ttstar}
\begin{equation}
C_{\mu_a}=  \frac{\partial}{\partial x_a}+B_a.
\end{equation}

If we focus on the dependence on a single twisted mass parameter and its angle, we get 
\begin{equation}
C_\mu= D_x+i\,\Phi,\qquad \overline{C}_{\bar\mu}=-D_x+i\,\Phi,
\end{equation}
with $D_x$ the anti--Hermitian part of $C_\mu$ and $i\,\Phi$ the Hermitian one, while, writing $\mu=(z+iy)/2$,
\begin{equation}
D_\mu=D_z-i\,D_y,\qquad  D_{\bar\mu}=D_z+i\,D_y.
 \end{equation}
The $tt^*$ equations become
\begin{gather}
[D_x,D_y]=[D_z,\Phi],\quad \text{and cyclic permutations of }x,y,z,
\end{gather}
which, seeing $\Phi$ as an (anti--Hermitian) adjoint Higgs fields, are identified with the Bogomolny monopole equations in  $\R^3$ with coordinates $x,y,z$. 

Finally, we can consider chiral operators which are twist fields for the flavor symmetry. In the LG examples discussed above, 
they would correspond, say, to exponentials $e^{\sum_a x_a Y_a}$. It should be clear that the action of such a chiral twist operator on 
a Bloch wave vacuum of parameters $x'_a$ would give a vacuum of parameters $x'_a + x_a$. 
In particular, this shows that the $x_a$ label the Hilbert space sectors ${\cal H}_{x_a}$ in which, as we go around the circle, the fields come back to themselves
up to a phase $\exp(2\pi ix_aQ_a)$, where $Q_a$ are the flavor symmetry charges.

Note that, since $x_a$ are characters of a symmetry, the $tt^*$ metric satisfies
 \begin{equation}
 \big\langle \overline{i;x_a}\,\big|\, j;y_a\big\rangle=G(x_a)_{i\bar\jmath} \sum_{k_a} \delta(x_a - y_a - k_a)
 \end{equation}
Sometimes we leave
the $x_a$ dependence implicit and not bother writing the subscript next to the ket.
\medskip

\paragraph{Doubly--periodic $tt^*$ geometries.} 
In the doubly--periodic case we have two Bloch angles, $\theta_a = 2 \pi x_a$ and $\tilde \theta_a = 2 \pi w_a$ for every mass parameter $\mu_a$. 
We can write
\begin{equation}
\begin{aligned}
C_{\mu_a}&=\frac{\partial}{\partial {x_a}}+ \Omega_{ab} \frac{\partial}{\partial {w_b}}+B_a= \frac{\partial}{\partial {\bar \lambda_a}} + B_a\\
 x_a &= \lambda_a + \bar \lambda_a \qquad w_a = \bar \Omega_{ab} \lambda_b + \Omega_{ab} \bar \lambda_b
\end{aligned}
\end{equation}
where the $n\times n$ matrix $B_a$
\begin{equation}
B_a\equiv \text{diag}\!\left(\partial_{\mu_a}W(Y)\bigg|_{Y=i\text{--th vacuum}\atop \text{in reference sheet}}\right),
\end{equation}
is independent of the $\lambda_b$, $\partial_{\lambda_b} B_a=0$.

It is useful to write 
\begin{equation}\label{firstdex}
C_{\mu_a}= D_{x_a}+\Omega_{ab}\,D_{w_b}\equiv D_{\bar\lambda_a}.
\end{equation}
Setting
\begin{equation}
D_{1,A}= \big(D_{\mu_a},-D_{\lambda_a}\big),\quad D_{2,A}=\big(D_{\bar\lambda_a}, D_{\bar\mu_a}\big),\qquad \begin{matrix}a=1,\dots,g,\\ A=1,\dots,2g,\end{matrix}
\end{equation}
the full set of $tt^*$ equations may be packed into the single equation
\begin{equation}\label{hyperholomorphic}
\big[D_{\alpha,A}, D_{\beta, B}\big]=\epsilon_{\alpha\beta}\, F_{AB}\qquad \text{where }  F_{AB}=F_{BA},
\end{equation}
which are the equations of a \textit{hyperholomorphic connection} on a hyperK\"ahler manifold (here $\R^{2g} \times T^{2g}$) also called hyperK\"ahler (or quaternionic) instanton \cite{salamon,nitta,verbitsky,bartocci}. Indeed, eqn.\eqref{hyperholomorphic} is equivalent to the statement that the curvature of the $tt^*$ connection $D$ is of type $(1,1)$ in all the complex structures of the hyperK\"ahler manifold. For $g=1$ these hyperholomorphic connections reduce to usual (anti)instantons in $\R^2\times T^2$, that is, to doubly--periodic instantons in $\R^4$.
\medskip

This may be seen more directly as follows. In complex structure $\zeta\in\C P^1$, the holomorphic coordinates on $\R^{2g}\times T^{2g}$ are
\begin{equation}\label{nesttolastdex}
u_a^{(\zeta)}= \mu_a-\bar\lambda_a/\zeta,\quad\text{and}\quad v_a^{(\zeta)}= \lambda_a+\bar\mu_a/\zeta.
\end{equation}
The flat $tt^*$ Lax connection with spectral parameter $\zeta$
\begin{equation}\label{lastdex}
\begin{aligned}
\nabla_{\mu_a}^{(\zeta)}& = D_{\mu_a}+\zeta\,C_{\mu_a}\equiv D_{\mu_a}+\zeta\,D_{\bar\lambda_a}\\
 \nabla_{\bar \mu_a}^{(\zeta)} &= D_{\bar \mu_a}+\frac{\overline{C}_{\mu_a}}{\zeta}\equiv D_{\bar\mu_a}-\frac{D_{\lambda_a}}{\zeta},
\end{aligned}
\end{equation}
annihilates, in this complex structure, all holomorphic coordinates $(u_a^{(\zeta)},v_a^{(\zeta)})$ and hence is the $(0,1)$ part, in complex structure  $\zeta$, of a connection $\mathcal{A}$ on $\R^{2g}\times T^{2g}$. The statement that the $tt^*$ Lax connection is flat for all $\zeta$ is then equivalent to the fact that the $(0,2)$ part of the curvature of $\mathcal{A}$ vanishes in all complex structures $\zeta$.
\medskip

The most general $tt^*$ geometry, depending on $N_s$ of standard parameters, $N_m$ twisted mass parameters and $N_d$ doubly-periodic twisted mass parameters is obtained considering such a hyperholomorphic connections  
which do not depend on some of the angular variables: we drop $2N_s+N_m$ angular variables and obtain a higher dimensional generalization of Hitchin, monopole and instanton equations. 
However, the $tt^*$ geometry has an additional requirement besides the condition that the connection is hyperholomorphic, namely the eq.(\ref{reality}) capturing the \emph{reality structure} \cite{ttstar}.

\paragraph{$tt^*$ geometries on $\R^g\times T^{2g}$.}

As  we shall discuss in section 
\ref{sec:geo3dimensions}, the typical $tt^*$ geometry of a 3d model is a variant of the periodic one discussed above in which the complex twisted mass parameters have the form
\begin{equation}
2\mu_a=z_a+i y_a,
\end{equation}
 where the $z_a$'s are real twisted mass parameters and the $y_a$ angular variables which appear on the same physical footing as the vacuum angles $x_a$. In fact, the  $\boldsymbol{S}$ operation 
 \begin{equation}
 \boldsymbol{S}:\quad x_a\rightarrow y_a,\quad y_a\rightarrow -x_a,
 \end{equation}
 should be a symmetry of the physics. To write more symmetric equations, we  add $g$ real variables $w_a$, which do not enter in the Berry connection, in order to complete the space $\R^g\times T^{2g}$ to the flat hyperK\"ahler space $\R^{2g}\times T^{2g}$ on which the $tt^*$ connection is hyperholomorphic (and invariant by translation in the $g$ directions $w_a$).
 We then reduce to a special case of the geometry described in eqns.\eqref{firstdex}--\eqref{lastdex} with holomorphic coordinates in  $\zeta=\infty$ complex structure
 \begin{equation}
 u_a^{(\infty)}\equiv \mu_a=z_a+i \,y_a,\qquad v_a^{(\infty)}\equiv\lambda_a=x_a+i w_a,
 \end{equation}
 (the parametrization being chosen to agree with our conventions for 3d models)

The fact that $\boldsymbol{S}$ is a symmetry of the physics means that it maps the $tt^*$ geometry into itself; in view of the discussion in  eqns.\eqref{firstdex}--\eqref{lastdex}, this means that the effect of $\boldsymbol{S}$ is to map the complex structure $\zeta$ into a complex structure $\tilde{\zeta}(\zeta)$.

To find the map $\zeta\mapsto\tilde\zeta$, we start with the 
$\boldsymbol{S}$--transformed complex coordinates
\begin{equation}
\tilde\mu_a\equiv\boldsymbol{S}(\mu_a)= z_a-i\, x_a,\qquad
\tilde\lambda_a\equiv\boldsymbol{S}(\mu_a)= y_a+i w_a,
\end{equation}
and define the $\boldsymbol{S}$--dual holomorphic coordinates in (dual) complex structure $\tilde\zeta$ as in eqn.\eqref{nesttolastdex},
\begin{equation}
\tilde u_a^{(\tilde\zeta)}=\tilde\mu_a-\overline{\tilde\lambda}_a/\tilde\zeta,\qquad 
\tilde v_a^{(\tilde\zeta)}=\tilde\lambda_a+\overline{\tilde\mu}_a/\tilde\zeta.
\end{equation}
The map $\zeta\mapsto \tilde\zeta(\zeta)$ is then defined by the condition that there exists two holomorphic functions, $f$ and $g$, such that
\begin{equation}
\tilde u_a^{(\tilde\zeta)}=f(u_a^{(\zeta)},v_a^{(\zeta)}),\qquad
\tilde v_a^{(\tilde\zeta)}=g(u_a^{(\zeta)},v_a^{(\zeta)}).
\end{equation}
$f$, $g$ are necessarily linear; writing $\tilde u_a^{(\tilde\zeta)}=\alpha_a u_a^{(\zeta)}+\beta_a v^{(\zeta)}_a$ and equating the coefficients of $x_a,y_a,w_a,z_a$,  one finds
\begin{equation}\label{tcay}
\tilde \zeta=\mathsf{C}(\zeta)\equiv \frac{1+i\,\zeta}{\zeta+i},
\end{equation}
which is the Cayley transform mapping the upper half--plane into the unit disk.
In particular, $\zeta = 1$ is a fixed point under this transformation, which is a rotation of $\pi/2$ of the twistor sphere 
around $\zeta = 1$.

From the discussion around eqn.\eqref{lastdex}, we see that a flat section, $\Pi^{(\zeta)}$, of the $tt^*$ Lax connection at spectral parameter $\zeta$, $\nabla^{(\zeta)}$, is a holomorphic section in complex structure $\zeta$; then, from the $\boldsymbol{S}$--dual point of view $\Pi^{(\zeta)}$ is holomorphic in the $\tilde\zeta$ complex structure, that is, a flat section of ${}^{\boldsymbol{S}}\nabla^{(\tilde\zeta)}$.  In particular, if $\Pi^{(\zeta)}(x_a,y_a,z_a)$ is a flat section of $\nabla^{(\zeta)}$, then
\begin{equation}\label{zetaoneS}
\boldsymbol{S}\Pi^{(\zeta)}(x_a,y_a,z_a)\equiv \Pi^{(\zeta)}(y_a,-x_a,z_a),
\end{equation}
is a flat section of $\nabla^{(\mathsf{C}^{-1}(\zeta))}$.

\paragraph{General ``non--flat'' $tt^*$ geometries.} As observed in \cite{GMN2d4d}, even for the case of surface defects in 4d gauge theories, the $tt^*$ equations reduce to the equations of a hyperholomorphic connection on a hyperK\"ahler manifold,
which is the Coulomb branch of the four-dimensional gauge theory compactified on a circle. In that case, though, the hyperK\"ahler manifold has a non--flat metric, and the $tt^*$ data has a more intricate dependence on the 
angular coordinates. A typical example of the $tt^*$ geometries which arise in this non--flat setup is associated to the moduli space $M$ of solutions of a Hitchin system on some Riemann surface $C$: the universal bundle on $M \times C$ 
supports a hyperholomorphic connection in the $M$ directions and a Hitchin system on the $C$ directions, and the two are compatible exactly as above: the (anti)holomorphic connection on $M$ in complex structure $\zeta$
commutes with the Lax connection of the Hitchin system with spectral parameter $\zeta$.

\subsection{A basic example of 2d periodic $tt^*$ geometry}\label{sec:basicexample}

The simplest and most basic example of periodic $tt^*$ geometry corresponds to the Landau--Ginzburg model
\begin{equation}\label{basicexample}
W(Y)=\mu\,Y - e^Y,
\end{equation} 
which may be seen as the mirror of a 2d chiral field \cite{HoriV} with a twisted complex mass
\begin{equation}
m_\text{twisted}=4\pi i\,\mu.
\end{equation}
 The exact $tt^*$ metric for this model is computed in Appendix A of \cite{CNV}. This $tt^*$ geometry is also a very simple case of the general 2d-4d structures analyzed in \cite{GMN2d4d}.\smallskip

We give a quick review of the $tt^*$ metric for this theory (with some extra detail) and then we shall compute the associated amplitudes ${\Pi_i}^a=\langle D_a|i\rangle$.

\subsubsection{$tt^*$ metric}

Taking the periodicity into account, this theory has a single vacuum at $Y = \log \mu$, as expected for a massive 2d chiral field. The $tt^*$ equations thus reduce to 
$U(1)$ monopole equations on $\R^2 \times S^1$, which can be solved in terms of a harmonic function. We expect the solution to be essentially independent on the phase of the mass, and the only singularities should occur when both the mass and the flavor Wilson lines are zero, so that the 2d chiral field has a zero-mode on the circle. Indeed, we will see momentarily that the correct solution to the $tt^*$ equations corresponds to a single Dirac monopole of charge $1$ placed at $\mu = \bar \mu = x = 0$. 
\footnote{The enthusiastic reader can check this result directly from the definition of the $tt^*$ data, by decomposing the 2d chiral field into 
KK modes on the circle and computing the contribution to the Berry's connection from each of these modes. The action for each KK mode is not periodic in $x$, 
and $n$-th KK mode gives a single Dirac monopole at $x = 2 \pi n$. Together they assemble the desired Dirac monopole solution on $\R^2 \times S^1$. }

It is interesting to describe in detail the relation between the monopole solution and the standard $tt^*$ data. 
In an unitary gauge, we would write 
\begin{align}\label{unitary}
 C_\mu &= \partial_x - i A_x + V \cr
- \bar C_{\bar \mu} &= \partial_ x - i A_x - V \cr
 D_\mu &= \partial_\mu - i A_\mu \cr
 D_{\bar \mu} &= \partial_{\bar \mu} - i A_{\bar \mu}
\end{align}
with $V$ being the Harmonic function, $A$ the associated monopole connection. 
Without loss of generality, we can split $V$ into an $x$-independent part and $x$-dependent part as
\begin{equation}
V(\mu, \bar \mu, x) = \frac{1}{2} v(\mu) + \frac{1}{2} \bar v(\bar \mu) + \frac{1}{2} \partial_x L(\mu, \bar \mu, x)
\end{equation}
for a periodic harmonic function $L(\mu, \bar \mu, x)$, and solve for the connection
\begin{align}
i A_\mu &= -\frac{1}{2}\partial_\mu a(\mu) + \frac{1}{2} \partial_\mu L \cr
 i A_{\bar \mu} &=+\frac{1}{2}\partial_{\bar \mu}\bar a(\bar \mu) - \frac{1}{2} \partial_{\bar \mu} L \cr
 i A_x &= -\frac{1}{2} v(\mu) + \frac{1}{2}\bar v(\bar \mu)
\end{align}
Thus
\begin{align}
C_\mu &= \partial_x + v + \frac{1}{2} \partial_x L \cr
- \bar C_{\bar \mu} &= \partial_ x - \bar v - \frac{1}{2} \partial_x L \cr
 D_\mu &= \partial_\mu +\frac{1}{2}\partial_\mu a- \frac{1}{2} \partial_\mu L \cr
 D_{\bar \mu} &= \partial_{\bar \mu} -\frac{1}{2}\partial_{\bar \mu}\bar a+\frac{1}{2} \partial_{\bar \mu} L
\end{align}

We can then go to the ``topological basis'' by the complexified gauge transformation with parameter $\frac{1}{2} L(\mu, \bar \mu, x)- \frac{1}{2} a(\mu) -\frac{1}{2} \bar a(\bar \mu)$:
\begin{align}\label{tftgauge1}
C_\mu &= \partial_x + v \cr
- \bar C_{\bar \mu} &= \partial_ x - \bar v - \partial_x L\cr
 D_\mu &= \partial_\mu +\partial_\mu a - \partial_\mu L \cr
 D_{\bar \mu} &= \partial_{\bar \mu} 
\end{align}

The gauge transformation parameter is directly related to the $tt^*$ metric \cite{ttstar}, which reduces in this case to a real positive function of $x$ and $|\mu|$, 
$G(x,|\mu|)$ of period $1$ in $x$: 
\begin{equation}
G(x,|\mu|) = e^{L(\mu, \bar \mu, x)- a(\mu) -\bar a(\bar \mu)}.
\end{equation}
Using\footnote{In a vacuum basis, the pairing $\eta$ is diagonal, proportional to the inverse determinant of the Hessian of the superpotential, see \cite{Vafa:1990mu}.}  the relation $\eta = \mu^{-1}$ the reality condition $|\mu|^{2}G(-x,|\mu|)\,G(x,|\mu|)=1$ tells us that $L$ is odd in $x$. 
Also, we find $a(\mu) = \frac{1}{2} \log \mu$ and $\log G = \log L - \frac{1}{2}\log |\mu|$. 

As $L$ is harmonic, 
\begin{equation}
\frac{1}{|\mu|}\frac{\partial\phantom{a}}{\partial\, |\mu|}\!\left(|\mu|\, \frac{\partial\phantom{a}}{\partial \,|\mu|}\, L\right)+ 4\,\frac{\partial}{\partial x}\!\left(\frac{\partial}{\partial x}L\right)=0,
\end{equation}
periodic and odd, it has an expansion in terms of Bessel--MacDonald functions of the form
\begin{equation}\label{ttspecialcase2d}
L(x,|\mu|)= \sum_{m=1}^\infty a_m\, \sin(2\pi m x)\, K_0(4 \pi m |\mu|),
\end{equation}
for certain coefficients $a_m$ which are determined by the boundary conditions. We may use either the UV or IR boundary conditions, getting the same $a_m$ \cite{CNV}. For instance, in the UV we must have the asymptotics as $|\mu|\rightarrow 0$
\begin{equation}\label{asymptoticsUV}
L(x,|\mu|)= -2\big(q(x)-1/2\big)\log|\mu|+\Lambda(x) + O\big(|\mu|\big),
\end{equation} 
 where $q(x)$ is the SCFT $U(1)$ charge of the chiral primary $e^{x Y}$ ($0\leq x <1$) at the UV fixed point, while the function $\Lambda(x)$ encodes the OPE coefficients at that fixed point \cite{ttstar,onclassification}. From the chiral ring relations we have $q(x)=x$. From the expansion
 \begin{equation}
 K_0(z)=-\log(z/2)-\gamma+O(z^2\log z)\qquad \text{as }z\sim 0.
\end{equation}
we get
\begin{equation}
(1-2x)\log|\mu|+\Lambda(x)=-\sum_m a_m\,\sin(2\pi m x) \Big(\log|\mu|+\log m+\log2\pi +\gamma\Big).
\end{equation}
Comparing the coefficients of $\log|\mu|$, we see that the $a_m$'s are just the Fourier coefficients of the first (periodic) Bernoulli polynomial, and hence
 \begin{equation}
 a_m= -\frac{2}{\pi}\, \frac{1}{m}.
 \end{equation}
 Then (for $0< x <1$)
 \begin{equation}
 \begin{split}
 \Lambda(x)&=(1-2x)\big(\log 2\pi +\gamma\big)+\frac{2}{\pi}\sum_{m\geq 1}\sin(2\pi m x)\,\frac{\log m}{m}=\\
 &= 2\log\Gamma(x)+\log\sin(\pi x)-\log\pi,
 \end{split}\end{equation}
 where the equality in the second line follows from Kummer's formula for the Fourier coefficients of the Gamma--function\cite{specialfunctions}. In particular, the UV OPE coefficients have the expected form \cite{ttstar,onclassification}.\medskip
 
As we have $v(\mu) = \log \mu$, we can recognize the periodic monopole solution
\begin{equation}\label{periodicmonopoles}
\begin{split}
V(\mu, \bar \mu, x) &= \log |\mu| - 2 \sum_{m=1}^\infty \, \cos(2\pi m x)\, K_0\big(4\pi m |\mu|\big) =\\
&= -\frac{1}{2} \sum_{n\in\Z} \left(\frac{1}{\sqrt{|2\mu|^2 + (x-n)^2}} -\kappa_n\right)-\gamma
\end{split}
\end{equation}
where $\kappa_n$ is some constant regulator (see eqn.\eqref{firstide}). 

It is convenient to give a representation of the solution $L(x,|\mu|)$ in terms of a convergent integral representation. From the equality (for $\mathrm{Re}\,z>0$)
\begin{equation}
K_0(z)=\frac{1}{2}\int\limits_0^\infty \frac{dt}{t}\,e^{-\tfrac{1}{2}\,z(t+t^{-1})}, 
\end{equation}
 we see that for $\mathrm{Re}\,\mu>0$
 \begin{equation}
 \begin{split}
 L(x,\mu,\bar\mu)&=-\frac{1}{\pi}\sum_{m=1}^\infty \frac{\sin(2\pi m x)}{m} \int\limits_0^\infty \frac{dt}{t}\, e^{- 2\pi m(\mu t+\bar\mu t^{-1})}=\\
 &= \frac{1}{2\pi i}\int\limits_0^\infty\frac{dt}{t}\, \log\!\left(\frac{1-e^{-2\pi(\mu t+\bar\mu t^{-1}- ix)}}{1-e^{-2\pi(\mu t+\bar\mu t^{-1}+ i x)}}\right).
 \end{split}
 \label{metricintegralrep}\end{equation}
 For $\mathrm{Re}\,\mu>0$ the integral is absolutely convergent.
 If $\mathrm{Re}\,\mu\not>0$ (and $\mu\neq 0$), just replace $\mu\rightarrow e^{i\alpha}\mu$ in such a way that $\mathrm{Re}(e^{i\alpha}\mu)>0$ (or, equivalently, rotate the integration contour).
 Notice that the expression \eqref{metricintegralrep} makes sense even for $\mu$ and $\bar\mu$ \emph{independent} complex variables (as long as $\mathrm{Re}\,\mu>0$ and  $\mathrm{Re}\,\bar\mu>0$). 
  

\medskip

2d $tt^*$ computes a second interesting physical quantities besides the metric, namely the CFIV `new index' $Q(x,|\mu|)$ \cite{CFIV}. Several explicit expression for the CFIV index of this model may be found in appendix \ref{2dcfiv}.

\subsubsection{The amplitude $\langle D_a|\phi(x) \rangle =\langle D_a|0\rangle_x$}
\label{thebasicbraneamplitude}
The equations for a flat section $\Pi$ of the $tt^*$ Lax connection look somewhat forbidding
\begin{align}
 \left(\partial_\mu + \zeta \partial_x\right) \log \Pi &= \partial_\mu L - \zeta v -\partial_\mu a \cr
\left(-\zeta \partial_{\bar \mu} +\partial_ x \right) \log \Pi &= \bar v + \partial_x L 
\end{align}
Observe that $\Pi$ is defined up to multiplication by an arbitrary function of $\zeta \mu -x - \zeta^{-1} \bar \mu$. 
This is related to the fact that any D-brane has infinitely many images, produced by shifts in the flavor grading of the Chan-Paton 
bundle. Starting from a single D-brane amplitude $\Pi_0$ one can produce a countable basis 
 \begin{equation}\label{prefactor}\Pi_k=e^{2\pi i k(\zeta \mu -x - \zeta^{-1} \bar \mu)}\,\Pi_0\qquad k\in \Z.\end{equation}
for the infinite-dimensional vector space of flat sections of the $tt^*$ Lax connection 

 Writing
 \begin{equation}\label{pi-phi}
 \log\Pi=\Phi-\frac{1}{2}\log\mu-\zeta \mu\big(\log\mu-1\big)-\zeta^{-1} \bar\mu\big(\log\bar\mu-1)+\mathrm{const.},
 \end{equation}
(we will fix the additive constant later by choosing a convenient overall normalization of $\Pi$)
we isolate the interesting part  
\begin{align}
 \left(\partial_\mu + \zeta \partial_x\right) \Phi &= \partial_\mu L  \cr
\left(-\zeta \partial_{\bar \mu} +\partial_ x \right) \Phi &=  \partial_x L 
\end{align}

In view of the expression
 \begin{equation}\label{LLLLL2}
 L=-\frac{1}{2\pi i}\sum_{m\neq 0} e^{2\pi i m x}\int_0^\infty \frac{1}{m\,t}\, e^{-2\pi |m| ((\mu\,t+\bar\mu\, t^{-1})}\,dt,
\end{equation}
 for $\mathrm{Re}\,\mu>0$ we look for a solution $\Phi$ of the form
 \begin{equation}\label{inrtegral}
 \Phi(x,\mu,\bar\mu)=\sum_{m\neq 0} e^{2\pi i m x}\int_0^\infty f_m(t)\; e^{-2\pi |m|(\mu\,t+\bar\mu\, t^{-1})}\, dt,
\end{equation}
for some functions $f_m(t)$ to be determined. Plugging this ansatz in the equations we get 
\begin{equation}
f_m(t)= \frac{i}{ 2\pi m } \frac{1}{ t -i\, \zeta\,\mathrm{sign}(m) }.
\end{equation}

 Then
\begin{equation}\label{Phiintegralform}
\Phi= \frac{1}{2\pi i}\int_0^\infty \frac{dt}{t-i \zeta} \;\log\!\Big(1-e^{-2\pi(\mu t+\bar\mu t^{-1}-ix)}\Big)-\frac{1}{2\pi i}\int_0^\infty \frac{dt}{t+i \zeta} \;\log\!\Big(1-e^{-2\pi(\mu t+\bar\mu t^{-1}+ix)}\Big).
\end{equation}
For $\mathrm{Re}\,\mu,\;\mathrm{Re}\,\bar\mu>0$ the integrals are absolutely convergent and define an analytic function of $\mu$ and $\bar\mu$ (seen as independent complex variables).

This expression has an important discontinuity along the imaginary $\zeta$ axis, where the poles cross the integration contours,
 and is analogous to the integral equations which 
gives the thimble brane amplitudes in the standard $tt^*$ case \cite{onclassification}. It is also a simple version of the integral equations which describe general 
2d-4d systems in \cite{GMN2d4d}. The discontinuity along the positive and negative imaginary axes are
\begin{equation}\label{stokes}
\pm \log\!\Big(1-e^{\pm 2\pi i (\zeta \mu- x -\zeta^{-1} \bar\mu)}\Big).
\end{equation} 
The two functions $\Pi_\pm$ defined by the analytic continuation from the positive and negative half-planes 
must correspond to the amplitudes for the thimble branes for the model. We will identify these branes momentarily.

The same discontinuities appear at fixed $\zeta$ as we vary the phase of $\mu$, as one has to rotate the integration contours 
while moving $\mu$ out of the $\mathrm{Re} \mu>0$ half-plane. 
Notice that the composition of the two discontinuities in $\Phi$ we encounter while rotating the phase of $\mu$ by $2 \pi$, i.e. 
$\pi i + 2 \pi i (\zeta \mu- x -\zeta^{-1} \bar\mu)$, cancel against the extra terms in the definition  \ref{pi-phi} of $\Pi$,
leaving only $$\Pi(e^{2 \pi i} \mu, e^{-2 \pi i} \bar \mu,x) = e^{2 \pi i x} \Pi(\mu, \bar \mu,x)$$
which is the gauge transformation which leaves the $tt^*$ data invariant. 
This equation is equivalent to the statement $q(x)=x$. Thus all the pieces conspire to make the 
sections $\Pi_\pm$ single-valued as functions of $\mu$ (but not of $\zeta$!). 

The function $\Phi$ is also periodic in $x$, and enjoys a number of interesting properties.
First of all, it satisfies the functional equation (for $\mathrm{Re}\,\zeta\mu,\; \mathrm{Re}\,\zeta^{-1}\bar\mu>0$)
\begin{equation}\label{property1}
\Phi(-x,\zeta^{-1}\bar\mu,\zeta\mu)-\Phi(x,\zeta\mu,\zeta^{-1}\bar\mu)= L(x,\mu,\bar\mu)
 \end{equation}
 which just says that the $tt^*$ metric can be computed out of the amplitude $\Pi(x)$ in the usual way. Second, for all integers $n\in\mathbb{N}$ it satisfies a `Gauss multiplication formula' of the same form as the one satisfied by $\log\Gamma(z)$
 \begin{equation}\label{property2}
 \Phi(0,n\zeta\mu,n\zeta^{-1}\bar\mu)= \sum_{k=0}^{n-1}\Phi(k/n, \zeta\mu,\zeta^{-1}\bar\mu).
 \end{equation}
Eqns.\eqref{property1}\eqref{property2} are shown in appendix \ref{prooffirsttwoide}.

\subsubsection{The limit $\bar\mu\rightarrow 0$ and brane identification} \label{sec:braneidentification}
 
Seeing the amplitude $\Pi(x,\mu,\bar\mu)$ as a function of independent complex variables $\mu$ and $\bar\mu$, it make sense to consider its form in the limit $\bar\mu\rightarrow 0$.  As discussed in section 2, this is the limit where we expect
$\Pi(x,\mu,0)$ to simplify, and satisfy a simple differential equation.   We will check to see how this
emerges in this section (eqn.(\ref{diffeq})).

The asymmetric limit $\bar \mu\rightarrow 0$ is also important 
to identify which kind of brane amplitude corresponds to each solution to the Lax equations,
and in particular to identify the unique solution which corresponds to a (correctly normalized) Dirichlet brane amplitude, $\langle D|x\rangle_\zeta$, and its relations with the Leftshetz thimble amplitudes.  We saw that the difference between the log of any two solutions, $\Pi_1, \Pi_2$, is a holomorphic function of $\zeta\mu- x-\bar\mu/\zeta$
\begin{equation}
\log \Pi_1-\log\Pi_2=f(\zeta\mu- x-\bar\mu/\zeta).
\end{equation}
In particular, two solutions which are equal at $\bar\mu=0$, are equal everywhere.
Therefore the identity of the corresponding boundary conditions is uniquely determined by comparing their $\bar \mu \to 0$ limit \cite{HIV},
with the period integrals of ${\rm exp }(-\zeta \beta W)$.  This limit can be alternatively
computed (assuming the correctness of our conjecture of the equivalence of this limit with supersymmetric partition functions
\cite{2dpart1,2dpart2})
with a direct localization computation for the partition function of the 2d chiral on a hemisphere \cite{Hori}. 

We are looking at 
\begin{equation}\label{muzerofirst}
\begin{split}
\log \Pi= &-\frac{1}{2}\log\mu-\zeta \mu\big(\log\mu-1\big)+\mathrm{const.} \\&+ \frac{1}{2\pi i}\int_0^\infty \frac{dt}{t-i \zeta} \;\log\!\Big(1-e^{-2\pi(\mu t-ix)}\Big)-\frac{1}{2\pi i}\int_0^\infty \frac{dt}{t+i \zeta} \;\log\!\Big(1-e^{-2\pi(\mu t+ix)}\Big).
\end{split}
\end{equation} 
We claim that choosing the additive constant to be $0$, the branch $\Pi_-$ of $\Pi$ in the negative $\zeta$ half plane becomes 
\begin{equation}\label{neumann} 
\Pi_- = \frac{1}{\sqrt{2 \pi}}\Gamma\big(- \zeta \mu+\{x\}\big)\mu^{-\{x\}} (-\zeta)^{\frac{1}{2}+\zeta \mu-\{x\}}
\end{equation}
and the branch $\Pi_+$ of $\Pi$ in the negative $\zeta$ half plane becomes
\begin{equation}\label{dirichlet}
\Pi_+ = \frac{\sqrt{2\pi}}{\Gamma\big(\zeta \mu+1-\{x\}\big)}\mu^{-\{x\}}\zeta^{\frac{1}{2}+\zeta \mu-\{x\}}
\end{equation}
where $\{x\}\equiv x-[x]$ is the fractional part, $0\leq \{x\}<1$. Note that these expressions are consistent with \eqref{property2} in view of the Gauss multiplication formula for $\Gamma(z)$ \cite{specialfunctions}.

A straightforward way to prove these identities is to observe that the right hand sides have the correct asymptotic behaviour 
at large $\zeta$, the correct discontinuities, and no zeroes or poles in the region where we want to equate them to 
the integral formula. Thus they must coincide with the result of the integral formula. 
By setting $x=0$ or $x=1/2$ in these identities we get well--known integral representations of $\log\Gamma(\mu)$ or, respectively, $\log\Gamma(\mu+1/2)-\tfrac{1}{2}\log\mu$ (see appendix \ref{appGamma2}). Setting $\mu=0$ in our identity produces a new proof of the Kummer formula (appendix \ref{appGamma3}). 

Comparing with a direct localization computation for the partition function of the 2d chiral on a hemisphere \cite{Hori},
we see that,
for all values of $\zeta$, the behavior \eqref{neumann} corresponds to a brane with Neumann boundary conditions and \eqref{dirichlet} to Dirichlet b.c.
We conclude that the thimble brane of the LG mirror corresponds to either Neumann or Dirichlet boundary conditions for the 2d chiral field.
The match with the localization computations is surprisingly detailed, especially if we turn off $x$ and identify $-\zeta \beta$ with $r \Lambda_0$ in \cite{Hori}.
 
Finally, we can compare the result to the expected integral expressions for the asymmetric conformal limits
$$\int_{D} e^{x Y- \zeta \mu Y + \zeta e^Y} dY$$
For  example, for $\zeta$ in the negative half-plane we can do the integral on the positive real $Y$ axis 
setting $t = - \zeta e^Y$, i.e. 
$$(-\zeta)^{-x + \zeta \mu}\int_{0}^\infty t^{x - \zeta \mu-1}e^{-t} dt = (-\zeta)^{-x + \zeta \mu} \Gamma(x - \zeta \mu)$$
which is as expected.

\subsection{A richer example}

After discussing a model which gives rise to a single periodic $U(1)$ Dirac monopole as a $tt^*$ geometry, it is naturally to seek a model associated to a 
single smooth $SU(2)$ monopole solution. It is not hard to guess the correct effective LG model: 
\begin{equation}\label{basicexampletwo}
W(Y)=\mu\,Y - e^{\frac{t}{2}+Y}+ e^{\frac{t}{2}-Y}.
\end{equation} 
We recognize this as the mirror of a $\C P^1$ sigma model \cite{HoriV} with $FI$ parameter $t$ and twisted mass $4 \pi i \mu$ for its $SU(2)$ flavor symmetry.  
We will come back to the standard $tt^*$ geometry in the $t$ cylinder momentarily.   Unlike the previous
example, it is not possible to solve this model explicitly.  Nevertheless we can predict properties
of the solution, based on the previous example, as well as general physical reasoning.

The model has two vacua, with opposite values of $W$, and will give rise to a rank $2$ bundle, with $SU(2)$ structure group.  
At large $|\mu|$ we get either of the vacua $Y \sim \pm\left(-\frac{t}{2}+\log \mu \right)$, and the two vacua are well-separated. The solution approaches an Abelian monopole of charge $\pm 1$.  On the other hand, $Y(\mu)$ does not have logarithmic singularities anywhere: there are never massless particles in the spectrum, 
and thus no Dirac singularities in the interior. This confirms that the $tt^*$ geometry for the $\mu$ parameter will be a smooth $SU(2)$ monopole. 
The parameter $t$ controls the constant part of the Higgs field and Wilson line at large $\mu$.

On the other hand, the $tt^*$ geometry for the $t$ parameter is well-known: the boundary conditions of the Hitchin system's Higgs field 
$C_t$ are controlled by 
\begin{equation}
\frac{1}{2} \mathrm{Tr}\, C_t^2= \left( \partial_t W \right)^2 = \frac{\mu^2}{4} - e^t
\end{equation}
Thus we have a standard regular singularity at the $t \to -\infty$ end of the cylinder, with residues $\pm \frac{\mu}{2}$ in the Higgs field and $\pm x$ in the connection.
We have the mildest irregular singularity at the $t \to \infty$ end of the cylinder. 

The $tt^*$ machinery predicts that the Lax connections for the BPS monopole connection associated to the $\mu$ direction
and for the Hitchin system in the $x$ direction will commute (for the same values of the spectral parameter). 
This fact may appear striking. It is useful to think about it in terms of an isomonodromic problem. For example, 
the Hitchin system has a unique solution for given $\mu$, $x$. Furthermore, up to conjugation, the monodromy data of the Lax connection 
with spectral parameter $\zeta$ only depends on the combination $\mu \zeta - x- \bar \mu \zeta^{-1}$
and it is annihilated by the combinations of derivatives $\partial_\mu + \zeta \partial_x$ and $\partial_{\bar \mu} - \zeta^{-1} \partial_x$. 
These facts are what make it possible to find connections $D_\mu + \zeta D_x$ and $D_{\bar \mu} - \zeta^{-1} \bar D_x$
which commute with the Hitchin Lax connections, and which become the Lax connection for the BPS monopole equation. 

The LG model has several interesting A-branes, which are mirror to the basic B-branes of the $\C P^1$ sigma model \cite{HIV}: 
we can have either a Dirichlet brane at the north or south pole, or a Neumann brane with $n + \frac{1}{2}$ units of world volume flux. 
The corresponding amplitudes where identified in \cite{GMN2d4d} with specific flat sections of the Hitchin system Lax connection.
The Dirichlet branes correspond to the monodromy eigenvectors at the regular singularity. The basic Neumann brane 
is the unique section which decreases exponentially approaching the irregular singularity. The whole tower of Neumann branes is obtained by transporting the 
basic one $n$ times around the cylinder. As the full BPS spectrum of the $\C P^1$ model is known, the actual brane amplitudes  
can be computed from the integral equations derived in \cite{dubrovin,onclassification}, corrected by the presence of twisted masses as in \cite{GMN2d4d}.

\subsection{Some doubly--periodic examples}
As we seek examples of well-defined doubly-periodic systems, it is natural to start from a simple, smooth doubly-periodic instanton solution \cite{Ford:2003vi} and 
work backwards to identify an effective LG model associated to it. The simplest choice would be a doubly-periodic $SU(2)$ instanton of minimal charge. 
The identification of the LG model is rather straightforward using the connection to the Nahm transform detailed in the next section. 
Here we can anticipate the answer: 
  \begin{equation}\label{doubly}
W= m_b \log \Theta(X+\frac{z}{2},\tau) - m_b\log \Theta(X-\frac{z}{2},\tau) - a X
\end{equation}
Here $X$ is the doubly-periodic LG field, $a$ the deformation parameter whose $tt^*$ geometry will reproduce the doubly-periodic instanton, 
$\Theta$ is the usual theta function and $m_b$, $z$ two extra parameters. 

The superpotential has discontinuities of the form $(n_1 + \tau n_2) a + (n_3 + z n_2) m_b$.  We will focus on the $tt^*$ geometry in $a$ first, and then 
extend it to $a, m_b$. The instanton is defined over the space parameterized by $a$ and the two angles $\theta_1$ and $\theta_2$ dual to the charges $n_1$ and $n_2$. 
The vacua are determined by 
\begin{equation}
a =m_b \frac{\Theta'(X+\frac{z}{2})}{\Theta(X+\frac{z}{2})}-m_b \frac{\Theta'(X-\frac{z}{2})}{\Theta(X-\frac{z}{2})}
\end{equation}
and $C_a$ is controlled by the critical value of $X$. 
At large $a$, 
\begin{equation}
X \sim \pm \left(\frac{z}{2} + \frac{m_b}{a} +\cdots \right),
\end{equation} 
and thus $z$ controls the large $a$ asymptotic value of the $SU(2)$ instanton connection on the $(\theta_1,\theta_2)$ torus direction and $m_b$ the first subleading coefficient. 

Because of the appearance of $m_b$ in the $n_2$ monodromy, the $C_{m_b}$ differential operator must include both the usual $\partial_{x_3}$ expected for a standard mass parameter, and 
an extra $z \partial_{x_2}$ which mixes it with the doubly-periodic instanton directions. Thus rather than a direct product of doubly-periodic instanton equations and 
periodic monopole equations, we get a slightly more general reduction of an eight-dimensional hyper-holomorphic connection down to a system over $\R^4 \times T^3$, where $T^3$ has a metric determined by $\tau$ and $z$. 

We can easily describe a system which behaves a bit better: 
 \begin{equation}\label{doublytwo}
 \begin{split}
W&= m_b \log \Theta(X+z,\tau) + m_b\log \Theta(X-z,\tau) -2 m_b\log \Theta(X,\tau) - a X =\\
&= m_b \log \left(\wp(X) - \wp(z) \right) - a X
\end{split}\end{equation}
This superpotential has only the standard $(n_1 + \tau n_2) a + n_3 m_b$ discontinuities, and thus we get three separate and compatible connections: 
an $SU(3)$ doubly-periodic instanton from the $a$ deformation, a rank $3$ periodic monopole from the $m_b$ deformation and a rank 3 Hitchin system 
from the $z$ deformation. 

The asymptotic form of $C_a$ for large $|a|$ in the three vacua is $X \sim z+m_b/a$, $X \sim -z + m_b/a$, $X \sim - 2 m_b/a$. 
The $tt^*$ geometry for $a$ should be smooth in the interior. 

In order to understand the other deformations, it is useful to massage a bit the chiral ring relation which follows from the superpotential.
We have 
\begin{equation}
m_b \wp'(X) = a \wp(X) - a \wp(z)
\end{equation}
Using the standard cubic relation for the Weierstrass function, we get
\begin{equation}
4 \wp(X)^3 - g_2 \wp(X)- g_3 = \frac{a^2}{m_b^2} \left(\wp(X) - \wp(z) \right)^2
\end{equation}
As the $C_{m_b}$ eigenvalues are the values of $\log \left(\wp(X) - \wp(z) \right)$, the above form of the chiral ring relation 
gives the holomorphic data of the periodic monopole solution. It appears to have logarithmic singularity, corresponding to a Dirac monopole singularity at $m_b=0$ of charges $1,1,-2$ 
and no logarithmic growth at infinity: there must be a smooth monopole configuration screening the Dirac singularity. 

The model has an interesting limit $z \to 0$, with constant $m_b z^2$: 
 \begin{equation}\label{doublytwo}
W= c \wp(X) - a X
\end{equation}
This is the basic building block for models considered in \cite{CV92b}, such as
  \begin{equation}\label{cpntwistedmasses}
W=\lambda\left(\sum_{a=1}^{N-1}\Big(\wp(Y_a)-m_a\,Y_a\Big)+\wp\big(-\sum\nolimits_{a=1}^N Y_a\big)\right).
\end{equation}

\subsection{Non--commutative $tt^*$ geometries}\label{non-com}
It is natural to wonder what would happen if we took a simpler version of the doubly-periodic examples, 
a superpotential involving a single $\theta$ function: 
\begin{equation}\label{nonc}
W= \log \Theta(X,\tau) - \mu \left(X+ \tau/2 + 1/2\right)
\end{equation}
This superpotential and the chiral ring relation 
\begin{equation}
\mu = \frac{\Theta'(X,\tau)}{\Theta(X,\tau)}
\end{equation}
only make sense if the parameter $\mu$ is taken to have a periodic imaginary part. 

This makes sense if $X$ is actually part of a 2d gauge multiplet and $\mu$ is 
the corresponding FI parameter. Indeed, in 2d the field strength of an $U(1)$ vector supermultiplet is a twisted chiral field $\Sigma$ with the real part of the $F$--term equal to the field strength 2--form. Hence the $F$--terms 
roughly take the form
\begin{equation}
i\int F^a\;\mathrm{Im}\,\partial_{\Sigma_a}W+\int d^2z\;D^a\;\mathrm{Re}\,\partial_{\Sigma_a}W,
\end{equation}
and the $\mathrm{Im}\,\partial_{\Sigma_a}W$ are field--dependent $\theta$--angles which need to be well--defined only up to shifts by integers. Indeed, the flux $\int F$ is quantized in multiples of $2\pi$, and the action is still well--defined mod $2\pi i$. Thus we may allow a (twisted) superpotential $W(\Sigma_a)$ such that 
$\partial_{\Sigma_a}W$ is defined up to integral multiples of $2\pi i$. 

Naively, one may thus expect the $tt^*$ geometry to be an instanton solution in $\R \times T^3$. 
The situation, though, is more complex than that. The images of a vacuum under the two translations of $X$ 
by $1$ or $\tau$ are associated to different values for $\mu$, as translations of $X$ by $\tau$ 
require a shift of $\mu$ by $2 \pi i$. Thus if we try to form Bloch wave vacua with angles $\theta_{1,2}$ 
as before, we cannot treat $\mu$ and $\theta_2$ as commuting variables. Rather, we need 
some Heisenberg commutation relation such that $e^{i n \theta_2}$ acts on $\mu$ by a shift of $2 \pi i n$.  

The natural guess is that, in situations such as this, the $tt^*$ equations may define a hyperholomorphic connection on a \emph{non--commutative} version of, say, $\R \times T^3$ where at least two torus directions do not commute among themselves. 
It turns out that such non--commutative $tt^*$ geometries are very common for 4 supercharge models arising from 4d gauge theories, as we shall see later in  
section \ref{4d}. In particular, the 4d theory with spectral curve \eqref{onechiral} may be modelled by a 2d theory with a superpotential $W$ such that
\begin{equation}
\exp\!\big(\partial_X W(X)\big)= \frac{\Theta(X+\mu^\prime/2,\tau)}{\Theta(X-\mu^\prime/2,\tau)}.
\end{equation} 
For $\mu^\prime$ small this gives
\begin{equation}\label{secondncmodel}
W(X)=\mu^\prime \log\Theta(X,\tau)+O({\mu^\prime}^2),
\end{equation}
whose $tt^*$ geometry may be meaningful only in the non--commutative framework.

Another situation where a non-commutative $tt^*$ geometry may appear is a 2d-4d system 
in the presence of Nekrasov deformation in the transverse plane to the defect
and/or a supersymmetric Melvin twist in the $tt^*$ compactification. The two are related because the 
Nekrasov deformation parameter behaves as a 2d twisted mass for the rotation (plus R-charge rotation) 
in the plane transverse to the defect, which is used to define the Melvin twist.
In such a situation, the electric and magnetic Wilson lines cease to be commutative variables. 
The corresponding non-commutative version of the $tt^*$ geometry should be related to the motivic 
Kontsevich-Soibelman wall-crossing formula. 
 
A full discussion of the non--commutative $tt^*$ geometries is outside the scope of the present paper. Here we limit ourselves to a general discussion of 
how non--commutative structures could possibly emerge from the standard $tt^*$ machinery.

\subsubsection{$tt^*$ geometry for the models \eqref{nonc}\eqref{secondncmodel}}

In these examples the chiral field $X$ takes values in a complex torus $E$ of periods $(1,\tau)$, that is, we periodically identify
\begin{equation}X\sim X+k+\tau m,\quad k,\, m\in\Z.\end{equation}
 For definiteness, we choose $\Theta\equiv \theta_3$ which vanishes at the point
$X_\mathrm{cr}=(1+\tau)/2$.
Since the superpotential is not univalued in $E$, to defined the $tt^*$ geometry we must lift the model to a cover where $W$ is well defined; in the process we get infinitely many copies of the single vacuum. However in this case there are more copies of the vacuum than just the lattice translates $X_0+(k+m\tau)$, $k,m\in\Z$. For instance for the model \eqref{nonc}, since $\mu$ is a periodic variable, 
the actual equation defining the classical vacua is \cite{nekrasovwitten}
\begin{equation}\label{vacua}
 \exp\!\Big[\partial_X W(X)\Big]=1.
\end{equation}
The \textsc{lhs} is a holomorphic function in $E\setminus (1+\tau)/2$ with an essential singularity at the point $X_\mathrm{cr.}=(1+\tau)/2$. 
By the Big Picard theorem, the equation
equation \eqref{vacua} has \emph{infinitely many} solutions in any open neighborhood of the point $(1+\tau)/2$. These solutions may
 be interpreted as cover copies of the vacuum due to the non--trivial 
monodromy around the point $X_{\mathrm{cr.}}$.

\paragraph{The monodromy action.} To be systematic, we consider $X$ as a field taking value in the K\"ahler 
manifold $K=E\setminus X_\mathrm{cr.}$ and go to its 
  universal cover
 $\widetilde{K}$ on which $W$ is defined as a univalued function by analytic continuation. Let $\mathsf{M}$ be the monodromy group of the cover $\widetilde{K}\rightarrow K$, which is identified with $\pi_1(K)$; 
we need to know how it
 is represented  on the vacuum bundle $\mathcal{V}\rightarrow  K$. Indeed, the monodromy group acts by symmetries just as in the ordinary periodic case.
For definiteness we choose $X=0$ as the base point, and consider the 
homotopy group of paths based at the origin, $\pi_1(K,0)$. This group is generated by three loops $u_1, u_2, \ell$ subject to a single relation
\begin{equation}\label{uuur}
 \ell=u_2^{-1}u_1^{-1} u_2u_1,
\end{equation}
where
\begin{equation}
 u_1=t \mod \Z+ \Z \tau,\qquad u_2= t\tau \mod \Z+\Z\tau,\qquad 0\leq t\leq 1,
\end{equation}
and $\ell$ is a loop which starts from the origin, go to the point $(1+\tau)/2$ along the segment connecting the two points, make a counter--clockwise loop around the 
point $(1+\tau)/2$ and then returns back to the origin along the segment.

One consequence of eqn.\eqref{uuur} is that --- if the monodromy along the loop $\ell$, $\mathsf{M}_\ell$, acts non--trivially on the vacuum bundle $\mathcal{V}$ ---
the two basic lattice translations $X\rightarrow X+1$ and $X\rightarrow X+\tau$ \emph{do not} commute. In the simplest periodic models we set the spectrum 
of the lattice translation operators to be $\exp(2\pi i x_i)$; in the present case, $\mathsf{M}_\ell\neq 1$ implies that the two translations cannot be diagonalized simultaneously on the vacua
and hence the vacuum angles $x_1$ and $x_\tau$ cannot be simultaneously defined. 

As in the standard periodic case, the action of the monodromy on the vacuum bundle
 is induced by the action of the monodromy on the superpotential $W$.
Hence, let us consider the monodromy action on $W$.
To encompass both models  \eqref{nonc}\eqref{secondncmodel} in a single computation, we consider the superpotential
\begin{equation}\label{WWWmuprime}
W(X)=\mu^\prime\, \log\Theta(X,\tau)-\mu\,X.
\end{equation} 
On $\widetilde{K}$ we introduce the meromorphic $\mathfrak{sl}(3,\C)$ connection (we set $\theta(x)\equiv \theta_3(\pi x)$)
$$A=\begin{pmatrix}
\ 0\ &d\left(\mu^\prime\frac{\theta^\prime}{\theta}-\mu\right) &\ 0\  \\
0 & 0 &\ dx\ \\
0 & 0 & 0
\end{pmatrix}$$
and look for solutions to 
\begin{equation}
d\Psi=\Psi A.
\end{equation}
A fundamental solution is
\begin{equation}
 \Psi= \begin{pmatrix}
        1 & \left(\frac{\mu^\prime\theta^\prime(x)}{\theta(x)}-\mu\right) & W\\
 0 & 1 & x\\
0 & 0 &1
       \end{pmatrix}
\end{equation}
with $W$ is as in eqn.\eqref{WWWmuprime}. The general solution is then given by $\mathsf{M}\Psi$ with $\mathsf{M}$ a constant matrix. Let $\gamma\in \pi_1(K,0)$ be a closed loop.
The analytic continuaion of the solution $\Psi$ along $\gamma$, $\Psi_\gamma$, is also a solution to the above linear problem, and hence there exists a constant 
$3\times 3$ matrix $\mathsf{M}_\gamma$ such that
\begin{equation}
 \Psi_\gamma =\mathsf{M}_\gamma\Psi.
\end{equation}
The matrices $\mathsf{M}_\gamma$ are upper triangular with 1's on the main diagonal. The map $\pi_1(K,0)\rightarrow SL(3,\C)$ given by
$\gamma\mapsto \mathsf{M}_\gamma$ is the monodromy representation we are interested in. Let us compute the monodromy representation of 
the generators $u_1, u_2,\ell$
\begin{align}
 &\mathsf{M}_{u_1}=\begin{pmatrix}1 & 0 & -\mu\\
                   0 & 1 & 1\\ 0 & 0 &1
                  \end{pmatrix}
&&\mathsf{M}_{u_2}=\begin{pmatrix}1 & -2\pi i\mu^\prime & -(\mu+i\pi \mu^\prime)\tau\\
                   0 & 1 & \tau\\ 0 & 0 &1
                  \end{pmatrix}&&
\mathsf{M}_{\ell}=\begin{pmatrix}1 & 0 & -2\pi i\,\mu^\prime\\
                   0 & 1 & 0\\ 0 & 0 &1
                  \end{pmatrix}.
\end{align}
One checks that these matrices satisfy the group relation \eqref{uuur}, and hence give a representation of the group $\pi_1(K)$.
The matrix $\mathsf{M}_\ell$ is a \textit{central element} of the monodromy group--algebra generated by $\mathsf{M}_{u_1}$,
$\mathsf{M}_{u_2}$. Hence we may diagonalize its action on the vacuum bundle introducing the $q$--vacua
\begin{equation}
 \mathsf{M}_\ell |q\rangle= q\, |q\rangle
\end{equation}
 (we use the same symbol to denote a monodromy matrix and the operator implementing it on $\mathcal{V}$). 
Then in the $q$ sector the cover group--algebra becomes identified with the quantum torus algebra $(\mathsf{M}_i\equiv \mathsf{M}_{u_i}$)
\begin{equation}\label{rrrrryu}
 \mathsf{M}_2\mathsf{M}_1=q\, \mathsf{M}_1\mathsf{M}_2.
\end{equation}

The vacuum bundle $\mathcal{V}$ over (the universal cover of) coupling constant space may be decomposed into $\mathsf{M}_\ell$--eigenbundles
\begin{equation}
 \mathcal{V}=\bigoplus_{q} \mathcal{V}_q.
\end{equation}
Since the $tt^*$ geometry is described by equations written in terms of commutators, and $\mathsf{M}_\ell$ is central and a symmetry, 
the $tt^*$ equations do not couple eigenbundles $\mathcal{V}_q$ with \emph{different} $q$. Hence we may fix $q$ and discuss the geometry in that sector.
In other words, we get a \emph{family} of $tt^*$ geometries labelled by the value of $q$.
In the vacuum eigenbundle with $q=1$ (if it exists at all), we see from eqn.\eqref{rrrrryu} that we may diagonalize simultaneously
 the lattice translation operators $\mathsf{M}_1$ and 
$\mathsf{M}_2$. Calling, as before, $\exp(2\pi i x_i)$ their respective eigenvalues, we get the standard commutative $tt^*$ geometry (triply--periodic instantons). 
If we deform the parameter $q$ away from its `classical' value $q\neq 1$, the lattice translation operators, $\mathsf{M}_1$ and $\mathsf{M}_2$,
 do not commute any longer, and  we get triply--periodic instanton on a \textit{non--commutative} deformation of the previous geometry, 
namely on the quantum torus obtained by deformation \emph{\'a la} Moyal of the usual commutative torus 
$$e^{2\pi ix_2}\,e^{2\pi i x_1}=q\, e^{2\pi i x_1}\,e^{2\pi i x_2}.$$ 

\paragraph{The value of $q$.} The obvious question at this point is what is the physically natural value of the non--commutativity parameter $q$. Although geometrically it makes sense to speak of generic $q\in \C^*$, we expect that the physical problem selects a definite value for $q$. Leaving a more complete analysis for future work, here we focus on the simplest  thimble amplitudes for the 4d theories modelled by the effective superpotential \eqref{secondncmodel}, in the UV asymmetric limit defined at the end of \S.\,\ref{sec:rev2d}. In this limit the vacuum wave functions may be identified with $\exp(-\zeta W)\,\xi_a$ where $\xi_a$ are closed forms dual to the Lefschetz thimble cycles $D_a$. If we define the branes $D_a$ so that the corresponding cycles are invariant under the monodromy along the path $\ell$, then the action of $\mathsf{M}_\ell$ on these vacua will be given by its action on the factor $\exp(-\zeta W)$ and hence
\begin{equation}\
q\, e^{-\zeta W}= \mathsf{M}_\ell\, e^{-\zeta W}\equiv e^{2\pi i \zeta \mu^\prime}\, e^{-\zeta W}.
\end{equation}
In particular, at $\zeta=1$ we get 
\begin{equation}\label{noncom}
q=e^{2\pi i \mu^\prime}.
\end{equation}
  As we will mention in section 8 this result
is in agreement with what one finds for the $tt^*$ geometry arising in 4d models.

\section{Spectral Lagrangian manifolds}\label{spectrallagrangianmanifold}
To the $tt^*$ geometry of any $(2,2)$ system there is associated a spectral Lagrangian manifold. The details vary slightly in the various cases so we treat them one at a time. 

\subsection{Ordinary models}

For an ordinary $(2,2)$ model (finitely many vacua, globally defined superpotential) the $tt^*$ equations for one complex coupling $t$ reduce to the Hitchin equations
\begin{equation}
\overline{D}_{\bar t}C_t=[D_t,\overline{D}_{\bar t}]+[C_t,\overline{C}_{\bar t}]=0,
\end{equation}
which, in particular, imply that the eigenvalues $\lambda(t)_j$ of the matrix $C_t$ are holomorphic functions of $t$. The spectral curve encodes the holomorphic functions $\lambda(t)_j$; it is simply the curve in $\C^2$
\begin{equation}\label{hitchinspectralcurve}
\det\!\big(C_t(t)-s\big)=0.
\end{equation}

In the case of several couplings $t_i$ ($i=1,2,\dots,g$), the $tt^*$ equations say that the various $C_i$'s \emph{commute} and are covariantly holomorphic, $\overline{D}_j C_i=0$. Then the $C_i$'s may be simultaneously diagonalized (more generally, simultaneously set in the Jordan canonical form) and moreover the corresponding eigenvalues depend holomorphically on the $t_j$'s.
The spectral manifold $\mathcal{L}$ encodes the $g$--tuples of eigenvalues of the $C_i$'s associated to a common eigenvector $\psi$, that is,
\begin{equation}\label{ordinarytt*spectral}
\mathcal{L}= \Big\{(s_1,\dots,s_g, t_1,\dots, t_g)\in \C^{2m}\;\Big|\; \exists\, \psi\not=0\ \text{s.t. }\big(s_i-C_i(t_j)\big)\psi=0\Big\}.
\end{equation}
Clearly $\mathcal{L}\subset \C^{2g}$ is a complex submanifold.
It is also a Lagrangian submanifold with respect to the holomorphic symplectic form \begin{equation}\omega=\sum_i ds_i\wedge dt_i.\end{equation}
To see this, notice that the spectral manifold is purely a property of the underlying holomorphic TFT. 
We may assume to be at a generic point in parameter space where the chiral ring $\mathcal{R}$ is semisimple. Then the eigenvalue of $C_i$ associated to the $k$--th indecomposable idempotent of $\mathcal{R}$ is simply $\partial C_{kk}/\partial t_i$ where $C$ is the $tt^*$ matrix introduced in ref.\cite{CV92b}. (In the particular case of a LG model, $C_{kk}$ is just $W^{(k)}$, the superpotential evaluated on the $k$--th classical vacuum configuration). Hence, locally on the $k$--th sheet, the equations of $\mathcal{L}$ take the form
\begin{equation}s_i=\frac{\partial}{\partial t_i} C_{kk},\end{equation}
and $\mathcal{L}$ is a Lagrangian submanifold. For a LG model it may be simply written as
\begin{equation}\label{specruvelg}
s_i= \partial W(Y_a,t_i)/\partial t_i\qquad \partial W(Y_a,t_i)/\partial Y_a=0. 
\end{equation}

The spectral manifold gives half of the spectral data which labels uniquely a solution of Hitchin's equations 
or of the higher-dimensional generalizations. The other half is a holomorphic line bundle on $\mathcal{L}$.
The line bundle can be defined as the eigenline associated to each point of the spectral manifold. 

\subsection{Periodic models}\label{specperiodicmodels}

Let us consider first the case in which we have a single triplet of parameters (a complex $t$ together with a vacuum angle $x$).
Just an in the ordinary case, the spectral curve $\mathcal{L}$ encodes the spectrum of the linear operator $C_t(t)$ which depends holomorphically on $t$. Hence the spectral curve is given by the same Hitchin formula as before, eqn.\eqref{hitchinspectralcurve}
\begin{equation}
\mathcal{L}:\ \ \mathrm{Det}\big[C_t(t)-s\big]=0.
\end{equation}
The only novelty is that now $C_t(t)$ is not a finite matrix, but rather a linear differential operator of the form
\begin{equation}
\frac{\partial}{\partial x}+B_t(t),
\end{equation}
and the matrix determinant gets replaced by a functional determinant in the Hilbert space $L^2(S^1,dx)\otimes \C^n$ of vector functions of period $1$. The expression
\begin{equation}\mathrm{Det}\big[C_t(t)-s\big]\equiv\mathrm{Det}\big[\partial_x+B_t(t)-s\big]\end{equation}
is simply the partition function, twisted by $(-1)^F$, of a system of one--dimensional free Dirac fermions with mass matrix $B_t(t)-s$. Hence the spectral curve has equation
\begin{equation}
\mathrm{Det}\big[\partial_x+B_t(t)-s\big]= \prod_j \left(e^{(\lambda_j(t)-s)/2}-e^{(s-\lambda_j(t))/2}\right)=0,
\end{equation}
where $\lambda_j(t)$ are the eigenvalues of $B_t(t)$. Usually one writes this equation in the form
\begin{equation}\label{periodiccurve}
\det\!\big[e^s-e^{B_t(t)}\big]=0.
\end{equation}
Since $B_t(t)$ (say for a periodic LG model) is a diagonal matrix whose $kk$--entry is $\partial_t W$ evaluated on the $k$--th (reference) classical vacuum (cfr.\! eqn.\eqref{whatLLLs}),  eqn.\eqref{periodiccurve} has the same form as the ordinary Hitchin curve \eqref{hitchinspectralcurve} but with all quantities exponentiated. 
This `exponentiation' is no mystery: the two formulae \eqref{hitchinspectralcurve} and \eqref{periodiccurve} are identical provided one keeps into proper account the role of the Hilbert space $L^2(S^1,dx)$.
Thus the spectral manifold is a Lagrangian sub manifold in $\C \times \C^*$. 
\medskip

The case of several triplets of couplings $t_i, x_i$, ($i=1,\dots,g$) is similar. The spectral manifold $\mathcal{L}$ is again given by the usual $tt^*$ equation \eqref{ordinarytt*spectral}, with the only specification that the $C_i$ are differential operators and $\psi$ is a non--zero eigenvector in
\begin{equation}\label{righthilbertspace}
\C^n\otimes L^2\big((S^1)^g,dx_1\wedge\cdots \wedge dx_g\big),
\end{equation} 
($n$ being the number of vacua in a reference sheet). The eigenvector equations for $\psi$ have the form
\begin{equation}
\Big(\partial_{x_r}+B_r(t)-s_r\Big)\psi=0,\qquad r=1,\dots,g,
\end{equation}
whose non--zero solutions are
\begin{equation}
\psi=\exp\!\Big(\sum\nolimits_r x_r\big(s_r-B_r)\Big)\psi_0,\qquad 0\neq\psi_0\in \C^n,
\end{equation}
(we have used the fact that the matrices $B_r$ commute).
The condition that $\psi$ belongs to the Hilbert space
\eqref{righthilbertspace} may be written in the form
\begin{equation}\label{periodictt*spectral}
\Big\{(e^{s_1},\dots,e^{s_g}, t_1,\dots,  t_g)\in (\C^*)^{g}\times \C^g\;\Big|\; \exists\, 0\neq\psi_0\in\C^n\ \text{s.t. }\big(e^{s_i}-e^{B_i(t_j)}\big)\psi_0=0\Big\}
\end{equation}
which is the same as the `exponentiation' of the spectral manifold equations. 

Eqn.\eqref{periodictt*spectral} gives the spectral manifold equations for the general periodic case. Again, $\mathcal{L}\subset (\C^*)^{m}\times \C^m$ is a complex submanifold which is also Lagrangian for the symplectic structure ($S_i\equiv e^{s_i}$)
\begin{equation}
\sum_i \frac{dS_i}{S_i}\wedge dt_i.
\end{equation}
In view of the definition of the matrices $B_r$, eqn.\eqref{whatLLLs}, the proof is the same as in the ordinary case, and will be omitted.

In particular, for a periodic LG model (with periodic couplings), eqn.\eqref{specruvelg} gets replaced by its `exponentiated' version
\begin{equation}
\exp\!\big(s_i\big)=\exp\!\Big({\partial W(Y_a,t_i)/\partial t_i}\Big)\qquad \partial W(Y_a,t_i)/\partial Y_a=0.
\end{equation}

In later sections we will encounter special periodic models which arise from the compactification of 3d gauge theories, for which the couplings $t_i$ are also periodic. In that case, the spectral manifold is naturally defined in $(\C^*)^{2g}$ rather than $(\C^*)^{g}\times \C^g$, with symplectic form ($S_i\equiv e^{s_i}$,$T_i\equiv e^{t_i}$)
\begin{equation}
\sum_i \frac{dS_i}{S_i}\wedge \frac{dT_i}{T_i}.
\end{equation}
We will denote these models as ``3d periodic models''.

\subsection{Doubly--periodic models}

The doubly--periodic case is similar, except that the $C_i$'s are now differential operators of the form
\begin{equation}\label{cidoublyperiodic}
C_i= \partial_{x_{1,i}}+\rho_i\,\partial_{x_{2,i}}+B_i.
\end{equation}
We consider first the case of just four parameters (a complex $t$ and two vacuum angles $x_1$ and $x_2$). Assuming $\mathrm{Im}\,\rho >0$, we introduce a complex coordinate $\zeta$ such that
\begin{equation}
(\rho-\bar\rho)\,\partial_\zeta =\partial_{x_{1}}+\rho\,\partial_{x_{2}},
\end{equation}
which takes values in a torus of periods $1$ and $\tau\equiv-\bar\rho$.
The spectral curve takes the form
\begin{equation}
\mathrm{Det}\left[\partial_\zeta+\frac{B-s}{\rho-\bar\rho}\right]=0.
\end{equation}
The \textsc{lhs} is now the partition function of a system of 2d chiral fermions on a torus of modulus $\tau$ coupled to a background gauge connection $A_\zeta=(B-s)/(\rho-\bar\rho)$. The spectral curve may be then written as 
\begin{equation}
\det\, \theta_1\!\!\left(\frac{s-B}{\rho-\bar\rho}\;\bigg| -\bar\rho\right)=0,
\end{equation} 
where $\theta_1(z\,|\,\tau)$ is the usual theta function.

In the general case the spectral manifold is
\begin{equation}
\mathcal{L}\equiv\Big\{(s_i, t_i)\in \mathcal{A}\times \C^g \;\Big|\; \exists\, 0\neq\psi_0\in\C^n\ \text{s.t. }\Theta(s_i-B_i)\psi_0=0\Big\}
\end{equation}
where $\mathcal{A}$ is the Abelian variety where the angular variables are valued in, and $\Theta$ is the basic theta--function for $\mathcal{A}$. All other cases (non--periodic, single periodic) may be obtained as degenerate limits of this expression. For instance, eqn.\eqref{periodiccurve} corresponds to $\mathcal{A}$ being an elliptic curve with an ordinary node, while
\eqref{hitchinspectralcurve} to an elliptic curve with a cusp. 

\subsection{Action of $\sp(2m,\mathbb{A})$ on the spectral manifolds}\label{actiononspectral}

In all cases the spectral manifold $\mathcal{L}$ is a (holomorphic) Lagrangian submanifold\footnote{\ To avoid misunderstandings, we stress that $\mathcal{L}$ is defined as a \emph{submanifold}, that is, as an abstract manifold together with a Lagrangian embedding $\mathcal{L}\xrightarrow{\iota} \mathcal{S}$; in particular, this means a definite choice of which coordinates we call $t_i$'s (\textit{i.e.}\! which coordinates are interpreted as couplings of the QFT).} of a holomorphic symplectic manifold $\mathcal{S}$ which is also an Abelian group. It makes sense to consider the action on $\mathcal{L}$ of  ambient symplectomorphisms
$\mathcal{U}\colon\mathcal{S}\rightarrow \mathcal{S}$. In order for the  transformed manifold $\mathcal{U}(\mathcal{L})$ to have a spectral interpretation\footnote{\ More precisely, we mean the following: when $\mathcal{U}$ is a group homomorphism, besides a symplectomorphism, one canonically identifies the Lagrangian submanifold $\mathcal{U}(\mathcal{L})$ as the spectral manifold of the $tt^*$ geometry of another supersymmetric QFT, thus inducing a group action on the field theories themselves. It would be interesting to see whether one can interpret in a similar way the action of more general symplectomorphism.},  $\mathcal{U}$ must be a group homomorphism of $\mathcal{S}$ as well as a symplectomorphism.

In the case of non--periodic models, $\mathcal{S}$ is the additive group $\C^{2g}$, and the group of symplectic homomorphisms is $\sp(2g,\C)$ acting on $(t_i,s_j)$ in the obvious linear way. For periodic models, 
the group of transformations compatible with the periodicities is much reduced. An important exception are 
3d periodic models, for which we can consider $\sp(2g,\Z)$ transformations, which preserve the periodicities of the 
$s_i$ and $t_i$ variables. 

For other periodic models, beyond the boring transformations 
of the form $t \to g t$, $s \to g^{-1} s$, we only have dualities between different types of spectral manifolds. 
For example, an $S$ transformation $s\to t$, $t \to -s$ may relate the spectral curves
for a Hitchin system on a cylinder and the spectral curve of a periodic monopole: the former has a periodic $t$ and non-periodic $s$, the latter has a periodic $s$ and a non-periodic $t$. 

The symplectic transformations will act both on the spectral manifold $\mathcal{L}$ and on the associated line bundle. Because of the one-to-one correspondence between the spectral data and the $tt^*$ geometry, 
one may suspect the symplectic action should lift to an action over the $tt^*$ geometries, and hence on the corresponding supersymmetric physical theories. 
Indeed, the lift coincides with the well-known notion of Nahm transform. We will discuss the Nahm transform and its relation to the $tt^*$ geometry in the next section. For now, we would like to examine a more direct physical interpretation of the symplectic action. 

Let's start with a standard $(2,2)$ LG model, defined by some superpotential $W(Y_a,t_i)$.
We can promote the parameters $t_i$ to chiral fields $P_i$, and consider a new LG model 
with superpotential 
\begin{equation}
W(Y_a,P_i, \tilde t_i) = W(Y_a,P_i) + \sum_i \tilde t_i P_i 
\end{equation} 
The F-term equations give us 
\begin{equation}
\tilde t_i + \partial_{P_i} W(Y_a,P_i) =0 \qquad \partial_{Y_a} W(Y_a,P_i)=0 \qquad \tilde s_i = P_i
\end{equation} 
Thus the spectral manifold of the new model is related to the spectral manifold of the old model by 
the basic symplectic transformation $\tilde s_i = t_i$, $\tilde t_i = - s_i$. 

This can be thought of as a functional Fourier transform at the level of chiral super-fields,
acting on path integrals as  
\begin{equation}
Z[t_i] \to \tilde Z[\tilde t_i]=\int DP^g e^{\int d^2 \theta \sum_i \tilde t_i P_i} Z[P_i]
\end{equation}
It is not hard to check that repeating this step, brings us back to the original theory, so this is an order
2 operation.

More generally, we can consider a model with superpotential 
\begin{equation}
W(Y_a,P_i, \tilde t_i) = W(Y_a,P_i) + \sum_{ij} A_{ij} \tilde t_i \tilde t_j +B_{ij}  \tilde t_i P_j  + C_{ij} P_i P_j
\end{equation} 
to obtain more general symplectic transformations. 

Inspired by the Fourier transform, we can describe the action of $\sp(2g,\C)$ on the $(2,2)$ theories in the following way. 
Let $\omega_s\equiv (q_1,q_2,\cdots,q_g,p^1,\cdots,p^g)$  be usual canonical operators acting in the Hilbert space $L^2(\R^g)$, and let
 \begin{equation}\label{elementofsp}
\mathcal{U}= \left(\begin{array}{c|c}
A & B\\\hline C & D\end{array}\right)\in \sp(2g,\C).
\end{equation} 
Since $\sp(2g,\C)$ is the complexification of $\mathrm{USp}(2g)$, the linear transformation
\begin{equation}
\omega_s\longmapsto \mathcal{U}_{st}\,\omega_t,
\end{equation}
is the complexification of an unitary transformation of $L^2(\R^g)$, and it is implemented by an invertible operator $U$. Consider its kernel in the Schroedinger representation
\begin{equation}
\exp\!\Big[{\kappa(q^\prime_i,q_j;\mathcal{U})}\Big]=\langle q_i^\prime\, |U|\,q_j\rangle.
\end{equation}

Then the action of  $\mathcal{U}\in \sp(2g,\C)$ on the space of $(2,2)$ theories (modulo $D$--terms) is given in terms of effective superpotentials as
\begin{equation}W(t_i)\longmapsto W_\mathcal{U}(t^\prime_i),
\end{equation}
where
\begin{multline}\label{abstractsction}
\exp\!\left(-\int d^2z\,d^2\theta\, W_\mathcal{U}(t^\prime_i)\right)=\\
=
\int\limits_{\text{TFT path}\atop\text{integral}} [dP_j]\;\exp\!\left(-\int d^2z\,d^2\theta\,\Big[W(P_j)-\kappa(t^\prime_i,P_j;\mathcal{U})\Big]\right).
\end{multline}

We claim that the spectral manifold $\mathcal{L}_\mathcal{U}$ of the transformed $(2,2)$ model $W_\mathcal{U}(t^\prime_i)$, as defined is precisely $\mathcal{U}(\mathcal{L})$, where $\mathcal{U}\colon\C^{2g}\rightarrow \C^{2g}$ is the linear map 
\begin{equation}
\begin{aligned}
t^\prime_i&= A_{ij} t_j+B_{ij}s_j,\\
s^\prime_i&=C_{ij} t_j+D_{ij}s_j.
\end{aligned}
\end{equation}

In this definition we do not specify the precise form of the D-terms 
for the new theory. Rather, we consider the action of the symplectic transformation on the space of $(2,2)$ QFTs modulo D-term deformations, including possibly integrating away 
some of the degrees of freedom if possible. In fact, the detailed form of the $D$--terms is irrelevant for $tt^*$, and we are interested not in the full effective action $S[P^i]$ but only in its topologically non--trivial part\footnote{\ We write the non--trivial part of the action as an $F$--term. Of course, it may be a twisted $F$--term as well. The examples in appendix A of \cite{onclassification} have actually twisted dual superpotentials. } $\int d^2\theta\, W(P^i)_\text{dual}$. 

It is not hard to extend this type of construction to the other types of theories with four super-charges we consider in this paper, by seeking transformations which reduce to the above symplectic transformations at the level of 
a low-energy $(2,2)$ LG description. The most important example are the periodic theories associated to 
Abelian flavor symmetries. If we gauge some flavor symmetries, we end up promoting the twisted masses $\mu_a$ 
to (twisted) chiral super fields $\sigma_a$ with linear (twisted) F-term couplings $\sum_a t_a \sigma_a$ 
to the FI terms. Thus we recover the symplectic transformation relating periodic monopole geometries and 
Hitchin systems on cylinders. 

In the special case of 3d periodic geometries, which arise from 3d ${\cal N}=2$ gauge theories compactified on a circle of finite size to a 2d theory with $(2,2)$ symmetry, the $\sp(2g,\Z)$ action on the spectral curve lifts all the way to Witten's $\sp(2g,\Z)$ \cite{Witten:2003ya} action on 3d SCFTs equipped with a $U(1)^g$ flavor symmetry, generated by the operations of gauging a flavor symmetry and of adding a background CS couplings. See \cite{DGGi} for a review and further references.

\subsection{A higher dimensional perspective} \label{higher}
The example of the 3d ${\cal N}=2$ gauge theories is actually very instructive. Although Witten's $\sp(2g,\Z)$ \cite{Witten:2003ya}  action can be defined directly in 3d terms, it is more elegantly described as the action of four-dimensional electric-magnetic duality on half-BPS boundary conditions for a free Abelian gauge theory with eight supercharges \cite{Gaiotto:2008ak}. 

It is simple and instructive to pursue this analogy here. Let's go back again to the 2d $(2,2)$ LG models
with some parameters $t_i$ which enter linearly in the superpotential $W(Y_a, P_i)$. 
This time, instead of promoting the $t_i$ to 2d chiral super fields, we can promote them to 
the boundary values of some free 3d hypermultiplets. 

More precisely, consider a set of free 3d hypermultiplets, decomposed into pairs of complex scalars $(P_i, \tilde P_i)$, 
rotated into each other by an $\sp(2g)$ flavor symmetry. The simplest half-BPS boundary condition $B$ for free hypers 
sets Dirichlet b.c. for the $\tilde P_i$ and Neumann for the $P_i$. The boundary value $P_i|_\partial$ of the $P_i$ 
behaves as a 2d chiral multiplet. If we add the 2d LG theory at the boundary and couple it to the 3d system through a superpotential $W(Y_a, P_i|_\partial)$
we obtain a deformed half-BPS boundary condition, which roughly sets $\tilde P_i = \partial_{t_i} W$. 
In other words, the boundary condition forces the hypermultiplet scalars $(P_i, \tilde P_i)|_\partial$ at the boundary 
to lie on the spectral Lagrangian manifold $\mathcal{L}$, with the identification $(P_i, \tilde P_i)|_\partial = (t_i,s_i)$ (see \cite{DGG} for an higher-dimensional version of this construction).

Up to D-terms, the map from 2d theories to half-BPS boundary conditions is invertible. Define a boundary condition $\tilde B$ 
by Dirichlet b.c. for the $P_i$ and Neumann for the $\tilde P_i$. If we put the 3d theory on a segment, 
with our boundary condition at one end and $\tilde B$ at the other end, we recover the original 2d theory. 
At this point, the symplectic action on 2d theories has an obvious interpretation in the language of half-BPS boundary conditions: 
it is the action of the (complexified) hypermultiplet flavor symmetries on half-BPS boundary conditions. 

It is easy to extend this to other situations:
\begin{itemize}
\item A 2d $(2,2)$ theory with Abelian flavor symmetry $U(1)^g$ can be transformed into a boundary condition for a free ${\cal N}=4$ 3d gauge theory: start with Neumann b.c. for the gauge fields and couple them to the 2d degrees of freedom at the boundary. 
The inverse operation involves a segment with Dirichlet b.c. for the gauge field. 
If we dualize the 3d gauge field we obtain an hypermultiplet valued in $\C \times \C^*$ and proceed as before. 
The duality transformation acts on boundary conditions as a gauging/ungauging of the Abelian flavor symmetry. 
This is related to the Nahm transform relating periodic monopoles and periodic Hitchin systems.  

\item A 3d ${\cal N}=2$ theory with Abelian flavor symmetry $U(1)^g$ can be transformed into a boundary condition for a free ${\cal N}=2$ 4d gauge theory. The bulk theory has an $\sp(2g,\Z)$ group of electric-magnetic duality transformations which acts on boundary conditions. 
This is related to the Nahm transform relating doubly periodic monopoles.  

\item A 4d ${\cal N}=1$ theory with Abelian flavor symmetry $U(1)^g$ can be transformed into a boundary condition for a free ${\cal N}=1$ 5d gauge theory. The 5d gauge theory can be dualized into a self-dual two-form. The duality transformation acts 
on boundary conditions as a gauging/ungauging of the Abelian flavor symmetry. 
This is related to the Nahm transform relating triply periodic monopoles and triply periodic instantons.  
\end{itemize}

Some of these examples we already encountered. Some we will encounter in the next sections. 
It is useful to point out that in this setup the bulk theory is always free and thus well-defined even in 5d. 

\subsection{Generalized Nahm's transform and the $tt^*$ geometry}\label{asaspecialinstance}

In the previous two sections we have seen  that the $tt^*$ equations for ordinary, periodic, and doubly--periodic systems are the higher dimensional generalizations of, respectively, Hitchin, monopole, and self--dual Yang--Mills equations. All these geometries get unified in the concept of hyperholomorphic connections on $U(N)$--bundles over a hyperK\"ahler manifold $\mathcal{M}$ \cite{salamon,nitta,verbitsky,bartocci}, which is possibly invariant under a suitable group of continuous isometries of $\mathcal{M}$, which reduces the number of coordinates (parameters) on which the geometry effectively depends, as well as of discrete isometries which lead to periodicities of various kinds.

An important tool in the theory of hyperholomorphic bundles and connections is the \emph{generalized Nahm transform} \cite{nahm,hitchinmonopoles,corrigan,braam,nakajima,nahmsurvey,bruzzobook}, which relates hyperholomorphic bundles on certain dual pairs of hyperK\"ahler manifolds $(\mathcal{M},\mathcal{X})$. The duality typically proceeds by defining a family of Dirac operators $\Dsl_x$ on $\mathcal{M}$ parameterized by a point $x \in \mathcal{X}$ and then constructing an hyperholomorphic connection on $\mathcal{X}$ from the kernel of the Dirac operators. A prototypical example of generalized Nahm transform is the Fourier--Mukai transform \cite{mukai,bruzzobook} where $\mathcal{M}$ and $\mathcal{X}$ are a dual pair of Abelian varieties.

Well known simple examples of the Nahm transformation relate monopole solutions on $\R^3$ to solutions of Nahm equations, periodic monopole solutions to solutions of Hitchin systems on a cylinder, instantons on $\R^4$ to solutions of algebraic equations, etc. In all cases where the spectral data can be defined, the generalized Nahm transform acts as a symplectic transformation 
on the spectral manifold. 

It is not hard to produce a long list of pairs of physical systems $(T_\mathcal{M},T_\mathcal{X})$ with four supercharges with the property that the corresponding $tt^*$ geometries are hyperholomorphic connections related by a Nahm transformation. This is particularly easy because many examples of Nahm transformations arise in well-known systems of intersecting D-branes in string theory. In all cases the two theories  $(T_\mathcal{M},T_\mathcal{X})$ are 
always related as we described above, by promoting some background couplings in one theory to dynamical degrees of freedom in the other theory. For example, if a periodic monopole geometry is associated to a $U(1)$ flavor symmetry of $T_\mathcal{M}$, then $T_\mathcal{X}$ will be obtained by gauging that flavor symmetry, and the $tt^*$ geometry associated to the corresponding FI parameter gives the dual solution of a Hitchin system on a cylinder. 

We would like to explain now briefly that the generalized Nahm transformation 
always coincides with the calculation of the $tt^*$ geometry for a certain physical system
and that the relation with a Fourier--Mukai transform also has a natural physical interpretation in the 
language of half-BPS boundary conditions for theories with eight supercharges. 
   
Much of the structure of the $tt^*$ geometry follows directly from general considerations about supersymmetric quantum mechanics (SQM) with four supercharges. In general, we have a $\Z_2$--graded Hilbert space $\mathcal{H}$, with grading operator $(-1)^F$, and a family of four {odd} Hermitian supercharges $\{Q_a(t)\}_{t\in\mathcal{X}}$, $a=1,...,4$, depending on F-term-type parameters $t$ taking value in some space $\mathcal{X}$. The $Q_a(t)$'s satisfy the \textsc{susy} algebra
\begin{equation}
\{Q_a(t),Q_b(t)\}=\delta_{ab}\,H(t).
\end{equation}
The $tt^*$ geometry computes the Berry connection on the bundle over $\mathcal{X}$ of the zero--eigenvectors of $H(t)$, which, as reviewed in the previous sections, is a hyperholomorphic connection (or a dimensional reduction thereof).

There is a simple interpretation of the Berry connection on the bundle of vacua. If we promote the parameters $t$ to dynamical superfields, with a very slow dynamics, the Berry connection encodes the effect of integrating away the original degrees of freedom in a Born-Oppenheimer approximation protected by supersymmetry. Remember that a massless Euclidean Dirac operator $\Dsl$, coupled to a 
gauge/gravitational background, in \emph{even} space--time dimensions $D=2m$, defines a SQM system with two supercharges $Q_1,Q_2$ under the dictionary
\begin{align}
&\gamma_5\leftrightarrow (-1)^F, && \Dsl\leftrightarrow Q_1, && i\gamma_5\Dsl \leftrightarrow Q_2, && \Dsl^2\leftrightarrow H.
\end{align}
The supersymmetry of this SQM system enhances to 4 supercharges precisely if $D=4n$, the gravitational background is hyperK\"ahler, and the gauge connection is \emph{hyperholomorphic} (in the particular case of $D=4$ this means (anti)self--dual). Indeed, under these conditions the Hamiltonian $\Dsl^2$ is invariant under a $\sp(2)$ $R$--symmetry, which geometrically corresponds to the centralizer of the holonomy group in $SO(D)$. Thus the $tt^*$ geometry encodes precisely the data required to define a low-energy supersymmetric dynamics on the parameter space 
$\mathcal{X}$ of the original theory. 

Parsing through the definitions of the generalized Nahm transform (or even of the standard Nahm transform)
makes it clear that the basic steps involving the Dirac operators $\Dsl_x$ simply coincides with the 
calculation of the Berry connection for the $N=4$ SQM associated to these Dirac operators. 
In other words, the Nahm transform emerges as expected from making the parameters of an $N=4$ SQM
dynamical. 

At this point, we can mimic our previous discussion by making the parameters $t$ dynamical not as 1d degrees of freedom,
but as boundary values of 2d degrees of freedom. We can add a direction to our system, and 
promote our 1d system with four supercharges to an half-BPS boundary of a 2d system with eight supercharges. 
We can consider a 2d $(4,4)$ non-linear sigma model with target space $\mathcal{X}$ defined on a half-space, and 
couple the boundary values of the 2d degrees of freedom $T_i$ to the original 1d $N=4$ SQM in the obvious way. 
This produces a half-BPS \footnote{More precisely a $(B,B,B)$ brane, a brane which is type B in each complex structure for the target.} brane ${\mathcal B}$ for the 2d $(4,4)$ non-linear sigma model.

This brane obviously captures the same protected information as the original 1d SQM. For example, we can consider the 2d theory on a segment, with ${\mathcal B}$ boundary conditions at one end and a $D0$ brane at the other end, i.e. Dirichlet boundary conditions $T_i = t_i$ at the other end. This quantum mechanical system has the same ground states and Berry connection as the original system. 

On the other hand, we can pick a different D-brane $\mathcal{D}$ at the other end of the segment, and thus find a different 
low energy $N=4$ quantum-mechanical system and a different $tt^*$ geometry associated to the pair $({\mathcal B},{\mathcal D})$.
For example, we could pick $\mathcal{D}$ to be a space-filling brane in $\mathcal{X}$. Then the 1d system is simply the SQM on 
$\mathcal{X}$. If $\mathcal{X}$ has a mirror $\mathcal{M}$, we can pick a family of branes $\mathcal{D}_m$ dual to 
D0 branes in $\mathcal{M}$ and thus obtain a family of SQM whose Berry connection is a hyperholomorphic connection on 
$\mathcal{M}$. 

This is just the action of mirror symmetry on half-BPS branes in the $(4,4)$ non-linear sigma model.
Mirror symmetry can be interpreted as a Fourier--Mukai transformation, with a kernel which 
defines a special BPS ``duality interface'' between the $\mathcal{X}$ and $\mathcal{M}$ non-linear sigma models. 

\subsection{An explicit example}

The above structures may be elementarily illustrated in a 1d $\mathcal{N}=4$ Landau--Ginzburg model which, as discussed above, may be identified with a Dirac operator coupled to a hyperholomorphic connection.
For simplicity we assume there is just one chiral field $Y$. 
Identifying the SQM Hilbert space with the space of square--integrable differential forms on $\C$, the supercharges in the Schroedinger representation are \cite{cec}
\begin{equation}
\bar Q=\bar\partial +\partial W\wedge,\qquad Q=\partial+\bar\partial \bar W\wedge
\end{equation} 
together with their Hermitian conjugates. The vacuum wave--functions are $1$--forms
\begin{equation}
\psi^j_1\; dY+\psi^j_2\; d\bar Y,
\end{equation}
and the solutions to the zero--energy Schroedinger equation $H\Psi=0$ may be identified with solutions of the negative--chirality Euclidean Dirac equation in $\R^4\simeq \C^2$,
\begin{equation}
\frac{1}{2}(1-\gamma_5)\Dsl \Psi=0,
\end{equation}
or, more explicitly,
\begin{equation}\label{DiracLG}
\begin{pmatrix}\partial_{\bar Y}  & \partial_{\bar Z} -\partial_YW\\
\partial_{Z}+\partial_{\bar Y}\bar W & -\partial_Y\end{pmatrix}
\begin{pmatrix}\psi^j_1\\ \psi^j_2\end{pmatrix}=0,
\end{equation}
which are invariant under the translations in the additional complex coordinate $Z$ (which may be assumed to take value in a compact torus).
The Dirac operator in eqn.\eqref{DiracLG} is coupled to a $U(1)$  connection on $\C^2$
\begin{equation}\label{whichconneU(1)}
A_Z=\partial_{\bar Y}\bar W,\qquad A_Y=0.
\end{equation}
which is self--dual. Indeed\footnote{\ Here $\Dsl^\pm =(1-\gamma_5)\Dsl/2$.},
\begin{equation}
\begin{split}
\Dsl^-\Dsl^+\equiv& \begin{pmatrix}\partial_{\bar Y}  & \partial_{\bar Z} -\partial_YW\\
\partial_{Z}+\partial_{\bar Y}\bar W & -\partial_Y\end{pmatrix}
\begin{pmatrix}-\partial_{Y}  & -\partial_{\bar Z} +\partial_YW\\
-\partial_{Z}-\partial_{\bar Y}\bar W & \partial_{\bar Y}\end{pmatrix}=\\
&= \Big(-\partial_{\bar Z}\partial_Z-\partial_{\bar Y}\partial_Y+|\partial_YW|^2\Big)\boldsymbol{1}_2.
\end{split}
\end{equation}

Given a family $A_\mu(t)$ of self--dual $U(N)$ connections depending on parameters $t_i$, the Nahm procedure requires to solve the chiral Dirac equation 
\begin{equation}
(1-\gamma_5)\gamma^\mu\big(\partial_\mu+A_\mu(t)\big)\Psi^j=0.
\end{equation}
 In the present LG example the connection is Abelian, $N=1$, and the $t_i$'s are the couplings in the superpotential $W$.
The normalizable zero--modes $\Psi^j$,which are automatically invariant by translation in the dumb coordinate $Z$, define a bundle over parameter space which is endowed with the natural induced connection. By definition, this is the Nahm transformed connection of the LG one \eqref{whichconneU(1)}.
Since the zero--modes $\Psi^j$ are precisely the \textsc{susy} vacua, the (translationally invariant) Nahm bundle is the \textsc{susy} vacuum bundle, whose rank $k$ is the Witten index of the LG model. Thus the Nahm connection coincides with the $tt^*$ one. By general $tt^*$ theory, the connection is hyperholomorphic and invariant by translation in half the directions.

Now we take this $tt^*$ geometry as the definition of a new $\mathcal{N}=4$ SQM system, by viewing the Dirac operators coupled to the $tt^*$ Berry connection as the new supercharges.  

For instance, for the one--dimensional family of LG models $W(Y)=W_0(Y)-PY$, parametrized by the coupling $P$, the supercharge corresponding to $\Dsl^-$ is
\begin{equation}\label{dualLGcharge}
\Dsl^-\Big|_{tt^*\atop \text{dual}}=\begin{pmatrix}\overline{D}_P\ &\ -C_P+Y\\
\overline{C}_P-\overline{Y}\ &\ -D_P
\end{pmatrix},
\end{equation}
where now $Y$ is a free parameter (a dual coupling). Note that the connection in eqn.\eqref{dualLGcharge} satisfies the $tt^*$ equations (with the same $tt^*$ metric) for all values of $Y$. The equation for the \textsc{susy} vacua of the dual theory
 \begin{equation}\label{duadiracq}
Q\Big|_{tt^*\atop \text{dual}}\,\Psi
\equiv \Dsl^-\Big|_{tt^*\atop \text{dual}}\,\Psi=0
 \end{equation} 
 has a single \emph{normalizable} solution which defines the vacuum line bundle $\mathcal{L}$ over the space $\C_Y\times \C_Z$ which is invariant by translation in the fictitious $Z$ direction. The $tt^*$ connection on $\mathcal{L}$
is just $J dW$, where $J$ is the quaternionic imaginary unit in $\mathbb{H}\simeq \C^2$. We have thus recovered the original LG model.

We close this subsection noticing that while eqn.\eqref{duadiracq} has (for any given value of $Y$) a single \emph{normalizable} solution, it has several physically interesting \emph{non--normalizable} solutions. Indeed,
let $\Pi$ be a $D$--brane amplitude of the original LG model with phase $\zeta=e^{i\theta}$. It is easy to check that the `right spinor'
\begin{equation}
\Psi=\begin{pmatrix}D_t \,\Pi\\ \overline{C}_t\, \Pi\end{pmatrix}\equiv -\zeta \begin{pmatrix}C_t \,\Pi\\ \overline{D}_t\, \Pi\end{pmatrix},
\end{equation}
satisfies the $tt^*$--dual chiral Dirac equation at $Y=0$
\begin{equation}
\Dsl^-\Big|_{tt^*\ \text{dual}}\Psi=0.
\end{equation}

\subsection{Review of the flat Nahm transform in $\R^{4n-k}\times T^k$}\label{reviewNahmflat}

The Nahm transform was  originally introduced as a generalization of the ADHM construction of $U(N)$ self--dual connections in $\R^4$ \cite{adhm}. One looks  for instantons in  the flat hyperK\"ahler space $\R^4$ which are invariant under a group of translations $\Lambda\subset \R^4$ \cite{nahmsurvey}. As a group, $\Lambda$ is isomorphic to $\R^k\times \Z^l$ for some $k$, $l$ with $k+l\leq 4$; the $\Lambda$--invariant instantons may be seen as field configurations of a $(4-k)$--dimensional theory which are periodic in $l$ directions, or equivalently theories defined on the quotient $(4-k)$--fold $M\equiv\R^4/\Lambda$ . It is well known that for $k=1$, $2$, $3$ and $4$,  the self--dual Yang Mills equations reduce, respectively, to monopole \cite{bogomolny}, Hitchin \cite{Hitchin}, Nahm \cite{nahm,hitchinmonopoles}, and the ADHM algebraic equations \cite{adhm}. The monopoles (resp. Hitchin, or Nahm fields) are then taken to be periodic in $l$ directions.

Let $\Lambda^\vee\subset (\R^4)^\vee$ be the dual group of $\Lambda$, \textit{i.e.}
\begin{equation}\Lambda^\vee=\{\alpha\in (\R^4)^\vee\;|\; \alpha(\lambda)\in\Z\ \ \forall\,\lambda\in\Lambda\}\simeq \R^{4-k-l}\times \Z^l.\end{equation}
The (flat) Nahm transform maps a $U(N)$ instanton on $\R^4$ invariant under $\Lambda$ into a $U(K)$ instanton on the dual $(\R^4)^\vee$ invariant under the dual group $\Lambda^\vee$ \cite{nahmsurvey}, which for $k=0,1,2,3,4$ concretely means a $l$--fold periodic solution to, respectively, the ADHM, Nahm, Hitchin, monopole, and YM self--dual equations. The dual solutions are allowed to have singularities of the appropriate kind \cite{nahmsurvey}.

Comparing with section \ref{section:periodictt*}, we see that the one coupling  $tt^*$ geometry corresponds to this $\R^4/\Lambda$ setting  with
\begin{itemize}
\item $\Lambda=\R^2$ for ordinary $(2,2)$ models;
\item $\Lambda=\R\times \Z$ for periodic models \cite{cherkiskapustin};
\item $\Lambda=\R\times \Z^2$ for 3d version of periodic models \cite{CW}
\item $\Lambda=\Z^2$ for doubly periodic models \cite{doublyperiodicinst}.
\end{itemize}

We review the $\R^4/\Lambda$ construction for $\Lambda$ a rank 4 lattice, so that $M$ is a torus $T^4$. All other cases, including the ones relevant for this paper, may be obtained from the $T^4$ one by sending some periods of the torus to either zero or infinity. The dual torus will be denoted as $\tilde T^4$ and its coordinates as $\tilde t_\mu$. 
By definition, the dual torus $\tilde T^4$, which can be viewed as T-dual of $T^4$, parametrizes the family of flat Abelian connections on the original $T^4$. Given a self--dual $U(N)$ connection $A$ on $T^4$ we may twist it by the flat $U(1)$ connections, forming the family of Dirac operators
\begin{equation}
\Dsl_{\tilde t} = \gamma^\mu\big(\partial_\mu+A_\mu+2\pi i \,\tilde t_\mu)
\end{equation} 
parametrized by points $\tilde t\in\tilde T^4$. The twisted connection is still self--dual, and $\Dsl_{\tilde t}$ may be seen as a supercharge of a $\mathcal{N}=4$ SQM system to which $tt^*$ geometry applies. Assuming all \textsc{susy} vacua have the same $(-1)^F$ grading, over the `coupling constant space' $\tilde T^4$ we have a vacuum bundle, of rank $K$ equal to the Witten index ($\equiv\mathrm{ind}\;\Dsl_{\tilde t}$) and whose Berry connection is hyperholomorphic, as we reviewed in the previous sections. This Berry connection is precisely the Nahm transform of $A_\mu$.

From the SQM interpretation, it is clear that the Nahm transformed connection has singularities of a rather standard form: the singularities appear at the loci in coupling constant space $\tilde T^4$ where the energy gap vanishes and the SQM vacuum states mix with the continuum. $tt^*$ is an IR description, and as all IR descriptions, should get in trouble at points where new light degrees of freedom appear.

All the above may be generalized to the higher dimensional case.
We have a pair of dual even--dimensional Abelian varieties $\mathcal{A}$ and $\mathcal{A}^\vee$ each one parametrizing the flat $U(1)$ connections of the other one.  $\mathcal{A}$, $\mathcal{A}^\vee$ are flat hyperK\"ahler manifolds. A hyperholomorphic connection on $\mathcal{A}$ may be twisted by the flat Abelian family parametrized by $\mathcal{A}^\vee$, giving a family of $\mathcal{N}=4$ SQM models whose $tt^*$ geometry defines a hyperholomorphic connection on $\mathcal{A}^\vee$ which is the Nahm transformed one. At the level of the correspondent coherent sheaves, it coincides with the  Fourier--Mukai transform \cite{mukai,bruzzobook}.

\subsection{Some examples from D-branes}
As highlighted in this section, much of the $tt^*$ structure only relies on an ${\cal N}=4$ super quantum mechanics. Some structure, of course, hinges on 
having a $(2,2)$ 2d theory: for example the spectral Lagrangian is tied to the twisted effective superpotential of the 2d theory in flat space. 
The notion of topological gauge for the $tt^*$ connection is also closely related to the existence of a 2d cigar geometry which maps chiral operators to 
states on the circle. Still, the structure which remains in a 1d setup is rather interesting, especially if we consider the generalization to 1d-3d systems,
\textit{i.e.}\! to half-BPS line defects in 3d ${\cal N}=4$ theories. Such defects preserve the same supersymmetry of an ${\cal N}=4$ SQM 
and may have flavor symmetries or parameters which give rise to a $tt^*$ geometry. This is essentially a dimensional reduction of the 
2d-4d systems reviewed in a previous section. 

The first obvious example is a massive 1d chiral field. There are three real mass parameters $m_i$ and the 2d calculations make it clear that the $tt^*$ geometry is a charge $1$ $U(1)$ 
Dirac monopole in $\R^3_m$. Notice that the Higgs field in the $tt^*$ monopole geometry is essentially the moment map for the flavor symmetry. This is why it diverges 
at $\vec m=0$, where the chiral field is massless. 

In order to obtain a smooth $SU(2)$ monopole solution we can look at a SQM with $\C P^1$ target, study the dependence on the $SU(2)_m$ flavor mass parameters $\vec m$. 
The theory has two vacua which when the mass parameter is turned on roughly corresponding to the north and south pole of $\C P^1$. 
At large $|\vec m|$, the dynamics in the two vacua is well approximated by a single free chiral of charge $\pm 1$. The theory has no non-compact directions at any value of $\vec  m$.
Thus the $tt^*$ geometry for $\vec  m$ is a smooth $SU(2)$ monopole with Abelian charge $(1,-1)$ at large $\vec m$, \textit{i.e.} a single smooth $SU(2)$ monopole. 
Notice that the asymptotic values of the Higgs field are given by the value of the moment map for (the Cartan sub algebra of) the $SU(2)$ flavor symmetry, 
which are $\pm t$, where $t$ is the FI/K\"ahler parameter for the $\C P^1$ theory. Thus $t$ controls the asymptotic values of the monopole geometry. 

Conversely, the $tt^*$ geometry for the FI/K\"ahler parameter $t$ is given by a solution of $SU(2)$ Nahm equations on $\R^+$,
which are the Nahm transform of a pair of $U(1)$ Dirac monopoles of charge $1$, at positions $\pm \frac{1}{2} \vec m$. 
At large $t$ the two vacua again correspond roughly to the north and south poles of $\C P^1$. If we use a GLSM description, the two vacua require the three 
scalar fields in the gauge multiplet to be equal to $\pm \frac{1}{2} \vec m$. 
We expect to find a solution of Nahm equations with a Nahm pole at the origin of $\R^+$, and 
constant diagonal vevs $( \frac{1}{2} \vec m, - \frac{1}{2} \vec m)$ at infinity. 

In order to generate more examples, we can look at the standard Hanany-Witten brane setup, 
with D3 branes stretched between NS5 branes, probed by a transverse D1 brane. 
Indeed, this setup gives 3d ${\cal N}=4$ field theories probed by 1d line defects, 
which can be interpreted as coupling the 3d theories to some 1d GLSMs or as 
the 1d version of Gukov-Witten monodromy defects. 

Depending on the choice of boundary conditions on the D1 brane, which may be realized concretely by having it end on 
a separate NS5 brane or D3 brane on a plane parallel to the system, one can get 1d defects with mass parameters corresponding to a motion parallel to the NS5 branes, 
of FI parameters corresponding to a motion along the D3 branes. 
For example, the 1d chiral can be engineered through a single semi-infinite D3 ending on a single NS5 brane,
with a D1 brane with fixed position $\vec m$ along the NS5 brane transverse directions. 
The $t$ geometry for the 1d $\C P^1$ model can be engineered by two semi-infinite D3 branes ending on a single NS5 brane,
with a D1 of fixed position $t$ along the D3 branes. Then $\vec m$ is the separation between the D3 branes.

In order to engineer the $\vec m$ $tt^*$ monopole geometry directly, we could look at a single D3 brane stretched between two NS5 branes, 
with a D1 probe of fixed position $\vec m$ along the NS5 brane transverse directions. Then $t$ is the separation between the NS5 branes. 
This setup realizes the smooth $SU(2)$ monopole geometry, but not through a $\C P^1$ 1d model. 
Instead, the physical interpretation of the brane system seems to be that of a single 1d chiral coupled to a 
3d $U(1)$ gauge field. It would be interesting to study this simple field theory model further: 
the prediction that such model should have two vacua, corresponding to the D1 ending on either NS5 brane, 
is reminiscent to a somewhat mysterious phenomenon which occur for certain surface defects \cite{GGS}. 

In general, the $tt^*$ geometry for the brane systems is recovered by S-duality: the D1 brane becomes an 
F1 string, whose endpoint explores the supersymmetric gauge fields on a system of intersecting D3 and D5 branes: 
the gauge fields on the D3 branes give the solutions of Nahm equations, the gauge fields on the D5 branes give the BPS monopole solutions. 

The world volume theory of the D1 brane could be interpreted as the 2d $(4,4)$ theory coupled to the 
1d system as in section \ref{higher}, with the choice of boundary conditions on the other end deciding which Nahm dual description 
emerges at the end. 

The brane construction reviewed in this section has obvious generalizations which are commonly used to describe higher-dimensional 
theories and defects:\begin{itemize}
\item A D2 probe of a D4-NS5 system engineers 2d theories or 2d-4d systems. The boundary conditions on the other end of the D2 probe 
correspond to gauging/ungauging a 2d flavor symmetry. The D2 probe theory is exactly the 3d free YM theory discussed in section \ref{higher}. 
Lift to M-theory gives the spectral data of the system.

\item A D3 probe of a D5-NS5 system (or a more general $(p,q)$ fivebrane web) 
engineers 3d theories or 3d-5d systems. The D3 probe theory is exactly the 4d free YM theory discussed in section \ref{higher}. 
T-duality together with a lift to M-theory produces the spectral data of the system. We will discuss this in detail in section \ref{Main}. 

\item A D4 probe of a D6-NS5 system (the Hanany-Zaffaroni setup \cite{Hanany:1997gh}) engineers 4d theories or 4d-6d systems. 
The D4 probe theory is exactly the 5d free YM theory discussed in section \ref{higher}. 
Double T-duality together with a lift to M-theory produces the spectral data of the system. Single T-duality gives a periodic $(p,q)$ fivebrane web.
We will discuss this in detail in section \ref{4d}.

\end{itemize}
\section{$tt^*$ geometry in 3 dimensions}
\label{sec:geo3dimensions}

In this section we would like to characterize the geometry of vacuum bundles
in theories in 3 dimensions, with ${\cal N}=2$ supersymmetry.  More precisely
we are interested in studying the vacuum geometry when the space is taken to be a flat $T^2$
with periodic boundary conditions for fermions, so as to preserve all supersymmetries.
Our strategy will be as follows.  We first clarify the structure of the parameter space
taking into account that the space is $T^2$.  We then view the 3d theory
as a special case of 2d ${\cal N}=(2,2)$ theories with infinitely many fields, and
use this relation to find the $tt^*$ geometry in 3 dimensions.  We shall see that they
correspond to generalized monopole equations.  We then show how this data
can be used to compute the partition function of the theory on infinitely elongated $S^3$ and $S^2\times S^1$
composed of two semi-infinite cigars joined in two different ways.

\subsection{The parameter space}

\begin{figure}[here!]
\centering
\includegraphics[width=.8\textwidth]{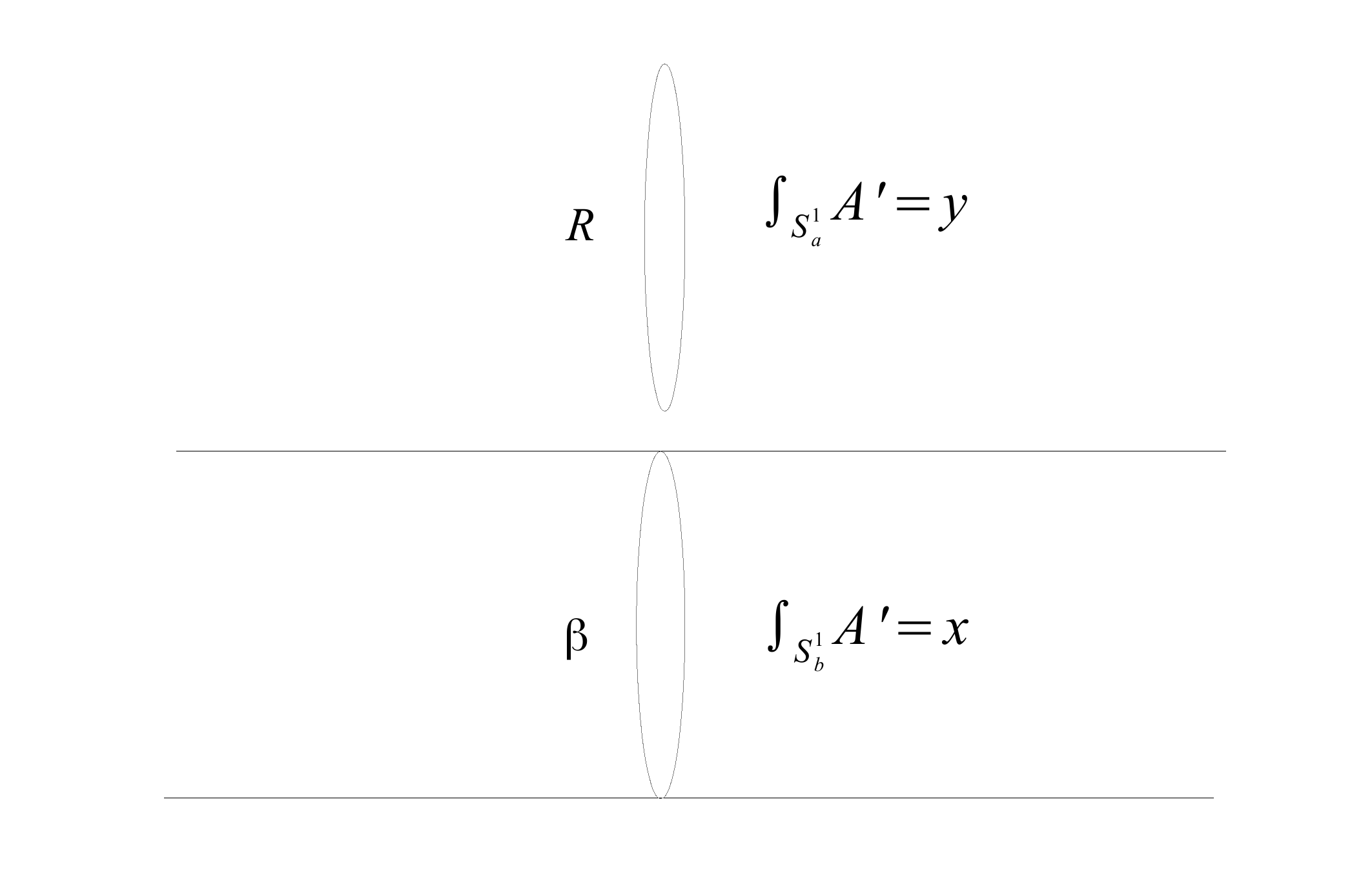}
\caption{In the 2+1 dimensional theory we take the space to be $T^2$ comprised of two circles $(S^1_a,S^1_b)$ of
lengths $(R,\beta)$ where we turn twisted by flavor symmetry by $y,x$ respectively (by turning on background
field $A'$ coupling to the flavor current).}
\label{fig:3d}
\end{figure}

Consider a 3d theory with a global $U(1)$ symmetry.  Furthermore we consider
the space to be a flat $T^2$.  In such a case we can associate a three parameter
deformations of the theory $(x,y,z)$ where $z\in \R$ denotes the twisted mass associated
to the $U(1)$ symmetry and $(x,y)\in T^2$ denote the fugacities 
for the $U(1)$ symmetries around the cycles of the $T^2$.  Another way of saying this is
to imagine weakly gauging this $U(1)$ symmetry.  In the ${\cal N}=2$ $U(1)$ vector multiplet
we have a $U(1)$ gauge field $A'$ and a scalar $\phi$ .  Then
$$ \langle \phi \rangle = z$$
$$\int _{S^1_a} A' =x,\quad \int_{S^1_b}A'=y,$$
where $S^1_{a,b}$ denote a basis for the two 1-cycles of $T^2$.  In the limit we turn off the gauge
coupling constant, we can view $(x,y,z)$ as parameters in the deformation space of the theory  (see Fig. \ref{fig:3d}).

If we have a rank $r$ flavor symmetry, the same argument, \textit{i.e.}, weak gauging and giving
vev to the adjoint $\phi$ in the Cartan subalgebra of the flavor group and to Wilson lines in the Cartan torus,
shows that we have a parameter space
$$(T^2\times \R)^r.$$
Taking into account the full symmetry of the problem amounts to dividing the above
space by the action of the Weyl group of the flavor symmetry.

Note that in presence of a $U(1)$ gauge symmetry we have, associated with it, a global
$U(1)$ symmetry (related to monopole number) where the $U(1)$ current is $J=*F$.
The twisted mass in this case corresponds to FI-term for the $U(1)$ gauge symmetry, while
the coupling constant corresponds to coupling the $U(1)$ gauge field $A$
to a background $U(1)$ gauge field $ A'$ with a  Chern-Simons interaction
$$\int  A'\wedge F$$
and the vevs of $A'$ along the two $S^1$'s give the $(x,y)$ parameters.  The FI parameter
plays the role of $m$.
At any rate, applying the logic of the previous discussion (as  a special case) to a theory which includes a gauge symmetry $U(1)^r$ will lead again to a parameter space $(T^2\times \R)^r$.

\subsection{Derivation of $3d$ $tt^*$ geometry from $2d$ perspective}

In this section we show how to derive the equations for $tt^*$ geometry for $3d$ by
viewing it from the $2d$ perspective.  What we will show is that from the $2d$ perspective
for each $U(1)$ symmetry the fundamental group of the parameter space receives an extra $\Z$,
since the 2d superpotential restricted to that sector will pick up an extra term $n t$ where $n\in \Z$
and $t$ is a 2d coupling associated to $U(1)$.  Once we show this, the structure of the
3d $tt^*$ falls in the class discussed in sections 3 where we obtained the generalized monopole 
equations.

The argument for this is as follows:  Suppose we have a $U(1)$ global symmetry
in 3d.  Consider compactifying the theory from 3d to 2d on $S^1_a$
with fugacity $y$ around the circle for the $U(1)$.   Then we obtain a ${\cal N}=(2,2)$ theory
in 2 dimensions, which includes a chiral deformation parameter given by
$$t=z+iy.$$
Note that $t$ takes values on a cylinder because $y$ is periodic.  On a space
$\R$, this 2d theory will in addition
have sectors ${\cal H}_n$ corresponding to $U(1)$ charge $n$.    The supersymmetry
algebra has a central term in this sector given by $nt$.  To see this, note that for a theory in 3 dimensions,
if we take the space to be $\R \times \R$ and consider a sector with flavor $U(1)$ charge $n$,
the fact that we have turned on the twisted mass parameter would have implied the central
charge to be $nz$.  Upon compactification of $\R$ to a circle $S^1_a$ turning on
fugacity $y$, given the holomorphic dependence of $W$ on $t,$ the central
charge in the supersymmetry algebra, which is the value of the superpotential in this sector, must be completed to
$nt$, as was to be shown.  Therefore we are in the category of $2d$ theories where
the vacua have a shift symmetry along which $W$ changes by an integer times a complex parameter
and, as we have already discussed, this leads to generalized monopole equations for the $tt^*$ geometry.
Note, in particular, that turning on the fugacity $x$ around the second circle $S^1_b$ corresponds to
weighing the $n$ vacua by 
$$|\alpha, n\rangle \rightarrow {\rm exp}(2\pi i xn) |\alpha,n\rangle.$$
In other words %
$$|\alpha, x\rangle =\sum_n {\rm exp}(2\pi i nx) |\alpha,n\rangle,$$
which is consistent with the definition of $x$--vacua discussed in section 3.  We therefore see that the $tt^*$ geometry
for 3d ${\cal N}=2$ theories corresponds to generalized monopole equations.

\subsection{Chiral algebra and line operators}

\begin{figure}
\centering
\includegraphics[width=.8\textwidth]{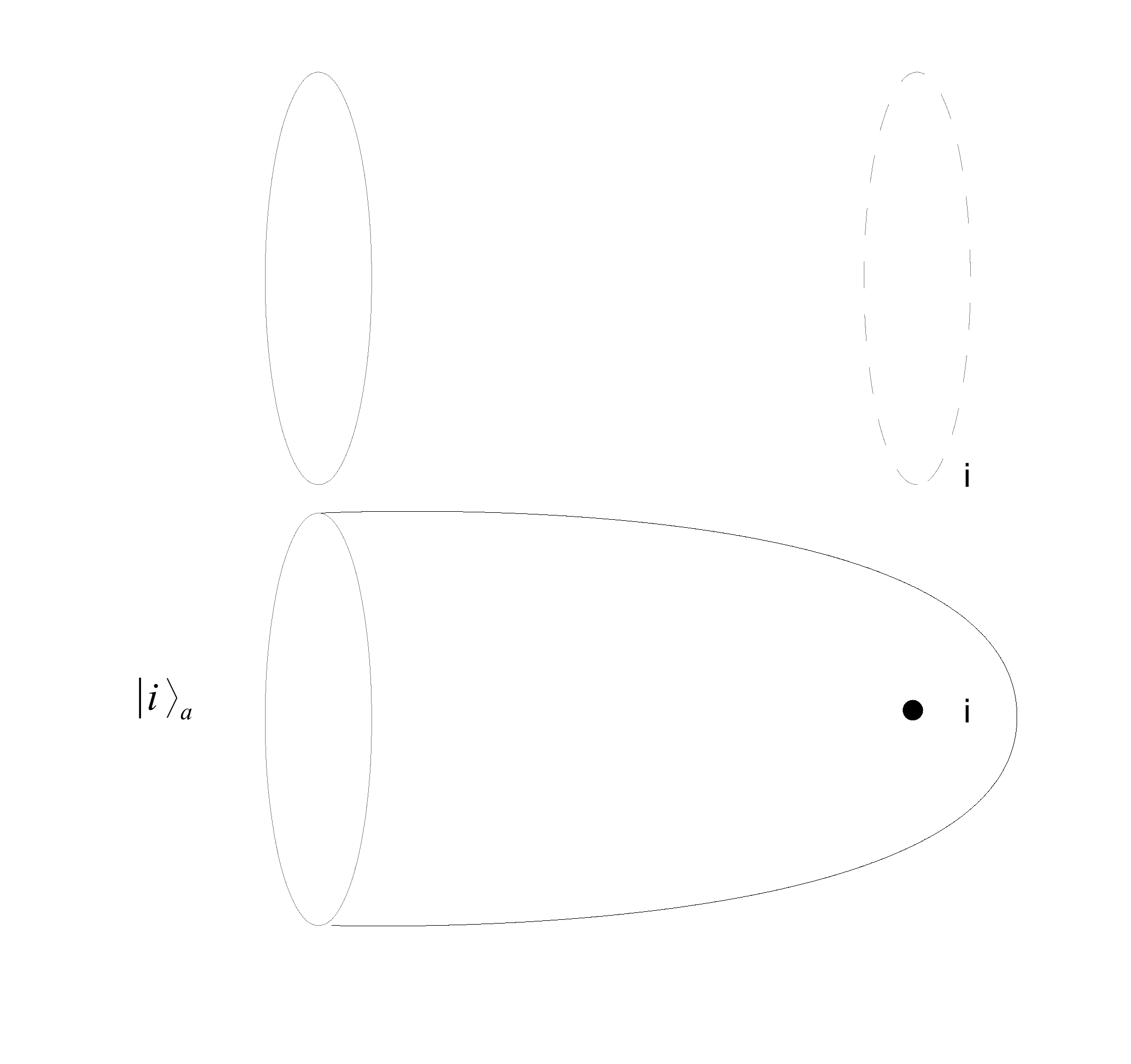}
\caption{Vacuum states in the 3d theory $|i\rangle _a$ can be obtained by doing a path-integral on an infinitely long
solid torus, which is equivalent to an infinitely long cigar times a circle.  The chiral fields in 2d are obtained
by wrapping the line operator along the circle $S^1_a$, \textit{i.e.}\! the circle in the solid torus which is not contracted.}
\label{fig:3dstate}
\end{figure}

Consider the 3d theory compactified on $S^1_a$, on a circle of size $R_a$.
This leads to an ${\cal N}=(2,2)$ theory in 2d.  Let us take a generic case
where we will have $n$ vacua with mass gap where $n$ is the Witten index of the theory.
From the 2d perspective we expect to have a chiral algebra with $n$ elements.  These
chiral fields should correspond to line operators from the 3d perspective wrapped around $S^1_a$.
Clearly they are localized over a point in 2d, so they could be in principle either point operators in 3d or line
operators.  The fact that their coupling involves $\int d^2\theta\ t^i\Phi_i$, and the imaginary part of $t$ is a global
parameter $y$ having to do with the holonomy around the $S^1_a$, shows that the operator must be a loop operator
wrapping $S^1_a$ and coupled to this global holonomy  (see Fig. \ref{fig:3dstate}).
Note that the algebra they form will depend on the radius $R_a$.

In the case of supersymmetric gauge theories, these line operators correspond to supersymmetric
Wilson lines.  See a nice discussion of them in \cite{kapustinrec}.  In particular
in the case of pure $N=2$ Chern-Simons gauge theory, where the theory is equivalent
to a topological theory, this algebra is isomorphic to the Verlinde algebra. We will return to this
discussion after considering the partition functions of these theories on spheres which we now turn to.

\subsection{Geometry of $T^2$ and partition functions on elongated $S^2\times S^1$ and $S^3$}

In this section we discuss the global interpretation of the partition functions computable using $tt^*$ geometry
in 3d.  The geometry of the space is captured by the two-torus $S^1_a\times S^1_b$.  For most of the discussion
we would be interested in a rectangular torus.  In particular if $\tau$ is the complex structure parameter for the
torus, we take $\tau_1=0, \tau_2=R_a/R_b$, in other words $\tau=i R_a/R_b$.  The reason for the choice
of rectangular torus is that if we set the $\tau_1 \not =0$ we would not have a reduction to a Lorentz-invariant
2d theory.  One can in principle also study this extension (which will induce some non-commutativity structure
from the 2d perspective if we are discussing any amplitude other than vacuum amplitudes), but for simplicity
we limit our discussion mainly to the rectangular case.
In addition to $\tau$ the geometry of the torus is characterized by its area $A=R_aR_b$.
Clearly there is an isomorphism of the theory which takes 
$$(\tau, A)\rightarrow (-1/\tau ,A),$$
by simply switching the role of the two circles.  

There are two inequivalent ways we can view this theory as a 2d theory, depending on whether
we take $S^1_a$ or $S^1_b$ as part of the spatial direction of the 2d theory. Of course, the geometry of the vacuum
bundle does not depend on this choice.  However, the $tt^*$ has more information than just the vacuum geometry:
It has also a choice of preferred sections for the vacuum bundle given by semi-infinite cigar cappings of the theory.
Let us take $S^1_b$ as part of the 2d spatial directions which are taken to form a semi-infinite cigar inside which the cycle $S^1_b$ shrinks.
Then the preferred choice of the vacuum bundles are labeled by chiral operators on the cigar:
$$|i\rangle_a$$
as discussed before for the general case in 2d (see Fig. \ref{fig:3dstate}). The label $a$ in the above state is to remind us that this is the
circle we have chosen not to shrink.  Moreover this corresponds to the fact that the line operators are wrapping
the $a$--cycle $S^1_a$.
  Similarly, we can consider the 2d theory obtained by viewing $S^1_a$ as part of the 2d spatial
dimensions and obtain the states  (see Fig. \ref{fig:3dbcycle}):

\begin{figure}
\centering
\includegraphics[width=.8\textwidth]{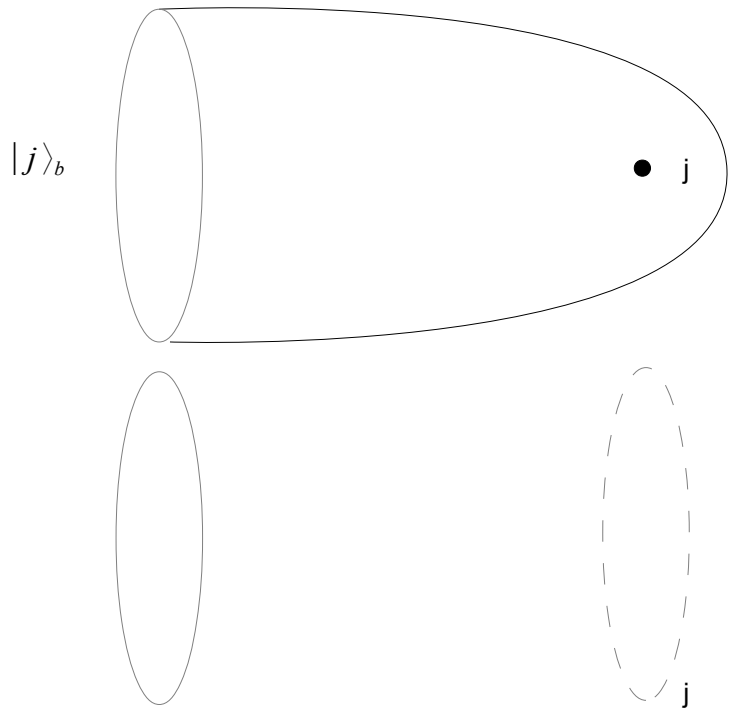}
\caption{The 3d vacuum states can be obtained by filling either of the two circles, leading
to two different bases for the vacua.  In this figure $|j\rangle_b$ denotes the state obtained by inserting
a line operator wrapped around the b-cycle.}
\label{fig:3dbcycle}
\end{figure}

$$|j\rangle_b.$$
  If we change $\tau\rightarrow -1/\tau$ we come back to the same theory.  In other words
  $|i(\tau)\rangle_{a}$ should be a linear combination of $|j(-1/\tau)\rangle_{b}$:
$$|i(\tau)\rangle_{a}={S_i}^j (\tau )|j(-1/\tau)\rangle_{b}$$
More precisely we have, restoring the $x,y$-dependence
$$|i(\tau),t=z+iy,x \rangle_{a}={S_i}^j (\tau )|j(-1/\tau),t'=z-ix,y\rangle_{b}.$$
${S_i}^j$ depends on all parameters, but here we are just exhibiting its
dependence on $\tau$. 
Note that ${S_i}^j$ satisfies
$${S_i}^j(\tau)\,{S_j}^k(-1/ \tau)={\delta_i}^k,$$
because this operation corresponds to $\pi$ rotation in 3d (and in particular takes
$(x,y)\rightarrow (-x,-y)$), and this acts trivially on the vacua (as can be seen by taking
the large area limit of a square torus and noting that this reflection can be generated by continuous
rotations for which the vacua are neutral).
We can also consider the D-brane boundary conditions, which are in 1-1 correspondence with the number
of vacua (in a massive phase).  Let $D_c$ denote one of the boundary states.   We then have
$$\Pi^{c(bb)}_i = {}_b\!\langle D_c |i \rangle_b.$$
Note that from the 3d perspective $\Pi^{c(bb)}_i$ is given by a path--integral in a space with the topology of a solid torus, whose boundary is a $T^2$ given
by the D-brane state $D_c$. It is useful to rewrite the boundary states $|D_d(-1/\tau)\rangle_b$ in terms of $|D_c(\tau)\rangle_a$.  First we have to recall that $D_c$ depends on $\zeta$ which determines which combination
of supercharges it preserves.  In going from $\tau$ to $-1/\tau$  the values of $\zeta$ also changes, as discussed in
eq.(\ref{tcay}): $ \tilde \zeta=\mathsf{C}(\zeta)\equiv \frac{1+i\,\zeta}{\zeta+i}$. 
Since the theories are the same, the boundary state should be a linear combination of one another.  In fact
as we have already noted these boundary states satisfy are sections of the Lax connection and therefore there
must exist a constant matrix ${E_{c}}^d$ such that
\begin{equation}\label{defEab}
|D^{\zeta}_c(\tau),t=z+iy,x\rangle_a ={E_c}^d \,|D^{\tilde \zeta}_d(-1/\tau),t=z-ix,y\rangle_b
\end{equation}
Note that repeating this operation is equivalent to a $Z_2$ spatial reflection.  This implies
that $E^4=1$ (using the fact that the ground states all have even fermion number).
  Indeed with a suitable
choice of basis (adapted from the point basis in the IR) it can be taken to be a diagonal matrix.
For simplicity of notation we will not explicitly write he corresponding phases.  Also,
we will not explicitly write the $\zeta$ in the definition of states.  Sometimes we choose one of 
the two preferred values $\zeta =\pm 1$ which are the fixed points of the transformation $\zeta\mapsto \mathsf{C}(\zeta)$.  We will return
to the significance of this choice later.

\begin{figure}
\centering
\includegraphics[width=.8\textwidth]{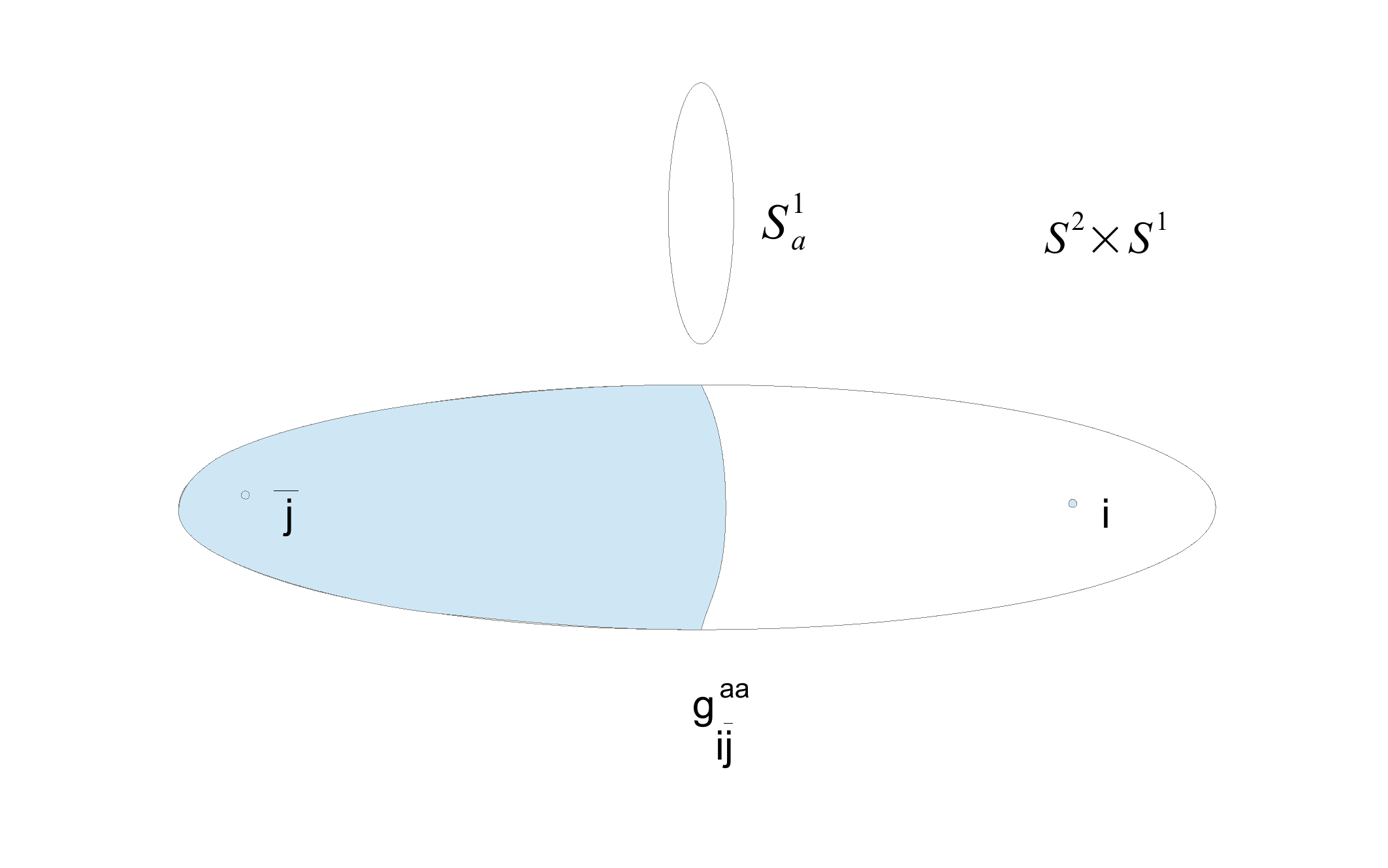}
\caption{The inner product on the Hilbert space restricted to the vacuum states can be represented
by the path-integral on $S^2\times S^1$ with infinitely elongated $S^2$, where the chiral and anti-chiral line operators
are inserted at the two ends.}
\label{fig:S2xS1}
\end{figure}

We can then compute, as in the general 2d case, the 2d topological metric $\eta$ and Hermitian metric $g$:
$$\eta_{ij}^{(aa)}= {}_a\!\langle j |i\rangle_a\qquad g_{i{\overline j}}^{(aa)}= {}_a\!\langle {\overline j} |i\rangle_a.$$
The topology of the space for both of these computations correspond to $S^2\times S^1_a$, where $S^2$
is a sphere with an infinitely elongated cylindrical neck.  The computation of $\eta$, which is a topological invariant,
can also be done for a finite size sphere.  For the computation of the Hermitian metric the infinite size
sphere is crucial.  We can also consider the partition function on $S^2\times S^1_b$ where the
role of $a,b$ are exchanged. 
Just as in the general 2d case, eqn.(\ref{deco}), we have (see \textit{e.g.}\! Fig. \ref{fig:S2xS1} for the metric $g$)
$$\eta_{ij}^{(aa)}={\hat \Pi}_j^{c(aa)}\Pi_i^{c(aa)}$$
$$g_{i{\overline j}}^{(aa)} ={\hat \Pi}^{c(aa)\dagger}_j\Pi_{ i}^{c(aa)}.$$

\begin{figure}
\centering
\includegraphics[width=1.1\textwidth]{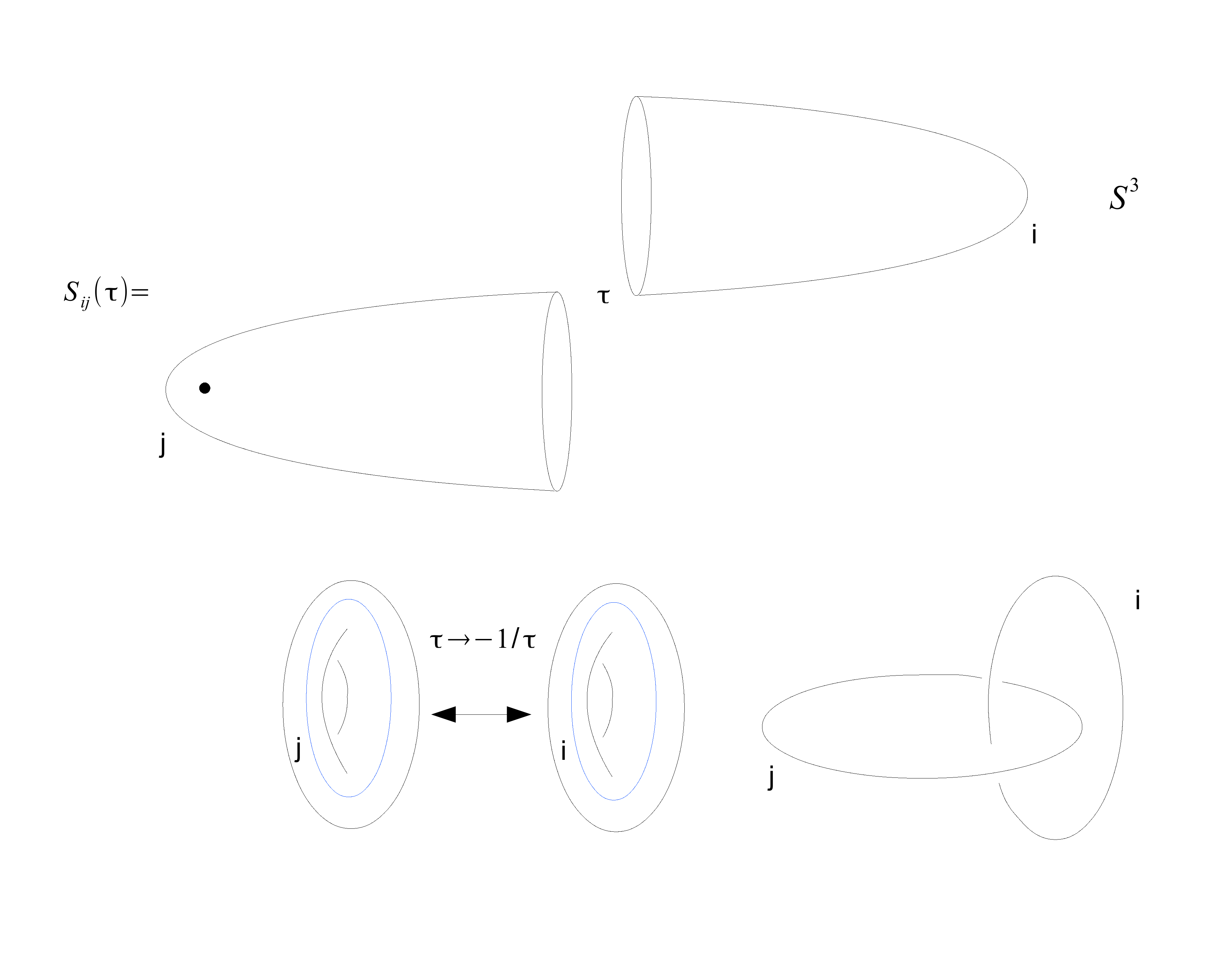}
\caption{The $S_{ij}$ can be viewed as the partition function on $S^3$ with line operators
inserted at the two ends.  This can be viewed as the Heegard decomposition of the $S^3$: the gluing
two solid tori each with a line operator inserted and whose boundaries are identified by the $\tau\rightarrow
-1/\tau$ transformation, exchanging the two cycles of $T^2$.}
\label{fig:FigureA}
\end{figure}

We can also consider capping different circles on the two sides, producing the 3d topology of $S^3$.
Notice that now there is no purely topological version, because $S^3$ does not admit an $SO(2)$ holonomy metric, and thus
the amplitude only makes sense when we consider a $S^3$ with an infinitely long flat neck.  Moreover, whether we choose the topological or the anti-topological theory on either side, the computation is  hard.  Let us then consider the inner product of the vacua thus obtained.
We define
$$S_{ji}(\tau)= {}_a\! \langle j(\tau)|i(-1/\tau) \rangle_b = {S_j}^k(\tau)  \;{}_b\!\langle k(-1/\tau) |i(-1/\tau)\rangle_b={S_j}^k (\tau)\, \eta_{ki}^{(bb)}.$$
The expressions of $S_{{\overline i} j}$ and $S_{{\overline i}{\overline j}}$ can be obtained from $S_{ij}$
using the reality matrix $M_{\overline i}^k$ discussed in the 2d context, so we will restrict our attention to $S_{ij}$.
Note that $S_{ij}$ can be viewed, from the 3d perspective, as the result of gluing two solid tori, each
with infinitely long necks, one of which has line operator $i$ inserted along its center corresponding to the $a$--cycle, and the other one with
the line operator $j$ inserted along the $b$--cycle.  In other words, topologically the two line operators
are Hopf linked.  This is familiar from the structure of Chern-Simons theory \cite{wittenCS}.  Of course
this is not accidental:  In the case of  $\mathcal{N}=2$ Chern-Simons theory with no matter, the theory is equivalent
to $\mathcal{N}=0$ Chern-Simons theory, for which the line operators are the Wilson loop observables.  In that context
$S_{ij}$ is the Hopf link invariant associated to loops indexed by the representations $i$ and $j$. This in turn
is the modular transformation matrix of the conformal blocks of the associated 2d RCFT.
 Unlike the topological
case, where $S_{ij}$ does not depend on any parameters, in the more general case we are considering $S_{ij}$ does
depend on parameters of the theory and in particular on $\tau$  (see Fig. \ref{fig:FigureA}).

In order to compute $S_{ij}$ we use the fact that we can compute $\Pi^{c(aa)}_i$ as discussed before.
Therefore it suffices to write $S_{ij}$ in terms of them.
We have
\begin{equation}\label{pitaupioneo}
\Pi^{c(aa)}_i (\tau)= {_a}\langle D_c(\tau)| i \rangle_a={S_i}^j(\tau)\,\langle D_c(-1/\tau)|j\rangle_b ={S_i}^j(\tau)\, \Pi^{c(aa)}_j(-1/\tau).
\end{equation}
In other words, we have
\begin{equation}\label{eq:sij}
{S_i}^j(\tau)=\Pi^{c(aa)}_i(\tau)  [\Pi^{{(aa)^{-1}}}(-1/\tau)]^{cj},
\end{equation}
leading to
$$S_{ij}=\Pi^{c(aa)}_i(\tau)  [\Pi^{{(aa)^{-1}}}(-1/\tau)]^{ck}\  \eta^{(bb)}_{kj}.$$
 The vacuum amplitude is given by
\begin{equation}\label{sphere}
S_{00}=\Pi^{c(aa)}_0(\tau)[\Pi^{{(aa)^{-1}}}(-1/\tau)]^{c{\hat 0}}  
\end{equation}
where $\hat 0$ denotes the spectral flow operator dual to the identity.
  Note that for the case of a single vacuum theory  we get
$$S_{00}=\Pi^{(aa)}(\tau) \Pi^{(aa)^{-1}}(-1/\tau).$$
This expression is similar to the expression of the partition functions for supersymmetric amplitudes
on ellipsoid $S^3_b$ for a theory of, say, free chiral theory, where instead of $\Pi^{(aa)}(\tau)$ one has the quantum
dilog with $\tau =b^2$ where $b$ is the squashing parameter.  As we will discuss later, this is not
accidental:  In a partial UV limit (similar to the ${\overline \beta}\rightarrow 0$ limit in 2d) the $\Pi^{(aa)}$ reduces to quantum dilog.  More generally we will argue in 
a later section that eqn.(\ref{sphere}) is consistent with the results of \cite{pasquettietal} in their computation
of the partition functions on $S^3_b$ in terms of sums over chiral blocks, which in the formula above is
the sum over $c$.

\subsection{Partition functions as gauge transformations}\label{sec:partitionsasgauge}

There is  a different (but equivalent) interpretation of the partition functions on infinitely elongated $S^2\times S^1$ and $S^3$ which is more convenient in actual computations. 

Let us consider first elongated $S^2\times S^1_a$. The elongated partition function is just the component of the $tt^*$ metric
$$g^{(aa)}_{0\bar 0}.$$
There are two natural trivializations of the vacuum bundle over $S^1_a\times S^1_b$ namely the ones given, respectively, by the topological and the anti--topological twisting on a cigar which caps the circle $S^1_b$.
The vacuum bundle Berry connections in these two natural   
trivializations read
\begin{equation}\label{tttyp}
\begin{cases}
D= \partial+g\partial g^{-1}\\
\overline{D}= \overline{\partial}
\end{cases}\qquad \text{and respectively}\qquad
\begin{cases}
D= \partial \\
\overline{D}= \overline{\partial}+g^{-1}\overline{\partial}g
\end{cases}
\end{equation}
where $g=g^{(aa)}$ is the $tt^*$ metric. We see that the $tt^*$ metric $g^{(aa)}$ is nothing else than the complexified gauge transformation mapping  
the Berry connection in the topological gauge to the one in the anti--topological gauge.

The same kind of identification holds for the quantity ${S_i}^j$ defined in eqn.\eqref{eq:sij}, and hence for the elongated $S^3$ partition function. Again we have two preferred trivialization of the same vacuum bundle given by the states $|i,\tau\rangle_a$ and $|j,-1/\tau\rangle_b$. In the first trivialization  the Berry connection $A$ has the form in the left part of eqn.\eqref{tttyp} with $g=g^{(aa)}$, while in the second one it is given by $\boldsymbol{S}A$, where $\boldsymbol{S}$ is the $\pi/2$ rotation acting as
$$\boldsymbol{S}\colon x\rightarrow y, \ y\rightarrow -x,\ \tau\rightarrow -1/\tau,$$
that is,
\begin{align*}
\boldsymbol{S}A_y(x,y,z,\beta, R)&=A_x(y,-x,z,R,\beta),\\
\boldsymbol{S}A_x(x,y,z,\beta, R)&=-A_y(y,-x,z,R,\beta),\\
\boldsymbol{S}A_z(x,y,z,\beta, R)&=A_z(y,-x,z,R,\beta).
\end{align*}
Since the connections $A$ and $\boldsymbol{S}A$ describe the same physical monopole in $x,y,z$ space, they are gauge equivalent, \textit{i.e.}\! there is a complex gauge transformation $S$ such that
\begin{equation}
\boldsymbol{S}A= S\,A\,S^{-1}+S\,dS^{-1},
\end{equation}
This matrix $S$ clearly coincides with the matrix ${S_j}^i$ defined in eqn.\eqref{eq:sij}.

Another way to see this identification, is to consider the brane amplitude $\Pi^c$ at $\zeta=\pm 1$. As discussed around eqn.\eqref{zetaoneS}, $\Pi^c$ and $\boldsymbol{S}\Pi^c$ satisfy the same Lax equations $\nabla\Pi^c=\bar\nabla\Pi^c=0$, and in fact both form a fundamental system of solutions of these linear  equations. Then they are linear combinations of one another with constant coefficients. More precisely, since they are written in different gauges,  we must have
\begin{equation}\label{piuspie}
{\Pi_i}^c= {U_i}^j\, \boldsymbol{S}{\Pi_j}^c\,
\end{equation} 
where $U$  is the (complexified) gauge transformation relating these two gauges.  Comparing with eqn.\eqref{pitaupioneo}, we get
\begin{equation}
{U_i}^j={S_i}^j,
\end{equation}
which is our identification.

This  identifications allows us to compute the partition function from the Berry connection without having to solve the Lax linear problem. In other words, we may read the partition function on the infinitely elongated $S^3$ directly from the $tt^*$ monopole configuration in $x,y,z$ space, without solving any additional partial differential equation.

The gauge viewpoint gives an alternative argument for the independence of the matrix $E$ mapping
D-branes at $\tau$ to $-1/\tau$, from all parameters, and how
it can be set to be the identity matrix. \emph{A priori,} the Lax equations imply eqn.\eqref{piuspie} in the weaker form $$\Pi_i^a={S_i}^j\, \boldsymbol{S}\Pi_j^b\, {E_b}^a,$$ where $E$ is a non--degenerate constant numerical matrix. In a theory with a mass--gap, rescaling the masses to infinity both connections $A$ and $\boldsymbol{S}A$ go to zero, hence $S\to 1$, while, using the point basis for the line operators and the corresponding thimble basis for branes, both $\Pi$ and $\boldsymbol{S}\Pi$  
approach the identity matrix. Then we remain with a diagonal $E$ matrix, which as discussed before satisfies
\begin{equation}
E^4=\boldsymbol{1}\qquad\text{diagonal in point/thimble basis, }\zeta=\pm1.
\end{equation}

\subsection{Massive limits and topological line operator algebra}\label{topological}
Consider 3d, ${\cal N}=2$ theories, which have a mass gap.  Such theories in the IR flow
to trivial theories with no non-trivial local correlation functions.  However, this does not mean the theory
is trivial:  It could still hold interesting topological non-local observables.  The simplest
examples of this kind are ${\cal N}=2$ pure Chern-Simons theories with no matter.  In such cases
the theory in the IR is locally trivial and the only non-trivial observables are the line operators associated with Wilson loops.
Supersymmetric Wilson loops are rigid in shape, but since this theory is equivalent to ${\cal N}=0$ Chern-Simons theory,
we can dispense with the condition of preserving supersymmetry and consider general Wilson loops, and use the topological
invariance of the theory to solve it, as was done by Witten \cite{wittenCS}.  

We would like to study this same phenomenon in the general case, and consider in addition the process of flow to the IR as well.
In fact, if we consider the Hilbert space of such a theory quantized in  $T^2$, the flow to the IR corresponds to changing
the area of $T^2$, while preserving its shape, given by the complex modulus $\tau$.  In other words, we would
be studying the flow
$$(\tau, A)\rightarrow (\tau, e^t A).$$
It is natural to conjecture that, for all such theories with mass gap, we always end up with a purely topological theory
in the IR, for which the line operators we have been studying play the role of non-trivial observables.  In particular it is natural
to conjecture that in this limit $S_{ij}$ will become independent of $\tau$ and satisfies the usual properties familiar
from the Verlinde algebra theory. 
 Moreover we conjecture that, as in the
case of the Verlinde algebra \cite{verlinealg}, the chiral ring becomes, in a suitable basis, an integral algebra whose multiplication table is given by 
positive integers
$${C_{jk}}^i={N_{jk}}^i\in {\Z}^+,$$
and that $S_{ij}$ diagonalizes the algebra which is equivalent to the statement that
$$\lambda_i^{l}={S_{il}\over S_{0l}}$$
satisfy the ring algebra
$$\lambda_i^{l}\lambda_j^{l}={N_{ij}}^k \,\lambda_k^{l},$$
where there is no sum in $l$ in the above formula, but there is a sum in $k$ in the RHS.

Let us try to see to what extent we can recover these structures in our context. 
Consider the ${\cal N}=2$ theory on $T^2$, and consider the set of line operators $\Phi_i$.  More
specifically, we will consider these line operators wrapped around the $a$ or the $b$ cycle of $T^2$ and
denote the corresponding operators by $\Phi_i^{(a)}$, $\Phi_i^{(b)}$.
These two sets of operators act on the ground states.  In general they have different spectrum,
because the radii are not equal.  Moreover the $(x,y)$ are not zero.  Let us therefore restrict
attention to the case $R_1=R_2=R$, \textit{i.e.}, $\tau=i$, and $A=R^2$.  Furthermore let us take
$x=y=0$.  For this particular case the spectrum of the two sets of operators is the same,
because spatial rotation by $\pi/2$ is a symmetry of this square torus, and represented
by a  unitary operator.  Let us denote this operator by $U$.  Note that on the ground states:

$$U|i\rangle_a=|i\rangle_b$$
which implies that
$$S_{ij}={}_a\langle i|j \rangle_b={}_a\langle i|U|j \rangle_a$$

In other words the matrix elements of $U$ can be identified with the matrix $S$.  We will
thus denote $U$ by $S$ from now on. 

From the fact that $U$ acts on line operators  taking the line operators around the $a$-cycle
to that on the $b$-cycle, we learn
$$S \Phi_i^{(a)} S^{-1}=\Phi_i^{(b)}.$$
Thus finding the $S$ matrix, for $\tau=i$ amounts to finding the change of basis involved
in going from the basis of vacua for the $a$-cycle to that for the $b$-cycle.  In particular, if we know how
$\Phi_i^{(b)}$ acts on the $|j\rangle_a$ we can compute the $S$-matrix (since the action of
$\Phi_i^{(a)}$ on $|j \rangle_a$ is known to be given by ${C_{ij}}^k|j\rangle_a$).

For the case where we are dealing with $U(1)$ gauge theories, the $\Phi_i^{(b)}$ corresponding
to the line operator in the fundamental representation can be identified
as the supersymmetric version of
$$\exp\!\left(i\int_b A\right).$$
The insertion of this operator in the cigar geometry $C$, for the topologically twisted theory, is equivalent to the insertion of the supersymmetrization of $\exp(i\int_C F)$.
This operator corresponds to changing the $\theta$-angle of the 2d theory by
$$\theta\rightarrow \theta +2 \pi .$$
In other words
$$\Phi^{(b)} \longleftrightarrow O_{\Delta \theta =2\pi}$$
where $O_{\Delta \theta =2\pi}$ is the operator changing the vacua by shifting $\theta $ by $2\pi$.  In 
other words, it is the holonomy of the $tt^*$ connection acting on the vacua as we go around one of the cycles of $T^2$ in the parameter space.  We do know the eigenvalues of the $O_{\Delta \theta =2\pi}$, but that turns out
not to be enough to fix the action of it on the vacua.  In particular, in principle this is a complicated operator,
which depends on solving the $tt^*$-geometry.  However, it turns out that in the $A\rightarrow \infty$ limit, \textit{i.e.}\! in the IR
limit, it is easy to fix this operator:  In this limit the classical vacua corresponding to point vacua do not mix
with each other, and so in the point basis, the action of $\theta \rightarrow \theta +2\pi$ is easy to find, as we will
see in the example section.  In particular we will find that in the IR limit we get the explicit form of $S$.  In this
way it is easy to check if the $S$ diagonalizes the ring algebra, and we shall see that this is indeed the case in the examples
we will consider.

Before going to that, we make some preliminary comment on the 3d brane amplitudes in general.

\subsection{Generalities of 3d brane amplitudes}\label{generalitiesonbrane}

We have already discussed the structure of the D-brane amplitudes in the 2d context and their
singularity structure as a function of the spectral parameter $\zeta$.  We can now describe how this structure changes in the 3d context. 
The standard essential singularity at $\zeta = 0,\infty$ is intimately connected to the presence of a compact direction in the 2d $tt^*$. 
If one were to look at a ``1d'' version of the $tt^*$ geometry, \textit{i.e.} say at solutions of Nahm equations or non-periodic monopoles, 
flat sections of the Lax connections would extend smoothly over the whole twistor sphere parameterized by $\zeta$. 

Conversely, suppose we want to study the 3d $tt^*$ geometry with a standard, BPS, Lorentz-invariant 3d boundary condition $B_{3d}$
and analyze the amplitude $\Pi[B_{3d}]$ in the usual 2d language. 
What is the 2d phase $\zeta$ associated to this problem? The 3d supercharges can be collected into complex 3d spinors of specific R-charge $\pm 1$, which we denote as $Q_\alpha^\pm$. A BPS boundary condition preserves a chiral half of the 3d supercharges,
of specific eigenvalue for the 3d gamma-matrix $\sigma^1$ in the direction orthogonal to the boundary. This corresponds to a specific value of $\zeta$. This is in agreement with the analysis in section \ref{section:periodictt*}. Indeed, consider a brane wrapped on a square torus $T^2$: under a $\pi/2$ rotation of $T^2$ a Lorentz--invariant brane amplitude should go to a brane amplitude of  the same kind, and this is possible only if $\zeta$ is a fixed point of the Cayley transform $\zeta\mapsto \mathsf{C}(\zeta)$ in eqn.\eqref{tcay}. In the conventions of section \ref{section:periodictt*}, where the periodic parameters $x_a,y_a$ are the imaginary parts of the holomorphic coordinates in the $\zeta=\infty$ complex structure, cfr.\!\! eqn.\eqref{nesttolastdex}, the two fixed points are $\zeta=\pm 1$.
Equivalently, the specific values of $\zeta$ corresponding to Lorentz--invariant branes may be obtained by requiring that the Stokes discontinuities, eqn.\eqref{stokes}, which are holomorphic functions of
 $$ x-i\, \tfrac{\zeta+\zeta^{-1}}{2}\,y$$
behaves correctly under $\pi/2$ rotations and hence are functions of $x+iy$ (resp.\! $x-iy$). This restriction on the values of $\zeta$ is also consistent with the analysis of the partition function on an infinitely elongated $S^3$ in the previous subsections which was based on $\zeta=\pm1$ brane amplitudes. The special properties of the $\zeta=\pm1$ amplitudes will be checked in an explicit example in \S.\,\ref{eqalitys00} below.
 
Upon compactification of the theory on a circle, it is probably possible to deform a Lorentz-invariant 3d boundary condition to a non-Lorentz invariant version which preserves 
a more general combination of the supercharges, and gives a flat connection for the spectral connection at general $\zeta$. 
This we expect, based on the fact that once we decompose the 3d theory, in terms of 2d data, such a generic parameter $\zeta$
emerges as a possibility in defining the brane amplitude.
As we may look at the 3d geometry as a 2d geometry in infinitely many ways, depending on which cycle of the torus we take as 
``internal'' and which one as 2d Euclidean time, we expect the essential singularities we encountered in 2d to appear at infinitely many locations.
If we identify $\zeta=\pm 1$ as the poles of the sphere, the essential singularities should appear at the equator, that is, for $\zeta$ on the imaginary axis (which, as already noted, is a fixed line for the Cayley transform \eqref{tcay}). 

On the other hand, from the point of view of the 2d theory which arises from compactification on a circle, the compactification of the 
3d theory on a cigar geometry also appears as a ``brane'', which preserves the supersymmetry corresponding to $\zeta = i$. 
There is actually a family of such ``branes'' which arise from a cigar with a line defect at the tip. Clearly, the amplitudes for such ``branes'' are closely related to 
the $S$ matrix defined above. We will illustrate this fact  in simple examples.  

\subsection{Comparison with susy partition functions on $S^2\times S^1$ and $S_b^3$}
Supersymmetric partition functions on $S^2\times S^1$ and $S_b^3$ have been computed recently \cite{pasquettietal,3dsquashed,pasq1}
in a variety of contexts.   Given that we have also been computing supersymmetric partition functions
on the same topoogies it is natural to ask the comparison between the two.  

The first point to notice is that they do not look to be the same:  The partition function computed
using $tt^*$ becomes supersymmetric {\it only in the limit} of infinitely elongated geometries.
This is not so for the supersymmetric partition functions on $S^3_b$ or $S^2\times S^1$ where
for finite metric the path-integral is supersymmetric. 

Indeed a similar question arises for 2d theories recently studied in \cite{exactK}.  In that context it was
found that at the conformal limit the $tt^*$ partition function for elongated $S^2$ \cite{2dpart1,2dpart2}, \textit{i.e.}, the amplitude $g_{0{\overline 0}}=\langle 0|{\overline 0}\rangle$, coincides with the supersymmetric partition function on $S^2$.  However, it was also
found that away from the conformal point the partition function on $S^2$ does not agree with the $tt^*$ partition function.
In that case, as we will argue, there is a limit of the $tt^*$ partition function which reproduces the simpler supersymmetric
partition function on $S^2$.  The same result works in the 3d case as well leading to the statement that
an asymmetric limit of the $tt^*$ partition functions lead to the supersymmetric partition functions.

Let us first discuss the case of 2d.
As discussed in section 2 eqn.(\ref{deco}) the $tt^*$ partition function on $S^2$, i.e. $g_{0\overline 0}$,
is given by
$$Z_{S^2}^{tt^*}=\Pi_a(t_i,{\overline t_i}) \Pi^{a*}(t_i,{\overline t_i}).$$
However the supersymmetric partition function on $S^2$ is made of blocks which are
holomorphic $\Pi$'s times anti-holomorphic $\Pi$'s.   This structure is true for $S^2_{tt^*}$ only
at the conformal point.  Aways from it, the answer is far more complicated.  However, we can consider the
asymmetric limit where we take the UV limit, corresponding to ${\overline \beta}\rightarrow 0$, with fixed $\beta$.
In this limit
$$\lim_{{\overline \beta}\rightarrow 0}\Pi_a(t_i,{\overline t_i})={\tilde \Pi}_a(t_i).$$
Moreover, as already discussed in section 2, in this limit $\Pi_a$ are given by
period integral with non-homogeneous $W$ (satisfying simple differential equations).
In this limit the partition function of the $tt^*$ agrees with the supersymmetric partition function on $S^2$:
$$Z_{S^2}={\tilde \Pi_a}(t_i){\tilde \Pi}^{a*}(\overline t_i)$$

Given this, it is natural to expect the same to work in the case of 3d.  Indeed, as has been found in \cite{pasquettietal}, the
partition function of supersymmetric theories in 3d decomposes into blocks, exactly as in eqn.(\ref{sphere})\footnote{\ The
appearance of inverse power in the $S^3$ partition function and its absence in \cite{pasquettietal} has to do with the choice of analytic continuations used there versus what we have here.  In our case $|q|<1$ whereas the two blocks used
in \cite{pasquettietal} used $|q|<1$ for one block and $|q|>1$ for the other block.}.

Indeed, as noted in \cite{pasquettietal}, the chiral blocks are solutions to the difference equations arising
from the ring relations satisfied by the line operators.  This is also the case for us, in the ${\overline \beta} \rightarrow 0$,
as follows from eqn.(\ref{diffeq}).  Therefore in the same limit as in the 2d case the 3d $tt^*$ geometry
should reduce to the supersymmetric partition functions on $S^3_b$ and $S^2\times S^1$.  We will verify
this expectation for the partition function of free chiral theory in section 6.

\section{Examples of $tt^*$ geometry in 3 dimensions}

The 3d $tt^*$ geometries should correspond to doubly-periodic solutions of the monopole equations, or their higher-dimensional 
generalizations. In this section we illustrate the correspondence in a number of examples. 
\subsection{Free 3d chiral multiplet}\label{sec:free3d}

The simplest example, of course, is a free 3d chiral multiplet of real twisted mass $m$, whose $tt^*$ geometry should give a $U(1)$ monopole solution 
on the space parameterized by $m$ and the flavor Wilson lines on the two cycles of $T^2$. 
If one of the two circles in the compact geometry is very small, we expect to recover the results for the 2d chiral field. 
This identifies the monopole solution as a doubly--periodic Dirac monopole of charge $1$. Indeed, 
the 3d free chiral field compactified on a circle of length $R_y$ (which in the previous section
we had simply called $R$) may be expanded in KK modes having 2d complex masses
 \begin{equation}\boldsymbol{m}_n=m+\frac{2\pi i}{R_y}(n+y)\qquad n\in\Z,
 \end{equation}
 where $y$ is the flavor Wilson like along the circle, which is a periodic variable of period 1.  
 The 2d mirror  is then described by the (twisted) superpotential of the form \cite{HoriV}
\begin{equation}\label{freeKK}
W(Y_n)=\sum_{n\in\Z} \Bigg(\frac{1}{2}\!\left(\frac{m}{2\pi}+i\,\frac{n+y}{R_y}\right)Y_n-e^{Y_n}\Bigg).
\end{equation}
Since the modes $Y_n$ are decoupled from each other, the $tt^*$ metric is simply the product of the metrics for each mode which, as described  in \S.\,\ref{sec:basicexample}, correspond to periodic monopole solutions.
The doubly--periodic monopole solution associated to the 3d free chiral is then the superposition of an infinite array of periodic Abelian monopole solutions, each corresponding to the contribution from a 2d KK mode.
Thus the harmonic function giving the Higgs field in the monopole solution is 
\begin{equation}
V_{\mathrm{chiral}}(m, x,y) =- \pi \sum_{n,k}\left(\frac{1}{\sqrt{m^2 + \frac{4\pi^2}{R_x^2} (x+k)^2+ \frac{4\pi^2}{R_y^2} (y+n)^2}} -\kappa_{k,n}\right)+\Lambda
\end{equation}
where $R_x\equiv\beta$ is the length of the $tt^*$ circle, $\kappa_{k,n}$ some constant regulator, and $\Lambda$ a constant (see appendix \ref{app:3dchiralmetric} for full details). In a natural basis (which in the $R_y\rightarrow 0$ limit reduces to the standard 2d `point' basis), the $tt^*$ metric is simply
\begin{equation}\label{opbasis}
G_\text{chiral}(m,x,y)=\exp\!\left(\frac{2}{R_x} \int\limits_0^x V_\text{chiral}(m,x^\prime,y)\,dx^\prime\right).
\end{equation} 
As discussed in section \ref{sec:geo3dimensions}, $G_\text{chiral}(m,x,y)$ may be interpreted as the partition function on the infinitely elongated $S^2\times S^1$ geometry
with flavor twist parameters $x,y$ around the equator of $S^2$ and the $S^1$, respectively
\begin{equation}
Z_{S^2\times S^1}=G_\text{chiral}(m,x,y).
\end{equation}
At large $|m|$, the harmonic function $V_\text{chiral}$ has a linear growth: 
\begin{equation}
V_\text{chiral} = \frac{R_x R_y}{2}\,|m|+O\!\Big(\!\exp\!\big[-\min(R_x,R_y)\,|m|\big]\Big).
\end{equation}  
This is slightly inconsistent: it corresponds to $\pm \frac{1}{2}$ units of flux for the gauge bundle on the $x$--$y$ torus. 
This is closely related to the $\Z_2$ anomaly for a 3d free chiral: depending on the sign of the mass $m$, integrating away a 3d chiral 
leaves a background Chern-Simons coupling of $\pm \frac{1}{2}$ for the flavor $U(1)$ symmetry. 
In general, we expect the slope of the Higgs field at large values of the masses, or the units of flux on the flavor Wilson line tori,
to coincide with the effective low energy background CS couplings for the corresponding flavor symmetries. 

Notice that the spectral data computed from $D_x + i D_y$ and $-D_m + V$ involves a holomorphic connection on the torus, which is 
covariantly constant in the $m$ direction. Thus the topological data of the holomorphic bundle, \textit{i.e.} the Chern class on $T^2$, is $m$-independent, 
and can only jump at the location of Dirac monopoles, by an amount equal to the Dirac monopole charge.  This explains the 
slopes we find at large $|m|$. 

A better defined choice (the ``tetrahedron theory'' in \cite{DGG}) is a theory $\Delta$ of a 3d chiral together with an additional background 
CS level of $-\frac{1}{2}$. This corresponds to the harmonic function 
\begin{equation}\label{vdeltaim}
V_{\Delta}(m, x,y) =- \frac{R_x R_y}{2}\, m - \pi \sum_{n,k}\left(\frac{1}{\sqrt{m^2 + \frac{4\pi^2}{R_x^2} (x+k)^2+ \frac{4\pi^2}{R_y^2} (y+n)^2}} -\kappa_{k,n}\right)+\Lambda
\end{equation}
which has coefficients $-1$ or $0$ for the linear growth at $m \to -\infty$ and $m \to \infty$ respectively, and corresponding effective CS couplings.

The harmonic function $V_\Delta$ has alternative representations which converge more rapidly than \eqref{vdeltaim} and are convenient to study particular limits
(see appendix \ref{app:3dchiralmetric}). For instance, we have the Fourier representation
\begin{equation}\label{foureirdelta}\begin{split}
V_\Delta(m,x,y) =& - R_x R_y\, m\, \Theta(-m) -\\
&-\frac{1}{2}\sum_{(k,\ell)\neq (0,0)}\, \frac{R_x R_y}{\sqrt{R_y^2\, \ell^2+ R_x^2\, k^2} }\;e^{2 \pi i k x+ 2 \pi i \ell y -\sqrt{R_y^2 \ell^2+ R_x^2 k^2}\; |m|}.
\end{split}\end{equation}

If we treat the $y$ direction as ``internal'' and the other two as the standard directions of 2d $tt^*$, we can assemble the 
monopole connection and Higgs field into the usual
$tt^*$ quantities\footnote{\ For convenience, we absorb the overall dependence of the $tt^*$ geometry on the length $R_x$ in the definition of $C_\mu$.}. Comparing wit eqn.\eqref{unitary}, we have
\begin{equation}
\begin{aligned}
C_\mu &= R_x\Big(\partial_x - i A_x \Big)+ V\\
-\bar C_{\bar \mu}&= R_x\Big(\partial_ x - i A_x\Big) - V \\
 D_\mu &= \partial_\mu - i A_\mu \\
 D_{\bar \mu} &= \partial_{\bar \mu} - i A_{\bar \mu},
 \end{aligned}
\end{equation}
where 
\begin{equation}
\mu = \frac{1}{4\pi}(m+2\pi i\,y/R_y).
\end{equation}
As for the 2d chiral model in \S.\,\ref{sec:basicexample}, we perform the complex gauge transformation to the standard  `point basis' topological gauge. 
From the definition 
\begin{equation}
V(\mu, \bar \mu, x) = \frac{R_x}{2}\, v(\mu) + \frac{R_x}{2}\, \bar v(\bar \mu) + \frac{R_x}{2} \,\partial_x L(\mu, \bar \mu, x)
\end{equation}
we find 
\begin{equation}\label{fourerL}
L_{\Delta}(m,x,y) =-\frac{1}{2\pi i}\sum_{k,\ell\in\Z\atop k\neq 0}\frac{R_y}{k\,\sqrt{R_x^2\,k^2+R_y^2\,\ell^2}}\; e^{2 \pi i k x+ 2 \pi i \ell y -\sqrt{R_y^2 \ell^2+ R_x^2 k^2}\; |m|} 
\end{equation}
and 
\begin{equation}
v_{\Delta}(\mu)  =   \log\! \left(1- e^{- 4\pi R_y \mu}  \right).
\end{equation}
This is a natural regularization of the $\sum_n \log\!\big(\mu+i \frac{n}{2R_y}\big)$ arising from the KK tower. 
By the same token, we propose 
\begin{equation}
a_{\Delta}(\mu)  =  \frac{1}{2}\log \!\left(1- e^{-4\pi R_y \mu}  \right).
\end{equation}

We can then go to the ``point topological basis'' by the complexified gauge transformation with parameter $\frac{1}{2} L(\mu, \bar \mu, x)-\frac{1}{2} a - \frac{1}{2} \bar a$ (cfr.\! eqn.\eqref{tftgauge1}):
\begin{equation}\label{3dcomplxgauge}
\begin{aligned}
\frac{1}{R_x} C_\mu &= \partial_x + v(\mu)  \\
 -\frac{1}{R_x} \bar C_{\bar \mu} &= \partial_ x - \bar v(\bar \mu) - \partial_x L(\mu, \bar \mu, x) \\
 D_\mu &= \partial_\mu +\partial_\mu a(\mu) - \partial_\mu L \\
 D_{\bar \mu} &= \partial_{\bar \mu} 
\end{aligned}
\end{equation}
We recognize that $v_{\Delta}(\mu) = \partial_\mu W_\Delta(\mu)$, where 
\begin{equation}W_\Delta\propto \mathrm{Li}_2(e^{- m R_y -2\pi i y})
\end{equation}
 is the twisted effective superpotential for a compactified 3d chiral multiplet with a $-1/2$ CS level. 

\subsubsection{Evaluation of the elongated $S^3$ partition function $Z_{S^3}=S_{00}$}

In this example with a single vacuum, the ${S_0}^0$, or equivalently the partition function on the infinitely
elongated $S^3$
should reduce, as discussed in \S.\,\ref{sec:partitionsasgauge}, to the gauge transformation which relates the Abelian monopole fields in the topological gauge \eqref{3dcomplxgauge}
to the same fields written in the $\boldsymbol{S}$--dual topological gauge based on the 2d $tt^*$ geometry for the opposite choice of ``internal'' circle, in which the parameter $\mu$ is replaced by its dual 
\begin{equation}
\mu_x =\boldsymbol{S}(\mu)\equiv \frac{1}{4\pi}\!\left(m-\frac{2\pi i\,x}{R_x}\right).
\end{equation}
To simplify the notation, we shall denote the effect of the action of $\boldsymbol{S}$ on any quantity by a tilde, that is, for all quantities $f$ we set
\begin{equation}\tilde f(x,y,m,R_x,R_y)=\boldsymbol{S}f(x,y,m,R_x,R_y)\equiv f(y,-x,m,R_y,R_x).
\end{equation}
Gauge invariant scalar quantities $s$ satisfy $\tilde s=s$; in particular, $\tilde V\equiv V$.

To compare the topological gauge with its $\boldsymbol{S}$--dual it is convenient to preliminary transform these two complex gauges in the corresponding unitary gauges by the inverse of the gauge transformation in eqn.\eqref{3dcomplxgauge} of imaginary parameter
\begin{equation}
\frac{1}{2}K\equiv \frac{1}{2}(L-a-\bar a),\quad\text{resp.}\quad \frac{1}{2}\tilde K\equiv \frac{1}{2}(\tilde L-\tilde a-\bar{\tilde a}).
\end{equation}
The two dual unitary connections $A$ and $\tilde A$ are, respectively,
\begin{equation}\label{gauheunit}
\begin{aligned}
A_m&= -\frac{R_y}{4\pi} \partial_{y} K  \\
A_{y}&= \frac{\pi}{R_y} \partial_{m} K \\
A_x&=\frac{i}{2}(v-\bar v) \\
V&=\frac{R_x}{2}(v+\bar v)+\frac{R_x}{2}\partial_x K,
\end{aligned}\quad\text{resp.}\qquad\ 
\begin{aligned}
\tilde A_m&= \frac{R_x}{4\pi} \partial_{x} \tilde K \\
\tilde A_y&=\frac{i}{2}(\tilde v-\bar{\tilde v})\\
\tilde A_{x}&= -\frac{\pi}{R_x} \partial_{m} \tilde K\\
V&=\frac{1}{2}(\tilde v+\bar{\tilde v})+\frac{R_y}{2}\partial_y \tilde K,
\end{aligned}\end{equation}
which satisfy the monopole equations
\begin{equation}
F=\tilde F=\frac{1}{2\pi}\ast dV.
\end{equation}
These two $U(1)$ connections are gauge equivalent. Hence there is a real function $\Lambda$ such that
\begin{equation}\label{lamblamb}
A-\tilde A= d\Lambda.
\end{equation}
The complete gauge transformation between the $\boldsymbol{S}$--dual topological gauge and the original one, which by the analysis in \S.\,\ref{sec:partitionsasgauge} is the infinitely elongated $S^3$ partition function, is then the composition of the above three complex gauge transformations, that is,
\begin{equation}\label{sspart}
S=\exp\!\left(-\frac{1}{2}K+i \Lambda +\frac{1}{2}\tilde K\right).
\end{equation}
To compute $\Lambda$ one starts from the known Fourier series for $K$
\begin{equation}\label{Kfouerie}
\begin{split}
K=L-a-\bar a&=\sum_{k,\ell} K(k,\ell;R_x,R_y)\;e^{2\pi i (k x+\ell y)}=\\
&=-\frac{1}{2\pi i}\sum_{k,\ell\in\Z\atop k\neq 0} \frac{R_y}{k\,\sqrt{R_x^2\,k^2+R_y^2\,\ell^2}}\; e^{2 \pi i k x+ 2 \pi i \ell y -\sqrt{R_y^2 \ell^2+ R_x^2 k^2}\; |m|} +\\
&\quad +\frac{1}{2}\sum_{\ell\geq 1}\frac{e^{-\ell R_ym}}{\ell}\big(e^{2\pi i \ell y}+e^{-2\pi i \ell y}\big),
\end{split}
\end{equation}
and the corresponding one for $\tilde K$ with coefficients
\begin{equation}
\tilde K(k,\ell;R_x,R_y)\equiv K(-\ell,k;R_y,R_x).
\end{equation}
Inserting these Fourier expansions in eqn.\eqref{gauheunit} one gets the expansions of the unitary connections $A$ and $\tilde A$, and then we may read the Fourier series for $\Lambda$ from eqn.\eqref{lamblamb}. One gets
\begin{multline}\label{fourierLambda}
\Lambda= -\frac{1}{4\pi}\sum_{k,\ell\neq 0} \frac{1}{k\ell}\;e^{2 \pi i k x+ 2 \pi i \ell y -\sqrt{R_y^2 \ell^2+ R_x^2 k^2}\, m} -\\
-\frac{i}{4}\Big(\log(1-e^{-R_y m+2\pi i y})-\log(1-e^{-R_y m-2\pi i y})+\\
+\log(1-e^{-R_x m-2\pi i x})-\log(1-e^{-R_x m+2\pi i x})\Big),
\end{multline}
(in writing this equation we assumed $m>0$).

As discussed in \S.\,\ref{sec:partitionsasgauge}, the partition function $S$ given by \eqref{sspart} should also be equal to $\tilde\Pi\, \Pi^{-1}$
where $\Pi$ are the $\zeta=\pm 1$ brane amplitudes. We shall check the validity of this relation after the computation of the amplitude $\Pi$.

\subsubsection{Branes for the 3d free chiral theory}\label{freechiraltheory3d}\def\bm{\boldsymbol{m}}
As we saw in \S.\,\ref{sec:braneidentification}, the function $\Phi$ defined in eqn.\eqref{Phiintegralform} corresponds in 2d to either the Neumann or Dirichlet brane amplitude depending on the value of $\zeta$.
That analysis is important for the 3d free chiral theory with real twisted mass $m$, compactified on a circle of length $R_y$, which may be seen as a 2d $(2,2)$ model with an infinite collection of decoupled KK modes
as in eqn.\eqref{freeKK}.
Again, since the modes do not interact, the brane amplitudes are given by an infinite product of the single mode amplitudes of section \ref{thebasicbraneamplitude}  with 2d twisted masses
\begin{equation}\label{wtwistedmasses}
4\pi \mu_n= m+\frac{2\pi i}{R_y}(n+y),\quad n\in\Z.
\end{equation}
To select a reasonable 
boundary condition for the 3d chiral field, we need to choose the boundary conditions of the individual 2d KK modes $Y_n$ in a coherent way. 
The most obvious choice is to seek either Dirichlet or Neumann b.c.\! for the 3d chiral field. This means selecting either Dirichlet b.c. \!for all the KK modes $Y_n$, or Neumann for all the $Y_n$. 
The complete 3d ``Dirichlet''/``Neumann'' amplitudes are
\begin{equation}\label{whatamplitude}
\begin{aligned}
\begin{matrix}\log \langle x|D;\zeta\rangle\\
\log \langle x|N;\zeta\rangle
\end{matrix}=&
-\frac{R_x}{4\pi\,R_y} \bigg(\zeta\,\mathrm{Li}_2(e^{-m\,R_y-2\pi i\,y})-
\zeta^{-1}\,\mathrm{Li}_2(e^{-m\,R_y+2\pi i\,y})\bigg)+\\
&- \frac{R_xR_y}{16\pi}\left[\zeta\left(m+\frac{2\pi i\,y}{R_y}\right)^2+\zeta^{-1}\left(m-\frac{2\pi i\,y}{R_y}\right)^2\right]-\\
&-\frac{1}{2} \log\!\left[ 2\sinh\!\left(\frac{1}{2}\big(m\,R_y+2\pi i\, y\big)\right)\right]+\begin{matrix}\boldsymbol{\Phi}_D(m,x,y,R_x,R_y;\zeta)\\
\boldsymbol{\Phi}_N(m,x,y,R_x,R_y;\zeta).\end{matrix}
\end{aligned}\end{equation}
where $\boldsymbol{\Phi}_D$, $\boldsymbol{\Phi}_N$ are the sums over all KK modes of the 2d functions $\Phi$  with, respectively, Dirichlet and Neumann b.c. 
However, from \S.\ref{generalitiesonbrane} we know that the physically interesting amplitudes, corresponding to proper Neumann/Dirichlet branes in the 3d sense, are the ones at fixed points of the Cayley transform $\mathsf{C}$, namely $\zeta=\pm1$. 

In order to write a sum over the KK modes having better convergence properties, it is convenient to rewrite the integral representation of the 2d thimble amplitude function $\Phi$ in a slightly more general form
\begin{equation}\label{Phiintegralform2}
\begin{split}
\Phi= &\frac{1}{2\pi i}\int_{L} \frac{dt}{t-i \zeta} \;\log\!\Big(1-e^{-2\pi(\mu t+\bar\mu t^{-1}-ix)}\Big)-\\
&\quad-\frac{1}{2\pi i}\int_{L} \frac{dt}{t+i \zeta} \;\log\!\Big(1-e^{-2\pi(\mu t+\bar\mu t^{-1}+ix)}\Big),
\end{split}
\end{equation}
where $L= e^{i\phi}\,\R_+$ is a ray in the complex plane such that: \textit{i)}
$\mathrm{Re}[t\,\mu]>0$ for $t\in L$ and \textit{ii)} the integrand has no pole in the angular sector $0\leq \arg t\leq \phi$.

Assuming the real mass $m$ in eqn.\eqref{wtwistedmasses} to be positive, and setting
\begin{equation}\label{defZz}
z=m R_x+2\pi i\frac{R_x}{R_y}y,
\end{equation}
we write the 3d function $\boldsymbol{\Phi}[\zeta]$ in  the form
\begin{equation}\label{eeexprr}
\begin{split}
\boldsymbol{\Phi}[\zeta]=&
 \frac{1}{2\pi i}\int_{L^-} \frac{dt}{t-i\, \zeta} \;\log\prod_{n<0}\Big(1-e^{2\pi i x-zt/2-\bar z t^{-1}/2-\pi i n\frac{R_x}{R_y}(t-t^{-1})}\Big)+\\
&+  \frac{1}{2\pi i}\int_{L^+} \frac{dt}{t-i\, \zeta} \;\log\prod_{n\geq 0}\Big(1-e^{2\pi i x-zt/2-\bar z t^{-1}/2-\pi i n\frac{R_x}{R_y}(t-t^{-1})}\Big)-\\
& -\frac{1}{2\pi i}\int_{L^-} \frac{dt}{t+i\, \zeta} \;\log\prod_{n<0}\Big(1-e^{-2\pi i x-zt/2-\bar z t^{-1}/2-\pi i n\frac{R_x}{R_y}(t-t^{-1})}\Big)-\\
& -\frac{1}{2\pi i}\int_{L^+} \frac{dt}{t+i\, \zeta} \;\log\prod_{n\geq 0}\Big(1-e^{-2\pi i x-zt/2-\bar z t^{-1}/2-\pi i n\frac{R_x}{R_y}(t-t^{-1})}\Big)-
\end{split}
\end{equation}
where $L^-$ is a ray in the upper--right quadrant and $L^+$ in the lower--right quadrant, and we assume that the poles at $t=\pm i\,\zeta$ are not in the angular sector bounded by $L^+,L^-$ and containing the positive real axis. Note that for the physical values, $\zeta=\pm 1$, the rays $L^\pm$ may be chosen arbitrarily in the respective quadrants.

From the discussion in \S.\,\ref{sec:braneidentification}, we know that this expression corresponds to a ``Dirichlet'' amplitude for $\mathrm{Re}\,\zeta<0$ and a ``Neumann''; amplitude for $\mathrm{Re}\,\zeta>0$
\begin{equation}
\boldsymbol{\Phi}[\zeta]=\begin{cases}\boldsymbol{\Phi}_N[\zeta] &\text{for } \mathrm{Re}\,\zeta>0\\
\boldsymbol{\Phi}_D[\zeta] &\text{for } \mathrm{Re}\,\zeta<0.
\end{cases}
\end{equation}
 To get the ``Neumann'' (resp.\! ``Dirichlet'') amplitude in the opposite half--plane one has to analytically continue the above expression, by deforming the contours while compensating the discontinuity each time one crosses a pole of the integrand.
  The physical amplitudes are then obtained by specializing the result to $\zeta=\pm1$.

The expression \eqref{eeexprr} may be written in a more suggestive form by 
introducing the compact quantum dilog function
\begin{equation}
\boldsymbol{\Psi}(z,q)\equiv (z\,q^{1/2};q)_\infty=\prod_{n=0}^\infty (1-z q^{n+1/2}),
\end{equation}
the product being convergent for $|q|<1$.
Then
\begin{equation}
\begin{aligned}\label{Phi3d2}
\boldsymbol{\Phi}[\zeta]&= \frac{1}{2\pi i}\int_{L^-} \frac{dt}{t-i\,\zeta}
 \;\log \boldsymbol{\Psi}(e^{2\pi ix -z t/2- \bar zt^{-1}/2- i\frac{\pi R_x}{2R_y}(t-t^{-1})},\; e^{i\pi \frac{R_x}{R_y}(t-t^{-1})})  \\
&+\frac{1}{2\pi i}\int_{L^+} \frac{dt}{t-i\,\zeta} \;\log\boldsymbol{\Psi}(e^{2\pi i x-z t/2-\bar zt^{-1}/2- i\frac{\pi R_x}{2R_y}(t-t^{-1})},e^{-i\pi \frac{R_x}{R_y}(t-t^{-1})})
\\
 &-\frac{1}{2\pi i}\int_{L^-} \frac{dt}{t+i\,\zeta} \;\log\boldsymbol{\Psi}(e^{2\pi i x-z t/2-\bar zt^{-1}/2-i \frac{\pi R_x}{2R_y}(t-t^{-1})},e^{i\pi \frac{R_x}{R_y}(t-t^{-1})})
\\
 &-\frac{1}{2\pi i}\int_{L^+} \frac{dt}{t+i\,\zeta} \;\log\boldsymbol{\Psi}(e^{-2\pi i x-z t/2-\bar zt^{-1}/2- i\frac{\pi R_x}{2R_y}(t-t^{-1})},e^{-i\pi \frac{R_x}{R_y}(t-t^{-1})}),
\end{aligned}
\end{equation}
all integrals being absolutely convergent for $L^\pm$ as above.

\paragraph{The asymmetric UV limit.} The asymmetric limit of the amplitudes $\langle x|N,\zeta\rangle$, $\langle x|D,\zeta\rangle$ as `$\bar\beta\rightarrow 0$' is given by a regularized sum of the asymmetric limit for each KK mode, and is computed in appendix \ref{appe:asym3d}. Not surprisingly, the limit is a quantum dilogarithms
\begin{equation}
\begin{split}
&\log\Pi_{3d}(\zeta=-1)=-\log\boldsymbol{\Psi}\big(e^{-m R_y-2\pi i y-4\pi x- 2\pi R_y/R_x};\; e^{-4\pi R_y/R_x}\big)-\\
&\qquad-\frac{R_y}{4 R_x}\left(m+\frac{2\pi i R_x}{R_y}y\right)-x \log\sinh\!\left[\frac{1}{2}\left(m R_y+2\pi i y\right)\right]+\mathrm{const.},
\end{split}
\end{equation}

Indeed, by the same argument as in \S.\,\ref{sec:braneidentification} in this limit the amplitude is holomorphic in $z=mR_x+2\pi i R_x y/R_y$ and it satisfies a difference equation of the form 
\begin{equation}\begin{split}
\Pi_{3d}(z+4\pi)&= \left(\prod_{n\in\Z} \left(\frac{z}{4\pi}+\frac{i R_x n}{2R_y}\right)\right)\Pi_{3d}(z)=\\
&=\Big(1-e^{-mR_y-2\pi i y}\Big)e^{(mR_y+2\pi i y)/2}\;\Pi_{3d}(z),
\end{split}\end{equation}
where the factor $e^{(mR_y+2\pi i y)/2}$ may be understood as arising from the $\Z_2$ anomaly of the free chiral at CS level zero.  Not only this result confirms that in this asymmetric limit we obtain the result for the partition function
of free chiral theory on $S^3_b$ and $S^2\times S^1$ which are made of quantum dilogs, when the twist parameters $x,y=0$, but it also predicts when
$x,y\not=0$ the result for these partition functions with twist line operators inserted at the two ends of the sphere.
 
 \subsubsection{The $\zeta=\pm 1$ amplitudes and the $S$--gauge transformation}\label{eqalitys00} 

In this subsection we check that the explicit expression \eqref{Phi3d2} for the brane amplitudes $\Pi[\zeta=\pm1]$ satisfies the expected relation with the elongated $S^3$ partition function, that is, the equality
\begin{equation}\label{strangererel}
\log\Pi[\zeta=\pm 1]-\log\tilde\Pi[\zeta=\pm 1]=\log S,
\end{equation}
where $S$ is the gauge transformation which, as described in \S.\,\ref{sec:partitionsasgauge}, gives the partition function on the elongated $S^3$.

We know explicitly the \textsc{rhs} of  eqn.\eqref{strangererel}  in the form of a double Fourier series
\begin{equation}
\log S=\sum_{k,\ell\in\Z} c(m; k,\ell)\; e^{2\pi i(k x+\ell y)},
\end{equation} 
while the \textsc{lhs} is known in the form of the integral representations \eqref{eeexprr}\eqref{Phi3d2}. 
The easiest way to check the validity of the equality \eqref{strangererel} is to compute the Fourier coefficients of the \textsc{lhs}, which is known to be a periodic function of $x,y$, and compare them with the $c(m;k,\ell)$'s.  
The Fourier coefficients $c(m;k,\ell)$ may be read from eqns.\eqref{strangererel}\eqref{Kfouerie}\eqref{fourierLambda}; for\footnote{\ The terms with $k\ell=0$ correspond to purely holomorphic gauge transformations which just change the holomorphic basis in the chiral ring $\mathcal{R}$ and hence are convention dependent.} $k\ell\neq0$ they are
 \begin{equation}\label{fourier}
 \begin{split}
- 4\pi i \,c(m;k,\ell)= &\left(\frac{R_x}{\ell\,\sqrt{R_x^2k^2+R_y^2\ell^2}}-\frac{1}{k\ell}+\frac{R_y}{k\,\sqrt{R_x^2k^2+R_y^2\ell^2}}\right)\times\\
&\qquad\qquad\quad\times
\exp\!\left(-\sqrt{R_x^2k^2+R_y^2\ell^2}\;|m|\right).
 \end{split}\end{equation}
 
The $k\ell\neq 0$ coefficients in the Fourier expansion of  $\log\Pi[\zeta=\pm1]$ coincide with the coefficients in the Fourier  series of the non--trivial part of the amplitude
\begin{equation}
\boldsymbol{\Phi}[\zeta=\pm1]=\sum_{k,\ell\in\Z} \boldsymbol{\Phi}_\pm(k,\ell)\, e^{2\pi i(kx+\ell y)}.
\end{equation}
We compute the coefficients $\boldsymbol{\Phi}_\pm(k,\ell)$ for $k>0$; the ones for $k<0$ are similar. The terms with $k>0$ arise from the first two integrals in eqn.\eqref{eeexprr}. Expanding in series the integrands, they become
\begin{multline}
-\frac{1}{2\pi i}\sum_{k\geq 1}\frac{e^{2\pi i k x}}{k}\sum_{n\geq 0}\int_{L^+} \frac{dt}{t-i\zeta}\:e^{-k z\, t/2 -k \bar z \,t^{-1}/2- \pi i \frac{R_x}{R_y}(t-t^{-1})kn}-\\
-\frac{1}{2\pi i}\sum_{k\geq 1}\frac{e^{2\pi i k x}}{k}\sum_{n< 0}\int_{L^-} \frac{dt}{t-i\zeta}\:e^{-k z\, t/2 -k \bar z \,t^{-1}/2- \pi i\frac{R_x}{R_y}(t-t^{-1})kn},
\end{multline}
which is already in the Fourier series form with respect to $x$. To get the double Fourier series one has to Poisson re--sum the KK modes. 
To do that, we deform the contours $L^\pm$ to their original position on the positive real axis. For $\zeta=\pm 1$ we get
\begin{equation*}
\begin{split}
&-\frac{1}{2\pi i}\sum_{k\geq 1}\frac{e^{2\pi i k x}}{k}\int_0^\infty \frac{dt}{t\mp i}\:e^{- k z\, t/2 - k \bar z \,t^{-1}/2}\sum_{n\in\Z}e^{- i\pi \frac{R_x}{R_y}(t-t^{-1})kn}=\\
&=-\frac{1}{2\pi i}\sum_{k\geq 1}\frac{e^{2\pi i k x}}{k}\sum_{\ell\in\Z}\int_0^\infty \frac{dt}{t\mp i}\:e^{- k z\, t/2 -k \bar z \;t^{-1}/2}\;\delta\!\left(\frac{R_x}{2\,R_y}k(t-t^{-1})-\ell\right)
=\\
&= -\frac{1}{2\pi i}\sum_{k\geq 1}\frac{e^{2\pi i k x}}{k}\sum_{\ell\in\Z}\int_0^\infty 
\frac{2R_y(t\pm i)\,dt}{R_xk(t+t^{-1})^2}\;e^{- k z\, t/2 -k \bar z \;t^{-1}/2}\;\delta\!\left(t-\frac{R_y\ell}{R_xk}-\sqrt{1+\frac{R_y^2\ell^2}{R_x^2k^2}}\right).
\end{split}
\end{equation*}
Using eqn.\eqref{defZz}, and recalling that we are assuming $m>0$,
the above expression becomes
\begin{equation}
-\frac{1}{4\pi i}\sum_{k\geq 1}\sum_{\ell\in\Z} \frac{R_y[\ell R_y\pm i k R_x+\sqrt{k^2R_x^2+\ell^2R_y^2}]}{k(k^2R_x^2+\ell^2R^2_y)}\;e^{2\pi i k x-2\pi i \ell y-|m|\sqrt{k^2R_x^2+\ell^2R_y^2}},
\end{equation}
so the $k\ell\neq 0$, $k>0$ Fourier coefficients are
\begin{equation}
\boldsymbol{\Phi}_\pm(k,\ell; R_x,R_y)= -\frac{1}{4\pi i}\,\frac{R_y[\pm i k R_x-\ell R_y+\sqrt{k^2R_x^2+\ell^2R_y^2}]}{k(k^2R_x^2+\ell^2R^2_y)}\;e^{-|m|\sqrt{k^2R_x^2+\ell^2R_y^2}}.
\end{equation}
For $k>0$, $\ell<0$ one has
\begin{equation}
\begin{split}
&\boldsymbol{\Phi}_\pm(k,\ell;R_y,R_x)-\boldsymbol{\Phi}_\pm(-\ell,k;R_x,R_y)=\\
&=-\frac{1}{4\pi i}\!\!\left(-\frac{1}{k\ell}+\frac{R_y}{k\sqrt{k^2R_x^2+\ell^2R_y^2}}+\frac{R_x}{\ell\sqrt{k^2R_x^2+\ell^2R_y^2}}\right)e^{-|m|\sqrt{k^2R_x^2+\ell^2R_y^2}}
,
\end{split}
\end{equation}
which, comparing with eqn.\eqref{fourier}, gives the equality \eqref{strangererel}.

 \subsection{The $\C P^1_0$ sigma model}
The next obvious step would be to seek a model which gives a smooth $SU(2)$ doubly-periodic monopole as the 3d $tt^*$ geometry. 
In 2d we used the mirror to the $\C P^1$ gauged linear sigma model for a similar purpose. 
It is natural to look at the 3d version of the same theory: a 3d $U(1)$ gauge theory coupled to two chiral multiplets of charge $1$. 
This theory has two flavor symmetries: an $SU(2)_m$ flavor symmetry with mass $m$, which rotates the chiral doublet, 
and an $U(1)_t$ ``topological'' flavor symmetry with mass parameter equal to the FI parameter $t$ for the theory. 
In order to define the theory fully, we need to select a Chern-Simons level for the theory. We select level $0$ for now. 

This theory happens to enjoy surprising mirror symmetry properties. These mirror symmetries are manifest in the branches of vacua which appear for special choices of the mass parameters. 
If we turn on a positive FI parameter and no $SU(2)$ mass $m$, the theory has a standard $\C P^1$ moduli space of vacua, where the chiral fields receive a vev controlled by the FI parameter. 
If we turn on a mass parameter $m$, we can integrate out the chirals and seek for a Coulomb branch for the theory.
As long as the Coulomb branch scalar $\sigma$ is in the interval $2|\sigma|<m$, integrating away the chirals of opposite flavor charge 
gives no net Chern-Simons coupling for the $U(1)$ gauge field, but produces a mixed CS coupling between the gauge and flavor symmetry, 
which shifts the effective FI parameter to $t + |m|$. If we tune the mass parameters so that $t = - |m|$, we find 
a Coulomb branch with the topology of $\C P^1$ XXX [refs?]. 
In conclusion, the theory has three $\C P^1$ branches of vacua, which appear along the rays 
\begin{align}
m&=0 \qquad t>0 \cr
t+m &=0 \quad m>0 \cr
t-m &=0 \quad m<0
\end{align}

The mirror symmetries of the theory coincide with the permutation group of the three branches, and act on the mass parameters as 
the Weyl group of $SU(3)$ acts on the Cartan generators $(\frac{2}{3} t, \frac{1}{2}m-\frac{1}{3} t, -\frac{1}{2}m -\frac{1}{3} t)$. 
Indeed, the mirror symmetries imply that the $U(1)_t \times SU(2)_m$ flavor group in the UV is promoted to an $SU(3)$ flavor group in the IR. 

The full $tt^*$ geometry should thus enjoy the same $S_3$ Weyl symmetry acting over the combined parameter space $\R^3_m \times \R^3_t$.
It is thus more natural to describe the $tt^*$ geometry as a bundle over $\R^3 \otimes \mathfrak{sl}(3)$.
The theory has two vacua, and thus the bundle will be of rank two. Inspection of the spectral data computed in the previous section
shows that the bundle has structure group $SU(2)$. The $S_3$ symmetry of the spectral data can be checked with some patience.

At fixed $t$, we can look at the bundle on $\R^3_m$: the asymptotic behaviour of the Higgs field at large $|m|$ is $\mathrm{diag}(|m|/2-t/2,-|m|/2+ t/2)$,
which is compatible with a single smooth doubly-periodic $SU(2)$ monopole. The $t$ parameter controls the constant subleading asymptotics 
of the Higgs field. The half-integral slope at large $|m|$ is consistent for an $SU(2)$ bundle: it corresponds to the minimal possible Chern class 
of an $SU(2)$ bundle on $T^2$. 

At fixed $m$, we can look at the bundle on $\R^3_t$: the asymptotic behaviour of the Higgs field at large positive $t$ is $\mathrm{diag}(t,-t)$,
at large negative $t$ (where the theory approaches a $\C P^1$ sigma model) is $\mathrm{diag}(m/2,-m/2)$. Thus at large negative $t$ the Higgs field goes to a constant
diagonal vev, controlled by the parameter $m$. If $m$ is set to zero, one finds instead a more complicated non-Abelian asymptotic behaviour, which is presumably associated to the 
low energy massless degrees of freedom of the $\C P^1$ sigma model. The asymptotics are again compatible with a single smooth doubly-periodic $SU(2)$ monopole.
The $m$ and $t$ monopole geometries differ by the choices of Chern classes for the $T^2$ bundle at infinity.  

It is also interesting to consider generalizations of this model with other Chern-Simons levels. We will do so in a later section, after we acquire some extra tools. 

\subsection{Main Example:  Codimension 2 Defects}\label{Main}
It turns out that the $tt^*$ geometries in 2, 3 and 4 dimensions, can all be exemplified in the context codimension
2 defects of 4, 5 and 6 dimensional theories supporting 4 supercharges, which arise in the context of geometric engineering \cite{KKV,KMV,AV}.  There are two equivalent
descriptions of this class of theories.  One starts either with  M-theory on a local Calabi-Yau threefold, or equivalently
\cite{LV}, with a network of $(p,q)$ 5-branes of type IIB \cite{Hana}.  This gives a theory in 5 dimensions.   One then considers
codimension 2 defects of this theory.  In the M-theory setup, this corresponds to wrapping M5 branes
over Lagrangian 3-cycles of CY, leading to a $3d\subset 5d$ defect, or in the $(p,q)$ web description
it can be viewed as D3 brane ending on the web.

This can lead to codimension 2 defects in 6 and 4 dimensions, and in particular
to 4 dimensional defect probes of $(2,0)$ and $(1,0)$ supersymmetric theories in 6d as follows:  Using M-theory/F-theory
duality, by restricting to elliptic CY, this would correspond to $4d\subset 6d$ defects \cite{HoIV,Mstring1,Mstring2}.
This is equivalent, in the $(p,q)$ 5-brane web, to requiring the space to be periodic in one of the directions
that the 5-branes wrap.  To obtain the 4 dimensional theories, one simply considers type IIA on the corresponding
Calabi-Yau, by compactifying the M-theory on the circle.  In this context the $(p,q)$ web becomes
the skeleton of the associated Seiberg-Witten curve of the theory, and the 2d defects are associated to surface defects
parameterized by points on the curve \cite{AV,AKV}.  This can also be described in purely
gauge theoretic terms \cite{GGS}. 

\begin{figure}
\centering
\includegraphics[width=.8\textwidth]{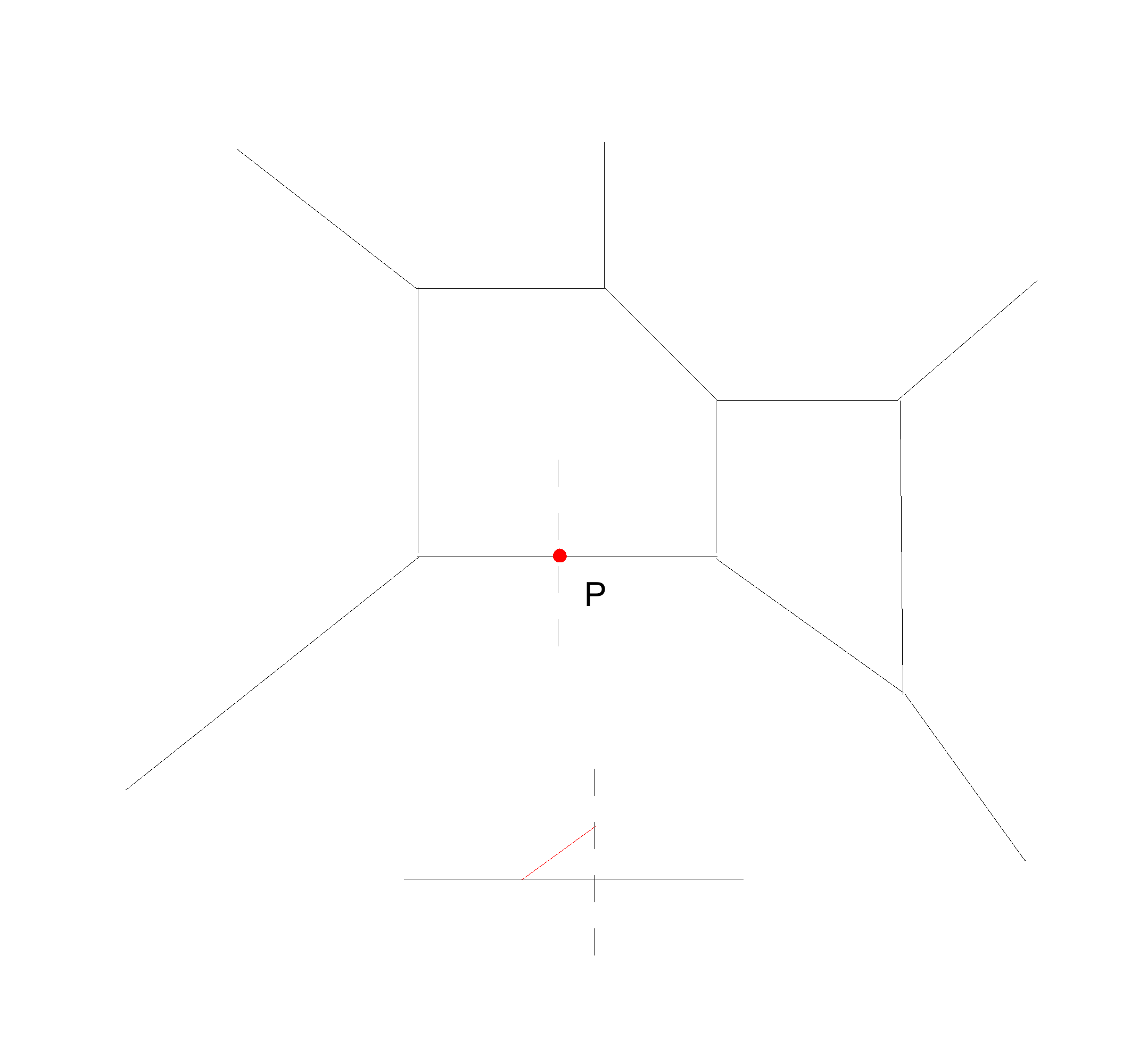}
\caption{A web of $(p,q)$ 5-branes engineers a 5d theory.  D3 branes (red line) suspended between the
web and a spectator brane (dashed line) gives rise to a 3d theory which can be viewed as a defect
of the 5d theory.  Changing the slope of the spectator brane corresponds to the $SL(2,\Z)$ action
on the 3d theory.}
\label{fig:web}
\end{figure}

In this section we focus on the $3d \subset 5d$ defects, and use the $(p,q)$ 5-brane web description, which is particularly
convenient for our purposes (see Fig. \ref{fig:web}).

\begin{figure}
\centering
\includegraphics[width=.8\textwidth]{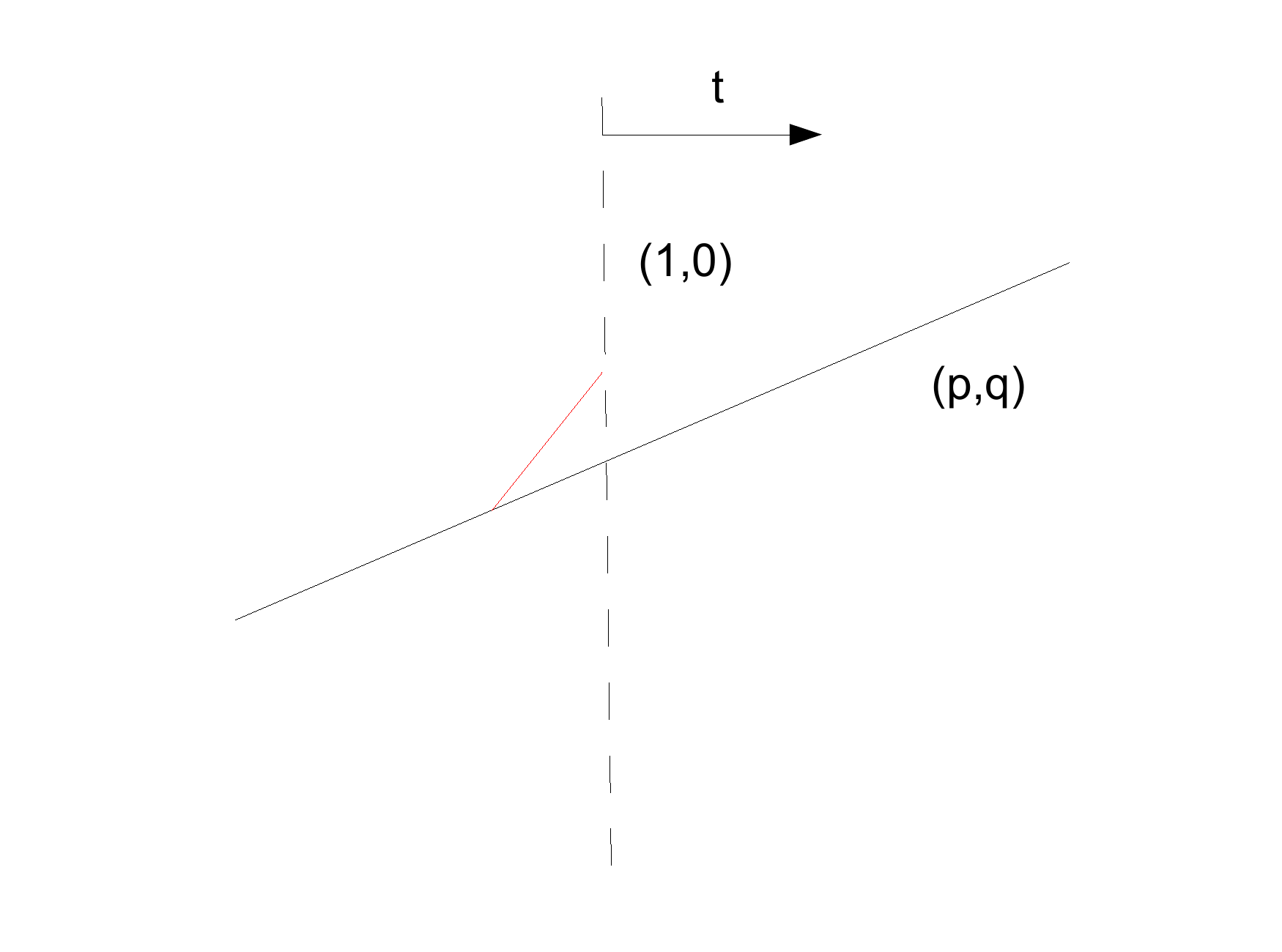}
\caption{The suspended D3 brane gives rise to a $U(1)$ Chern-Simons theory at level $v\wedge w=q$.
Moving the spectator brane by $t$ corresponds to changing the FI term by $t$.}
\label{fig:brane}
\end{figure}

The slope of the $(p,q)$ 5-brane is $p/q$ due to supersymmetry (at type IIB coupling constant $\tau=i$).
  We consider the 3d theory obtained
by having an extra D3 brane ending on the web. To make the theory dynamical we need
a finite length D3 brane and for this purpose we need an extra spectator brane (these were originally introduced in 
\cite{AKV} in the M-theory context and was related to framing of the associate knot invariants).  In particular if we have a $v=(1,0)$ brane and a $w=(p,q)$ brane with
a D3 brane stretched between them we get an ${\cal N}=2$ supersymmetric Chern-Simons
theory in 3d with CS level $v\wedge w= q$ with the associated monopole flavor symmetry with CS level at level $p$ \cite{CCV}  (see Fig. \ref{fig:brane}).

  In particular the $SL(2,{\Z})$ action of Witten \cite{W3} on the space of 3d theories with $U(1)$ flavor
symmetry corresponds to $SL(2,{\Z})$ action on the spectator brane, where the $T$ operation adds a unit background CS coupling and 
the $S$ operation gauges the flavor symmetry.  
For definiteness we will take the spectator brane to be a $(1,0)$, and act by $SL(2,{\Z})$ on the
rest of the web.

Consider the case of $(p,q)=(1,k)$.  This is a pure $U(1)_k$ Chern-Simons theory. 
The 3d theory is massive and the $\sigma$ field is frozen at $k\sigma = t$.  The changing of $t$ corresponds
to moving the spectator brane (see Fig. \ref{fig:brane}).  In this context the above relation gets interpreted as follows:
$(t,s)$ can be viewed as the $(x,y)$ component of the D3-brane, which is at the intersection of the projection
of the two 5-branes on the plane.

We can also consider compactifications of these theories on a circle.  The corresponding web geometry becomes the Seiberg-Witten curve
which generically takes the form
$$f(s,t)=\sum c_{n,m}\,e^{nt+ms} =\sum c_{mn}\, T^n S^m=0$$
where $S=e^s, \ T=e^t$.  In particular the points $(n,m)\in\Z^2$ such that $c_{m,n}\neq 0$ form a convex polygon.  Moreover the semi-infinite 5-branes correspond to pairs of adjacent points on the edges of the polygon.  If $(n_1,m_1)$ and $(n_2,m_2)$ are two adjacent points on the edge of the polygon, there is a $(p,q)$
5-brane with 
$$(p,q)\wedge (n_1-n_2,m_1-m_2)=p(m_1-m_2)-q(n_1-n_2)=0$$ 
(for a recent discussion see \cite{Vrec}).
Moreover, from the 3d probe theory we get a 2d theory with $(2,2)$ supersymmetry.
As was shown in \cite{AV} the corresponding Seiberg-Witten curve can be interpreted
as the spectral curve of the 2d theory in the following sense:  The 2d theory has $t$ as a parameter
and has a field $\Sigma$.  Moreover $f(s,t)=0$ corresponds to the spectrum of $\Sigma=s$ for
the fixed value of $t$.  In other words there is a superpotential $W(\Sigma,t)=W_0(\Sigma)+t\Sigma $ which satisfies
$$\Sigma=-\partial_{t} W $$ 
and 
$$\partial_\Sigma W_0+t =0\ \Longleftrightarrow\  f(s,t)=0.$$
Another way of saying this is that locally solving $t(s)$ using $f(s,t)=0$ leads to solving for a branch of $W$
$$W(\Sigma)=\int^{\Sigma} t(s)\, ds -t\Sigma. $$
In other words, the spectral geometry of the line operators of the 3d theory wrapped around the circle
is the SW curve.
Note that there are in general multiple vacua.  For example, considering the $U(1)_k$ theory discussed
above, upon compactification on a circle we find $k$ vacua, where 
the spectral curve becomes 
\begin{equation}
e^{k s} = e^{t}.
\end{equation}
In other words the 3d loop operator $S$ satisfies the relation $S^k=T$.  Defining ${\tilde S}=S/(T^{1/k})$
we see that
$${\tilde S}^k=1$$
We recognize this as the Verlinde algebra of $U(1)_k$ \cite{verlinealg}.  Indeed
this is the familiar result for the loop operator of a $U(1)$ Chern-Simons theory at level $k$ \cite{wittenCS},
where $\tilde S$ is equivalent to the Wilson loop operator wrapped around the circle, in the fundamental representation of $U(1)$.   

In fact we can do more:  We can suspend $N$ D3 branes between the 5-branes.  In this case we would get an ${\cal N}=2$
$U(N)$ Chern-Simons gauge theory at level $k$.  In this context the field ${\tilde S}$ should be viewed as an $N\times N$ matrix valued
loop operator.  The relation ${\tilde S}^k=1$ still holds.  This means that we choose $N$ eigenvalues at $k$-th roots of unity.
Using the fact that gauge symmetry acts as permutation of the eigenvalues we see that the number of inequivalent
vacua are now given by 
$${k(k+1)...(k+N-1)\over {N!}}$$
which is the same as the dimension of the Verlinde algebra for $U(N)$ conformal theory at level $k$.
Indeed the resulting ring of the line operators is isomorphic to the Verlinde ring.

We now wish to study the $tt^*$ geometry of these 3d systems.   Having a single probe will
lead to 3d monopole systems on $T^2\times \R$, \textit{i.e.}\! doubly periodic monopole $SU(n)$ systems
if we have $n$ vacua\footnote{The construction can probably be generalized to other classical groups by a judicious use of orbifolds.}.
Indeed such a system was already studied in \cite{CW} and in particular it was noted there that the
spectral curve associated to the doubly--periodic monopole equations are captured
by the SW curve of the above physical system.
Here we are finding a physical explanation of why the corresponding web appeared
as part of a solution to the monopole equations.
We now give a brief review of their results. We refer to reference \cite{CW} for more details.  

The basic idea is that the moduli spaces of doubly-periodic monopoles on $\left(\R \times T^2 \right)_t$ are labelled by the 
coefficients $Q_\pm$ of the linear growth of the Higgs field at large $|t|$, by the constant subleading coefficients $M_\pm$ 
in the Higgs field at large $|t|$, and by two sets of angles $p_\pm$ and $q_\pm$ which from our point of view combine with $M_\pm$ to give other doubly-periodic deformation directions of the $tt^*$ geometry.
Other parameters are the locations $t_i$ in $\left(\R \times T^2 \right)_t$ of the Dirac monopole singularities. These give other doubly-periodic  deformation directions of the $tt^*$ geometry.
A certain linear combination of these parameters is redundant: a translation of $\left(\R \times T^2 \right)_t$ will in general shift the $r_i$ and possibly the $(M,p,q)$ by multiples of the $Q$.

The spectral curve for the doubly-periodic geometry is then given by an equation of the general form 
\begin{equation}
\sum c_{n,m} e^{n t} e^{m s} =0
\end{equation}
where the $(n,m)$ integer points for non-zero $c_{n,m}$ form a convex Newton polygon in the plane. 
The shape of the polygon encodes the $Q_\pm$ coefficients and the coefficients on the boundary of the 
polygon encode the (complexified) $M_\pm$ and $t_i$ data. The coefficients of the interior coefficients are moduli of the 
periodic monopole configuration. More precisely, the monopole moduli space is parameterized by a choice of spectral curve 
with given $(M,p,q,t_i)$ and of a line bundle on it.  Each interior point of the Newton polygon gives two complex parameters: 
a coefficient in the spectral curve and a modulus for the line bundle. Indeed, the monopole moduli space is an hyperK\"aler manifold. 

It is now clear that the same geometry is describing the $tt^*$ solutions of 3d theory on the probes of our 5-brane web system
compactified on $T^2$, where the real Coulomb branch moduli of the bulk 5d theory combine with the gauge Wilson lines and the dual photons 
to give the hyperK\"ahler geometry of the doubly-periodic monopole moduli space. The mass deformation parameters correspond to the $M_\pm$ and $t_i$ 
parameters.   As already noted, the probes are a $D3$ brane segment stretched from the $(p,q)$ brane web to a separate $(1,0)$ brane 
lying on a plane parallel to the plane of the web and the position of the $(1,0)$ brane the $D3$ brane ends on becomes the (FI) mass parameter $t$ and the 
$tt^*$ geometry corresponding to the $t$ deformation becomes the doubly-periodic monopole geometry. 

We can now reinterpret our previous examples as brane webs, and then add a few more. 

A single 3d chiral multiplet, or better the $T_\Delta$ theory, can be engineered by a web including a $(-1,0)$, a $(0,1)$ and a $(1,-1)$ fivebranes
coming together to a point \cite{CCV}. This configuration is rigid. The obvious $\Z_3$ symmetry generated by the $ST$ $SL(2,\mathbb{Z})$ duality transformation
corresponds to the basic mirror symmetries of the $T_\Delta$ theory \cite{DGG}. To be precise, if the $D3$ brane probe ends on an $(1,0)$ brane parallel to the web we get the description of the theory as an $U(1)$ CS theory at level $1/2$, coupled to a single chiral of charge $1$. 

The slopes of the fivebranes, $-1$ for negative $t$ and $0$ for positive $t$ match the background CS couplings for the $U(1)_t$ flavor symmetry and the 
asymptotic values of the Higgs field in the monopole solution. The spectral curve is 
\begin{equation}
e^{s} = 1-e^{-t}
\end{equation}
This is also consistent with the relation we found for the loop operator associated with a chiral field with mass parameter $t$, which we obtain by ungauging the $U(1)$, by converting the spectator $(1,0)$ brane to a $(0,1)$ brane.

Next, we can look at an $U(1)_{k-\frac{1}{2}}$ Chern-Simons theory coupled to a chiral of charge $1$. 
For large negative $t$ we have two branches of vacua in flat space: either $\sigma =0$ and the chiral gets a vev or we integrate away the chiral and we have an effective $CS$ level $k-1$ and $(k-1) \sigma = t$. 
For large positive $t$ we have one branch of vacua only: we integrate away the chiral and have an effective $CS$ level 
$k$, with $k \sigma =t$. Thus we expect a brane system with a $(1,k)$ fivebrane, a $(0,1)$ fivebrane and a $(1,k-1)$ fivebranes. 

As our next example we consider the brane description for the 3d $\C P^1$ gauge theory with twisted mass. 
In order to describe the algebra of the wrapped loop operators when we compactly the theory to 2d on $S^1$ we view the 3d model as a 2d model with infinite towers of KK modes. The (twisted) superpotential is
\begin{equation}
W=\sum_{n\in\Z}\Big[e^{Y^+_n}+e^{Y^-_n}-\big(i\, n+t_2+\Sigma\big)Y^+_n-\big(i\,n-t_2+\Sigma\big)Y^-_n\Big]+2\pi \,t_1 \Sigma.
\end{equation}
There are two distinct vacua satisfying
\begin{equation}
\left\{\begin{aligned} &\cosh(2\pi \Sigma_0)=\cosh(2\pi t_2)+\frac{1}{2}\,e^{2\pi t_1}\\
& Y^\pm_n=\log(\Sigma_0+i\,n\pm t_2).
\end{aligned}\right.
\end{equation}
One has
\begin{align}
&C_1=\frac{\partial}{\partial x_1}+2\pi\Sigma_0 && S_1=e^{-2\pi\Sigma_0}\\
&C_2=\frac{\partial}{\partial x_2}+\log\frac{\sinh \pi(\Sigma_0-t_2)}{\sinh \pi(\Sigma_0+t_2)} && S_2=\frac{\sinh \pi(\Sigma_0+t_2)}{\sinh \pi(\Sigma_0-t_2)},
\end{align}
from which we get the equations for the spectral curve $\mathcal{L}$
\begin{align}
&S_1+S_1^{-1}=T_2+T_2^{-1}+T_1 && (1-S_1T_2)S_2=T_2-S_1.
\end{align}
It is easy to check that $\mathcal{L}$ is indeed a Lagrangian submanifold of $(\C^*)^4$, as expected.

\begin{figure}
\centering
\includegraphics[width=.8\textwidth]{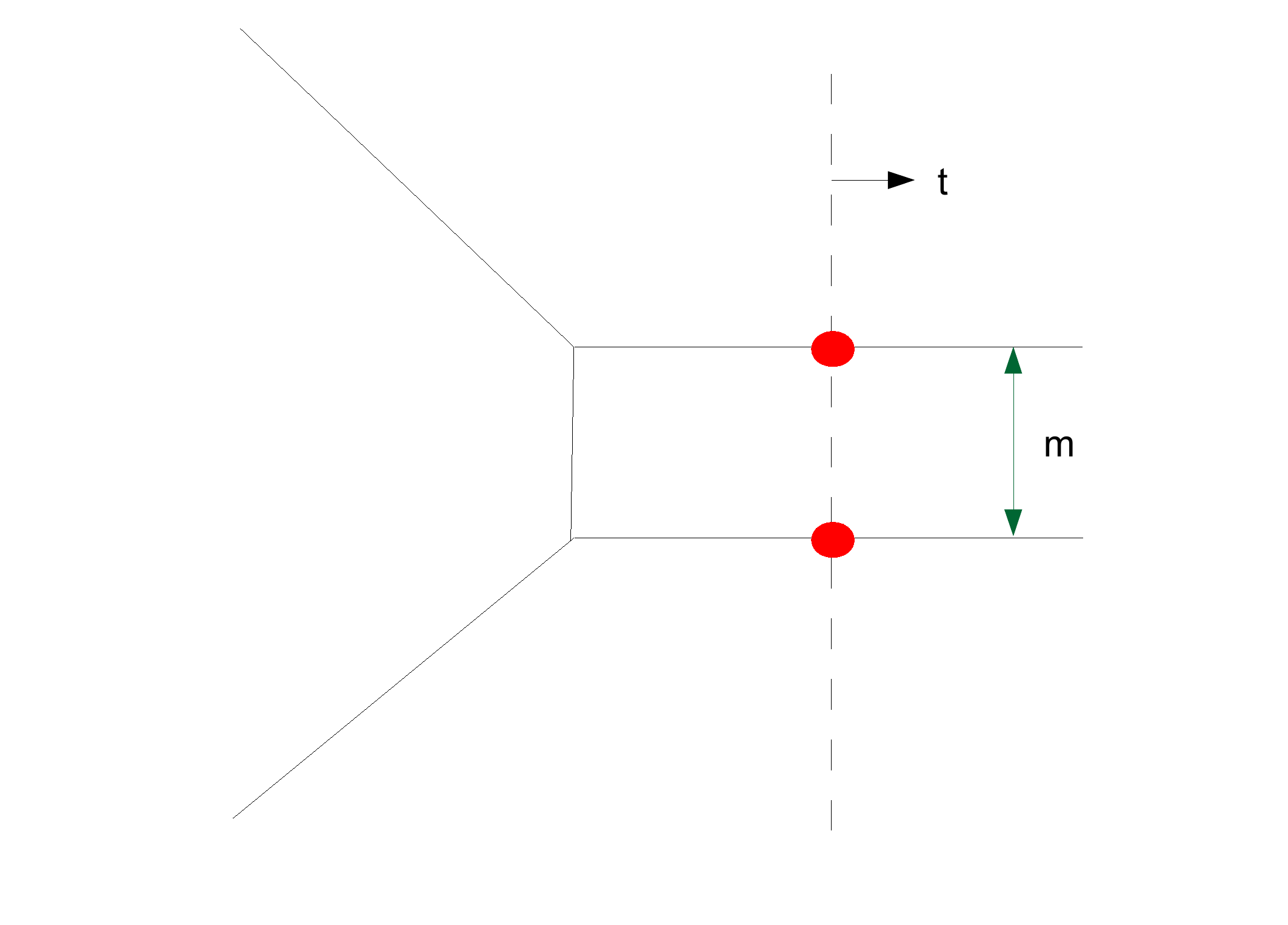}
\caption{The web geometry which leads to $\C P^1$ gauge theory at CS level 0.  The separation
of the horizontal lines is controlled by the mass parameter associated to flavor symmetry rotating
the two flavors in opposite directions.  The movement of the spectator brane corresponds to changing
the FI-parameter $t$.}
\label{fig:cp1}
\end{figure}

For the $t$ geometry for the $\C P^1$ gauge theory, where we fix the mass parameter but vary $t$ by moving the spectator brane,  the first equation would need to be the spectral curve, where $S_1,T_1$ define the curve and $T_2$ is a parameter.
In other words, the Newton polygon can be taken to include $(0,-1),(0,0),(0,1)$ and a $(-1,0)$. Thus we need a $(1,1)$ brane, a $(-1,1)$ brane and two $(0,-1)$ branes (see Fig. \ref{fig:cp1}).

The parallel $(0,-1)$ branes give rise to the $SU(2)_m$ flavor symmetry, and their separation is the parameter $m$. 
The spectral curve is the expected (where $t_2 = m/2 + i\pi$, $t_1= t$). 
\begin{equation}
e^s + e^{-s} = c_{0,0} + e^{t}
\end{equation}
where $c_{0,0} = - e^{m/2} - e^{-m/2}$.\footnote{
We could also seek a five-brane geometry which would reproduce directly the $m$ geometry for the $\C P^1$ gauge theory with zero CS level.
Although it is straightforward put the spectral data in the correct form, up to a small redefinition $m \to 2t_m$,
the spectral curve 
\begin{equation}
e^{s_m} + e^{- s_m} =-e^{2 t_m -t} + e^{t_m}  + 2 e^{-t} + e^{- t_m} - e^{-2 t_m -t}
\end{equation}
is non-generic: the corresponding brane system has normalizable moduli, and really engineers a more complex 3d-5d system. 
This is an important cautionary tale, which was encountered before in the context of 2d-4d systems. A given 3d theory may not have enough deformations to  
reproduce all moduli of a doubly-periodic monopole geometry, but rather it may produce some (usually somewhat special) slice of that 
moduli space. In 2d-4d examples, that slice is often a singular locus in the full moduli space. The physical interpretation is that the brane construction produces a larger theory. If we restrict the Coulomb branch moduli to the values which correspond to the original system's spectral curve, 
a Higgs branch may open up and an RG flow to the original system may become available by moving along the Higgs branch. 
It would be interesting to verify if the same picture holds in the 3d-5d setup.
It is also interesting to see if one can embed the D-model setup in \cite{link} into string theory.
If so, one can engineers arbitrary spectral geometries in higher dimensions in that way.}

\begin{figure}
\centering
\includegraphics[width=.8\textwidth]{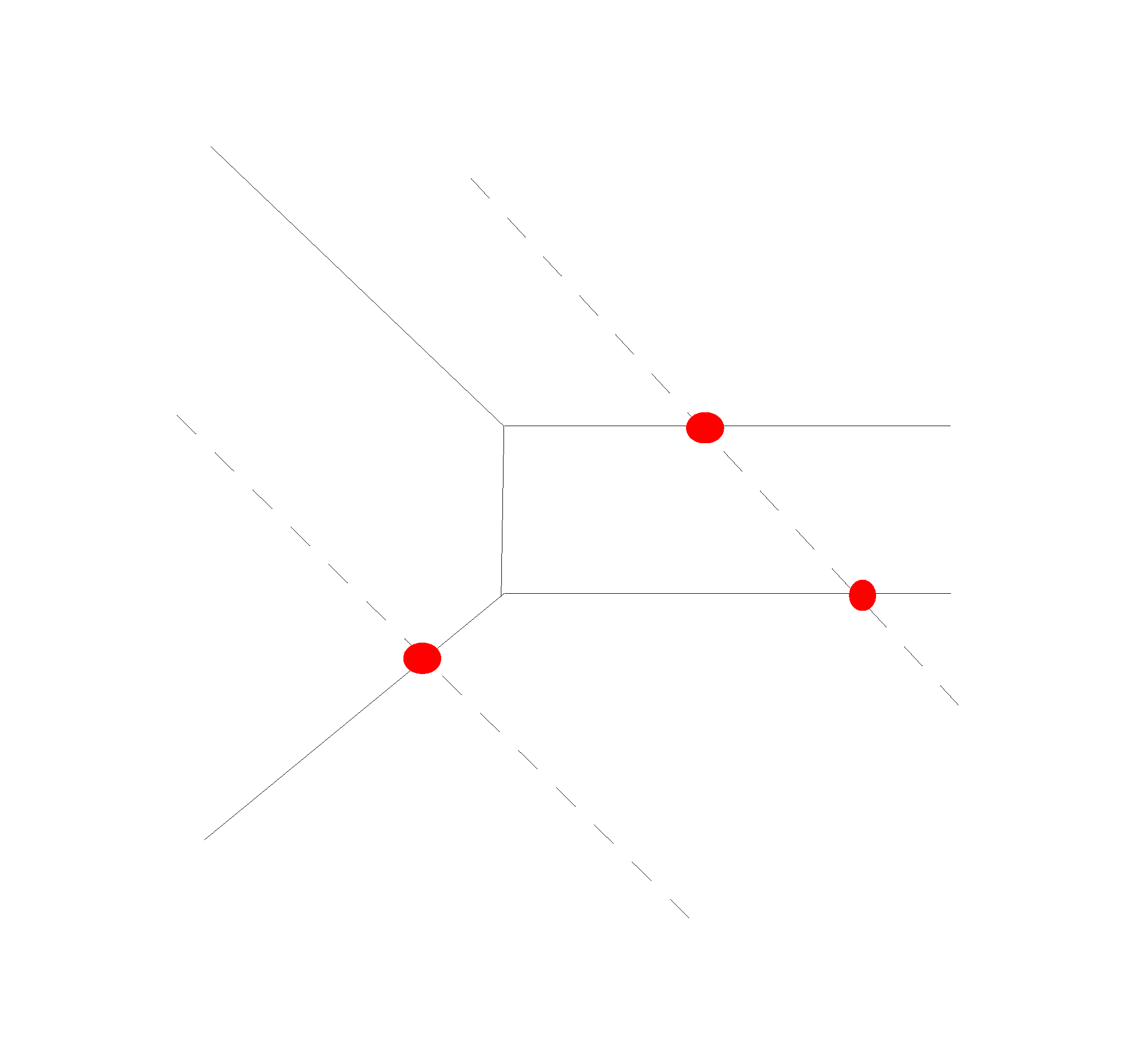}
\caption{The web geometry which leads to $\C P^1$ gauge theory at CS level 1.
Depending on the sign of the FI parameter $t$ the 3d theory has 1 or 2 vacua.
The one corresponding to 1 has degeneracy 2, in the sense that if we compactify the theory
it splits into two distinct vacua.  Having a vacuum geometry which splits upon compactification is
a signature of non-trivial topological structure in the IR.}
\label{fig:cp11}
\end{figure}

A final example is a $\C P^1$ theory with an extra CS coupling of $1$. At large negative $t$ we still have a $\C P^1$ sigma model, 
but for large positive $t$ we now have a single branch with effective CS coupling $2$. At $t=0$ a semi-infinite Coulomb branch opens up.
The brane system involves the same two $(0,1)$ fivebranes, a $(1,0)$ brane and an $(1,2)$ fivebrane. 
For this case, it is more convenient to rotate the spectator brane to achieve the CS coupling 1, and not rotate
the entire web.  In particular we take the spectator brane to be a $(1,-1)$ brane instead of $(1,1)$ brane, and
use the same brane as the one for CS level 0 (see Fig. \ref{fig:cp11}).

\subsection{Loop operator algebras as deformed Verlinde Algebra}

As discussed in \S.\ref{topological}, we expect that for the massive 3d theories, in the infrared limit the theory become topological.  In this section we give some examples of this and point out that these give a mass deformation structure
to the Verlinde algebra, which would be potentially interesting for topological phases of matter.

Let us go back to the two 5brane system with branes $(1,0),(1,k)$.   This has
$k$ vacua in 2d, given by $S^k=e^t$.  As already noted, suspending
a D3 brane between them leads to a $U(1)_k$ Chern-Simons theory, and the ${\cal N}=2$ loop algebra
is isomorphic to the Verlinde algebra.  Let us check how the S matrix computed in the ${\cal N}=2$ context
match up with that of the S-matrix of the Verlinde algebra.  The S-matrix intertwines
loop operators wrapping each of the two cycles of $T^2$.  Let us denote the generators of the two loop operators by $S_a$
and $S_b$:
$$S S_a S^{-1}=S_b$$
 Let us use a basis of vacua adapted to the $S_a$, where it acts (after suitably normalizing it) as
 $$S_a|n\rangle =\omega^n |n\rangle$$
 where $\omega$ is a primitive $k$-th root of unity.
  As discussed in \S.\ref{topological}, 
the action of $S_b$ for a $U(1)$ gauge theory is the same as the action of the holonomy of $tt^*$ by going
through a path where $\theta\rightarrow \theta+ 2\pi$, with $\theta$ the imaginary part of $t$.  In the IR, this
holonomy can be computed easily in the point basis, and it corresponds to the permutation of the $k$ vacua.  In other words
$$S_b|n\rangle =|n+1\rangle.$$
Since the S-matrix intertwines between them we learn that
$$S_{ij}={1\over \sqrt k}\, \omega^{ij}.$$
which agrees with the expected form of the S matrix for the Verlinde algebra of $U(1)_k$.
This analysis can be extended to the case in which, instead of just one suspended D3 brane, we have $N$ of them, giving the S-matrix for $U(N)_k$ CS theory.  We leave checking the details to the reader.

Instead, we will focus on asking how such a structure gets realized in our models.  Consider in particular
the $\C P^1$ gauge theory at level $1$ (see Fig. \ref{fig:cp11}).  This is a particularly interesting case, and is a special instance of the theories studied in \cite{kapustinrec}, involving $U(N)_{k/2}$ coupled to $k$ fundamental chiral fields,
to explain the relation observed by Gepner \cite{Gepner} between the Verlinde algebra for $U(N)_k$ and the quantum cohomology ring for Grassmannian $Gr(N,k)$ \cite{Intriligator,WittenV}.
As already discussed, for $t\ll0$ we expect to get a pure $U(1)$ CS theory at level 2,
and thus the considerations of the previous discussion applies; in particular, in the IR we get
the same structure as the 2d Verlinde algebra.  On the other hand, one may ask what S-matrix structure
do we get for $t\gg 0$.  In this case the theory is the $\C P^1$ sigma model.  Let us also assume that in addition we
have a mass parameter, and ask how the S-matrix behaves in this regime.  Let us go to
a basis in which the $S_a$ operator is diagonal, and given by the intersection of the 5-brane
with the brane web.  This corresponds to two points on the 3d web, which become infinitely far
away in the IR.  Moreover, it is also clear that $S_b$ which corresponds to the $\theta\rightarrow \theta +2\pi$
is also diagonal in this basis, because the 3d vacua do not get permuted.  This implies that in this regime of parameters
the $S$ matrix becomes trivial, \textit{i.e.}\! the identity operator.   This suggest that there is no non-trivial topological
degrees of freedom in this regime of parameters.
The $tt^*$ geometry for the doubly periodic
system thus interpolates between a trivial S-matrix in one regime of parameters, to the non-trivial S-matrix
(corresponding to that of the Verlinde algebra) in a different regime.  This is indeed exciting and is worth
studying further.  

The general structure which emerges from this discussion is that by looking at the 3d vacua
we can determine if in the IR, upon compactification on an $S^1$, we get a topologically non-trivial
theory or not.  In particular, if the 3d vacua reflect the degeneracy of the compactified theory, then
the theory becomes trivial.  This is the case when the projection of the spectator 5-brane with $(p,q)$ type $v$ to the
5-brane plane intersects the web in as many points as the vacua, which in turn is the
case if the product of $v\wedge w_i= \pm1$ for each 5-brane $w_i$ it intersects. Otherwise at each intersection point
we get the structure of a $U(1)_{v\wedge w}$ Verlinde algebra.  Moreover if we consider having $N$ suspended
D3 branes the S-matrix in the IR will have the structure of the Verlinde algebra for
$$\prod_i U(N)_{v\wedge w_i}.$$
Clearly we have found a beautiful interplay between deformations of 2d RCFT's and geometry, captured
by doubly periodic monopole equations, which should be further studied, especially in view of application
to topological phases of matter.

\subsection{Class R three--dimensional theories associated to three-manifolds}
There is a rich class of three dimensional Abelian Chern-Simons matter theories which can be obtained from a product of $m$ $T_\Delta$ theories, 
by acting with an arbitrary $\sp(2m,\mathbb{Z})$ transformation and adding certain superpotential couplings described in \cite{DGG,CCV,tangles}.  
The main point of interest of this class of theories is that there is a large network of mirror symmetries relating different UV theories in the class, and 
the space of equivalence classes of IR SCFTs, dubbed ``class R'' in \cite{DGG}, seems to have a rich structure. 

The spectral data for the $tt^*$ geometry of a class R theory $T$, which coincides with the parameter space of supersymmetric vacua 
${\cal L}[T]$ discussed in \cite{DGG}, is invariant under the mirror symmetries and is presented as the image of the product ${\cal L}[T_\Delta]^{m}$ of parameter spaces of the individual chiral multiplets 
under the $\sp(2m,\mathbb{Z})$ transformation and a toric symplectic quotient of $(\mathbb{C}^*)^{2m}$ determined by the choices of super-potentials. 
At the level of the $tt^*$ doubly-periodic geometry itself, the $\sp(2m,\mathbb{Z})$ transformation is the Nahm transform discussed in a previous section. 
The symplectic quotient is simply the restriction of the monopole data to a linear subspace in $\R^{3m}$, the locus where one sets to zero 
the mass parameters and flavor Wilson lines for the flavor symmetries broken by the superpotential terms. 

There is a subset of class R theories $T_M$ which are associated to certain decorated three-manifolds $M$: the data of the theory is constructed from a triangulation of the 
three manifold $M$, and the mirror symmetries insure invariance under 2-3 moves which relate different triangulations of the same manifold. 
Thus the final 3d SCFT only depends on the choice of manifold $M$, and so will the corresponding $tt^*$ geometry. 
The construction is designed in such a way that the parameter space ${\cal L}[T_M]$ coincides with the space of flat $SL(2,\C)$ connections on $M$. 
A typical example of $M$ could be a knot complement in $S^3$. It would be interesting to find a similar geometric relation between the three-dimensional geometry 
$M$ and the $tt^*$ geometry of $T_M$.

\section{$tt^*$ geometry in 4 dimensions}
In this section we discuss the $tt^*$ in 4 dimensions.  The structure of the argument is very similar
to that of the 3d case, except that in this case we have 2 distinct possibilities:  We can discuss either
flavor symmetries, which correspond to line operators, or 2-form symmetries which couple
to conserved anti-symmetric 2-form.  These arise in particular in theories with $U(1)$ gauge factors
where we consider $B\wedge F$ terms as well as the FI term.
 We will see that in the case of flavor symmetries the parameter space is $(T^3)^r$ where
$r$ is the rank of the flavor symmetry group.  In the case of 2-form symmetries, we find that the parameter
space is $(T^3\times R)^r$ where $r$ is the number of 2-form symmetries.
Furthermore the derivation of the 4d $tt^*$ geometries proceed as in 3d case.  We see that
for the case of flavor symmetries the theory has sectors indexed by an integer $n$ where $W$ has
a central charge $n \mu$ for a complex parameter $\mu$.  In the case of 2-form symmetries we see that
there are sectors labeled by a pair of integers $(n_1,n_2)$ for which $W$ shifts by $(n_1+n_2\rho) \mu$.

\subsection{The case of flavor symmetries}
Consider a theory in $4d$ where we take the space to be a flat torus $T^3$ with periodic
boundary condition for fermions, preserving all supersymmetry.  Let us assume this theory has a flavor symmetry of rank $r$.
We can turn on fugacities for the rank $r$ flavor group in the Cartan of the flavor group along each circle.
Therefore the parameter space is
$$T^{3r},$$
modulo the action of the Weyl group\footnote{\ This is the most general flavor twisting in $T^3$ whenever the flavor group is an Abelian group times a product of simple groups of isotype $A_{N-1}$ and $C_n$ which have all dual Coxeter labels equal $1$. For more general flavor groups $G_F$, the space $T^{3r}/\mathrm{Weyl}$ gets replaced by the moduli space $\mathcal{M}_3$ of communing triples in $G_F$ which is a disconnected space (see \cite{morgan,wittencomtriples}). Restricting ourselves to the $tt^*$ geometry of the largest connected component of $\mathcal{M}_3$,  we reduce back to the situation discussed in the text. }.  In order to develop the $tt^*$ geometry for this theory, consider
the first step, where we compactify the theory on a circle down to 3 dimensions.  Then we get a theory
with a flavor group of rank $r$.  Moreover the twisted mass parameters of this 3d theory is identified
with the fugacity of the flavor group around the circle.  So, unlike the generic flavor group in 3d where the
corresponding twisted mass parameter is parameterized by $\R$, the fugacities are periodic.  This is 
the only difference from a generic 3d  theory with flavor symmetry.  Therefore the $tt^*$ geometry
is the same as in the generic case, namely the generalized monopole equations in $3r$ dimensions.  Here
the parameter space is the compact $T^{3r}$.  The chiral operators of the 2d case now correspond
to surface operators in the internal geometry (to see this note that the twist operators are codimension 2-operators, which is a surface operator in 4 dimension).

\subsection{The case of 2-form symmetries}
This is the case where the theory has a conserved anti-symmetric 2-form `current' $J_{\mu\nu}$:
$$\partial_\mu J_{\mu\nu}=d*J=0.$$
This couples to a background 2-form tensor field $B_{\mu \nu}$: 
$$\int d^4x\; B \wedge *J$$
which we take to be flat.  In the ${\cal N}=1$ supersymmetric case the background tensor
field is part of an ${\cal N}=1$ tensor multiplet, which includes in addition a real scalar field $\phi$ whose constant
vev deforms the theory.  A generic way this structure appears is when we have a $U(1)$ gauge symmetry.
In that case $J=*F$, which is conserved because $d*J=dF=0$.  The coupling to the background $B$
field corresponds to a $\int d^4x\, B\wedge F$ term and the vev of the scalar field $\phi$ corresponds to the FI parameter
for the $U(1)$ field. For each such 2-form symmetry, we have, in addition to the choice of the
vev of $\phi$ which is generically parameterized by\footnote{We will discuss
some examples where the 4d theory is a probe in a 6d $(1,0)$ theory where the
vev is parameterized by an $S^1$, instead of $\R$.  In such a case
we get $T^4$ as the parameter space.} $\R$, we choose a 2-form $B$ on $T^3$ which is
periodic (assuming, as is typically the case, that the integrals of $*J=F$ are quantized), which then is
parametrized again by a $T^3$. Thus, altogether, we get the parameter space $T^3\times \R$.
 If we have $r$ such 2-form symmetries this gives the parameter space becomes
$$(T^3\times \R)^r.$$
The 4 dimensional ${\cal N}=1$ supersymmetric theories do admit BPS strings, with the central
term being controlled by the scalar vev (which in the $U(1)$ gauge theory case
corresponds to FI-term) in the tensor multiplet.  Let us call this real parameter $\mu$, which denotes
the tension of the string.
Now consider compactifying the theory on $T^2$ to 2-dimensions with a complex structure $\rho$.    The parameter $\mu$
gets complexified by the component of the $B_{12}$ along the $T^2$.  Let us call this $x_{12}$, \textit{i.e.}
$$\mu\rightarrow \mu+i\,x_{12}.$$
In particular the strings will have a BPS tension proportional to $\mu$.
However now we have in addition a more refined sector
in the 2d theory which will be labeled by a pair of integers $n_1,n_2$ depending on wrapping
number of the string around the two cycles.   Then the norm of the central term in this sector will be the length of the
string times the tension, \textit{i.e.},
$$W_{n_1,n_2}=\mu (n_1 R_1 +i n_2 R_2)$$
where we have taken the $T^2$ to be a rectangular torus of radius $R_1,R_2$.  Redefining $\hat \mu =R_1 \mu$,
we have
$$W_{n_1,n_2}={\hat \mu}(n_1+\rho \,n_2)$$
where $\rho$ denotes the complex structure parameter for $T^2$.  We are thus in the same situation as doubly
periodic $W$'s discussed in  section 3.  As discussed there, the $tt^*$ geometry in that case
become that of self-dual Yang-Mills and its generalizations to higher dimensions, corresponding to hyper-holomorphic
connections.

The chiral operators of the 2d theory will now correspond to surface operators wrapped around the $T^2$ fiber
over each point in the 2d theory.  The fact that they are surface operators follows from the fact
that they couple to $\mu$ whose imaginary part includes the expectation value of $B$ along the $T^2$ which
can be gauged away locally, and is only accessed by operators wrapping the entire $T^2$.

\subsection{Partition functions on elongated $S^3\times S^1$ and $S^2\times T^2$}
Just as in the 3d case, we recall that the $tt^*$ geometry has more information than just the vacuum bundle
and in particular it has a preferred basis of vacua corresponding to chiral operators, coming from the
topologically twisted path integral on semi-infinite cigar.  Consider a rectangular $T^3$ geometry and choose one
of the circles to be the circle we was to contract inside the cigar  (see Fig. \ref{fig:FigureB}):

\begin{figure}[here!]
\centering
\includegraphics[width=.8\textwidth]{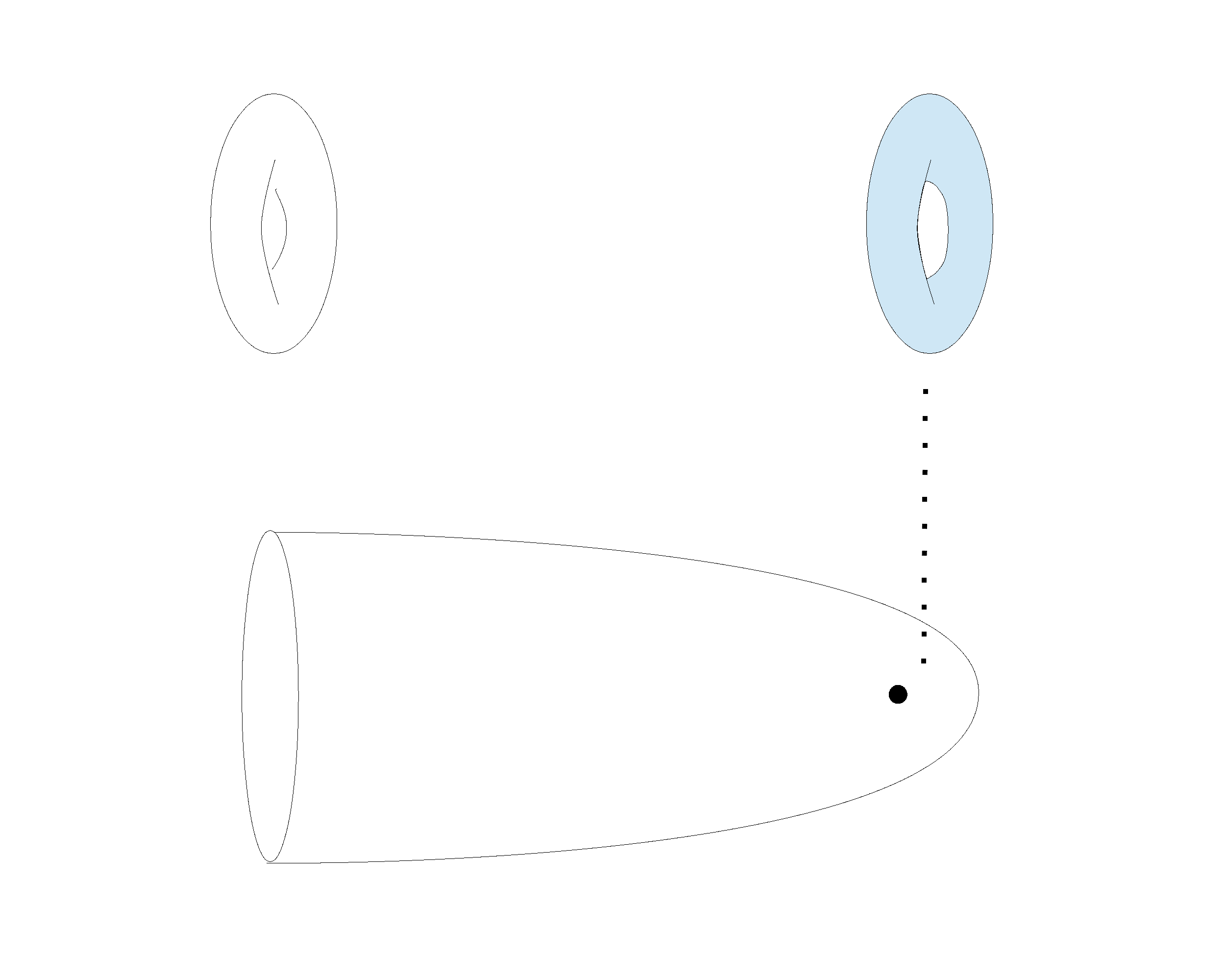}
\caption{The states of the $4d$ theory on $T^3$ can be obtained by doing the path integral
on an infinite cigar times $T^2$, with surface operator, wrapping $T^2$ being inserted at the tip
of the cigar.}
\label{fig:FigureB}
\end{figure}

   We can consider D-brane boundary conditions
and we can compute this, as before, in terms of $\Pi^a_i$.  Or we can consider capping another circle obtaining
a compact geometry.
 Moreover we have three inequivalent
choices to cap the other circle.  If we choose the same circle to contract on the other cigar as well, we would
be computing the usual $g_{i{\overline j}}$ metric and $\eta_{ij}$ of the 2d theory depending on whether both
cigars are topological, or one is topological and the other anti-topological.  The path-integral
for this configuration will have the topology of infinitely elongated $T^2\times S^2$.  On the other hand
if we contract one of the other circles on the second cigar, we will get something which has the topology
of infinitely elongated $S^1\times S^3$.  As explained in the context of the 3d problem these can all be computed.
In this case the analog of $S$-transformation will be played by a non-abelian discrete subgroup of $SO(3,{\Z})\subset SL(3,{\Z})$, generated
by $\pi/2$ rotations of $12,23,31$ planes.

\subsection{Gauging and ungauging}
As discussed in the context of Nahm transformation, we expect that making the flavor symmetry dynamical
has the effect of mapping the $tt^*$ geometry to its Nahm transform.  In the context of a 4d theory with a $U(1)$
flavor symmetry, as already discussed, we expect to get monopole equations on $T^3$.  On the other hand
for a gauge $U(1)$ symmetry we expect to get self-dual connections on $\R\times T^3$.  Indeed the two are Nahm
transforms of one another.  In particular, if we consider the Fourier-Mukai transform of the self-dual
connection on $\R\times T^3$ we expect a T-dual geometry, which gets rid of one dimension
(given by $\R$) and maps $T^3$ to the dual $T^3$, which is indeed the expected geometry for the dual system. 

\section{Examples of $tt^*$ geometry in 4 dimensions} \label{4d}
In this section we would like to discuss some examples of $tt^*$ geometries which arise from four-dimensional ${\cal N}=1$ theories compactified to 2d $(2,2)$ on a torus of complex structure $\tau$ and area $A$.
Although the analysis is not conceptually different from the 3d and 2d examples which appeared in the previous sections, the existence of various anomalies in four dimensions 
field theories complicate our work. 

The basic example of a single free chiral multiplet in four dimensions illustrates well the situation. The KK reduction on a 2-torus gives us a double tower of 2d chiral multiplets, 
of masses $\mu_{k,n} = \mu + \frac{2 \pi}{R_z}\left(k + \tau n\right)$, 
where $k,n$ are the KK momenta and $R_z \mu = \theta_3 + \tau \theta_2$ is the complex combination of the two flavor Wilson lines on $T^2_{\tau}$, 
which behaves as a twisted mass parameter for the theory reduced to two dimensions.
Correspondingly, we should expect the $tt^*$ geometry to be a triply-periodic $U(1)$ BPS Dirac monopole solution, defined on the $T^3$ parameterized by the three flavor Wilson lines
$\theta_{1,2,3}$, the standard $\theta_1 = 2 \pi x$ and the internal $\theta_2 =  2 \pi y$ and $\theta_3 = 2 \pi z$. 

There is an obvious problem with that: there are no (non--trivial) single-valued harmonic functions on a compact space. In other words, we can assemble the triply-periodic array of 
Dirac monopoles, but we cannot make the solution fully periodic in the three Wilson lines. For simplicity, let's take momentarily $\tau = i \frac{R_z}{R_y}$ and 
assemble a periodic array of the doubly-periodic monopole solutions we encountered in 3d. 
Formally this will correspond to a harmonic function of the form
\begin{gather}
V_{4d}(x,y,z)=v(z)+\!\!\!\!\sum_{(k,\ell)\neq(0,0)}\!\! \!V(k,\ell)\;e^{2\pi i k x+2\pi i\ell y}\cosh\!\!\left[\frac{2\pi}{R_z}\sqrt{R_x^2k^2+R_y^2\ell^2}\!\left(z-\frac{1}{2}\right)\right]\\
\intertext{where}
V(k,\ell)=-\frac{R_xR_y}{2\,\sqrt{R_x^2k^2+R_y^2\ell^2}\;\sinh\!\Big[\frac{\pi}{R_z} \sqrt{R_x^2k^2+R_y^2\ell^2}\Big]}.
\end{gather}
The harmonic function $V_{4d}(x,y,z)$ has a source which corresponds to the periodic array of doubly-periodic Dirac monopoles 
if and only if the first derivative of the zero-mode $v(z)$ has discontinuity $\frac{2 \pi R_x R_y}{R_z}$ at all integer $z$. 
This is impossible for a function which is both harmonic and periodic; indeed harmonicity requires something like 
\begin{equation}
v(z) = \frac{2 \pi}{R_z} \left( (k+\frac{1}{2}) z - \frac{1}{2} k (k+1) \right)\qquad \qquad k < z < k+1.
\end{equation}
Thus the Higgs field fails to be periodic by $V(z+1) - V(z) = \frac{2 \pi}{R_z}(z+ \frac{1}{2})$. Correspondingly, the field strength of the gauge connection on the $x$--$y$ torus will 
not be periodic in the $z$ direction, but rather the total flux will increase by one as $z \to z + 1$. 
Of course, we can make slightly different choices to sacrifice periodicity, say, in the $y$ direction and keep periodicity in the $z$ direction. 

The lack of periodicity is also visible from the twisted effective superpotential for a 4d chiral multiplet compactified to 2d,  
or better its first derivative. 
Indeed, the contribution of $tt^*$ $B$--matrix \eqref{whatLLLs}, written as a sum of the corresponding matrices $B_{n,m}$ for the decoupled $Y_{n,m}$ KK modes, is\footnote{\ The sum is not absolutely convergent, and hence the order of summation matters (in particular, different orders lead to functions which fail to be periodic in different directions). Here the symmetric Eisenstein order convention is implied.}
\begin{equation}
B= \sum_{n,m\in\Z} \log\!\left[\frac{z+n}{2R_z}+i\,\frac{y+m}{R_y}\right]\xrightarrow{\ \text{regularization}\ }\log\Theta\!\left(z+y\tau,\; \tau=i\frac{R_z}{R_y}\right),
\end{equation}
where
\begin{equation}
\Theta(w,\, \tau)\equiv\theta_1(\pi w\,|\,\tau)=2\sum_{n=0}^\infty(-1)^n\,q^{\frac{1}{2}(n+\frac{1}{2})^2}\sin\!\big((2n+1)\pi w\big),\quad q=\exp(2\pi i\tau).
\end{equation}
Under a translation $y\to y+1$, the $ \log \Theta(z+y \tau, \tau)$ shifts by
$$i\pi-2\pi i(z+y\tau)-\pi i \tau.$$
This corresponds to the choice of $V$ which fails to be periodic in the $y$ direction. 

Finally, we can express the problem in terms of the spectral curve 
\begin{equation}
e^p = \Theta(z+y\tau, \tau)
\end{equation}
which is not a well-defined curve in $\C^* \times T^2$. 

The relation to the anomaly in the 4d flavor symmetry of a single chiral field becomes a bit more obvious if we imagine 
a collection of 4d chiral multiplets, having charges $q_i\in\Z$ under the flavor symmetry. 
The $B_\mu$ matrix (equal to the value of the derivative of the effective superpotential with respect to $\mu\equiv z+y\tau$ on the reference vacuum, cfr. \S.\,4 satisfies 
\begin{equation}
B_\mu = \sum_i q_i\, \log \Theta\big(q_i(z+y\tau), \tau\big).
\end{equation}
Under a translation $y \to y + 1$ it shifts by 
\begin{equation}\label{dangerousshift}
- \pi i (2\mu +\tau) \sum_i q_i^3 + i \pi \sum_i q_i^2 
\end{equation}
Thus the coefficient of the dangerous shift linear in $\mu=z+y\tau$ is the coefficient of the total $U(1)^3$ anomaly. 
Eqn.\eqref{dangerousshift} is equivalent to the statement that 
$$\exp B_\mu \equiv \exp\!\Big(\partial_\mu W|_{\text{vacuum}}\Big)$$
is a section of a line bundle $L$ over the elliptic curve of period $\tau$ with Chern class
\begin{equation}
c_1(L)=\sum_iq_i^3.
\end{equation}
In particular, 
the $U(1)^3$ anomaly coefficient measures the failure to commute of  the two translations $T_z\colon z\to z+1$ and $T_y\colon y\to y+1$.
 
More generally, if we look at multiple $U(1)$ flavor symmetries, with 4d chirals of charges $q_{i,a}$ under the $a$-th flavor symmetry, 
the coefficient of $\mu_c$ in the discontinuity of $B_{\mu_a}\equiv \partial_{\mu_a} {\cal W}\big|_\text{vacuum}$ under $\mu_b \to \mu_b + \tau$ equal to 
\begin{equation}
-2 \pi i \sum_i q_{i,a} q_{i,b} q_{i,c}.
\end{equation}
Thus any mixed anomaly between the $U(1)$ flavor symmetries will cause trouble with the periodicity of the $tt^*$ geometry. 
We will encounter similar statements for non-Abelian flavor symmetries, by considering their Cartan subgroup.

For example, consider a theory of two 4d chirals. The theory has a non-anomalous ``vector'' flavor symmetry which rotates a chiral 
in one direction, and the other chiral in the opposite direction and an anomalous ``axial'' symmetry which rotates them in the same direction. 
If we do not turn on a Wilson line for the axial symmetry, the $tt^*$ geometry for the vector symmetry is well defined, but trivial, as the contribution of the two chirals essentially cancels out. 
On the other  hand, if we allow a generic fixed flavor Wilson line $\mu'$ for the 
``axial'' flavor symmetry and study the $tt^*$ geometry for the vector symmetry we still have some trouble, although less serious: 
the harmonic function $V$ is not periodic, but it shifts by a constant (\textit{i.e.}\! $e^{\partial_\mu W|_\text{vacuum}}$ is a section of a topologically trivial line bundle).  Now the $tt^*$ connection has a curvature $F=\ast dV$ which is strictly periodic, and hence well--defined (up to gauge transformations).

Correspondingly, the spectral curve 
\begin{equation}\label{onechiral}
e^p = \frac{\Theta(\mu+ \mu', \tau)}{\Theta(-\mu+ \mu', \tau)}
\end{equation}
is not a well-defined curve in $\C^* \times T^2$, as $p$ is multi-valued by $4 \pi i \mu'$. 
Unlike the case with a single chiral field, this spectral curve can still make sense as a curve in a non-trivial $\C^*$ bundle over $T^2$, 
and the $tt^*$ monopole geometry can make sense if we think about the Higgs field as a periodic scalar field whose profile is a section of an affine bundle 
over $T^3$. 

\begin{figure}
\centering
\includegraphics[width=.8\textwidth]{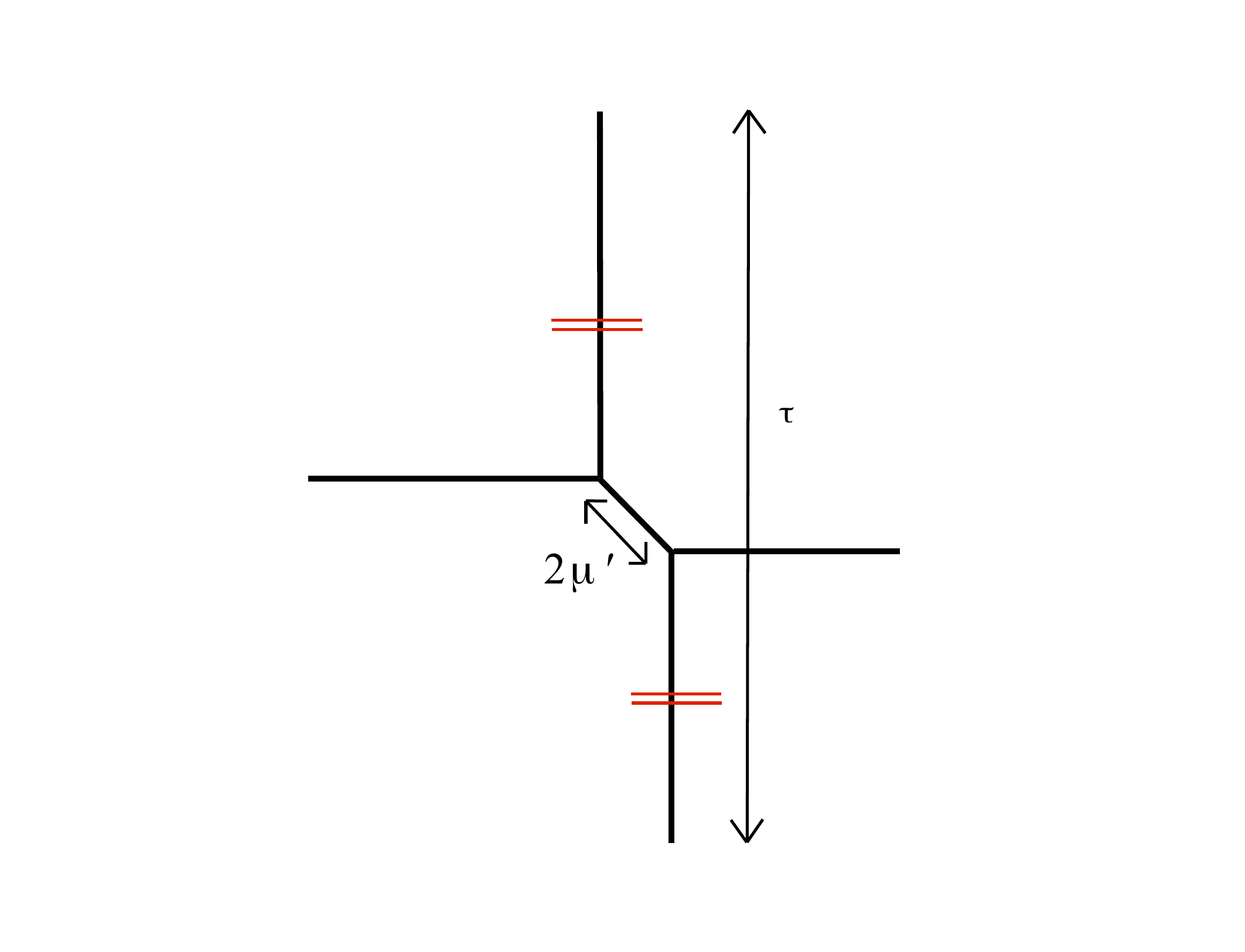}
\caption{A Single M5 brane in the presence of a Taub-NUT, is dual, after
compactification on a circle, to this 5-brane web diagram on a cylinder with circumference $\tau$.
The mass parameter $m=2 \mu'$ is induced from the R-twist around the compactified circle
leading to ${\cal N}=2^*$ theory in 5d with adjoint mass $m$.}
\label{fig:1M5}
\end{figure}

To gain more insight into this case it is useful to consider the 5-brane construction of the last section associated with this
geometry by viewing the 4d theories as probes of 6d (2,0) or (1,0) theories.  As discussed in \cite{HoIV} the 5-brane geometries which 
are on a cylinder, instead of a plane, are equivalent to 6d theories.  Moreover as noted in \cite{Mstring1,Mstring2,HI} the 6d theories
can be viewed either as circle compactification of 6d gauge theories or (1,0) SCFT's theories.   For example consider the
brane geometry given by  Fig. \ref{fig:1M5}.
This corresponds to an M5 brane geometry compactified on a circle with a twist around the circle
corresponding to a mass $m=2\mu'$, which in 5d becomes the $U(1),\  {\cal N}=2^*$ theory where
$\mu'$ is the mass parameter for the adjoint field.  Note that the plane geometry is twisted,
in that as we go around the vertical direction, we shift along the horizontal direction
by an amount $\mu'$.   Indeed, upon compactification on another circle, this gives rise to an
${\cal N}=2$ theory in $d=4$ with the Seiberg-Witten curve given by \cite{HoIV}
$$e^p\, \Theta(-x+\mu',\tau)-\Theta(x+ \mu', \tau)=0$$
where this is an equation for  $x$ and $p$.

\begin{figure}
\centering
\includegraphics[width=.8\textwidth]{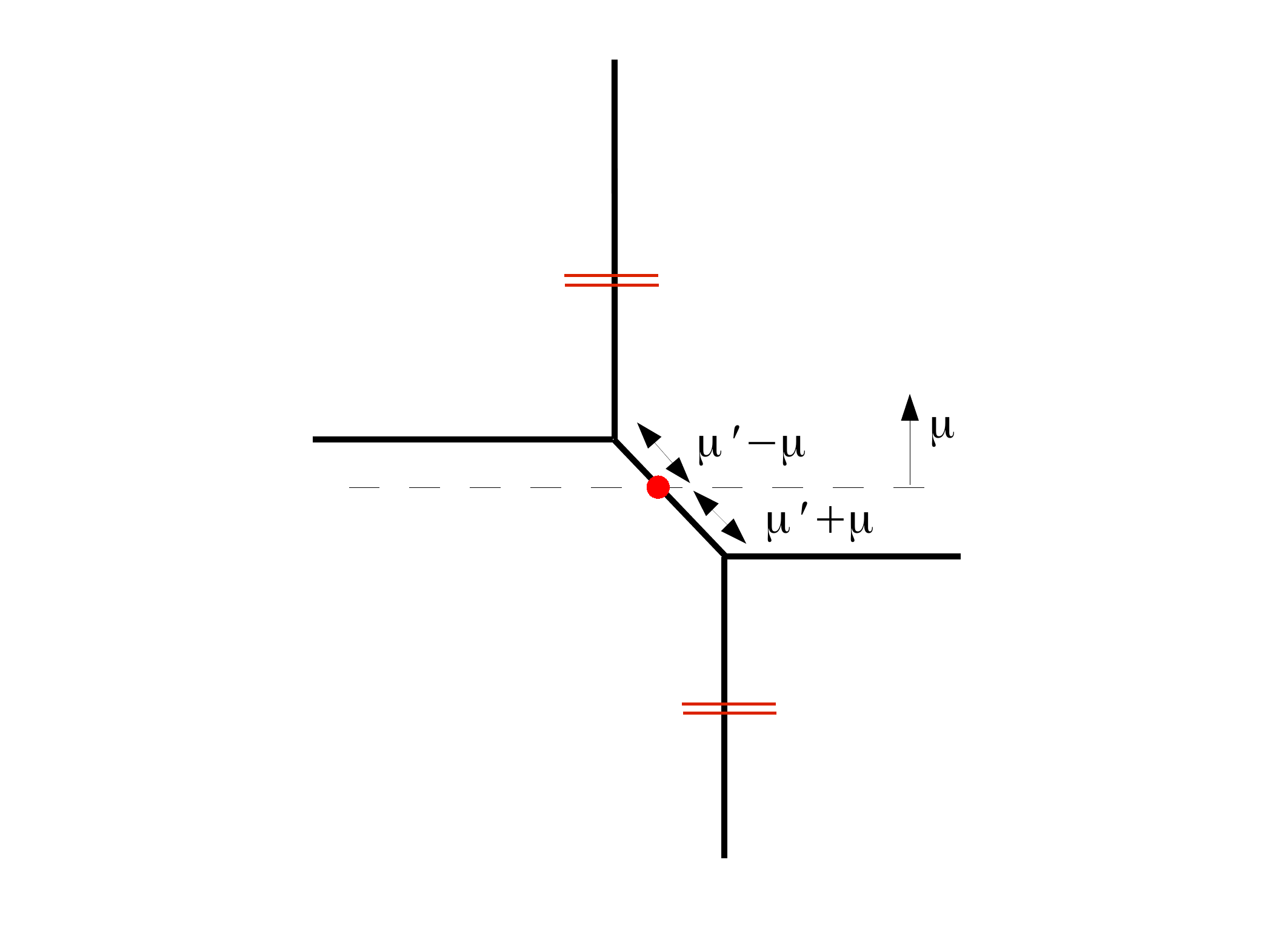}
\caption{This web diagram is dual to 4d theory with two chiral fields
of masses $\mu'+\mu$ and $\mu'-\mu$.  The variable $\mu$ which changes
with the position of the spectator brane is a periodic variable.}
\label{fig:1M5probe}
\end{figure}

 The 4d probe of M5 brane, upon compactification on a circle,
corresponds to 3d theories corresponding to the suspended branes, where the
suspended brane ends on the web at $x=\mu$.  This gives rise to a theory
with two chiral fields of masses $\mu+\mu'$ and $-\mu+\mu'$.  See Fig. \ref{fig:1M5probe}.

\begin{figure}
\centering
\includegraphics[width=.8\textwidth]{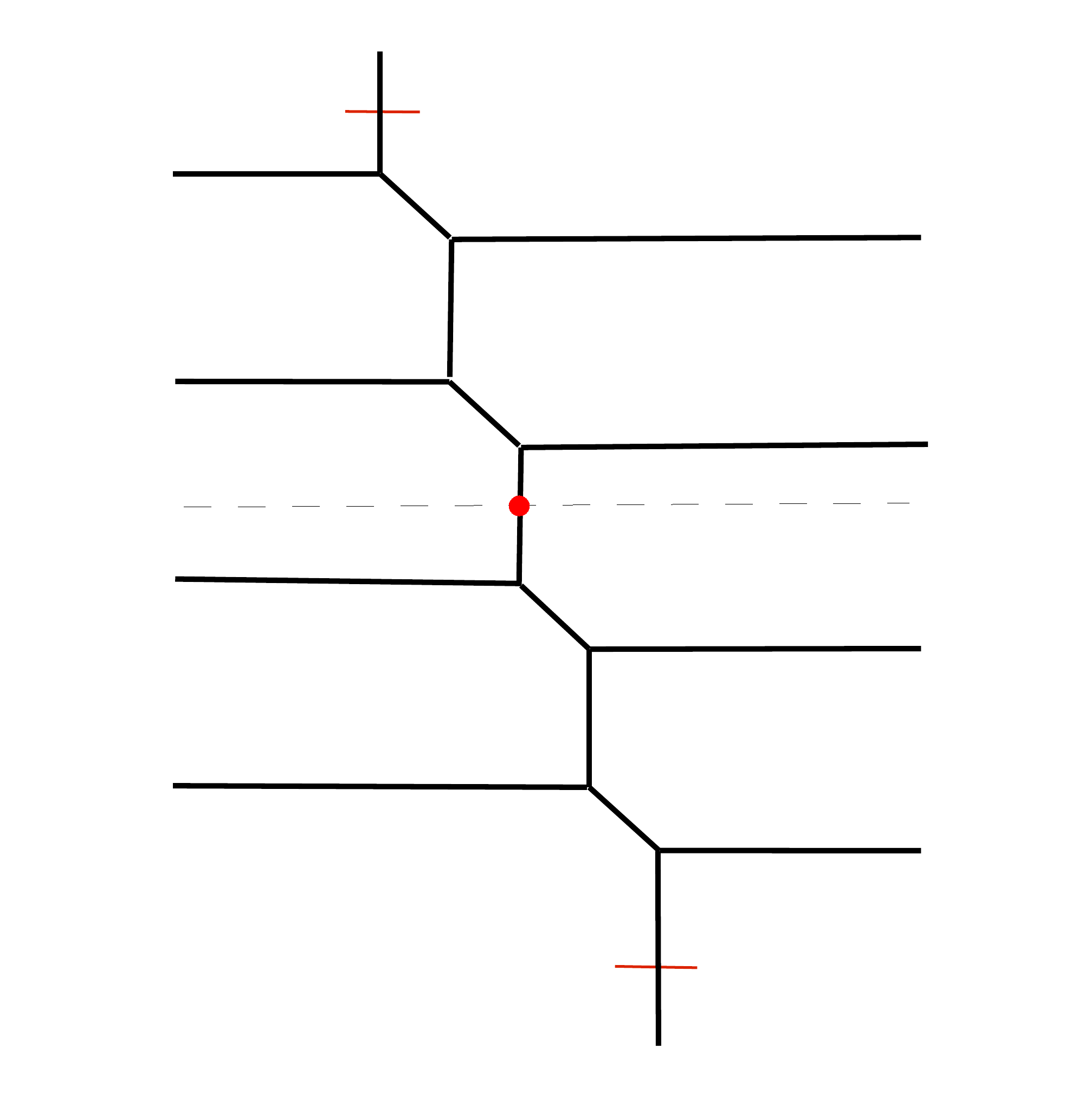}
\caption{This web diagram includes a 4d subsector with $2n=8$ chiral
fields of charges $(+1^4,-1^4)$ under the $U(1)$ flavor symmetry.}
\label{fig:An}
\end{figure}

  In fact this geometry
gives rise to the spectral curve eq.(\ref{onechiral}), and this is because
the brane probe theory supports two chiral fields (reflected in this case by the two
1-branes stretched between the web and the D3 brane). The fact that the $\mu$ parameter is periodic
(while in the 3d case it took values in $\R$)
is simply a reflection of the fact that the vertical direction is periodic, and as the spectator goes
around the vertical direction by $\tau$ it comes back to the original position, thus giving a parameter
space $T^3$.
 The lack of periodicity of the Higgs field, \textit{i.e.}\! the shift in the horizontal direction as we
go around the vertical direction, will give rise to a generalized $tt^*$ geometry which as discussed in section 3
corresponds to the Nahm transform of  hyperholomorphic connections on a non-commutative
space.   Moreover as found for this model in eq.(\ref{noncom}) the non-commutativity parameter is proportional
to $\mu'$.
The non-commutativity would disappear, and we get the ordinary $tt^*$ commutative geometry, if we set
$\mu'=0$ and avoid turning on Wilson lines for anomalous flavor symmetries.
We can also consider more chiral fields which corresponds to a 4d probe of the $(1,0)$ theory given by
an M5 brane probing an $A_{n-1}$ singularity. See Fig. \ref{fig:An}.
If we use $n$ chirals of charge $1$, $n$ of charge $-1$  we get a spectral curve: 
\begin{equation}
e^p = \prod_a \frac{\Theta(\mu- \mu_a, \tau)}{\Theta(-\mu+ \mu'_a, \tau)}
\end{equation}
Again, if we want to avoid the anomalous flavor symmetry we can turn off the axial flavor
Wilson lines and set $\sum_a \mu_a -\sum_a \mu'_a =0$.  Without loss of generality we can
set  $\sum_a \mu_a =\sum_a \mu'_a =0$ by shifting $\mu$ if necessary.
This is a sensible spectral curve, and corresponds to a triply-periodic collection of $U(1)$ Dirac monopoles
$n$ of charge $1$ and $n$ of charge $-1$, at positions $\vec \theta_a$ and $\vec \theta_a'$ on $T^3$ constrained to  
satisfy $\sum_a \vec \theta_a =\sum_a \vec \theta'_a =0$.

\begin{figure}
\centering
\includegraphics[width=.8\textwidth]{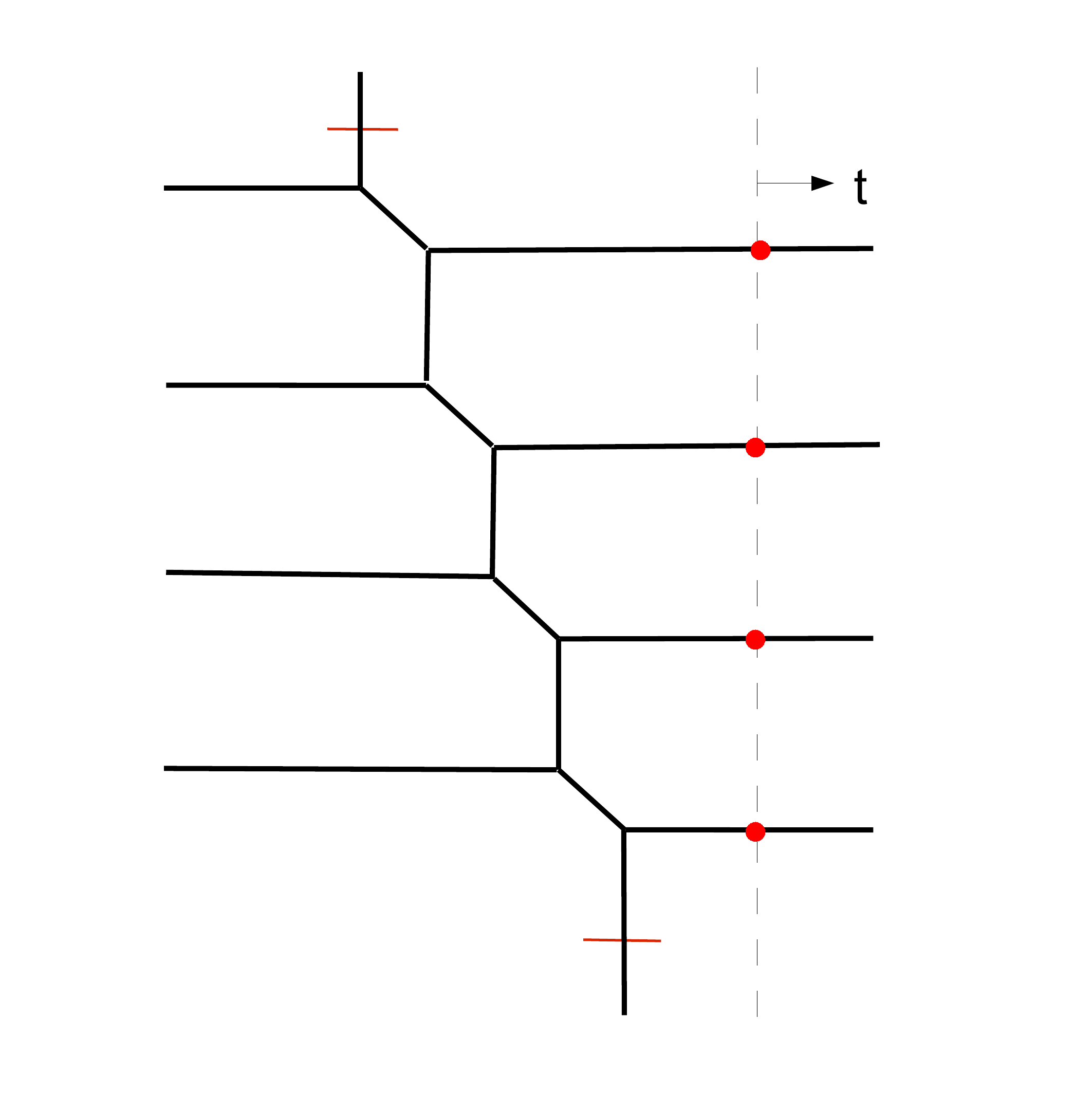}
\caption{This web diagram together with the probe engineers SQCD with 4 flavors.  The theory
has $4$ vacua.  The FI parameter of the $U(1)$ theory is controlled by $t$, the position of the
spectator brane. The rank of the group depends on how many branes we suspend between
the spectator brane and the web.  For $U(n_c)$ we need $n_c$ suspended branes.}
\label{fig:SQED}
\end{figure}

The obvious next step is to gauge the $U(1)$ flavor symmetry, to get ${\cal N}=1$ SQED with $n$ flavors. Although the theory has a Landau pole, the 
$tt^*$ geometry is oblivious to the 4d gauge coupling, and should thus be relatively well-defined. 
As this is simply a Nahm transform of the previous problem, the spectral curve is
\begin{equation}
e^t = \prod_a \frac{\Theta(\sigma- \mu_a, \tau)}{\Theta(-\sigma+ \mu'_a, \tau)}
\end{equation}
where $t$ is the (complexified) FI parameter. 
We can better write the equation as 
\begin{equation}
e^t\prod_a\Theta(-\sigma+ \mu'_a, \tau) - \prod_a \Theta(\sigma- \mu_a, \tau)=0
\end{equation}
This equation is a degree $n$ theta function on the torus, and has $n$ zeroes $\sigma^*_i$, 
which represent the gauge Wilson lines in the $n$ vacua of the theory compactified on $T^2$. 
In terms of brane diagrams, the spectator brane is now oriented vertically.  See Fig. \ref{fig:SQED}.

Geometrically, the $tt^*$ geometry is a triply-periodic instanton geometry, \textit{i.e.} an instanton in $\R_t \times T^3$. 
The complexified FI parameter $t$ is a coordinate on $\R_t \times S^1$, while by definition the zeroes $\sigma^*_i$ characterize 
the holomorphic $SU(n)$ bundle on the remaining $T^2$ directions. Thus the $\mu'_a$ and $\mu_a$ parameters 
label the holomorphic $SU(n)$ bundle at large positive and large negative $t$ respectively.
This instanton solution appears to be rigid. The spectral curve, for example, has no moduli. 

\subsection{SQCD}
Much as it happens for the Grassmanian GLSM \cite{onclassification} (see also the review
in \cite{GGS}), the twisted chiral ring relations for 
an $U(n_c)$ four-dimensional gauge theory with $n_f$ flavors are closely related to the ones for a $U(1)$ theory,
and can be engineered as in the case of $U(1)$ SQED discussed above, by
taking $n_c$ suspended D-branes.  In this case
each of the Wilson lines $\sigma_i$ in the Cartan of $U(n_c)$ must solve 
\begin{equation}
e^t \prod_a\Theta(-\sigma_i+ \mu'_a, \tau) - \prod_a \Theta(\sigma_i- \mu_a, \tau)=0
\end{equation}
with the extra constraints that they should be distinct solutions. 

Thus the ${n_f \choose n_c}$ vacua of the system coincide with the possible choices of $n_c$ distinct roots of 
the above degree $n_f$ theta function.\footnote{\ The remaining $n-n_c$ roots are nothing else but the Wilson lines of the Seiberg-dual gauge group.}
It is pretty clear that the $tt^*$ geometry should thus be the rank ${n_f \choose n_c}$ triply-periodic
instanton obtained as the $n_c$-th exterior power of the rank $n_f$ bundle described above. 

In order to describe the $tt^*$ geometry for the true SQCD theory, \textit{i.e.}\! a $SU(n_c)$ gauge theory with $n_f$ flavors, 
we need to ``ungauge'' the diagonal $U(1)$ gauge symmetry, \textit{i.e.}\! do a Nahm transform of the rank ${n_f \choose n_c}$ triply-periodic
instanton to some triply-periodic monopole geometry for the vector $U(1)$ flavor symmetry.

\subsection{4d probes of more general 6d (1,0) SCFT's}

\begin{figure}
\centering
\includegraphics[width=.8\textwidth]{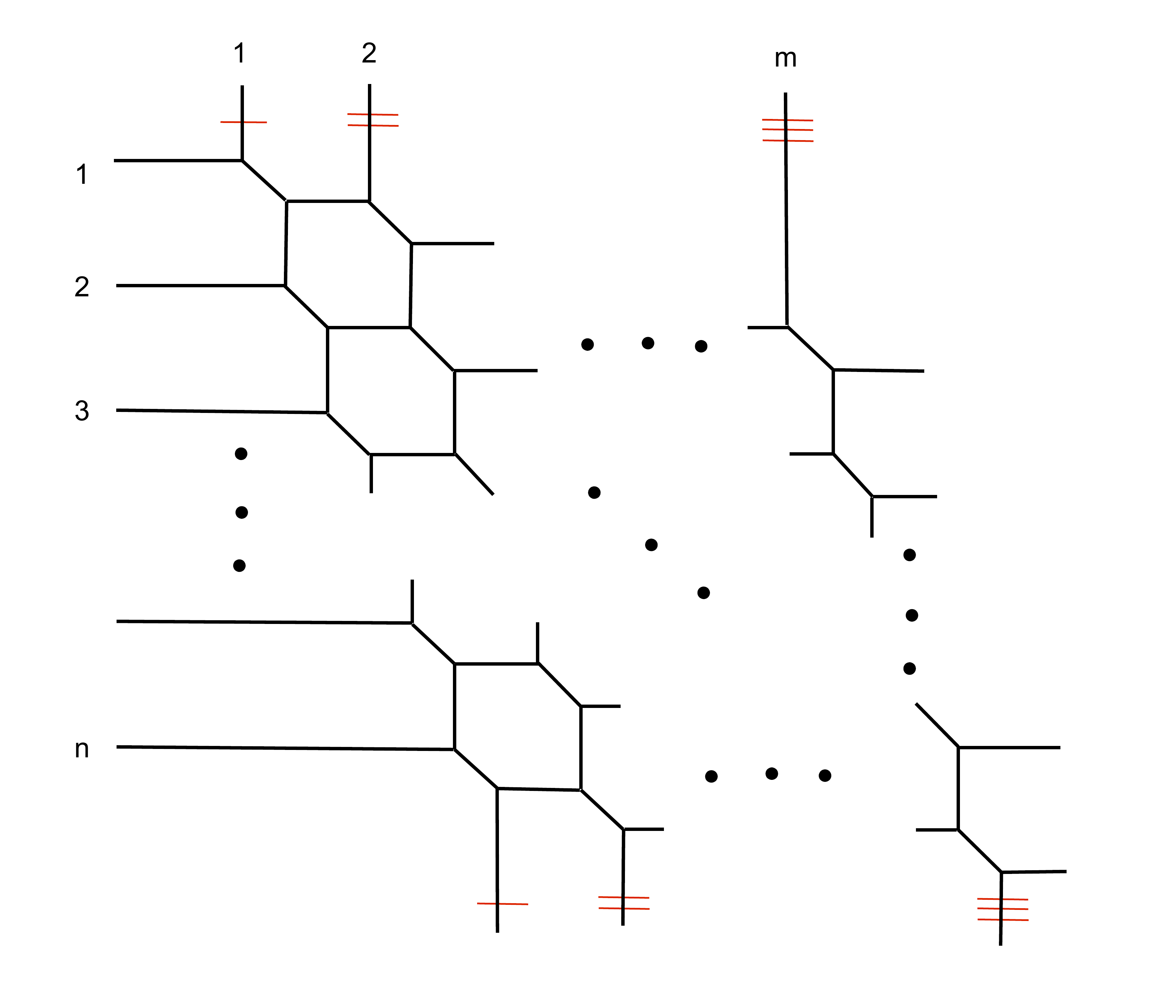}
\caption{This web diagram is a circle compactification of the 6d (1,0) SCFT of $m$ M5 branes in the presence of
$A_{n-1}$ singularity.}
\label{fig:nm}
\end{figure}

It is natural to consider more general singly periodic web diagrams. See Fig. \ref{fig:nm}.
  In this case we have
$n$ horizontal directions on the web broken by $m$ vertical directions.  This 5d theory corresponds to
compactifications of the 6d $(1,0)$ SCFT, given by $m$ parallel M5 branes probing an
$A_{n-1}$ singularity, on a circle \cite{Mstring2,HI}.  The distance between the vertical
lines relate to the separation of M5 branes.   Taking the spectator 5-brane in the vertical
direction and suspending a D3 brane, gives rise in 6d, to a 4d theory with ${\cal N}=1$.  The corresponding
theory will have $n$ vacua, giving rise to $tt^*$ geometry of $SU(n)$ instantons on $\R\times T^3$ with instanton number $m$.
The corresponding spectral curve would be given by 
\begin{equation}
\sum_{k=0}^{m} a_k\, e^{kt} \prod_{a=1}^n \Theta(-\sigma+ \mu_a^k, \tau) =0,
\end{equation}
which gives us a more-general $SU(n)$ instanton geometry on $\R_t \times T^3$, with boundary conditions at large $|t|$ still controlled by 
the $\mu_a^0$ and $\mu_a^n$ parameters.\footnote{ The $(n-1)m+(m+1)-1=nm$ complex normalizable moduli
coming from the choices of the other $\mu_a^i$, the $a_k$, and getting rid of one overall normalization for
the equation,  combine with the moduli of the line bundle over the spectral curve to give a moduli space of solutions of hyperK\"ahler dimension $m n$. }

Strictly speaking, it is not obvious that the $U(1)$ gauge group associated to the FI parameter $t$ will survive the field-theory limit. 
It is more-likely that a well-defined co-dimension two defect in the $(1,0)$ 6d theories would support an $U(1)$ flavor symmetry in its world volume, 
and that the triply-periodic instanton $tt^*$ geometry is the result of gauging that $U(1)$ flavor symmetry. 
In order to describe the $tt^*$ geometry of the original defect, we should do a Nahm transform back to 
a triply-periodic monopole geometry, and re-interprete the spectral curve 
\begin{equation}
\sum_{k=0}^{m} a_k\, e^{kp} \prod_{a=1}^n \Theta(-\sigma+ \mu_a^k, \tau) =0
\end{equation}
as the spectral curve for a $U(N)$ triply-periodic monopole solution on $T^3_m$, in the presence of $n$ Dirac monopoles of charge $1$ and $n$ of charge $-1$.

\begin{figure}
\centering
\includegraphics[width=.8\textwidth]{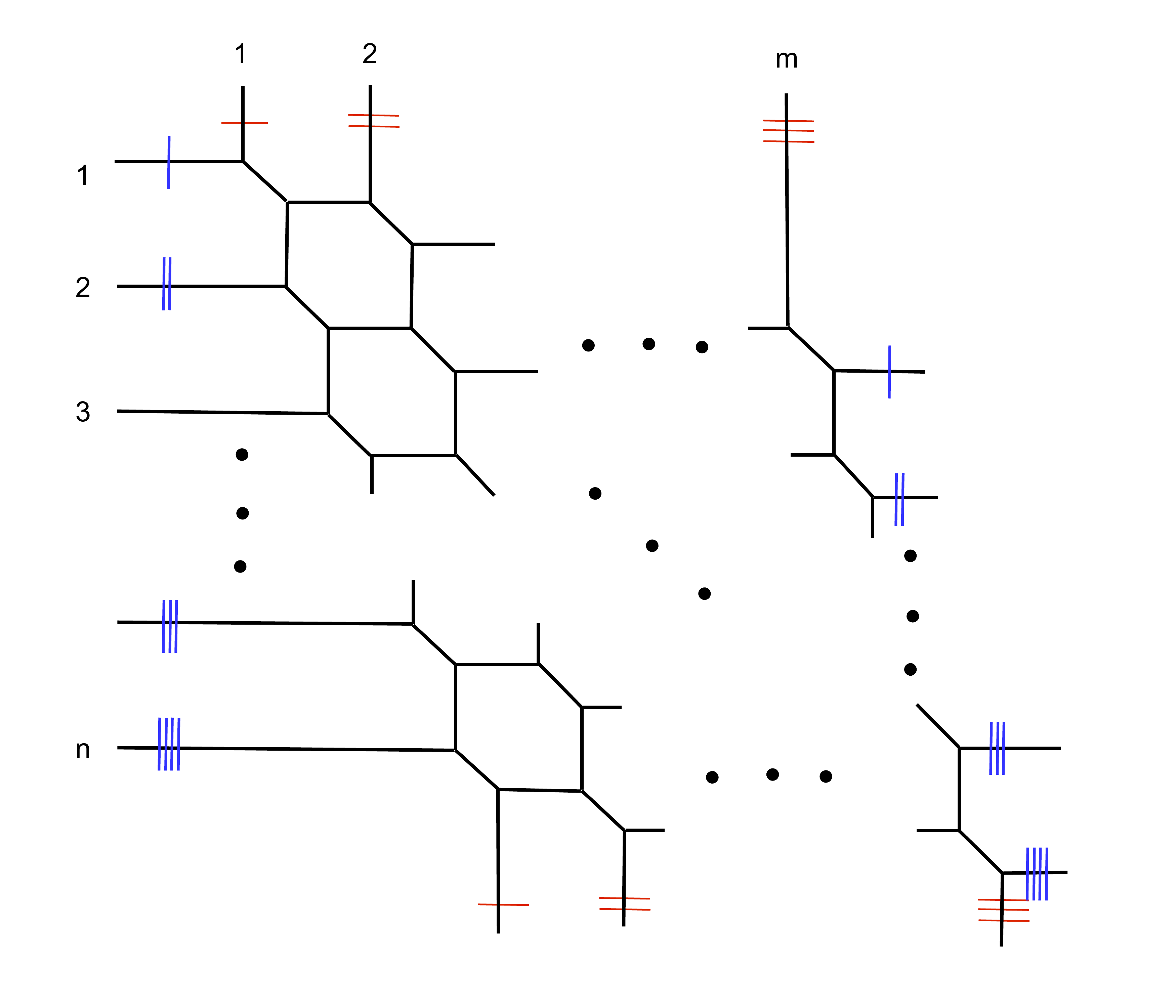}
\caption{This web diagram is a circle compactification of the little 6d string theory of $m$ M5 branes in the presence of
$A_{n-1}$ singularity with one transverse circular geometry.  Adding a spectator brane and a probe
to this geometry realizes the $tt^*$ geometry associated to $SU(n)$ instantons with instanton number
$m$ on (non-commutative) $T^4$.}
\label{fig:nml}
\end{figure}

It is also natural to ask if we can get instantons on $T^4$, by having a `periodic' version of FI parameter $t$.
This is indeed possible, because we can consider the doubly periodic brane geometry, \textit{i.e.}\! 5-branes
not on a cylinder but on a $T^2$.   See Fig. \ref{fig:nml}.

As noted in \cite{Mstring1} this geometry will engineer the little string theories.  More specifically
this corresponds to $m$ M5-branes probing an $A_{n-1}$ singularity where one transverse
dimension to the M5 branes has been compactified on the circle.  This would then lead to the
parameter space being $T^4$ and the $tt^*$ geometry would correspond to $SU(n)$ instantons
of instanton number $m$ on $T^4$.\footnote{Exchanging the vertical and the horizontal
direction will map this to $SU(m)$ instantons of instanton number $n$, which is an instance of
Fourier-Mukai/Nahm transformation.}  The corresponding spectral curves will involve level
$(n,m)$ genus 2 $\Theta$ functions
as discussed in \cite{HoIV}.

As already discussed, to obtain the conventional $tt^*$ geometry we had to turn off anomalous flavor
Wilson lines.  Moreover we have argued that when we turn them on the
$tt^*$ geometry becomes non-commutative.  Given the unusual nature of this result
it is interesting to note that we can get a confirmation of our results from a different perspective
from the elegant work \cite{GM,GMS}.  In particular they show that the moduli space of non-commutative
instantons on $T^4$ is given by the moduli space of the $(1,0)$ superconformal theories we have discussed.
Moreover they show that the spectral curve for such instantons are precisely the associated
Seiberg-Witten curves.  The non-commutativity is mapped to horizontal and vertical shifts
as we go around the cycles of the plane of the 5-branes.  This agrees with what we expected in that when we have
4d flavor symmetries which would have anomalies (if gauged)
we get non-commutative versions of $tt^*$ geometry.  Note that in these contexts the value of non-commutative parameters are not part of the
moduli space of the $tt^*$ geometry.  They are fixed background values.  Moreover turning on the angular
parts of the mass parameters, we expect to get 3 non-commutativity parameter for each periodic
direction of the 5brane plane in agreement with the results of \cite{GM,GMS}.

There is a further modification of this setup, which is worth mentioning. 
On the periodic fivebrane web picture, it corresponds to having the bundles of semi-infinite fivebranes 
end on groups of D7 branes, as in \cite{Benini:2009gi}.

Alternatively, if we T-dualize to a system of D6 branes crossing NS5 branes \cite{Hanany:1997gh}, we can consider the full Hanany-Zaffaroni setup which includes $D8$ branes, 
to describe a somewhat larger class of $(1,0)$ theories which can be interpreted, as in lower dimensional cases, as ``Higgs branch descendants'' of the 
6d theories described above, \textit{i.e.}\! sit at the bottom of an RG flow initiated by turning on some special Higgs branch vevs. 
The spectral curve is a slight generalization of the above:
\begin{align}
&e^{(N+1) \tau} \prod_a\Theta(-\sigma+ \mu'_a, \tau)^{n'_a}  + \cr &\sum_{k=1}^N e^{(N+1-k) \tau} \prod_a\Theta(-\sigma+ \mu'_a, \tau)^{\mathrm{max}(n'_a-k,0)}  \Theta(\sigma- \mu_a, \tau)^{\mathrm{max}(n_a+k-N-1,0)}\Theta_n^{(k)}(\sigma,\tau)+\cr & \prod_a \Theta(\sigma- \mu_a, \tau)^{n_a}=0
\end{align}
which is constructed in such a way to describe $U(N)$ triply-periodic monopole solution on $T^3_m$, in the presence of Dirac monopoles of charge $n_a$ and  $-n'_a$.

\section{Line operators and the CFIV index}

The $tt^*$ geometry in 2 dimensions led, in particular, to the calculation of a new supersymmetric index, 
the CFIV index \cite{CFIV}, given by\footnote{\ The peculiar overall factor $1/2$ in the \textsc{rhs} is an artifact of the choices of normalization of the charges. The Fermi number $F$ is normalized in such a way that the supercharges have Fermi number $\pm 1$. Then the odd superspace coordinates $\theta$ also have charges $\pm 1$. Instead the $Q_{ab}$ is normalized in such a way that, at criticality, the difference between its maximal and minimal eigenvalues is $\hat c$. Effectively, this is the same as assigning axial charge $1$ to the superpotential $W$, and hence charges $\pm 1/2$ to the $\theta$'s. Therefore $Q_{ab}$ is $1/2$ the CFIV index. }
\begin{equation}\label{q5}
Q_{ab}=\lim_{L\rightarrow\infty}{i\beta \over 2 L}\,\mathrm{Tr}_{ab}(-1)^F F\, e^{-\beta H}
\end{equation}
where the space is taken to be a segment of length $L$ with boundary conditions $a$, $b$ at the two ends, and we take the infinite volume limit $L\rightarrow \infty$.   $Q_{ab}$ can be identified with the $tt^*$ connection in the direction
of RG flow, in a suitable gauge.
 
$Q_{ab}$ is an index in the sense that it depends
only on a finite number of parameters in the theory (F-terms) and is insensitive to all the others. In the limit $\beta\rightarrow\infty$ it becomes an index in the ordinary sense, which counts the net number of short supersymmetry representations in the Hilbert space sector specified by the boundary conditions $a$, $b$.
Furthermore, one can exchange space and Euclidean time and relate $Q_{ab}$ to the expectation value of
the axial R-symmetry charge $Q^5$ (using the 2d fact that $j^5_\mu=*j^F_\mu$), which is broken away from the conformal point:
$$Q_{ab}=\frac{1}{2}\,\langle a| Q^5|b\rangle.$$
This can be interpreted as the action of the operator 
$$Q^5=\int_{S^1_\beta} \Big(\,j^F_\beta+Q\text{--exact}\Big)$$ on the ground states:
\begin{equation}\label{opinsert}
Q_{ab}=\frac{1}{2}\,\Big\langle a\,\Big| \oint_{S^1_\beta}  j^F_\beta\; \Big|\,b\Big\rangle
\end{equation}
where by $j^F_\beta$ we mean the component of the Fermion number R-current $j^F$ in the direction of
$S^1_\beta$.
At the conformal point, this corresponds to the spectrum of the R-charges of the Ramond ground states,
which by spectral flow, gives the spectrum of chiral operators in the theory.  Twice the highest eigenvalue
of $Q^5$ corresponds to $\hat c$, the central charge of the $\mathcal{N}=2$ theory.
Away from the conformal point, even though $Q^5$ is no longer conserved, one
can still compute its spectrum restricted to the ground states, and  it was shown \cite{CFIV} that the entire spectrum of $Q_{ij}$ is monotonically decreasing as we
flow to the infrared.  Applied to the highest eigenvalue of $Q^5$ as one flows from one fixed point
to another, this leads to the statement that along RG flow $\hat c$ decreases.

It is natural to ask what are the physical implications of the CFIV index, applied to theories which
arise from 3 or 4 dimensions.  In this section we take some preliminary steps in this direction trying to find the physical meaning of this quantity.  Moreover we compute it explicitly for the case of free chiral fields in 3 dimensions with a twisted mass.

\subsection{CFIV index and 3d theories}

Consider a 3d theory, compactified on a rectangular torus with periods $R,\beta$.  We can view
this as a 2d theory on a spatial circle of length $\beta$, by viewing the 3d fields as an infinite
tower of KK modes arising from compactification on a circle of length $R$.
The conserved 2d R-symmetry which corresponds to the fermion number $F$ will lift
in the 3d context to the conserved 3d R-symmetry which is present for all 3d theories with ${\cal N}=2$.
We can thus interpret the expression \eqref{q5} as computing the same quantity, except that
the space is now two dimensional, comprising of $\R^1\times S^1$, where the length of $\R^1$ is taken
to be $L$ and we take $L\rightarrow \infty$, and the length of $S^1$ is $R$.  Moreover the answer can now depend
on $x_i,y_i$, where the $y_i$'s correspond to the imaginary part of the 2d coupling parameters, and $x_i$'s can
be viewed as an additional insertion of flavor fugacities around the $\beta$ circle:
\begin{equation}\label{3dq5}
Q_{ab}(x_i,y_i,m_i, \rho)=\lim_{L\to\infty} {i\beta \over 2L}\,\mathrm{Tr}_{ab}\Big[(-1)^F F\, e^{-\beta H +2\pi i x_i f_i}\Big]
\end{equation}
where we have separated out the imaginary piece $y_i$ from the real part $m_i$ of the twisted mass parameters,
and set $\rho =\beta/R=-i\tau$; $f_i$ denote the $i$-th flavor charge.
We will compute this quantity for the case of the 3d free chiral model and verify this
Hilbert space interpretation.

Moreover, we can also look at this quantity from the perspective of the dual channel.  Namely we can
consider the Hilbert space of the 3d theory on a $T^2$ with periods $(\beta,R)$ and flavor Wilson lines $(x_i,y_i)$
around the two cycles. In this context we have
\begin{equation}\label{3dt2}
2\,Q_{ab}(x_i,y_i,\mu_i, \rho)=\Big\langle a\,\Big| \oint_{S^1_\beta}\oint_{S^1_R}  j^F_\beta\; \Big|\,b\Big\rangle_{(\beta_{x_i},R_{y_i})}\equiv \beta R\; 
\big\langle a\,\big|\, j^F_\beta\, \big|\,b\big\rangle_{(\beta_{x_i},R_{y_i})}. 
\end{equation}
Of course, by interchanging the role of the two circles, the CFIV index also computes the vacuum matrix elements of the other component of the current $j^F_R$ (the third component, $j^F_L$, has vanishing matrix elements between vacua). 

\subsection{Possible interpretations of $Q$ in terms of line operators}
In the case of 2d, at the conformal fixed point the (differences) in the spectrum of $Q$
determine the dimension of chiral fields.  In particular in that case we have
$$\Phi_i (0) {\overline \Phi_i }(z)\sim \frac{A}{ |z|^{2 Q_i}},\qquad z\sim 0,$$
where $Q_i$ denote the charge of the chiral field $\Phi_i$, and $A$ can be read from the $\beta\to0$ behavior of the $tt^*$ metric \cite{ttstar}.  It is natural to ask if a similar statement holds
for the case of 3d theories at their conformal limit where $m_i=0$.  In this case, as already mentioned, the chiral fields are replaced by line operators,
and so the question would be: How the partition function of the theory depends on separation $|z|$ of a line operator
and its conjugate?  

Consider the line operators wrapping the $S^1_R$.  From another perspective, this can be interpreted
as a particle defect.  A natural question in this context would be how the energy of the system depends
on the separation between a line operator and it conjugate, \textit{i.e.}\! the Casimir energy of line operator/ anti-line operator system.
If $Q$ is related to such an energy, as we increase $R\rightarrow \infty$ for a fixed $\beta$, \textit{i.e.} as we take $\rho \rightarrow 0$,
$Q$ should grow linearly in $R$ since the energy $E(R,|z|)$ should be proportional to the spatial size $R$ of the system, up to finite size corrections. 
So we would expect $Q/R$ to have a finite limit as $R\rightarrow \infty$.  As we shall see in the explicit
example below, this is indeed the case.  It would be interesting to see if $Q/R$ in this limit is related
to the Casimir energy of pairs of conjugate line operators.  It would also be interesting
to connect the conformal limit of this computation to the cusp anomalous dimension for line operators
(see e.g. \cite{malda}).

\subsection{On the CFIV index for the 3d chiral model}

We consider the free chiral model in 3d with twisted mass $m$ which upon reduction on a circle of radius $R$ is
equivalent to the 2d (2,2) LG model \eqref{freeKK} with
$$
W(Y_n)=\sum_{n\in\Z} \Bigg(\frac{1}{2}\!\left(\frac{m}{2\pi}+i\,\frac{n+y}{R}\right)Y_n-e^{Y_n}\Bigg).
$$
We further consider putting the theory on a $tt^*$ circle of length $\beta$. We set $\rho=\beta/R=-i\tau$. We write $x$ for the vacuum angle with period $1$, $y$ for the second angle (also with period $1$) associated with the imaginary part of the 2d twisted mass, and set $z=\beta\,m/2\pi$; then the 3d real mass (made dimensionless by multiplying it by $\beta$) is $2\pi z$.

The CFIV index for this model is the sum over the KK modes of the CFIV index for the 2d single--mode theory. In appendix \ref{2dcfiv} we present several convenient expressions of this last index. From, say, eqn.\eqref{QBessel} we have
\begin{equation}\label{newindex}
\begin{split}
&Q(x,y,z,\rho)=\\
&=-\frac{1}{\pi}\sum_{n\in\Z}\sum_{k\geq 1} \frac{\sin(2\pi kx)}{k}\;\Big(2\pi k\big|z+i(y+n)\rho\big|\Big)\; K_1\!\Big(2\pi k\big|z+i(y+n)\rho\big|\Big).
\end{split}
\end{equation}
See eqn.\eqref{exactexpression} below for the Poisson--resummed expression of $Q(x,y,z,\rho)$ as a double Fourier series in the two periodic variables $x,y$.
\bigskip 

$Q(x,y,z,\rho)$ has two interesting limits. One is $R\rightarrow 0$ at $\beta$ fixed, that is, $\rho\rightarrow \infty$, while keeping $y\rho$ fixed. In this case all terms in the sum over the KK modes $n$, except for the zero mode $n=0$, vanish exponentially, and we get back the $2d$ expression with complex twisted mass $m+2\pi i y/R$.

The second one is the opposite limit $R\rightarrow \infty$. Before computing it, let us list the physical properties we expect the answer to have. 

\paragraph{Physical expectations as $R\to\infty$.} The $tt^*$ amplitudes in this limit are correlators of line operators wrapped on a cycle of large length $R$. Since the CFIV index is believed to give the values of some kind of \emph{extensive} quantity, like  the Casimir energy of the vacuum states on $T^2$ created by the line operators, we expect that, asymptotically for large $R$ and fixed $\beta$, $Q$ becomes proportional to $R$
\begin{equation}
Q(x,y,z,\beta/R)\bigg|_{R\rightarrow\infty} =\frac{R}{\beta}\cdot f(x,y,z)+\text{finite--size corrections},
\end{equation}  
where $f(x,y,z)/\beta$ is the \emph{finite} linear density of the said extensive quantity, which should scale with the temperature $\beta^{-1}$ by dimensional considerations. We also expect the function $f(x,y,z)$ to be $y$--independent, since a finite flavor twist over a circle of infinite length should not affect 
the value of the local density of an extensive quantity. Moreover, as a function of $x$, $f(x,y,z)$ should be periodic of period $1$ and odd (this last condition reflects consistency with CPT). Finally, since the mass of the physical particle in 3d is $2\pi |z|/\beta$, and the CFIV index is a Hilbert space trace of the form (\ref{3dq5}), for \emph{large} values of the mass $m=2\pi |z|/\beta$ the density $Q/R$ should have the standard thermodynamical expression
\begin{equation}\label{physicalpredictions}
\begin{split}
\frac{1}{\beta}\, f(x,y,z)&\sim \frac{i\beta}{2}\Big(e^{2\pi ix}-e^{-2\pi ix}\Big) \int \frac{d^2p}{(2\pi)^2}\,e^{-\beta\sqrt{p^2+m^2}}=\\
&=\frac{i}{4\pi}\Big(e^{2\pi ix}-e^{-2\pi ix}\Big) \left(\frac{1}{\beta}+m\right)e^{-\beta m},
\end{split}
\end{equation}
where the factors $e^{\pm 2\pi ix}$ arise from the role of $e^{2\pi ix}$ as particle number fugacity, and their relative combination is fixed by PCT and reality of $f(x,y,z)$. In fact, the \textsc{rhs} turns out to be the \emph{exact} expression of the terms proportional to $e^{\pm 2\pi i x}$ in the Fourier expansion of $f(x,y,z)$. This means that multiparticle states of total flavor charge $\pm1$ do not contribute to $Q$. This last statement is exact for all $R$, not just for $R$ large. Indeed, the term proportional to $e^{2\pi ix}$ in eqn.\eqref{newindex} is 
\begin{equation}
\begin{split}
\frac{i\beta}{2\pi}\,e^{2\pi i x}\sum_{n\in\Z}&\Big(\big|m+2\pi i(y+n)/R\big|\Big)\; K_1\!\Big(\beta \big|m+2\pi i(y+n)/R\big|\Big)\equiv\\
&\equiv\frac{i\beta}{2}\,e^{2\pi i x}\sum_{n\in\Z}\int\limits_{-\infty}^{+\infty}\frac{dp}{2\pi}\;e^{-\beta\sqrt{p^2+(2\pi)^2(n+y)^2/R^2+m^2}},
\end{split}
\end{equation}
which is the partition function of a particle of mass $m$ in an infinite cylinder of circumference $R$ and holonomy $\exp(2\pi i y)$.

More generally, given that this theory has only one physically distinct vacuum, we may think of computing the trace in eqn.\eqref{q5} by inserting a complete set of intermediate states, which may be taken to be free particle states; this implies that the coefficient of $e^{2\pi i k x}$ in the Fourier expansion of $Q$ (and hence of $f$) should be $O(e^{-\beta |k|\,|m|})$ for large $|m|$. Again, this is manifestly true for the expression \eqref{newindex}.


\paragraph{The function $f(x,y,z)$.} Before showing that these expectations are correct, and giving an explicit formula for $f(x,y,z)$, we rewrite the expression \eqref{newindex} in a more compact and illuminating form.
We start form the following integral representation of $K_1(w)$
\begin{equation}
K_1(w)=\frac{1}{w}\,\int_0^\infty dt\; e^{-t-w^2/4t}\qquad |\mathrm{arg}\,w|<\pi/4.
\end{equation}
Plugging this formula into \eqref{newindex} we get
\begin{equation}
Q(x,y,z,\rho)=-\frac{1}{\pi}\sum_{k\geq 1} \frac{\sin(2\pi kx)}{k}\;\int\limits_0^\infty dt \;e^{-t}\,\left(\sum_{n\in\Z}e^{-4\pi^2k^2(z^2+(y+n)^2\rho^2)/4t}\right).
\end{equation} 
The expression inside the big parenthesis is a $\theta_3$ function. To get a rapidly convergent expression for $\rho$ small we have just to express this function in terms of the $\theta$--function for the inverse period using its modular transformation properties. More precisely, we have
\begin{equation}
\begin{split}
\sum_{n\in\Z}e^{-4\pi^2k^2(z^2+(y+n)^2\rho^2)/4t}&=e^{-\pi^2k^2(z^2+y^2\rho^2)/t}\;\sum_{n\in\Z} \Big(e^{-2\pi^2 k^2\rho^2/t}\Big)^{n^2/2}\;\Big(e^{-2\pi^2k^2\rho^2 y/t}\Big)^n=\\
&=e^{-\pi^2k^2(z^2+y^2\rho^2)/t}\;\theta_3\big(i\pi^2k^2\rho^2\, y/t\;|\; i\pi k^2\rho^2/t\big),
\end{split}
\end{equation}
where
\begin{equation}
\theta_3(w|\sigma)= \sum_{n\in\Z} q^{n^2/2}\,e^{2 i n w},\qquad q=e^{2\pi i \sigma}.
\end{equation}
Thus
\begin{equation}\begin{split}
&Q(x,y,z,\rho)=\\
&=-\frac{1}{\pi}\sum_{k\geq 1} \frac{\sin(2\pi kx)}{k}\;\int\limits_0^\infty dt \;e^{-t}\,e^{-\pi^2k^2(z^2+y^2\rho^2)/t}\;\,\theta_3\big(i\pi^2k^2\rho^2\, y/t\;|\; i\pi k^2\rho^2/t\big).
\end{split}\end{equation}

The simplest way to compute the $\rho\rightarrow 0$ (\textit{i.e.}\! the $R\rightarrow\infty$) limit is to replace the theta function by its $S$--modular transform
\begin{equation}
\theta_3\big(w\,|\,\sigma\big)= (-i\sigma)^{-1/2}\;\exp\!\big(-i w^2/\pi\sigma\big)\; \theta_3\big(-w/\sigma\,|\,-1/\sigma\big),
\end{equation}
which gives
\begin{equation}\label{dualexppre}\begin{split}
&Q(x,y,z,\rho)=\\
&=-\frac{1}{(\pi)^{3/2}\,\rho}\sum_{k\geq 1} \frac{\sin(2\pi kx)}{k^2}\;\int\limits_0^\infty dt\;\sqrt{t} \;e^{-t-\pi^2k^2 z^2/t}\;\,\theta_3\!\left(-\pi y \;\Big|\; \frac{it}{\pi k^2\rho^2}\right).
\end{split}\end{equation}

The function $f(x,y,z)$ is defined by taking first the infinite size limit $R\rightarrow \infty$ at fixed $x,y,z$. In the limit $\rho\rightarrow 0$ ($R\rightarrow\infty$) the theta function in the integrand of \eqref{dualexppre} may be replaced by its asymptotic expression, which is just $1$. Thus, since $\rho=\beta/R$, 
for large $R$
\begin{equation}
Q(x,y,z,\beta/R)\approx -\frac{R}{(\pi)^{3/2}\,\beta}\sum_{k\geq 1} \frac{\sin(2\pi kx)}{k^2}\;\int\limits_0^\infty dt\;\sqrt{t} \;e^{-t-\pi^2k^2 z^2/t},
\end{equation}
Now,
\begin{equation}
\int\limits_0^\infty \exp\!\Big(-t-\frac{w^2}{4t}\Big)\,\sqrt{t}\;dt=\frac{1}{\sqrt{2}}\,w^{3/2}\; K_{3/2}(w)\equiv\frac{\sqrt{\pi}}{2}\,e^{-w}\left(1+w\right)\end{equation}
and
\begin{equation}\label{leadingQ}
Q(x,y,z,\beta/R)\approx -\frac{R}{\beta}\,\frac{1}{2\pi}\sum_{k\geq 1} \frac{\sin(2\pi kx)}{k^2}\;e^{-2\pi k |z|}\Big(1+2\pi k |z|\Big),
\end{equation}
which is of the expected form with
\begin{equation}\label{yyyrr}
\begin{split}
f(x,y,z)=&\frac{1}{4\pi i }\Big(\mathrm{Li}_2\big(e^{-2\pi (|z|+ix)}\big)-\mathrm{Li}_2\big(e^{-2\pi (|z|-ix)}\big)\Big)+\\
&+\frac{|z|}{2 i} \Big[\log\big(1-e^{-2\pi(|z|-ix)}\big)-\log\big(1-e^{-2\pi(|z|+ix)}\big)\Big].
\end{split}
\end{equation}
This expression exactly matches the physical predictions \eqref{physicalpredictions}. In particular, $f(x,y,z)$ is independent of $y$, periodic and odd in $x$, and has a leading behavior for large $|m|$ as expected, eqn.\eqref{physicalpredictions}. As already mentioned, the coefficient of $e^{\pm 2\pi i x}$ in the Fourier expansion of $f(x,y,z)$ are exactly given by eqn.\eqref{physicalpredictions}.

In particular, at $z=0$ we have
\begin{equation}
 f(x,y,0)=-\frac{1}{2\pi}\,\Pi(x),
\end{equation}
where $\Pi(x)$ is the Lobachevsky function (a.k.a.\! the Clausen integral), which expresses \emph{inter alia} the volume of the ideal tetrahedra in hyperbolic 3--space. One has 
\begin{equation}
\Pi(x)=\sum_{m\geq 1}\frac{\sin(2\pi mx)}{m^2}= \frac{\mathrm{Li}_2(e^{2\pi ix})-\mathrm{Li}_2(e^{-2\pi i x})}{2i}=-\int\limits_0^{2\pi x}\log\!\Big(2\sin\!\big(s/2\big)\Big)ds.
\end{equation}

\paragraph{The double Fourier series for $Q$.} We can write alternative  expressions for the 3d chiral CFIV index which are more convenient for computing sub--leading corrections to  the large $R$ behaviour \eqref{leadingQ}.
We start from eqn.\eqref{dualexppre} 
\begin{equation}\label{dualexppre2}\begin{split}
&Q(x,y,z,\rho)=\\
&=-\frac{1}{(\pi)^{3/2}\,\rho}\sum_{k\geq 1} \frac{\sin(2\pi kx)}{k^2}\;\int\limits_0^\infty dt\;\sqrt{t} \;e^{-t-\pi^2k^2 z^2/t}\;\,\theta_3\!\left(-\pi y\;\Big|\; \frac{it}{\pi k^2\rho^2}\right)=\\
&= -\frac{1}{(\pi)^{3/2}\,\rho}\sum_{k\geq 1} \frac{\sin(2\pi kx)}{k^2}\;\int\limits_0^\infty dt\;\sqrt{t} \;e^{-t-\pi^2k^2 z^2/t}\;\,\sum_{n\in\Z}e^{- t n^2/k^2\rho^2}\, e^{-2\pi i n y}.
\end{split}\end{equation}
The integral in $t$ may be computed using
\begin{equation}
 \int\limits_0^\infty dt\,\sqrt{t}\; e^{-a t-b/t}= \frac{\sqrt{\pi}}{2 \,a^{3/2}}\big(1+2\sqrt{ab}\big)\, e^{-2\sqrt{ab}}\qquad \mathrm{Re}\,a,\; \mathrm{Re}\,b>0.
\end{equation}
This allows us to rewrite the CFIV index explicitly as a double Fourier series in $x$ and $y$:
\begin{equation}\label{exactexpression}
\begin{split}
Q(x,y,z,\rho)
=\frac{1}{\rho}\, f(x,0,z)
- \frac{\rho}{\pi} &\sum_{k\geq 1\atop n\geq 1} \frac{k\,\sin(2\pi k x)\, \cos(2\pi n y)}{(n^2+k^2\rho^2)^{3/2}}\big[\rho+2\pi |z|\sqrt{k^2\rho^2+n^2}\big]\;\times\\
&\qquad\times\exp\!\Big(-2\pi |z|\,\sqrt{n^2/\rho^2+k^2}\Big),
\end{split}\end{equation}
which, of course, matches with the result one would obtain starting with the known double Fourier series for the $tt^*$ metric of the 3d chiral field.

\paragraph{Finite--size corrections.} The first term in the \textsc{rhs} of the exact equation \eqref{exactexpression} is the previous asymptotic behavior as $\rho\rightarrow 0$ (keeping fixed $2\pi |z|/\rho\equiv |m| R$). One may go on and compute the corrections in powers of $\rho^2$. The first correction is $O(\rho^2)$
\begin{equation}
\begin{split}
Q-\frac{1}{\rho}\,f(x,y,z)=&-\frac{\rho^2}{2\pi}\Big[\mathrm{Li}_3(e^{-|m|R+2\pi y})+\mathrm{Li}_3(e^{-|m|R-2\pi y})+\\
&\ +|m|R\,\mathrm{Li}_2(e^{-|m|R+2\pi y})+|m|R\,\mathrm{Li}_2(e^{-|m|R-2\pi y})\Big]+O(\rho^4).
\end{split}
\end{equation}

\subsection{General 3d $\mathcal{N}=2$ models: large mass asymptotics}

From the previous physical discussion, we expect that the limit
\begin{equation}
\lim_{R\to\infty} \frac{1}{R}\,Q(R)
\end{equation}
exists for all 3d $\mathcal{N}=2$ models compactified to 2d on a circle of length $R$. This will correspond to the energy per unit length of the system described at the beginning of the section. This fact may be checked explicitly for large twisted masses/FI parameters. Indeed, in this limit the $tt^*$ equations linearize (this being just the statement that a non--Abelian monopole looks Abelian far away from its sources) and we get
\begin{equation}\label{IRasy}
\begin{split}
Q_{ab}\approx& \frac{i R}{2}\bigg(\Delta_{ab}\!\int \frac{d^2p}{(2\pi)^2}\, e^{-\beta \sqrt{p^2+m_{ab}^2}}\ +\\
&+\sum_{\boldsymbol{k}\in\Z^r\atop \text{primitive} }
\Delta(\boldsymbol{k})_{ab}\, e^{2\pi i \boldsymbol{k}\cdot \boldsymbol{x}}\! \int\frac{d^2p}{(2\pi)^2}\, e^{-\beta\sqrt{p^2+m_{ab}(\boldsymbol{k})^2}}+\\
&+ \sum_{\ell\geq 2}\sum_{\boldsymbol{k}\in\Z^r\atop \text{primitive} }c(\ell, \boldsymbol{k})\, e^{-\ell\beta m_{ab}(\boldsymbol{k})}\,e^{2\pi i \ell \boldsymbol{k}\cdot\boldsymbol{x}}+\text{sub--leading} \bigg)
\end{split}
\end{equation}
where $\boldsymbol{x}=(x_1,\dots,x_r)$ are the flavor chemical potentials, $m_{ab}(\boldsymbol{k})$ (resp.\! $m_{ab}$) is the mass of the lightest particle in the 3d Hilbert space sector  $\mathcal{H}_{ab}$ having flavor charges $\boldsymbol{k}$ (resp.\! being flavor neutral);  the coefficients $c(\ell,\boldsymbol{k})$ are polynomially bounded in terms of the masses, and hence the terms in the last line are to be thought of as `subleading'. The coefficients $\Delta_{ab}$ and $\Delta(\boldsymbol{k})_{ab}$ are integers which satisfy the PCT conditions
\begin{equation}
\Delta_{ba}=-\Delta_{ab},\qquad \Delta(\boldsymbol{k})_{ba}=-\Delta(-\boldsymbol{k})_{ab},
\end{equation}
and count the net multiplicities, in the sense of the CFIV susy index, of 3d BPS particles having the corresponding quantum numbers. Note that the sum in eqn.\eqref{IRasy} is over the primitive flavor charge vectors only. In writing the above equation we made use of a genericity assumption, namely that there are no accidental alignments in the 2d effective central charge complex plane.

The bottom line is that is we may read the 3d BPS spectrum of a $\mathcal{N}=2$ theory from the asymptotical behavior  at infinity of the associated (higher dimensional) $tt^*$ monopole fields. In turn this spectrum may be related, through the CFIV index, with the scaling behavior with the separation $\Delta$ of an extensive energy--like quantity associated with a configuration of two parallel line operators placed at the distance $\Delta$.   

\section*{Acknowledgements}

We would like to thank M. Aganagic, U. Bruzzo, C. Cordova, T. Dimofte, B. Dubrovin, T. Dumitrescu, J. Gomis, S. Gukov, B. Haghighat, J. Heckman, D. Jafferis, J. Maldacena,
S. Sachdev, M. Verbitsky and E. Witten for useful discussions.   We would also like to thank the SCGP for hospitality during the 11th Simons Workshop on math
and physics, where this work was initiated.

The work of C.V. is supported in part by NSF grant PHY-1067976.

\newpage
\appendix

\section{Proofs of identities used in \S.\ref{sec:basicexample}}\label{appednix31}

\subsection{Eqns.\eqref{property1} and \eqref{property2}}\label{prooffirsttwoide}
We set $\zeta=1$ by rescaling $\mu$ and $\bar\mu$. Then from eqn.\eqref{Phiintegralform}
\begin{equation}
 \Phi(x,\mu,\bar\mu)=\frac{1}{2\pi i}\left(\int_0^\infty \frac{dt}{t-i}\,\sum_{m\geq 1}\frac{e^{-2\pi m(\mu t+ \bar\mu t^{-1}+i x)}}{m}-\Big(i\leftrightarrow -i\Big)\right),
\end{equation}
and perform the change of variables $t=s^{-1}$
\begin{equation}
 \begin{split}
\Phi(x,\mu,\bar\mu)&=\frac{1}{2\pi i}\left(\int_0^\infty \frac{i\,ds}{s(s+i)}\,\sum_{m\geq 1}^\infty \frac{e^{-2\pi m(\bar\mu s+\mu s^{-1}+ix)}}{m}-\Big(i\leftrightarrow -i\Big)\right)=\\
&= \frac{1}{2\pi i}\left(\int\limits_0^\infty ds\,\left(\frac{1}{s}-\frac{1}{s+i}\right)\,\sum_{m\geq 1} \frac{e^{-2\pi m(\bar\mu s+\mu s^{-1}+ix)}}{m}-\Big(i\leftrightarrow -i\Big)\right)=\\
&=\frac{1}{2\pi i}\left(\int_0^\infty \frac{ds}{s-i}\,\sum_{m\geq 1} \frac{e^{-2\pi m(\bar\mu s+\mu s^{-1}-ix)}}{m}-\Big(i\leftrightarrow -i\Big)\right)+\\
&\quad+\frac{1}{2\pi i}\left(\sum_{m\geq 1}\frac{2i \sin(2\pi m x)}{m}\int_0^\infty \frac{ds}{s}\; e^{-2\pi m(\bar\mu s+\mu s^{-1})}\right).
 \end{split}
\end{equation}
Comparing with eqn.\eqref{LLLLL2} we get the desired functional equation for $\Phi$
\begin{equation}\label{desfunceqn}
\Phi(x,\mu,\bar\mu)-\Phi(-x,\bar\mu,\mu)= L(x,\mu,\bar\mu).
\end{equation}
As stated in the main body of the text, this is equivalent to the equation for the $tt^*$ metric in terms of the amplitudes $\Pi$ (at, say, $\zeta=1$)
\begin{equation}\label{tttyyui}
\log\Pi_\mathrm{can}-\log\Pi_\mathrm{can}^*= \log G-\log|\eta|\equiv L,
\end{equation}
where $\Pi_\mathrm{can}$ are the amplitudes in the canonical base (in which $\eta=1$), that is,
\begin{equation}
\log\Pi_\mathrm{can}=\log\Pi+\frac{1}{2}\log\mu.
\end{equation}
Then, in view of eqn.\eqref{pi-phi}, eqn.\eqref{tttyyui} reduces to \eqref{desfunceqn} since 
$\Phi(x,\mu,\bar\mu)^*=\Phi(-x,\bar\mu,\mu)$ on the physical slice where $\bar\mu=\mu^*$.

In view of eqn.\eqref{Phiintegralform}, eqn.\eqref{property2} follows from the obvious identity, valid for all natural numbers $n$,
\begin{equation}
\log\!\Big(1-e^{-2\pi n(\mu t+\bar\mu t^{-1}+i x)}\Big)=\sum_{k=0}^{n-1}
\log\!\Big(1-e^{-2\pi (\mu t+\bar\mu t^{-1}+i x+i k/n)}\Big).
\end{equation}

\subsection{The asymmetric UV limit $\bar\mu\rightarrow 0$}\label{appGamma1}

Again we set $\zeta=1$ by a redefinition of the $\mu,\bar\mu$, and we assume first the redefined $\mu,\bar\mu$ to satisfy $\mathrm{Re}\,\mu,\;\mathrm{Re}\,\bar\mu>0$.
From eqn.\eqref{pi-phi} we have
\begin{equation}\label{appone}
\Big(\partial_\mu-\partial_x\Big)\log\Pi(x,\mu,\bar\mu)= \log\mu -\frac{1}{2\mu}+\Big(\partial_\mu-\partial_x\Big)\Phi(x,\mu,0),
\end{equation}
while from eqn.\eqref{Phiintegralform},
\begin{equation}
2\pi i\, \Phi(x,\mu,0)= \int_0^\infty \frac{dt}{t-i}\;\log\!\Big(1-e^{-2\pi(\mu t+ix)}\Big)-\Big(i\leftrightarrow -i\Big),
\end{equation}
from which we get
\begin{equation}\begin{split}
\big(\partial_\mu-\partial_x\big)\Phi(x,\mu,0)&=-i\int_0^\infty
\frac{dt}{1-e^{-2\pi(\mu t+ix)}}-\Big(i\leftrightarrow -i\Big)=\\
&= 2 \sum_{n\geq 1} \sin(2\pi n x)\int_0^\infty dt\; e^{-2\pi n\mu t}=\\
&=\frac{1}{\pi \mu}\sum_{m\geq 1}\frac{\sin(2\pi n x)}{n}=  
\frac{1}{\mu}\left(\frac{1}{2}-x\right).
 \end{split}
\end{equation}
Plugging this result in \eqref{appone}, we get
\begin{equation}
\Big(\partial_\mu-\partial_x\Big)\log\Pi(x,\mu,0)=\log\mu-\frac{x}{\mu}.
\end{equation}
This equation has the general solution
\begin{equation}
\log\Pi(x,\mu,0)=-x\log\mu+f(x+\mu),
\end{equation}
for some function $f(w)$. To fix $f(w)$ it is enough to compute $\log\Pi(x,\mu,0)$ for a fixed value of $x$ (and all $\mu$). We shall compute it for $x$ a half--integer.

\subsubsection{$\log\Pi(x,\mu,0)$ for $x\in\tfrac{1}{2}\Z$}\label{appGamma2}

For $x\in\Z$ we have
\begin{equation}
 \begin{split}
  \Phi(x\in\Z,\mu,\bar\mu=0)&=\frac{1}{2\pi i}\left(\int_0^\infty \frac{dt}{t-i} \sum_{m\geq 1} \frac{e^{-2\pi m\mu t}}{m}-\Big(i\leftrightarrow -i\Big)\right)=\\
&=\frac{1}{\pi} \int_0^\infty \frac{dt}{t^2+1} \sum_{m=1}^\infty \frac{e^{-2\pi m \mu t}}{m}=\\
&= \frac{1}{\pi}\int_0^\infty dt\;\left(\sum_{m=1}^\infty\frac{e^{-2\pi m\mu t}}{m}\right) \frac{d}{dt} \arctan(t)=\\
&=-\frac{1}{\pi}\int_0^\infty \arctan(t)\, dt\; \frac{d}{dt}\sum_{m=1}^\infty\frac{e^{-2\pi m\mu t}}{m}=\\
&=2\mu \int_0^\infty dt\;\frac{\arctan(t)}{e^{2\pi \mu t}-1}.
 \end{split}
\end{equation}
By a change of variables, we rewrite the last expression in the form
\begin{equation}
 2\int_0^\infty dt \; \frac{\arctan(t/\mu)}{e^{2\pi t}-1}.
\end{equation}
Then from eqn.\eqref{muzerofirst}, we have
\begin{equation}\label{appp2}
\log\Pi(x\in\Z,\mu,0)=\left(\mu-\frac{1}{2}\right)\log\mu-\mu+\mathrm{const.}+2\int_0^\infty dt \; \frac{\arctan(t/\mu)}{e^{2\pi t}-1}.
\end{equation}
Now we invoke Binet's formula
\begin{equation}\label{appeapp}
 \begin{split}
  \log\Gamma(z)=(z-1/2)\log z-z +\frac{1}{2}\log (2\pi)+2 \int_0^\infty \frac{\arctan(t/z)}{e^{2\pi t}-1}\,dt,
 \end{split}
\end{equation}
to conclude that, choosing the constant in \eqref{appeapp} to be zero, for all $\mu$ with $\mathrm{Re}\,\mu>0$, we have
\begin{equation}\label{pppaa5}
 \Pi(x\in\Z,\mu,\bar\mu=0)=\frac{\Gamma(\mu)}{\sqrt{2\pi}},
\end{equation}
as claimed.\medskip

For $x\in\tfrac{1}{2}+\Z$, 
\begin{equation}
\log\Pi(x=1/2,\mu,0)=\left(\mu-\frac{1}{2}\right)\log\mu-\mu+\Phi(1/2,\mu,0),
\end{equation}
while, from identity \eqref{property2}
\begin{equation}
\Phi(1/2,\mu,0)=\Phi(0,2\mu,0)-\Phi(0,\mu,0)\end{equation}
which in view of eqns.\eqref{muzerofirst}\eqref{pppaa5} is equivalent to
\begin{equation}
\log\Pi(x=1/2,\mu,0)=\log\Gamma(2\mu)-\log\Gamma(\mu)-\frac{1}{2}\log\mu -(2\mu-1/2)\log 2.
\end{equation}
Using the Gamma function identity
\begin{equation}
 \Gamma(2z)= \pi^{-1/2}\, 2^{2z-1}\, \Gamma(z)\,\Gamma(z+1/2),
\end{equation}
this becomes
\begin{equation}
\log\Pi(x=1/2,\mu,0)=\log\Gamma(\mu+1/2)-\frac{1}{2}\log\mu-\log\sqrt{2\pi},
\end{equation}
as claimed.

\subsubsection{Kummer formula}\label{appGamma3}
For $\epsilon\sim 0$, one has
\begin{equation}\label{kkum1}
\begin{split}
\log\Pi(x,\epsilon,0)+x\,\log\epsilon=&\log\sqrt{2\pi}+\left(x-\frac{1}{2}\right)\log\epsilon+\frac{1}{\pi}\sum_{m\geq 1}\frac{\cos(2\pi m x)}{m}\int_0^\infty \frac{dt}{t^2+1}-\\
&-\frac{1}{\pi}\sum_{m\geq 1}\frac{\sin(2\pi m x)}{m}\int_0^\infty \frac{t\,dt}{t^2+1}\,e^{-2\pi m \epsilon t}
\end{split}\end{equation}
One has
\begin{gather}
\int_0^\infty \frac{t\,dt}{t^2+1}\, e^{-2\pi m\, \epsilon\, t}= -\log(2\pi m \,\epsilon)-\gamma+O(\epsilon)\\
\sum_{m\geq 1}\frac{\cos(2\pi m x)}{m}=-\log(2\sin \pi x)\\
\sum_{m\geq 1}\frac{\sin(2\pi m x)}{m}=\frac{\pi}{2}(1-2x),
\end{gather}
and eqn.\eqref{kkum1} becomes
\begin{equation}
\begin{split}
\log\Pi(x,\epsilon,0)+x\,\log\epsilon=&\log\sqrt{2\pi}-\frac{1}{2}\log(2\sin\pi x)+\left(\frac{1}{2}-x\right)\Big(\log2\pi+\gamma\Big)+\\
&+\frac{1}{\pi}\sum_{m\geq 1}\frac{\sin(2\pi m x)}{m}\,\log m
+O(\epsilon)\end{split}
\end{equation}
According to our results, the limit as $\epsilon\rightarrow 0$ of the \textsc{lhs} is $\log\Gamma(x)$ (for $0<x<1$). The resulting expression for $\log\Gamma(x)$ is the celebrated Kummer formula.

\subsubsection{The $\zeta=1$ thimble brane function $\Phi(x,\mu,\bar\mu)$ for $\mathrm{Re}\,\mu \gtrless0$}\label{contoursmany}

In \S.\ref{thebasicbraneamplitude} we have saw that, for $\mathrm{Re}\,\mu>0$, the solution to the $tt^*$ brane amplitude for the model $W=\mu\,Y-e^Y$, which corresponds to the basic Lefshetz thimble brane, may be written in the form
\begin{equation}\label{remupositive}
\begin{split}
\Phi(x,\mu,\bar\mu)=\int\limits_\R ds\,f(s)\equiv& - \int\limits_\R \frac{ds}{2\pi} \Bigg(
\frac{\log\!\big[1-\exp(-2\pi\mu\, e^s-2\pi \bar\mu\ e^{-s}-2\pi i\, x)\big]}{e^{-s}+i}+\\
&+\frac{\log\!\big[1-\exp(-2\pi\mu\, e^s-2\pi \bar\mu\, e^{-s}+2\pi i\, x)\big]}{e^{-s}-i}\Bigg),
\end{split}
\end{equation}
the integral being evaluated along the real axis $\R\subset\C$. The integrand $f(s)$ has poles at
\begin{equation}
s=s_k\equiv \frac{i\pi}{2}+k\, i \pi.
\end{equation}

%
In \S.\ref{freechiraltheory3d}, we need the expression for the function $\Phi(x,\mu,\bar\mu)$ valid in the region $\mathrm{Re}\,\mu<0$, where eqn.\eqref{remupositive} does not apply since the integral does not converge.

The integral of the meromorphic function $f(s)$ along a contour $\gamma\subset\C$, produces a solution $\Phi(x,\mu,\bar\mu)_\gamma$ to the $tt^*$ brane amplitude equations provided:
\begin{itemize}
\item the integral $\int_\gamma f(s)\,ds$ is convergent;
\item in the physical region $\bar\mu=(\mu)^*$, the function  $\Phi(x,\mu,\bar\mu)_\gamma$ satisfies the reality condition
\begin{equation}\label{apprealitycondt}
\Phi(x,\mu,\bar\mu)_\gamma-\Phi(-x,\mu,\bar\mu)_\gamma^\ast=L(x,\mu,\bar\mu).
\end{equation}
\end{itemize}
For $\mathrm{Re}\,\mu>0$ the function $\Phi(x,\mu,\bar\mu)_\R$ defined by the \textsc{rhs} of  \eqref{remupositive} satisfies both requirements by the functional equation \eqref{desfunceqn}, together with the identity $\Phi(x,\mu,\bar\mu)_\R^\ast=\Phi(x,\bar\mu,\mu)_\R$ valid in the physical region.

The Lefshetz thimble amplitude for $\mathrm{Re}\,\mu<0$  is given by the integral along the line $\mathrm{Im}\,\pi$ parallel to the real axis. 
This contour defines the function $\Phi(x,\mu,\bar\mu)_{\R+i\pi}$. 
From the symmetry of the integrand one see that
\begin{equation}\label{negativethinmle}
\Phi(x,\mu,\bar\mu)_{\R+i\pi}=-\Phi(1-x,-\mu,-\bar\mu)_\R,
\end{equation}
where both sides are well--defined for $\mathrm{Re}\,\mu<0$ and satisfy the reality condition \eqref{apprealitycondt}.
However, $\Phi_{\R+i\pi}$ is related to \emph{some} amplitude of the form $\langle x|D^\prime\rangle$, whereas we need to compute for $\mathrm{Re}\,\mu<0$ the amplitude $\Pi\equiv\langle x|D\rangle$ for the \emph{same} basic brane $|D\rangle$ which is associated to the function $\Phi_\R$ for $\mathrm{Re}\,\mu>0$.

\subsubsection{The limit $\bar\mu\rightarrow 0$ of $\Pi$ in the negative half--plane}

We already know that the $\bar\mu=0$ limit of the $\zeta=1$ Lefschetz thimble amplitude in the positive half--plane $\mathrm{Re}\,\mu>0$ is
\begin{gather}\label{originalamplitude}
\log\Pi(x,\mu,0)_\R=\log\Gamma(\mu+x)-x\log\mu-\log\sqrt{2\pi},\qquad 0\leq x\leq 1\\
\text{i.e.}\quad \Phi(x,\mu,0)_\R=\log\Gamma(\mu+x)-(\mu+x-1/2)\log\mu+\mu-\log\sqrt{2\pi}.
\end{gather}
While, from eqn.\eqref{negativethinmle},
\begin{equation}
\begin{split}
\log\Pi(x,\mu,0)_{\text{negative}\atop \text{half--plane}}=-&\Phi(1-x,-\mu,0)_\R+\left(\mu-\frac{1}{2}\right)\log\mu -\mu.
\end{split}\end{equation}
Inserting eqn.\eqref{originalamplitude}, we get
\begin{equation}\label{asympblue2}
\log\Pi(x,\mu,0)_{\text{negative}\atop \text{half--plane}}=
-\log\Gamma(1-x-\mu)-x\log\mu +i\pi(x+\mu-1/2)+\log\sqrt{2\pi},
\end{equation}
which is the expression used in the text.

\subsection{Explicit expressions for the CFIV new susy index}\label{2dcfiv}

From the $tt^*$ metric, we can read the `new susy index' which in the present case reads (we set $M=2 \beta|\mu|$)
\begin{equation}\label{QBessel}
Q(x,M)\equiv -\frac{M}{2}\,\frac{\partial L}{\partial M}=-\frac{1}{\pi} \sum_{m\geq 1}\frac{\sin(2\pi m x)}{m}\,(2\pi m M)\,K_1(2\pi mM),
\end{equation}
where $K_1(z)$ is the Bessel--MacDonald function of index 1. 
This expression in particularly useful for large $M$ (IR limit), where the terms in the sum after the first one are negligible. Alternatively we have the Poisson re--summed expressions in terms of a sum over monopole contributions, see eqn.\eqref{periodicmonopoles}.

The Poisson re--summed expression for $L(x,M)\equiv\log G(x,M)+\log|\mu|$ reads
\begin{equation}\label{Lpoisson}\begin{split}
L(x,M&)=(1-2x)\Big(\log M+\gamma\Big)
-\\
&-\sum_{k\in\Z}\log\!\left(\frac{\sqrt{M^2+(x-k)^2}+x-k}{\sqrt{M^2+(k-1/2)^2}+(1/2-k)}
\,\exp\!\Big[{-(x-1/2) |k|}\Big]\!\right)\!.\end{split}\end{equation}
Then
\begin{equation}\label{Qpoisson}
Q(x,M)=\left(x-\frac{1}{2}\right)+\frac{1}{2}\sum_{k\geq 1}\left[\frac{(k-x)}{\sqrt{M^2+(k-x)^2}}+\frac{(1-k-x)}{\sqrt{M^2+(1-k-x)^2}}\right]
\end{equation}
from which we recover the UV result $Q(x,0)=x-1/2$. From the periodic $U(1)$ monopole point of view, $Q(x,M)$ is the component $r\, A_\theta$ of the Abelian connection $A$ in cylindric coordinates $(r=M,\theta, x)$ in a suitable gauge.

Besides the two series representations  of $Q(x,M)$, \eqref{QBessel} and \eqref{Qpoisson}, we give two convenient integral representations. The first one, for $0<x<1$,  is
\begin{equation}
Q(x,M)
=\left(x-\frac{1}{2}\right)-\frac{M^2}{4}\int\limits^{+\infty}_{-\infty}\frac{d\xi}{(\xi^2+M^2)^{3/2}}\;\Big( \big[ x+\xi \big]+\big[ x-\xi \big]\Big),
\end{equation} where $[z]$ is the integral part of the real variable $z$. This formula may be obtained  replacing in the integrand $M^2 (\xi^2+M^2)^{-3/2}$ with $d[\xi (\xi^2+M^2)^{-1/2}]/d\xi$ and integrating by parts: one gets the series \eqref{Qpoisson}; instead writing $[z]=z-\{z\}$ and plugging in the Fourier series expansion of $\{z\}$ one gets the Bessel series \eqref{QBessel}. 

The second one is particularly convenient for checking monotonicity properties
\begin{equation}
Q(x,M)=-\sin(2\pi x)\,M \int_0^\infty \frac{e^{-\pi M(t+t^{-1})}\; dt}{1-2\cos(2\pi x)\, e^{-\pi M(t+t^{-1})}+e^{-2\pi M(t+t^{-1})}}.
\end{equation}

\section{Technicalities for the 3d chiral (\S.\ref{sec:free3d})}\label{app:3dchiral}

\subsection{Details on the $tt^*$ metric for the 3d free chiral multiplet}\label{app:3dchiralmetric}

As in \S., we identify the 3d free chiral multiplet with twisted real mass
$m$, compactified on a circle of length $R_y$, as the 2d $(2,2)$ model with superpotential
\begin{equation}\label{thhhismod}
W(Y_n)=\sum_{n\in\Z} \bigg(e^{Y_n}-\frac{i}{2}\Big(\frac{m}{2\pi}+i\frac{n+y}{R_y}\Big)Y_n\bigg).\end{equation}
The length of the $tt^*$ circle is $R_x$, and we set $\rho=R_x/R_y$. $x$ is the $tt^*$ vacuum angle with period $1$ and $y$ is also a periodic variable of period 1. We write $z= m R_x/2\pi\in\R$.
The $tt^*$ quantities for  the model \eqref{thhhismod} will be denoted by the boldface version of the symbols used in \S.\ref{thebasicbraneamplitude} and appendix \ref{appednix31} to denote the corresponding quantity for the 2d model obtained by neglecting all non--zero KK modes.

Since the various KK modes do not interact,
\emph{formally} we have
\begin{equation}
\log\boldsymbol{G}(x,y,z,\rho)=``\sum_{n\in\Z}\log G\big(x,|i z-(n+y)\rho|/2\big)'',
\end{equation}
and we have to give a precise meaning to this expression. The 2d $tt^*$ metric $G$ depends on the chosen basis for the chiral ring $\mathcal{R}$. In \S. \ref{thebasicbraneamplitude}  the metric was written in the so--called `point' basis; in the chiral operator basis would read
\begin{equation}\label{changeofbasis}
G(x)_{\text{operator}\atop\text{basis}}\equiv \langle \overline{e^{xY}}\;|\; e^{xY}\rangle= e^{x \log|\mu|^2}\,G(x)_{\text{point}\atop\text{basis}}.
\end{equation}
Instead of summing the series for $\log G$, it is  convenient to sum the series for its derivative with respect to $x$; still at the pure formal level, we have
\begin{equation}\label{formallymetric}
\begin{split}
\frac{\partial}{\partial x}&\log \boldsymbol{G}(x,y,z,\rho)=
\sum_{n\in\Z}\frac{\partial}{\partial x}\log G(x,|iz +(y-n)\rho|/2)=\\
&=\sum_{n\in\Z}\Big(\log\frac{|iz+(y-n)\rho|^2}{4}-4 \sum_{m\geq 1}\cos(2\pi m x)\, K_0\big(2\pi m |iz+(y-n)\rho|\big)\Big),
\end{split}
\end{equation}
where the first term in the large parenthesis in the \textsc{rhs} is the effect of the change of basis \eqref{changeofbasis} and the other terms are as in eqn.\eqref{ttspecialcase2d}. One has the identity \cite{gradstein,linton}
\begin{equation}\label{firstide}
\begin{split}
&\log\frac{z^2+y^2}{4}-4\sum_{k=1}^\infty\cos(2\pi k x)\; K_0(2\pi k\sqrt{z^2+y^2})=\\
&=-\frac{1}{\sqrt{z^2+y^2+x^2}}-\sum_{k\in\Z\atop k\neq 0}\left(\frac{1}{\sqrt{z^2+y^2+(x-k)^2}}-\frac{1}{|k|}\right)-2\gamma,
\end{split}
\end{equation}
and hence the \textsc{rhs} of eqn.\eqref{formallymetric} may be seen as a \emph{regularized} version of the formal expression
\begin{equation}
-\sum_{n\in\Z^2}\sum_{k\in\Z}\frac{1}{\sqrt{z^2+(y-n)^2\rho^2+(x-k)^2}}+\mathrm{const},
\end{equation}
that is, literally, the potential $\boldsymbol{V}$ for a doubly--periodic array of $U(1)$ monopoles located at $(x,y\rho, z)=(k,n\rho, 0)_{(k,n)\in\Z^2}\subset\R^3$.
For a convenient choice of the additive constant, the regularized version is \cite{linton}
\begin{equation}\label{boldV}\begin{split}
\boldsymbol{V}(x,y,z,\rho)=\; &  \Lambda-
\frac{1}{\sqrt{z^2+y^2\rho^2+x^2}}-\\
&-\sum_{(k,\ell)\in\Z^2\atop (k,\ell_\neq (0,0)}\left(\frac{1}{\sqrt{z^2+(y-\ell)^2\rho^2+(x-k)^2}}-\frac{1}{\sqrt{\ell^2\rho^2+k^2}}\right),
\end{split}\end{equation}
where the constant $\Lambda$ is
\begin{equation}
\Lambda= 2\Big(\log \frac{4\pi}{\rho} -\gamma\Big)-8\sum_{k=1}^\infty\sum_{\ell=1}^\infty K_0(2\pi m n\rho).
\end{equation}

The function $\boldsymbol{V}$ in eqn.\eqref{boldV} is harmonic and doubly periodic, hence solves the $tt^*$ equations. In order to identify it with the correctly normalized $tt^*$ metric, we send the KK radius $R_y$ to zero. In this limit we should recover the 2d answer of \S. \ref{thebasicbraneamplitude}; indeed, we claim that
\begin{equation}
\lim_{\rho\rightarrow\infty} \boldsymbol{V}= \frac{\partial}{\partial x}\log G_{2d},
\end{equation}
where $G_{2d}$ is the 2d $tt^*$ metric \emph{but} in the original point basis. This follows from the Newman expression for the function $\boldsymbol{V}$ \cite{linton,newman}
\begin{equation}\label{boldV2}
\boldsymbol{V}(x,y,z,\rho)=\frac{2}{\rho}\, \log\!\big(2 |\sin\pi(y+i x/\rho)|\big)-4\sum_{\ell\in\Z}\sum_{k=1}^\infty \cos(2\pi k x)\, K_0\!\big(2\pi k \sqrt{z^2+(y-\ell)^2\rho^2}\big),
\end{equation}
As $\rho\rightarrow \infty$  (keeping $y\rho$ fixed) the first term vanishes as do all terms with $\ell\neq0$; we remain with the 2d expression.

Besides \eqref{boldV}\eqref{boldV2} there is a third equivalent expression of the function $\boldsymbol{V}(x,y,z,\rho)$ which is useful (for additional representations of $\boldsymbol{V}$ in terms of Ewald sums and heat kernels see \cite{linton})
\begin{equation}
\boldsymbol{V}(x,y,z,\rho)= 2\pi\,\frac{|z|}{\rho}-\sum_{(a,b)\in\Z^2\atop (a,b)\neq(0,0)}\frac{1}{\sqrt{a^2 \rho^2+b^2}}\;\exp\!\big(2\pi i\,a x+2\pi i\,b y-2\pi\sqrt{a^2+b^2/\rho^2}|z|\big)
\end{equation}

To get the expressions used in \S.\ref{sec:free3d}, one has to use $z=R_x m/2\pi$, $\rho=R_x/R_y$ and perform an overall rescaling to get the standard normalization of the monopole potential
\begin{equation}\label{Vvariousexp}
\begin{split}
V(x&,y,m,R_x,R_y)= \frac{R_x}{2}\;\boldsymbol{V}(x,y,R_x m/2\pi, R_x/R_y)\\
&=\frac{2\Lambda}{R_x}-\pi \sum_{(k,\ell)\in\Z^2\atop (k,\ell_\neq (0,0)}\left(\frac{1}{\sqrt{m^2+\frac{4\pi^2}{R_y}(y-\ell)^2+\frac{4\pi^2}{R_x}^2(x-k)^2}}-\frac{1}{\sqrt{\frac{4\pi^2}{R_y^2}\ell^2+\frac{4\pi^2}{R_x^2}k^2}}\right)\\
&=\frac{R_x\,R_y}{2}\,|m|-\sum_{(a,b)\in\Z^2\atop (a,b)\neq(0,0)}\frac{R_xR_y}{2\,\sqrt{a^2 R_x^2+b^2 R_y^2}}\;\exp\!\big(2\pi i\,a x+2\pi i\, b y-\sqrt{a^2 R_x^2+b^2 R_y^2}\,|m|\big).
\end{split}
\end{equation}

From the above computations we get for the 3d $tt^*$ metric (in a basis which reduces to the 2d point basis as $R_y\rightarrow 0$) we get
\begin{equation}
\boldsymbol{G}(x,y,m,R_x,R_y)=\exp\!\left(\frac{2}{R_x}\int\limits_0^x V(x^\prime, y,m, R_x,R_y)\,dx^\prime\right),
\end{equation} 
which is eqn.\eqref{opbasis}.

To get the analogue expressions for tetrahedron theory we have just to shift $V$ by a term linear in $m$, as explained in the main body of the paper.

 \subsection{The asymmetric UV limit for the 3d brane amplitudes}\label{appe:asym3d}

The amplitudes for the 3d chiral model may be written as a product on the KK modes:
\begin{equation}
\log\Pi_{3d}(m,x,y;\zeta)=\sum_{n\in\Z} \log\Pi_{2d}\!\left(\mu=\frac{m R_x}{4\pi}+\frac{i R_x}{2R_y}(y+n);\zeta\right).
\end{equation}
The asymmetric limit of the 3d amplitudes are then the product of the 2d asymmetric limit. Here we limit ourselves to the case $\zeta=-1$ corresponding to Neumann b.c., the extension to $\zeta=+1$ being straightforward.  From eqn.\eqref{neumann}, we have (for $0\leq x\leq 1$)
\begin{multline}\label{3dasassuminasymlim}
\log\Pi_{3d}(\bm;\zeta=-1)\Big|_{\text{asymmetric}\atop\text{UV limit}}=\\
=\sum_{n\in\Z}\log\Gamma\!\left(\frac{\bm}{4\pi}+x+\frac{i R_x}{2R_y}n\right)-x \sum_{n\in\Z}\log\!\left(\frac{\bm}{4\pi}+\frac{iR_x}{2R_y}n\right),
\end{multline}
where
\begin{equation}
\bm= m R_x+\frac{2\pi i R_x}{R_y}y,
\end{equation}
is the complexified twisted mass measured  in units of the inverse radius $R^{-1}_x$.

We insert in the expression \eqref{3dasassuminasymlim} the Weierstrass representation of the Gamma function
\begin{equation}
\log\Gamma(z)=-\gamma\, z -\log z-\sum_{m=1}^\infty\left[\log\!\left(1+\frac{z}{m}\right)-\frac{z}{m}\right].
\end{equation}
The idea is to invert the order of summation in $m$ and $n$; 
unfortunately, the expression is not absolutely convergent (recall that in the asymmetric limit we have infinitely many massless $2d$ fields), and this inversion is not legitimate. However, if we take three derivatives with  
respect to $\bm$ or $x$ the double series becomes absolutely convergent and the inversion of the summations will be allowed. Hence the result is well--defined, without further prescriptions, up to a quadratic polynomial in $\bf$ and $x$ (related with the specification of the background field CS level). 

With this warning, we perform the inverted--order sum formally,  using the symmetric $\zeta$--regularized sums 
\begin{gather}\sum_{n\in\Z}1=1+2\zeta(0)=0\quad \quad\quad\sum_{n\in\Z}n=\zeta(-1)-\zeta(-1)=0,\\
\sum_{n\neq 0}\log n= i\pi\, \zeta(0)-2\,\zeta^\prime(0)=-i\pi/2 +\log 2\pi,
\end{gather} as well the identity
\begin{equation}
\log s+\sum_{n=1}^\infty \log\!\left(1+\frac{s^2}{n^2}\right)=\log\sinh(\pi s)-\log\pi.
\end{equation}
We get
\begin{equation}
\begin{split}
\log\Pi_{3d}(\zeta=-1)=&-\log\boldsymbol{\Psi}\big(e^{-m R_y-2\pi i y-4\pi x- 2\pi R_y/R_x};\; e^{-4\pi R_y/R_x}\big)-\\
&-\frac{R_y}{4 R_x}\left(m+\frac{2\pi i R_x}{R_y}y\right)-x \log\sinh\left[\frac{1}{2}\left(m R_y+2\pi i y\right)\right]+\mathrm{const.},
\end{split}
\end{equation}
where the constant is independent of $m, y$.


\bibliography{references}

\end{document}